\documentclass[aps,twocolumn,nofootinbib,groupedaddress,superscriptaddress,longbibliography,notitlepage]{revtex4-2}

\usepackage{amsmath,amssymb}
\usepackage[normalem]{ulem}
\usepackage{graphicx}
\usepackage{multirow}
\usepackage{mathtools}
\usepackage{dsfont}
\usepackage{amsfonts}
\usepackage{xcolor}
\usepackage{url}
\usepackage[colorlinks=true,linkcolor=magenta,citecolor=magenta,hypertexnames=false]{hyperref}
\newcommand{\nocontentsline}[3]{}
\newcommand{\tocless}[2]{\bgroup\let\addcontentsline=\nocontentsline#1{#2}\egroup}
\def\ba#1\ea{\begin{align}#1\end{align}}
\def\bg#1\eg{\begin{gather}#1\end{gather}}
\def\bpm{\begin{pmatrix}}
\def\epm{\end{pmatrix}}

\newcommand{\nn}{\nonumber \\ }
\newcommand{\bb}[1]{{\mathbf #1}}
\newcommand{\bs}[1]{{\boldsymbol #1}}
\newcommand{\bx}{\bb x}
\newcommand{\bk}{\bb k}
\newcommand{\bK}{\bb K}

\newcommand{\td}[1]{\widetilde{#1}}
\newcommand{\cm}{\overline}
\newcommand{\mc}[1]{\mathcal{#1}}
\newcommand{\mf}[1]{\mathfrak{#1}}

\newcommand{\om}{\omega}

\newcommand{\vph}{\varphi}
\newcommand{\vep}{\varepsilon}
\newcommand{\ep}{\epsilon}
\newcommand{\Z}{\mathbb{Z}}

\newcommand{\N}{\mathbb{N}}

\newcommand{\trm}[1]{\textrm{#1}}

\newcommand{\ket}[1]{|#1\rangle}

\newcommand{\brk}[2]{\langle#1|#2\rangle}

\newcommand{\EBR}{BR}

\newcommand{\nW}[1]{m[#1]}
\newcommand{\nR}[1]{n(#1)}

\newcommand{\eqv}{\Leftrightarrow}

\newcommand{\nSimeq}{\mathrel{\ooalign{$\simeq$\cr\hidewidth$/$\hidewidth}}}
\allowdisplaybreaks

\begin{document}
\title{Building blocks of topological band theory for photonic crystals}
%%%%%%

%%%%%%Authors
\author{Yoonseok \surname{Hwang}}
\affiliation{Department of Physics, University of Illinois Urbana-Champaign, Illinois 61801, USA}
\affiliation{Anthony J. Leggett Institute for Condensed Matter Theory, University of Illinois Urbana-Champaign, Illinois 61801, USA}
\affiliation{Blackett Laboratory, Imperial College London, London SW7 2AZ, United Kingdom}

\author{Vaibhav \surname{Gupta}}
\affiliation{Department of Physics, University of Illinois Urbana-Champaign, Illinois 61801, USA}
\affiliation{Anthony J. Leggett Institute for Condensed Matter Theory, University of Illinois Urbana-Champaign, Illinois 61801, USA}

\author{Antonio Morales-P\'erez}
\affiliation{Donostia International Physics Center, Paseo Manuel de Lardizabal 4, 20018 Donostia-San Sebastian, Spain}
\affiliation{Material and Applied Physics Department, University of the Basque Country (UPV/EHU), Donostia-San Sebastián, Spain}

\author{Chiara Devescovi}
\affiliation{Donostia International Physics Center, Paseo Manuel de Lardizabal 4, 20018 Donostia-San Sebastian, Spain}
\affiliation{Institute for Theoretical Physics, ETH Zurich, Zurich, Switzerland}

\author{Mikel García-Díez}
\affiliation{Donostia International Physics Center, Paseo Manuel de Lardizabal 4, 20018 Donostia-San Sebastian, Spain}
\affiliation{Physics Department, University of the Basque Country (UPV/EHU), Bilbao, Spain}

\author{Juan L. Ma\~nes}
\affiliation{Physics Department, University of the Basque Country (UPV/EHU), Bilbao, Spain}
\affiliation{EHU Quantum Center, University of the Basque Country UPV/EHU, 48940 Leioa, Spain}

\author{Maia G. Vergniory}
\affiliation{Donostia International Physics Center, Paseo Manuel de Lardizabal 4, 20018 Donostia-San Sebastian, Spain}
\affiliation{Département de Physique et Institut Quantique, Université de Sherbrooke, Sherbrooke, QC J1K 2R1 Canada}

\author{Aitzol Garc\'ia-Etxarri}
\affiliation{Donostia International Physics Center, Paseo Manuel de Lardizabal 4, 20018 Donostia-San Sebastian, Spain}
\affiliation{IKERBASQUE, Basque Foundation for Science, Mar\'ia D\'iaz de Haro 3, 48013 Bilbao, Spain}

\author{Barry \surname{Bradlyn}}
\affiliation{Department of Physics, University of Illinois Urbana-Champaign, Illinois 61801, USA}
\affiliation{Anthony J. Leggett Institute for Condensed Matter Theory, University of Illinois Urbana-Champaign, Illinois 61801, USA}
%%%%%%

%%%%%%
\begin{abstract}
We derive a framework for classifying topological bands in three-dimensional photonic band structures, where the zero frequency polarization singularity implied by Maxwell's equations complicates the direct application of existing symmetry-based approaches.
Building on recent advances in the regularization of photonic bands, we use the recently introduced concept of stable real-space invariants (SRSIs) to show how photonic band structures can be unambiguously characterized in terms of equivalence classes of band representations.
We classify topologically trivial photonic bands using SRSIs, treating them as the fundamental building blocks of 3D photonic band structures.
This means that if certain bands cannot be constructed from these building blocks, they are necessarily topological.
Furthermore, we distinguish between photonic and electronic band structures by analyzing which SRSI values are allowed in systems with and without polarization singularity.
We also explore the impact of the polarization singularity on the behavior of Wilson loops, providing new insights into the topological classification of 3D photonic systems.
\end{abstract}
%%%%%%

%%%%%%
\maketitle

%\let\oldaddcontentsline\addcontentsline
%\renewcommand{\addcontentsline}[3]{}
%%%%%%

%%%%%%
\section{Introduction}
\label{sec:intro}
%%%%%%
In the last decade, topological phases in condensed matter systems have attracted significant attention due to their unique properties, including the presence of exotic boundary states and bulk responses~\cite{hasan2010colloquium,chiu2016classification}.
In particular, the study of topological phases in photonic systems has emerged as an exciting direction, offering both new route for exploring band topology with a high degree of control, as well as a promising platform for realizing lossless and directional communication, enhanced light-matter interaction, and quantum information processing~~\cite{haldane2008possible,lu2014topological,ozawa2019topological}.
Photonic crystals~\cite{yablonovitch1987inhibited,john1987strong}, which have periodic dielectric structures, have proven to be a useful platform for simulating and studying topological phases.
The ability to engineer photonic band structures allows for precise control of the topological properties of light, which can lead to novel photonic devices and applications.

There has been significant recent progress in establishing topological band theory for electronic systems, including approaches such as topological quantum chemistry (TQC)~\cite{bradlyn2017topological,vergniory2019complete,elcoro2021magnetic}, symmetry-indicator (SI) methods~\cite{po2017symmetry,watanabe2018structure,tang2019comprehensive}, and other similar works~\cite{kruthoff2017topological,song2018quantitative,zhang2019catalogue}. 
Despite this, the particular constraints of Maxwell's equations present obstacles to applying these techniques directly to topological phases in photonic crystals.
To properly characterize and classify the topology of photonic band structures, a key challenge lies in addressing the polarization singularity of Maxwell's equations at zero frequency~\cite{datta1993effective,krokhin2002long,wolff2013generation,watanabe2018space,christensen2022location,wang2023non}.
This singularity impedes the direct application of conventional topological band theory methods such as TQC and SIs.

To understand the origin of the polarization singularity and how it complicates the topological classification of photonic bands, we can consider Maxwell's equations in three-dimensional (3D) photonic crystal.
In a periodic structure, as the momentum $\bk$ and frequency $\om$ go to zero, the transversality condition $\nabla \cdot \bb D(\bb r) = 0$ for the displacement field $\bb D(\bb r)$ reduces to $\bk \cdot \bb D_\bk = 0$ for the Fourier component $\bb D_\bk$~\cite{datta1993effective,krokhin2002long}.
This implies that there are two physical modes which are connected to zero frequency and momentum $(\om, \bk) = \bb 0$, and which are transverse ($T$) with polarization vectors orthogonal to $\bk$. 
This implies that the field eigenmodes cannot be smoothly continued to $\bk=0$ (the $\Gamma$ point) near zero frequency, due to the nonzero Euler characteristic of the sphere surrounding $\Gamma$.

This lack of analyticity in the eigenstates as a function of $\bk$ means that a naive application of TQC and SI methods to photonic bands is destined to fail.
To use the SI method to analyze a given band structure, the multiplicities of irreducible representations (irreps) of the little group at the high-symmetry momenta (HSM) in the Brillouin zone must be known.
However, since the symmetry transformation of photonic modes is determined by their polarization vector, the irrep multiplicities are not well-defined for the bands connected to the singular point $(\bk,\omega)=(\Gamma,0)$.
To address this issue, Ref.~\cite{christensen2022location} showed that the polarization singularity could be resolved by introducing auxiliary modes in addition to the physical $T$ modes.
One of the auxiliary bands corresponds to the longitudinal ($L$) mode that regularizes the polarization singularity.
This approach allows for the consistent assignment of irrep multiplicities and symmetry indicators to photonic bands~\cite{christensen2022location} and enables the construction of a transversality-enforced tight-binding model that accurately reproduces the topological properties of photonic bands~\cite{morales2025transversality,devescovi2024axion} by placing the auxiliary bands at unphysical imaginary frequencies.
In particular, Ref.~\cite{christensen2022location} exploited the additivity of regularized photonic bands to infer the allowed irrep multiplicities of physical bands, providing an indirect but systematic characterization of physical bands.

Nevertheless, the polarization singularity obstructs the construction of real-space localized wave functions~\cite{albert2000generalized,busch2003wannier} which would be analogous to Wannier orbitals for electronic bands.
Wannier orbitals and the (elementary) band representations induced from them serve as the building blocks of electronic band structures, defining trivial topology.
This raises the pressing question of how to identify the analogous building blocks for topologically trivial photonic band structures.
Given the unique role of polarization in 3D photonic crystals, defining band topology for photonic bands requires a more careful approach compared to their electronic counterparts.

In this work, we introduce a general framework for quantifying topologically trivial bands in photonic crystals, allowing us to capture their relative topology and define topological bands as well.
Our approach is based on regularizing the polarization singularity by introducing auxiliary bands and using stable real-space invariants (SRSIs)~\cite{hwang2025stable} to capture the topological equivalence of trivial photonic bands, independent of the choice of auxiliary bands.
Using this method, we classify the trivial building blocks of photonic bands and quantify them with SRSIs.
Furthermore, our approach enables a quantitative comparison of trivial electronic and photonic bands based on the allowed values of SRSIs in any space group.
Based on the theory of polyhedra, we further show that trivial photonic bands form a finitely generated set, enabling a systematic classification directly at the level of physical bands.
We also investigate the physical properties of trivial photonic bands, such as the Wilson loop spectrum, which highlight the unique impact of polarization singularity on photonic bands, in contrast to electronic cases, using both ab initio calculations and the tight-binding model method.
%%%%%%

%%%%%%
\section{Regularization of singularity and trivial photonic bands}
\label{sec:regular}
%%%%%%
We begin by reviewing the regularization scheme for the polarization singularity, which involves introducing auxiliary bands, as described in Refs.~\cite{christensen2022location,morales2025transversality}.
For the regularized photonic and auxiliary bands, we will define symmetry-data vectors and (elementary) band representations (EBRs) to identify topologically trivial photonic bands, analogous to the framework of TQC.
Crucially, this will allow us to disentangle the topological characteristics of the polarization singularity (which are common to all photonic band structures) from other topological invariants of the lowest photonic bands.

\begin{figure}[t]
\centering
\includegraphics[width=0.48\textwidth]{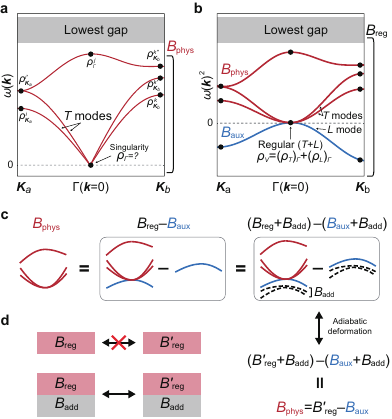}
\caption{{\bf Polarization singularity and regularization of photonic bands.}
\textbf{a} Below the lowest gap, a set of bands $\mc{B}_{\rm phys}$ is connected to the transverse ($T$) modes around zero frequency ($\om=0$) and zero momentum (at $\Gamma$ in the Brillouin zone), where the polarization singularity is located.
At high-symmetry momenta, the little-group irreps $\rho^i_\bK$ are assigned, except at the singularity.
\textbf{b} The band structure can be represented by $\om(\bk)^2$, the square for frequency.
Introducing auxiliary bands $\mc{B}_{\rm aux}$ regularizes the singularity and defines the regularized bands $\mc{B}_{\rm reg}$.
One of $\mc{B}_{\rm aux}$ corresponds to the longitudinal ($L$) mode around the singular point.
The auxiliary bands have negative $\om(\bk)^2$, indicating their auxiliary nature.
$\mc{B}_{\trm reg,aux}$ can be assigned little-group irreps, $\rho_V$ and $(\rho_L)_\Gamma$, at the singularity.
\textbf{c} The physical bands $\mc{B}_{\rm phys}$ are considered a formal difference between $\mc{B}_{\rm reg}$ and $\mc{B}_{\rm aux}$.
The physical description of $\mc{B}_{\rm phys}$ must remain equivalent regardless of the choice of $\mc{B}_{\rm aux}$, including any additional introduction of auxiliary bands $\mc{B}_{\rm add}$.
\textbf{d} Two topologically distinct regularized bands, $\mc{B}_{\rm reg}$ and $\mc{B}'_{\rm reg}$, regularize $\mc{B}_{\rm phys}$ in a topologically equivalent manner if they can be adiabatically deformed to each other by adding trivial bands, i.e. $\mc{B}_{\rm reg} + \mc{B}_{\rm add} \simeq \mc{B}'_{\rm reg} + \mc{B}_{\rm add}$.
}
\label{fig:reg}
\end{figure}

To begin with a simpler case without singularity, let us consider a set of photonic bands $\mc{B}$ below the lowest gap (see.~Fig.~\ref{fig:reg}\textbf{a}).
To define the topology of $\mc{B}$, we use the symmetry-data vector $\bb v[\mc{B}]$, which encodes the multiplicities of little-group irreps at HSM, or maximal $k$-vectors.
Recall that the little group $G_\bK$ consists of elements that leave the momentum $\bK$ invariant, up to reciprocal lattice vectors.
The little-group irreps are determined by the way the displacement fields $\bb D_\bK$ transforms under symmetry elements of $G_\bK$.
The symmetry-data vector $\bb v[\mc{B}]$ is defined as
\ba
\bb v [\mc{B}]
=& \big[ \nR{\rho_{\bK_1}^1}, \nR{\rho_{\bK_1}^2}, \dots, \nR{\rho_{\bK_1}^{N_{\bK_1}}}; \nn
& \nR{\rho_{\bK_2}^1}, \dots \nR{\rho_{\bK_2}^{N_{\bK_2}}}, \dots; \nR{\rho_{\bK_M}^{N_{\bK_M}}} \big]^T,
\label{eq:symvec}
\ea
by counting the multiplicity of each irrep $\rho^i_\bK$ ($i=1,\dots,N_\bK$) at maximal $k$-vectors $\bK = \bK_1, \dots, \bK_M$.
See Fig.~\ref{fig:reg}\textbf{a} for an example.
The total length of $\bb v [\mc{B}]$ is $N_{irrep} = \sum_{a=1}^M N_{\bK_a}$.

By comparing $\bb v[\mc{B}]$ with the symmetry-data vector of a linear combination of (E)BRs, we can determine whether $\mc{B}$ is topologically trivial or nontrivial.
The BRs are the representations of the space group (SG) that are induced from the real-space localized orbitals~\cite{zak1981band,bradlyn2017topological}, and hence any topologically trivial set of bands must transform as a BR.
Each band representation is determined by a set of site-symmetry group representations at the Wyckoff positions (WP) within the unit cell at which the orbitals are localized, where the site-symmetry group consists of elements that leave the WP invariant.
Each BR determines a symmetry-data vector, so we define a matrix $\EBR$, where each column represents the symmetry-data vector of a corresponding BR.
The (E)BRs for all SGs are listed in the BANDREP tool~\cite{bilbao_EBR}, available via the Bilbao Crystallographic Server (BCS) \cite{aroyo2006bilbao1,aroyo2006bilbao2,vergniory2017graph,elcoro2017double}.
By defining the list of BR multiplicities or orbitals as an $N_{\rm orb}$-dimensional vector $\bb m$, the symmetry-data vector and BR multiplicity vector can be related as
\bg
\bb v [\mc{B}] = \EBR \cdot \bb m.
\label{eq:map_m_v}
\eg
Since the matrix $\EBR$ is not invertible, multiple solutions for $\bb m$ may exist.
If $\bb m$ cannot be chosen to be a vector with nonnegative integers, i.e. $\bb m \notin \N_0^{N_{\rm orb}}$, the corresponding bands are necessarily topological.
On the other hand, for trivial band structures, $\bb m$ must satisfy $\bb m \in \N_0^{N_{\rm orb}}$.

For physical photonic bands $\mc{B}_{\rm phys}$ below the lowest frequency gap, as shown in Fig.~\ref{fig:reg}\textbf{a}, however, the displacement fields and their polarizations cannot be defined due to the singularity at zero frequency.
This obstructs the assignment of little-group irreps at $\Gamma$, and so prevents a determination of BR multiplicities and the symmetry-data vector.
This issue can be resolved by introducing topologically trivial, auxiliary bands $\mc{B}_{\rm aux}$~\cite{christensen2022location}, which includes the $L$ mode, as shown in Fig.~\ref{fig:reg}\textbf{b}.
When combined with two $T$ modes connected to $\mc{B}_{\rm phys}$, this removes the singularity.
To gain intuition for this process, we can examine the long-wavelength properties of the displacement field. While $O(3)$ symmetry is emergent in isotropic media, the vectorial nature of the displacement field $\bb D$ under the little group $G_\Gamma$ at $\Gamma$ can be more generally justified from $k \cdot p$ theory.
In particular, the $T$ and $L$ modes of $\bb D$ transform as components of a three-dimensional vector representation $\rho_V$ of $G_\Gamma$, which may be reducible in anisotropic media but remains well-defined at each $\bk$ (Thus, our convention for assigning little-group irreps applies to the displacement field).
The representation $(\rho_L)_\Gamma$ of $L$ mode at $\Gamma$ can always be chosen as the trivial irrep $\rho_{\rm triv}$ of $G_\Gamma$ in any space group~\cite{christensen2022location}.
This assignment is consistent with the transformation properties of a longitudinal vector aligned with $\hat{\bk}$, which transforms as a vector under $O(3)$ and also depends on the wavevector $\bk$.
Under a symmetry operation, the transformation of the vector and that of its $\bk$-dependence cancel out, justifying its identification with the trivial irrep at $\Gamma$.
This means that the irreps of $T$ modes can be assigned as $(\rho_T)_\Gamma = \rho_V - \rho_{\rm triv}$~\cite{watanabe2018space,christensen2022location} (We will provide a clear example of how $\rho_{V,\rm triv}$ are assigned in Sec.~\ref{sec:model}).
With well-defined irrep at $\Gamma$, the regularization allows the definition of symmetry-data vectors $\bb v [\mc{B}_{\rm reg}]$ for the regularized bands $\mc{B}_{\rm reg} = \mc{B}_{\rm phys} + \mc{B}_{\rm aux}$.

With this regularization scheme, the physical bands can be understood as a formal difference between the regularized and auxiliary bands, i.e.
\bg
\mc{B}_{\rm phys} = \mc{B}_{\rm reg}-\mc{B}_{\rm aux},
\label{eq:B_phys}
\eg
as shown in Fig.~\ref{fig:reg}\textbf{c}.
Thus, band-theoretical properties of physical bands can be understood for those of a pair $(\mc{B}_{\rm reg}, \mc{B}_{\rm aux})$.
In particular, because of the definition of $(\rho_T)_\Gamma$, $\mc{B}_{\rm reg}$ must contain $\rho_V$, while $\mc{B}_{\rm aux}$ must include $\rho_{\rm triv}$ (see Fig.~\ref{fig:reg}\textbf{b}).
When both $\mc{B}_{\rm reg,aux}$ are topologically trivial, they allow well-defined, nonnegative BR multiplicity vectors $\bb m_{\rm reg,aux} \in \N_0^{N_{\rm orb}}$ (Note that $\mc{B}_{\rm aux}$ is trivial by definition).
In this case, we define $\mc{B}_{\rm phys}$ as a set of {\it trivial photonic bands}, and the BR multiplicity vector of $\mc{B}_{\rm phys}$ is then defined as:
\bg
\bb m_{\rm phys} = \bb m_{\rm reg} - \bb m_{\rm aux},
\label{eq:m_phys}
\eg
where $\bb m_{\rm phys}$ can take general integer values, in contrast to the nonnegative integer values for $\bb m_{\rm reg,aux}$.
%%%%%%

%%%%%%
%%%%%%

%%%%%%
\section{Stable equivalence and stable real-space invariants}
\label{sec:rsi}
%%%%%%
Our primary goal is to classify trivial photonic bands (since nontrivial topology is implied by not being trivial) and to develop a systematic method to quantify their characteristics.
For trivial photonic bands, we assigned the BR multiplicity vector $\bb m_{\rm phys}$ in Eq.~\eqref{eq:m_phys}.
However, $\bb m_{\rm phys}$ alone does not serve as a reliable topological invariants for two main reasons.

First, even for regular band structures, the configuration of site-symmetry irreps that defines the BR multiplicity vector can change through adiabatic processes, continuous deformations between configurations of site-symmetry irreps that neither close the band gap nor break the system's symmetry.
This means that different BR multiplicity vectors can represent the same band topology.

Second, for photonic bands below the lowest gap, the structure in Eq.~\eqref{eq:B_phys} introduces more complex adiabatic processes.
Since the choice of auxiliary bands is not unique, additional trivial bands $\mc{B}_{\rm add}$ can be introduced to $\mc{B}_{\rm aux}$.
Thus, both pairs $(\mc{B}_{\rm reg},\mc{B}_{\rm aux})$ and $(\mc{B}_{\rm reg}+\mc{B}_{\rm add},\mc{B}_{\rm aux}+\mc{B}_{\rm add})$ describe the same physical photonic bands $\mc{B}_{\rm phys}$, as shown in Fig.~\ref{fig:reg}\textbf{c}.

Now, suppose that $\mc{B}_{\rm phys}$ can be represented by a pair $(\mc{B}'_{\rm reg},\mc{B}_{\rm aux})$, where $\mc{B}'_{\rm reg}$ and $\mc{B}_{\rm reg}$ are topologically distinct and cannot be adiabatically deformed to each other, i.e. $\mc{B}_{\rm reg} \nSimeq \mc{B}'_{\rm reg}$ where $\mc{B}_1 \simeq \mc{B}_2$ ($\mc{B}_1 \nSimeq \mc{B}_2$) means that the bands $\mc{B}_1$ and $\mc{B}_2$ can (cannot) be adiabatically deformed into each other meaning that they are (are not) topologically equivalent~\cite{bradlyn2017topological,hwang2025stable}.
However, it may be the case that when additional bands are introduced suitably so that $\mc{B}_{\rm reg} + \mc{B}_{\rm add} \simeq \mc{B}'_{\rm reg} + \mc{B}_{\rm add}$, as illustrated in Fig.~\ref{fig:reg}\textbf{d}.
In these cases we say that $\mc{B}_{\rm reg}$ and $\mc{B}'_{\rm reg}$ are {\it stably equivalent}.
Thus, despite the topological distinction between $\mc{B}_{\rm reg}$ and $\mc{B}'_{\rm reg}$, the inclusion of additional auxiliary bands followed by a series of adiabatic deformations, as shown in Figs.~\ref{fig:reg}\textbf{c,d}, enforce that both pairs $(\mc{B}_{\rm reg},\mc{B}_{\rm aux})$ and $(\mc{B}'_{\rm reg},\mc{B}_{\rm aux})$ characterize the same physical bands $\mc{B}_{\rm phys}$ whenever $\mc{B}_{\rm reg}$ and $\mc{B}'_{\rm reg}$ are stably equivalent.
%

%%%%%%
\begin{figure}[b!]
\centering
\includegraphics[width=0.45\textwidth]{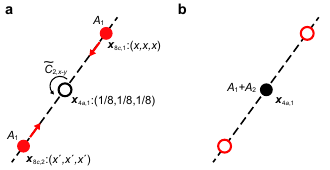}
\caption{{\bf Adiabatic deformation process in SG $P4_332$.}
The site-symmetry irreps $A_1$ and $A_2$ at WP $4a$ can be adiabatically deformed into the $A_1$ irrep at WP $8c$.
Note that $\td{C}_{2,x-y}$ maps $\bx_{8c,1}$ to $\bx_{8c,2}$, and $x'=-x+1/4$.
}
\label{fig:adia}
\end{figure}
%%%%%%

Stable equivalence of band representations is captured by stable real-space invariants (SRSIs) introduced in Ref.~\cite{hwang2025stable}.
There it was proved that if two configurations of site-symmetry irreps have the same SRSIs, they are stably equivalent.
The SRSIs are topological invariants defined by the BR multiplicity vector, and they have been computed for all 230 space groups (SGs)~\cite{hwang2025stable}.
For photonic crystals (or electronic systems without spin-orbit coupling), $\Z$-valued and $\Z_2$-valued SRSIs, which take values in integers and $\{0,1\}$ respectively, can be defined.
%

%%%%%%
{
\renewcommand{\arraystretch}{1.4}
\begin{table}[t!]
\centering
\begin{minipage}{0.48\textwidth}
\caption{
WPs and site-symmetry irreps in the SG $P4_332$ (No. 212).
The first column lists WPs with their representative positions.
The second and third columns denote the generators of site-symmetry group $G_W$ and site-symmetry irreps $(\rho)_W$ at the representative position, respectively.
When we denote $(\rho)_W$, the multiplicity of the WP is omitted for simplicity.
Note that $\td{C}'_{2,x-y}=\{2_{1\cm{1}0}|5/4,5/4,5/4\}$, and $x$, $y$, and $z$ are free real parameters.}
\label{table:sg212}
\end{minipage}
\begin{tabular*}{0.48\textwidth}{@{\extracolsep{\fill}}c| c c}
\hline \hline
WP $W$ & $G_W$ & $(\rho)_W$ \\
\hline
$4a(1/8,1/8,1/8)$ & $C_{3,111},\td{C}_{2,x-y}$ & $(A_1)_{a},(A_2)_{a},(E)_{a}$
\\
$4b(5/8,5/8,5/8)$ & $C_{3,111},\td{C}'_{2,x-y}$ & $(A_1)_{b},(A_2)_{b},(E)_{b}$ \\
$8c(x,x,x)$ & $C_{3,111}$ & $(A_1)_{c},({}^1E{}^2E)_{c}$ \\
$12d(y,1/4-y,1/8)$ & $\td{C}_{2,x-y}$ & $(A)_{d},(B)_{d}$
\\
$24e(x,y,z)$ & $\{E|\bb 0\}$ & $(A)_{e}$ \\
\hline \hline
\end{tabular*}
\end{table}}
%%%%%%

To illustrate this, we focus on the SG $P4_332$ (No. 212), generated by the three-fold rotation $C_{3,111} = \{ 3_{111}| \bb 0 \}$, two-fold screw rotation $\td{C}_{2,x-y} = \{ 2_{1\cm{1}0}| 1/4,1/4,1/4 \}$, four-fold screw rotation $\td{C}_{4z} = \{ 4_{001}| 3/4,1/4,3/4 \}$, time-reversal symmetry $\mc{T}$, and lattice translations $\{E|\bb a \in \Z^3\}$
Note that the point-group elements $4_{001}$, $3_{111}$, $2_{1\cm{1}0}$, $E$ transform $(x,y,z)$ to $(-y,x,z)$, $(z,x,y)$, $(-y,-x,-z)$, $(x,y,z)$, respectively.

Several WPs, $4a$, $4b$, $8c$, $12d$, and $24e$, are located within a unit cell.
Table~\ref{table:sg212} summarizes the types of WPs $W$ with their representative positions, site-symmetry group $G_W$, and site-symmetry irreps $(\rho)_W$.
In SG $P4_332$, there are three $\Z$-valued SRSIs, $\bs \theta_\Z = (\theta_1,\theta_2,\theta_3)$, and a single $\Z_2$-valued SRSI $\theta_4^{(2)}$.
They can be expressed in terms of the BR multiplicities as~\cite{hwang2025stable}
\ba
\theta_1 =& \nW{(A_1)_a} + \nW{(A_1)_b} + \nW{(A_1)_c} \nn
& + \nW{(A)_d} + \nW{(A)_e},
\nn
\theta_2 =& \nW{(A_2)_a} + \nW{(A_2)_b} + \nW{(A_1)_c} \nn
& + \nW{(B)_d} + \nW{(A)_e},
\nn
\theta_3 =& \nW{(E)_a} + \nW{(E)_b} + 2 \nW{({}^1E{}^2E)_c} \nn
& + \nW{(A)_d} + \nW{(B)_d} + 2\nW{(A)_e},
\nn
\theta^{(2)}_4 =& \nW{(A_1)_a} + \nW{(A_2)_a} + \nW{(E)_a} \pmod 2.
\label{eq:sg212_rsi}
\ea
in terms of the BR multiplicity vector $\bb m$.
Note that $\theta_1 + \theta_2 + 2 \theta_3 = \nu/4$, where $\nu$ is the total number of bands that are being considered, or equivalently the total dimension of all site-symmetry irreps per unit cell.
Additionally, $\nu \in 4\Z$ in the SG $P4_332$ due to the symmetry protection or compatibility relation of bands.

To understand how the SRSIs are left invariant under adiabatic processes, let us focus on a key example of the adiabatic deformation that involves transforming the $(A_1)_c$ irrep at the $8c$ WP into the $(A_1)_a$ and $(A_2)_a$ irreps at the $4a$ WP.
(See Fig.~\ref{fig:adia} for schematic illustration.)
The $4a$ WP consists of 4 positions: the point $\bx_{4a,1} = (1/8,1/8,1/8)$ and its symmetry-related locations. The site-symmetry group of $\bx_{4a,1}$, denoted as $G_{4a,1}$, is isomorphic to the point group $32$, i.e. $G_{4a,1} \simeq 32$ generated by $C_{3,111}$ and $\td{C}_{2,x-y}$.
The $8c$ WP consists of 8 positions, each with a site-symmetry group isomorphic to the point group $3$.
For a representative position $\bx_{8c,1} = (x,x,x)$, $G_{8c,1} \simeq 3$ generated by $C_{3,111}$.
At the $8c$ WP position $\bx_{8c,1}=(x,x,x)$, the $A_1$ irrep has a $C_{3,111}$ eigenvalue of $1$.
Due to the screw rotation $\td{C}_{2,x-y}$, $\bx_{8c,1}$ maps to another symmetry-related position $\bx_{8c,2}=(x',x',x')$, where $x' = -x + 1/4$, as illustrated in Fig.~\ref{fig:adia}\textbf{a}.
By taking symmetric and antisymmetric linear combinations of the $A_1$ orbitals at $\bx_{8c,1}$ and $\bx_{8c,2}$, we obtain orbitals localized at $\bx_{4a,1}$.
Both combinations have the same $C_{3,111}$ eigenvalue $1$ but the opposite $\td{C}_{2,x-y}$ eigenvalues.
Thus, they correspond to the $A_1$ and $A_2$ irreps at the $4a$ WP.
This means that by tuning the parameter $x$ to $1/8$, the $A_1$ irrep at $8c$ can be smoothly deformed to the direct sum $A_1 \oplus A_2$ representation at $4a$, as shown in Fig.~\ref{fig:adia}\textbf{b}.
This adiabatic process is written as $(A_1+A_2)_a \eqv (A_1)_c$, where $(\rho)_W$ denotes the site-symmetry irrep $\rho$ at a WP $W$.

The adiabatic process $(A_1 + A_2)_a \eqv (A_1)_c$ changes the BR multiplicity vector $\bb m$, but the SRSIs in Eq.~\eqref{eq:sg212_rsi} remain unchanged.
By construction, the SRSIs do not change under any adiabatic process in the SG $P4_332$.
Moreover, SRSIs also capture the stable equivalence between two configurations of site-symmetry irreps.
For instance, $2(A_1)_a$ and $2(A_1)_b$ are not adiabatically deformable to each other.
However, since both have the same SRSI values, $\bs \theta = (\theta_1, \theta_2, \theta_3, \theta^{(2)}_4)=(2,0,0,0)$, they can be deformed into each other if additional trivial bands that correspond to $({}^1E{}^2E)_c$ are included (see the Methods).
This defines the stable equivalence, between $2(A_1)_a$ and $2(A_1)_b$.
By capturing the stable equivalence between configurations of site-symmetry irreps, SRSIs can be used to characterize topology of trivial photonic bands.
A concrete example illustrating how different choices of $(\mc B_{\rm reg}, \mc B_{\rm aux})$ lead to the same physical bands and identical SRSIs of physical photonic bands is also provided in the Methods.
%%%%%%

%%%%%%
\section{SRSI classification of trivial photonic bands}
\label{sec:classify}
%%%%%%
Equipped with the SRSIs, we now classify the trivial photonic bands by finding a mapping between SRSIs and symmetry-data vector.
We will first review how we can construct a one-to-one mapping between the $\Z$-valued SRSIs $\bs \theta_\Z$ and the symmetry-data vector in any space group, while imposing a constraint from the polarization singularity on this mapping.
Here, we focus on symmetry-indicated topology, i.e. those topological properties that are determined solely by the symmetry-data vector.

To classify trivial photonic bands, we proceed in the following steps.
First, for a given space group, we define the $\Z$-valued SRSIs $\bs \theta_\Z$ and the symmetry-data vector $\bb v[\mc{B}]$, which encodes the multiplicities of little-group irreps at HSM.
Second, we construct a linear map from $\bs \theta_\Z$ to $\bb v[\mc{B}]$.
Next, using this map allows us to translate physical constraints into {\it linear inequalities} on $\bs \theta_\Z$, including those imposed by the polarization singularity, which determines the irrep structure $(\rho_T)_\Gamma$.
The allowed SRSIs form the lattice $Lat_{\bs \theta_\Z, ph}$, defined as the set of integer-valued $\bs \theta_\Z$ satisfying these inequality constraints.
While the construction and solution of the inequalities can be carried out algorithmically using the Hilbert basis method (see the Supplementary Material (SM)~\cite{supple}), in the main text we construct and visualize this set directly as the lattice $Lat_{\bs \theta_\Z, ph}$.

To set this up, for a given set of bands $\mc{B}$ with BR multiplicity $\bb m$, we express the $\Z$-valued SRSIs in Eq.~\eqref{eq:sg212_rsi} as $\bs \theta_\Z = \Delta_\Z \cdot \bb m$, by introducing 3-by-11 matrix $\Delta_\Z$ ($\bb m$ has a length of 11, and its basis for site-symmetry irreps follows the ordering in the third column in Table~\ref{table:sg212}).
We invert this equation as $\bb m = \Delta_\Z^\ddagger \cdot \bs \theta_\Z + \bb m_{ker}$, by introducing a pseudoinverse $\Delta_\Z^\ddagger$ of $\Delta_\Z$.
Here $\bb m_{ker}$ is a generic vector in the kernel of $\Delta_\Z$, such that $\Delta_\Z \cdot \bb m_{ker} = \bb 0$.
Note that $\Z$-valued SRSIs can always be defined such that $\Delta_\Z^\ddagger$ is integer-valued~\cite{hwang2025stable}, as is the case in our study.
By combining this with Eq.~\eqref{eq:map_m_v}, we obtain
\bg
\bb v [\mc{B}]
= \EBR \cdot \Delta_\Z^\ddagger \cdot \bs \theta_\Z
\label{eq:map_v_rsi}
\eg
Here, we used the fact that $\EBR \cdot \bb m_{ker} = \bb 0$, which can be shown by explicit computation (In fact, this holds in any SG, as shown in Ref.~\cite{hwang2025stable}).
Since both $\EBR$ and $\Delta_\Z^\ddagger$ are integer-valued, Eq.~\eqref{eq:map_v_rsi} ensures that integer-valued SRSI vectors are mapped to integer-valued symmetry data vectors.

The symmetry-data vector $\bb v [\mc{B}]$ is a list of multiplicities of all little-group irreps defined at HSM, including $\Gamma=(0,0,0)$, $R=(\pi,\pi,\pi)$, $M=(\pi,\pi,0)$, and $X=(0,\pi,0)$.
Not all the irrep multiplicities are independent due to the compatibility relations.
In particular, compatibility relations enforce that $\nR{\Gamma_3} = \nR{R_1R_2}$, $\nR{\Gamma_4} = \nR{M_2M_3}$, $\nR{\Gamma_5} = \nR{M_1M_4}$, and $\nR{M_5} = \nR{X_1} = \nR{X_2}$.
This allows us to define the reduced symmetry-data vector:
\ba
& {\bb v}' [\mc{B}] = [\nR{\Gamma_1}, \nR{\Gamma_2}, \nR{\Gamma_3}, \nR{\Gamma_4}, \nR{\Gamma_5}, \nR{R_3}, \nR{M_5}]
\nn
& = (\theta_1, \theta_2, \theta_3, \theta_2 + \theta_3, \theta_1 + \theta_3, \theta_1 + \theta_2 + \theta_3, \theta_1 + \theta_2 + 2\theta_3)^T
\nn
& := \mc{M} \cdot \bs \theta_\Z
\label{eq:sg212_symvec_map}
\ea
with dimension $N_{BZ}=7$.
Note that this reduction is always possible in any space group, since symmetry-data vectors generated in the form $\bb v = \EBR \cdot \bb m$ [as in Eq.~\eqref{eq:map_v_rsi}] automatically satisfy all compatibility relations~\cite{po2017symmetry,elcoro2020application}.
In Eq.~\eqref{eq:sg212_symvec_map}, we introduced a integer-valued submatrix $\mc{M}$ of $\EBR \cdot \Delta_\Z^\ddagger$ in Eq.~\eqref{eq:map_v_rsi},
\bg
\mc{M} = \bpm 1 & 0 & 0 & 0 & 1 & 1 & 1 \\ 0 & 1 & 0 & 1 & 0 & 1 & 1 \\
0 & 0 & 1 & 1 & 1 & 1 & 2 \epm^T.
\label{eq:sg212_map_mat}
\eg

\begin{figure}[t]
\centering
\includegraphics[width=0.48\textwidth]{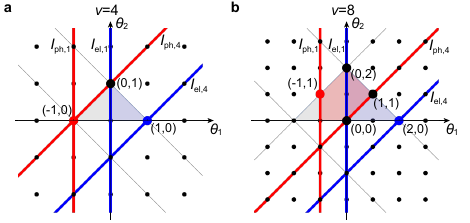}
\caption{{\bf SRSIs of photonic and electronic bands allowed in SG $P4_332$.}
Three $\Z$-valued SRSIs, $\bs \theta_\Z = (\theta_1, \theta_2, \theta_3)$, are determined by symmetry-data vector, as in Eq.~\eqref{eq:sg212_symvec_map}.
The allowed values of $\bs \theta_\Z$ for the number of bands \textbf{a} $\nu=4$ and \textbf{b} $\nu=8$ are shown.
Note that $\nu = 4(\theta_1+\theta_2+2\theta_3)$.
Gray, red, and blue regions represent $M_{co}$, $M_{ph}$, and $M_{el}$, respectively, defined by inequalities $I_{ph,i}$ and $I_{el,i}$ $(i=1,\dots,7)$.
Red (blue) dots in $M_{ph}$ ($M_{el}$) are the allowed SRSIs $\bs \theta_\Z$, which give physically allowed, integer-valued symmetry-data vectors.
The larger black dots represent SRSIs allowed in both electronic and photonic band structures.
Inequalities $I_{{ph},1,4}$ ($I_{{el},1,4}$) hold at the red (blue) lines.
}
\label{fig:rsi_lattice}
\end{figure}

We now seek to constrain the allowed values of $\bs \theta_\Z$ that can arise from physically realizable, trivial photonic bands.
These constraints are derived by linearly mapping $\bs \theta_\Z$ to a symmetry-data vector via Eq.~\eqref{eq:sg212_symvec_map}, whose components represent multiplicities of little-group irreps.
Physical band structures must correspond to nonnegative irrep multiplicities, except where modified by physical considerations such as the polarization singularity.
For photonic bands $\mc{B}_{\rm phys}$ below the lowest gap, the expression of $(\rho_T)_\Gamma = \rho_V - \rho_{\rm triv}$, representing the irreps assigned for $T$ modes, enforces certain conditions on the symmetry-data vector $\bb v'[\mc{B}_{\rm phys}]$.
In SG $P4_332$, the little group at $\Gamma$ is the point group 432, with vector representation $\Gamma_4$ and trivial representation $\Gamma_1$.
Thus, we assign $(\rho_T)_\Gamma = - \Gamma_1 + \Gamma_4$, and this condition can be encapsulated by two inequalities $I_{ph, 1}$ and $I_{ph, 4}$:
\ba
I_{ph,1}:& \, \nR{\Gamma_1} = v'_1[\mc{B}_{\rm phys}] \ge -1, \nn
I_{ph,4}:& \, \nR{\Gamma_4} = v'_4[\mc{B}_{\rm phys}] \ge 1,
\label{eq:sg212_ineq_ph1}
\ea
where $v'_i [\mc{B}_{\rm phys}]$ represents the $i$th components of $(\bb v'[\mc{B}_{\rm phys}])_i$.
For other little-group irreps not constrained by polarization singularity, $I_{ph, j}: \, v'_j [\mc{B}_{\rm phys}] \ge 0$ for $j=1,\dots,N_{BZ}=7$ except for $j=1$ and 4.
The inequalities $I_{ph,1,\dots,7}$ can be expressed in terms of $\theta_{1,2}$ with a fixed number of bands $\nu$, by recalling that $\nu/4 = \theta_1 + \theta_2 + 2\theta_3$.
For example, when $\nu=4$, we have:
\bg
\theta_1 \ge -1, \quad
\theta_2 \ge 0, \quad
\theta_1 + \theta_2 \le 1, \quad
-\theta_1 + \theta_2 \ge 1, \nn
\theta_1 - \theta_2 \ge -1, \quad
\theta_1 + \theta_2 \ge -1, \quad
\nu \ge 0.
\label{eq:sg212_ineq_example}
\eg
In this case, $I_{ph,1}$ and $I_{ph,4}$ are represented as $\theta_1 \ge -1$ and $-\theta_1+\theta_2 \ge 1$, respectively.
Note that the last inequality from $I_{ph,7}$ is trivially satisfied for $\nu = 4$.
The equality holds on the red lines in Fig.~\ref{fig:rsi_lattice}\textbf{a}.

Contrary to the singular photonic bands, regular band structures $\mc{B}$, such as electronic bands (and photonic bands above the lowest gap), are not subject to polarization constraints.
Each little-group irrep multiplicity is a nonnegative integer, and thus
\bg
I_{el, i}: \, v'_i [\mc{B}] \ge 0 \quad i=1, \dots N_{BZ}
\label{eq:sg212_ineq_el}
\eg
holds.
Since $I_{ph, j}$ and $I_{el, j}$ ($j \ne 1, 4$) are identical and satisfied in both electronic and photonic bands, we collectively refer to them as $\{I_{co}\}$.
This defines a region $M_{co}$, which is a set of ($\theta_1,\theta_2,\theta_3$) that satisfies $\{I_{co}\}$.
The subregion $M_{ph}$ ($M_{el}$) is further constrained by $\{I_{ph, 1}, I_{ph, 4}\}$ ($\{I_{el, 1}, I_{el, 4}\}$).
For fixed $\nu=4$ and $8$, the regions $M_{ph,el,co}$ are shown in Fig.~\ref{fig:rsi_lattice}.

We can now classify trivial photonic bands by defining the lattice of allowed SRSI values as the intersection between $M_{ph}$ and a region $Z$:
\bg
Lat_{\bs \theta_\Z, ph} = M_{ph} \cap Z.
\label{eq:lat_ph_triv}
\eg
Here, $Z = \{\bs \theta_\Z | \bs \theta_\Z \in \Z^{N_{\theta_\Z}} \}$ where $N_{\theta_\Z}$ is the number of $\Z$-valued SRSIs allowed in the SG.
The lattice of $Lat_{\bs \theta_\Z, el}$ SRSIs allowed for electronic bands is defined similarly.
%

%%%%%%
%%%%%%

The lattice of SRSIs and the corresponding symmetry-data vectors for photonic and electronic bands can be compared quantitatively.
To illustrate this, let us consider the cases of (i) $\nu=4$ and (ii) $\nu=8$, treating $\theta_{1,2}$ as variables (Recall that $\theta_3=\nu/8-\theta_1/2-\theta_2/2$, and $\nu \in 4\Z$ in SG $P4_332$).
For each case, the allowed $\Z$-valued SRSI values in $Lat_{\bs \theta_\Z, ph,el}$ are shown in Figs.~\ref{fig:rsi_lattice}\textbf{a,b}, respectively.
Specifically, for $\nu=4$, we have $
Lat_{\bs \theta_\Z, ph} = \{(0,1,0),(-1,0,1)\}$.
These SRSI values correspond to a set of bands with little group irreps,
\ba
& [(\rho_T)_\Gamma + \Gamma_1 + \Gamma_2, R_3, M_2 M_3 + M_5, X_1 + X_2]
\nn
& [(\rho_T)_\Gamma + \Gamma_3, R_1 R_2, M_2 M_3 + M_5, X_1 + X_2],
\label{eq:sg212_symvec_ph}
\ea
respectively.
For electronic bands, we find that $Lat_{\bs \theta_\Z, el} = \{(0,1,0),(1,0,0)\}$ with
\ba
& (\Gamma_2 + \Gamma_4, R_3, M_2 M_3+M_5, X_1 + X_2)
\nn
& (\Gamma_1 + \Gamma_5, R_3, M_1 M_4 + M_5, X_1 + X_2).
\label{eq:sg212_symvec_el}
\ea
%

%%%%%%
\begin{figure*}[t]
\centering
\includegraphics[width=0.9\textwidth]{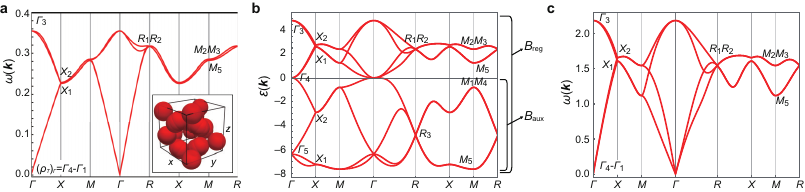}
\caption{
{\bf Ab-initio and tight-binding models in SG $P4_332$.}
\textbf{a} Frequency spectrum of the SG $P4_332$ photonic crystal built by the MPB.
Only the lowest four bands, isolated from higher bands, are shown, as described in the Methods section.
The inset displays the unit cell structure of photonic crystal.
The box shows the boundary of primitive unit cell $x,y,z \in [-0.5, 0.5]$.
\textbf{b-c} Band structures in the tight-binding models.
\textbf{b} Energy spectrum for physical and auxiliary bands.
Regularized bands $\mc{B}_{\rm reg}$ corresponds to the EBR induced from $(E)_a$, while the auxiliary bands $\mc{B}_{\rm aux}$ are chosen to have a symmetry-data vector identical to that induced from $(A_1)_a$.
All irreps of the auxiliary bands except $\Gamma_1$ have negative $\vep(\bk)$.
\textbf{c} Frequency spectrum for physical bands corresponding to $\mc{B}_{\rm phys}$ in Eq.~\eqref{eq:sg212_irrep_model}.
Note that $\Gamma=(0,0,0)$, $X=(0,\pi,0)$, $M=(\pi,\pi,0)$, and $R=(\pi,\pi,\pi)$.
}
\label{fig:sg212_band}
\end{figure*}
%%%%%%

By comparing $Lat_{\bs \theta_\Z, ph}$ and $Lat_{\bs \theta_\Z, el}$, we conclude that $\bs \theta_\Z = (-1,0,1)$ can only occur in photonic band structures.
Additionally, we observe that the presence of $\Gamma_3$ is unique to photonic band structures when $\nu=4$, and the band degeneracies at $\Gamma$ differ between photonic and electronic bands, even when their $\Z$-valued SRSIs match, as in the case of $\bs \theta_\Z = (0,1,0)$.
This equality of $\bs \theta_\Z$ implies that both band structures share the same symmetry-data vector, and thus the same multiplicities of little-group irreps.
Nevertheless, the actual irrep content at $\Gamma$ differs.
For instance, the photonic band includes $(\rho_T)_\Gamma$, which is absent in the electronic case.
The resolution of this apparent contradiction lies in the fact that, unlike for electronic bands, photonic band symmetry data vectors need not be entirely nonnegative.
While we use SRSIs as explicit invariants, our formalism inherently accounts for such distinctions: the definition of physical photonic bands relies on a regularization scheme, which implicitly fixes the structure of $(\rho_T)_\Gamma$.
Despite these differences, photonic bands satisfy the compatibility relations just like electronic bands.
This is because the photonic bands are constructed as a formal difference between regularized and auxiliary bands, each of which individually respects the compatibility relations.

We note that the lattice $Lat_{\bs \theta_\Z, ph/el}$ obtained above can also be computed systematically using the Hilbert basis method, which solves the linear inequalities on $\bs \theta_\Z$ under the integrality constraint.
Each solution $\bs \theta_\Z$ can be expressed as a sum of two parts:
\bg
\bs \theta_\Z = \bb h_c + \sum_{A=1}^{D_r} n_A \, \bb h_{r,A}, \quad n_A \in \N,
\eg
where the $\bb h_{r,A}$ are generators of the recession cone (unbounded directions), and $\bb h_c$ belongs to the finite set of compact solutions, i.e. those solutions that come as close as possible to saturating the inequalities.
For both photonic and electronic bands, the recession cone is generated by the same three vectors: $\bb h_{r,1} = (1,0,0)$, $\bb h_{r,2} = (0,1,0)$, and $\bb h_{r,3} = (0,0,1)$.
The key difference lies in the compact part: while the electronic case has only the trivial compact solution $(0,0,0)$, the photonic case admits two additional solutions, $\bb h_c = \{(-1,0,1), (0,1,0)\}$.
Together with the cone generators, these compact parts define the complete lattice of allowed SRSIs.
Further details and a worked example are provided in the SM~\cite{supple}.

Finally, we outline how the above framework can be extended to stable topological photonic bands.
For any given photonic band structure, including both trivial and topological cases, a symmetry-data vector can be defined from the irrep content at HSMs, with $(\rho_T)_\Gamma$ and any other regular little-group irreps.
From this symmetry-data vector, the corresponding SRSIs $\bs \theta_\Z$ can be computed.
When these SRSIs fall outside the allowed region $Lat_{\bs \theta_\Z, ph}$ for trivial bands, the band structure must be topological.
Then, we can define a denser lattice $\td{Lat}_{\bs \theta_\Z, ph}$ of all SRSIs consistent with symmetry and integrality of symmetry-data vector, including both trivial and stable topological bands.
Unlike the case of trivial bands, where $\bs \theta_\Z$ must be integral, Ref.~\cite{hwang2025stable} shows that stable topological bands exhibit fractional $\bs \theta_\Z \notin \Z^{N_{\theta_\Z}}$, while still yielding a well-defined integer-valued symmetry-data vector.
This motivates defining $\td{Lat}_{\bs \theta_\Z, ph} = M_{ph} \cap \td{Z}$, where $\td{Z} = \{ \bs \theta_\Z | \bb v (\bs \theta_\Z) \in \Z^{N_{\bb v}} \}$ ensures that the resulting $N_{\bb v}$-dimensional symmetry-data vector $\bb v(\bs \theta_\Z) = BR \cdot \Delta^\ddagger \cdot \bs \theta_\Z$ is integer-valued.
Consequently, topological band structures yield fractional SRSIs that belong to $\td{Lat}_{\bs \theta_\Z, ph}$ but not to $Lat_{\bs \theta_\Z, ph}$.
%%%%%%

%%%%%%
\section{Trivial photonic bands and Wilson loops in SG $P4_332$}
\label{sec:model}
%%%%%%
Earlier, we demonstrated that bands with $\bs \theta_\Z = (-1,0,1)$ in SG $P4_332$ can only be realized in photonic band structure when the number of bands is $\nu=4$.
The corresponding little-group irrep content is:
\bg
\mc{B}_{\rm phys}: [(\rho_T)_\Gamma + \Gamma_3, R_1 R_2, M_2 M_3 + M_5, X_1 + X_2].
\label{eq:sg212_irrep_model}
\eg
Here, we will demonstrate that this band structure can be modeled using tight-binding (TB) methods based on our SRSI analysis.
We will examine the effects of the polarization singularity on physical quantities, with a focus on the Wilson loop spectrum.
Before constructing the TB model, we first present the ab-initio calculation.
We constructed a photonic crystal in the SG $P4_332$ and computed the eigenspectra using the MIT Photonic Bands (MPB) package~\cite{johnson2001block-iterative}.
For this, we create a primitive unit cell with a lattice constant $a=1$, filled with a non-magnetic ($\mu = 1$), homogeneous, isotropic medium with a dielectric constant $\epsilon = 11$.
Then, we carve out spheres of radius $r = 0.2$, centered at the $24e$ Wyckoff positions (WPs) (Thus, $\epsilon =1 $ inside the spheres).
Specifically, the spheres are centered at $(0,-0.4,0.2)$ and other symmetry-related locations.
The unit cell structure is shown in the inset of Fig.~\ref{fig:sg212_band}\textbf{a}, and more details are provided in the Methods.
The frequency spectrum is shown in Fig.~\ref{fig:sg212_band}\textbf{a}, where the lowest four bands are separated from higher bands by the lowest frequency gap.
For those bands, the irreps at HSM exactly match those in Eq.~\eqref{eq:sg212_irrep_model}.

Now, we realize the same band structure with a TB model by introducing auxiliary bands, as in Refs.~\cite{morales2025transversality,devescovi2024axion}.
We will use SRSIs to determine the site symmetry irreps for the pair $(\mc{B}_{\rm reg},\mc{B}_{\rm aux})$ that describes the physical bands.
To do this, let us assume without loss of generality that the site-symmetry irreps corresponding to $\mc{B}_{\rm reg,aux}$ are located only at maximal WPs, $4a$ and $4b$.
Note that any irrep at a nonmaximal WP can always be moved to maximal WPs while preserving all the symmetries of system.
With this assumption, the three $\Z$-valued SRSIs in Eq.~\eqref{eq:sg212_rsi} are simplified as $\theta_1 = \nW{(A_1)_a} + \nW{(A_1)_b}$, $\theta_2 = \nW{(A_2)_a} + \nW{(A_2)_b}$, and $\theta_3 = \nW{(E)_a} + \nW{(E)_b}$.
One can immediately find the solutions for $(\theta_1,\theta_2,\theta_3)=(-1,0,1)$.

Consider a specific solution where $\nW{(A_1)_a}=-1$, $\nW{(E)_a}=1$, and zero for all other irreps.
Since the site-symmetry irreps $(A_1)_a$ and $(A_1)_b$ induce identical symmetry-data vectors, the physical band in Eq.~\eqref{eq:sg212_irrep_model} can be represented equivalently as $(E)_a - (A_1)_{a/b}$.
The little-group irrep content of physical bands is compatible with this decomposition, as since the irreps induced from $(E)_a$ and $(A_1)_{a/b}$ are
\ba
(E)_a:& \, (\Gamma_3 + \Gamma_4 + \Gamma_5, R_1 R_2 + R_3, M_1 M_4 \nn
&+ M_2 M_3 + 2M_5, 2X_1 + 2X_2),
\nn
(A_1)_{a/b}:& \, (\Gamma_1 + \Gamma_5, R_3, M_1 M_4 + M_5, X_1 + X_2).
\label{eq:sg212_tb_symvec_aux}
\ea
With this decomposition of the symmetry-data vectors for $\mc{B}_{\rm reg, aux}$ established, we fix the regularized bands to correspond to the EBR induced from $(E)_a$.
The auxiliary bands are then chosen such that their symmetry-data vector coincides with that induced from $(A_1)_a$, or equivalently $(A_1)_b$.

We construct the TB model $H_{TB}(\bk)$ with basis orbitals $(E)_a$, where the complete set of bands represents the regularized bands [See the SM~\cite{supple} for the details and explicit form of $H_{TB}(\bk)$].
To realize physical bands in Eq.~\eqref{eq:sg212_irrep_model}, we choose hopping parameters to assign positive energy $\om (\bk)^2$ to irreps corresponding to $(E)_a-(A_1)_{a/b}$ and negative energy to those irreps corresponding to $(A_1)_{a/b}$ at every HSM, as shown in Fig.~\ref{fig:sg212_band}\textbf{b}.
Additionally, the irrep $(\rho_T)_\Gamma$ at $\Gamma$ has zero energy.
When we define the eigenvalue of $H_{TB}(\bk)$ as $\vep(\bk)$, the physical bands have nonnegative eigenvalue $\vep(\bk) \ge 0$.
Thus, the frequency spectrum with $\om(\bk)=\sqrt{\vep(\bk)}$ can be used to simulate the photonic crystal.
Figs.~\ref{fig:sg212_band}\textbf{b,c} display the band structures of $H_{TB}(\bk)$ defined by $\vep(\bk)$ and $\om(\bk)$, respectively.

To study how the polarization singularity affects the Wilson loop behavior in photonic band structure, we computed the $k_z$-directed Wilson loop $W_{k_z} (\phi;k_\rho)$.
(See the Methods for the definition of Wilson loop.)
Here, $k_\rho$ and $\phi$ are the radius and azimuthal angle in the cylindrical coordinates.
The Wilson loop spectra, $\{\Theta_{k_z}(\phi;k_\rho)\} = {\rm Spec}[-i \log W_{k_z} (\phi;k_\rho)]$, is obtained from the phases of $W_{k_z} (\phi;k_\rho)$ eigenvalues, by changing $\phi$ but fixing $k_\rho$.

\begin{figure}[t]
\centering
\includegraphics[width=0.48\textwidth]{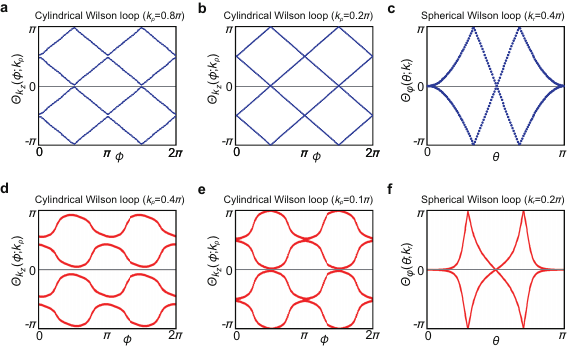}
\caption{{\bf Wilson loop spectra and windings.}
The cylindrical and spherical Wilson loop spectra for the $\bb H$ fields in the ab-initio model(\textbf{a-c}) and for the tight-binding model (\textbf{d-f}).
\textbf{a,b} The spectrum of cylindrical Wilson loop becomes gapless and exhibits helical winding as $k_\rho$ approaches 0.
Note that due to the small scale of inversion symmetry breaking in the ab initio structure, the radius of the cylinder for which the the ab-initio cylindrical Wilson loop spectra clearly does not wind is larger than in the tight binding model. We refer the reader to the SM~\cite{supple} for a detailed discussion.
\textbf{c} The spectrum of spherical Wilson loop exhibits winding structure corresponding to $|\mf{e}|=2$ for the Euler number $\mf{e}$.
\textbf{d-f} We observe qualitatively similar behavior in the Wilson loop spectra for the tight-binding model.
}
\label{fig:sg212_wilson}
\end{figure}

Now, let us consider the limit where the radius $k_\rho$ goes to zero.
If the band structure were regular around the $\Gamma$ such as for electronic bands, the Wilson loop would be required to converge to a $\phi$-independent value $\lim_{k_\rho \to 0}W_{k_z}(\phi;k_\rho)\equiv W_{k_z}(0)$.
However, for photonic bands in both ab-initio and TB models, the Wilson loop spectra becomes gapless and winds as a function of $\phi$ even as $k_\rho \to 0$, as shown in Figs.~\ref{fig:sg212_wilson}\textbf{a,b} and \textbf{d,e}.
In our model, we show in the SM~\cite{supple} that the effective Hamiltonian around the singularity enforces that $
\lim_{k_\rho \to 0} \, W_{k_z} (\phi;k_\rho)$ can be approximated to
\bg
\bpm 
0 & e^{-i \frac{\pi}{4}} & 0 & 0 \\
-e^{i \frac{\pi}{4} + 2i\phi} & 0 & 0 & 0 \\
0 & 0 & 0 & -e^{-i \frac{\pi}{4}} \\
0 & 0 & e^{i \frac{\pi}{4} - 2i\phi} & 0
\epm.
\eg
Thus, we conclude that
\bg
\lim_{k_\rho \to 0} \, \{\Theta_{k_z}(\phi;k_\rho)\}
\simeq \left( -\phi-\frac{\pi}{2}, -\phi+\frac{\pi}{2}, \phi-\frac{\pi}{2}, \phi+\frac{\pi}{2} \right),
\eg
which is compatible with the winding pattern observed in the Wilson loop spectra in Fig.~\ref{fig:sg212_wilson}.
%

%%%%%%
%%%%%%

In both our TB and ab-initio models, the cylindrical Wilson-loop spectrum exhibits winding as $k_\rho \to 0$, reflecting the influence of the polarization singularity at $\Gamma$.
While this behavior is characteristic of both our tight-binding and ab initio models, it is not generic, as it depends on the symmetry data.
To more robustly capture the topological constraint imposed by the singularity, we consider a spherical Wilson loop $W_\vph (\theta; k_r)$ that encloses $\Gamma$.
This Wilson loop exhibits the Euler number $\mf{e}$~\cite{bzduvsek2017robust,ahn2018band} of the transverse polarization vectors.
In spherical coordinates with $(k_r,\theta,\vph)$, we first fix $k_r$, and then compute $\vph$-directed Wilson loop by varying $\theta$.
The resulting spectrum, $\{\Theta_{\vph}(\theta;k_r)\}= {\rm Spec}[-i \log W_{\vph} (\theta;k_r)]$, is shown in Figs.~\ref{fig:sg212_wilson}\textbf{c,f}, for both the tight-binding and ab-initio models.
This winding can be interpreted as resulting from the nonzero Euler number $|\mf{e}|=2$, protected by emergent transversality (in the photonic case) or effective $I\mc{T}$ symmetry (in the tight-binding model) at small $k_r$ and $\omega$, where $I$ and $\mc{T}$ are inversion and time-reversal symmetries.
This emergent transversality is a consequence of the long-wavelength homogenization of Maxwell's equations in photonic crystals~\cite{datta1993effective,krokhin2002long}, while we explicitly show the emergence of effective $I\mc{T}$ symmetry in our tight-binding model in SM~\cite{supple} (Note that $|\mf{e}|$ is equal to the number of times $\Theta_\vph (\theta;k_r)$ crosses $\pi$~\cite{ahn2018band}).
Thus, using spherical and cylindrical Wilson loops, we demonstrated the distinction between photonic and electronic band structures, along with the different possible values of SRSIs.
%%%%%%

%%%%%%
\section{Discussion}
\label{sec:discuss}
%%%%%%
In this work, we introduced a comprehensive framework to classify and quantify trivial photonic bands in three-dimensional photonic crystals, addressing the challenges arising from the polarization singularity.
Our approach, based on the regularization of singularity introduced in Ref.~\cite{christensen2022location} and the stable real-space invariants (SRSIs) introduced in Ref.~\cite{hwang2025stable}, allows for the characterization of photonic bands in a way that overcomes the limitations of traditional band theory approaches: in particular, the SRSIs provide a well-defined topological characterization even in the presence of the singularity and independently of the choice of auxiliary bands in regularization.
We demonstrated that, for systems in SG $P4_332$, SRSIs are powerful tools for classifying trivial photonic bands and capturing the stable equivalence inherent in the classification problem.
This framework also allows for quantitative comparisons between photonic and electronic bands by employing the same SRSI formalism.
We emphasize that our approach can be applied to any space group, and we further demonstrate its effectiveness in another space group, $P432$ (No. 207), as shown in the SM~\cite{supple}.

Importantly, we highlighted the unique role of the polarization singularity in photonic bands, demonstrating how it constrains both band structure and Wilson loop behavior in ways distinct from electronic systems.
A key insight is that the polarization singularity imposes physical constraints on the symmetry-data vector of photonic bands, and consequently, on the SRSIs.
This enables a consistent classification of the topology of photonic bands based on their SRSI values.
The presence of the singularity also manifests in the Wilson loop behavior, as we observed winding structure in the cylindrical Wilson loop spectrum, which is a direct consequence of the polarization singularity.

We have characterized photonic band structures with $\Z$-valued SRSIs by focusing on their one-to-one relationship with the symmetry-data vector, which implies we have focused on the symmetry-indicated phases.
Our formalism naturally extends to non-symmetry-indicated phases by additional consideration of $\Z_n$-valued SRSIs.
Since the polarization singularity only imposes constraints on the symmetry-data vector, and thus on the $\Z$-valued SRSIs, the corresponding band structures can also be assigned $\Z_n$-valued SRSIs based on the choices of $\mc{B}_{\rm reg}$ and $\mc{B}_{\rm aux}$.
For instance, in the SG $P4_332$, for the band structures with $(\theta_1,\theta_2,\theta_3)=(-1,0,1)$, a $\Z_2$-valued SRSI $\theta_4^{(2)}$ in Eq.~\eqref{eq:sg212_rsi} allows two topologically distinct trivial band structures with $\theta_4^{(2)}=0$ and 1, with EBR decomposition $(E)_a - (A_1)_a$ and $(E)_a - (A_1)_b$, respectively.
While we have focused on properties determined by the symmetry-data vector and thus $\Z$-valued SRSIs, this illustrates that $\Z_n$-valued SRSIs can encode additional distinctions beyond symmetry indicators.

Our framework opens several directions for further studies.
First, the methods we developed can be extended to classify topological photonic bands with fractional SRSIs, going beyond the trivial case.
This is evident from the fact that we classified the building blocks of trivial photonic bands using integer-valued SRSIs.
Our framework separates genuine band topology from singular behavior at $\Gamma$, by using auxiliary bands only to define symmetry data, while computing SRSIs from the physical bands alone.
Additionally, studying the possibility of assigning irrep choices for the longitudinal ($L$) mode~\cite{christensen2022location} instead of trivial irrep can be an interesting direction for future studies.
While we used physical considerations to motivate the choice of the trivial irrep for the auxiliary $L$ mode in this work, our framework is flexible enough to accommodate alternative irrep choices.
Also, several distinct photonic bands with the same symmetry-data vector and $\Z$-valued SRSIs can differ in $\Z_n$-valued SRSIs.
As it has been shown how Wilson loops can distinguish between electronic bands with different $\Z_n$SRSIs~\cite{cano2022topology}, examining whether and how such distinctions can be meaningfully assigned and detected in photonic band structures could be an interesting further study.
Moreover, naive Wilson-loop diagnostics of surface band structures~\cite{fidkowski2011model,huang2012entanglement} may be ambiguous in the presence of polarization singularities.
As we saw in SG $P4_332$, a polarization singularity at $\Gamma$ can induce apparent winding in the Wilson-loop spectrum, falsely indicating nontrivial surface band connectivity even when the band is topologically trivial.
Understanding such ambiguities may offer deeper insights into photonic band structures.
Finally, we anticipate that our framework will find applications in model-building for photonic band structures, particularly as it has already been applied to construct tight-binding models in Ref.~\cite{morales2025transversality} and this work.
Future work could also explore the extension of these ideas to lower-dimensional systems or other excitations beyond photons, such as phonons.
In such systems, for example the lowest bands of 2D photonic crystals or the acoustic phonons in 3D solids~\cite{park2021topological,christensen2022location}, zero frequency modes at $\Gamma$ form a well-defined (regular) vector irrep at $\Gamma$ and do not exhibit a polarization singularity.
As such, the SRSI framework can be applied directly without regularization, but with the added requirement that the SRSI constraints incorporate the vector-irrep condition at $\Gamma$.
%%%%%%

%Methods

%%%%%%
\section*{Methods}
%%%%%%
Here, we summarize the derivation and properties of stable real-space invariants.
We also provide the details on the ab-initio calculations for simulating the photonic bands, and the review of Wilson loop method.
%%%%%%

%%%%%%
\subsection{Stable real-space invariant}
%%%%%%
The stable real-space invariants (SRSIs) can be obtained by considering all possible adiabatic deformation of site-symmetry irreps in a given space group.
For a clear explanation of algorithm obtaining the SRSIs, let us focus on the SG $P4_332$ (No. 212), which is studied in the main text.
In this SG, the following forms the basis for all possible adiabatic processes between the site-symmetry irreps:
\bg
(A_1 + A_2)_{a} \eqv (A_1)_{c}, \quad
2(E)_{a} \eqv ({}^1E{}^{2}E)_{c},
\nn
(A_1 + A_2)_{b} \eqv (A_1)_{c}, \quad
2(E)_{b} \eqv ({}^1E{}^{2}E)_{c},
\nn
(A_1 + E)_a \eqv (A)_d, \quad
(A_2 + E)_a \eqv (B)_d,
\nn
(A_1 + E)_b \eqv (A)_d, \quad
(A_2 + E)_b \eqv (B)_d,
\nn
(A + {}^1E{}^{2}E)_c \eqv (A)_e, \quad
(A + B)_d \eqv (A)_e,
\label{eq:sg212_adia}
\eg
where $(\rho_1+\rho_2)_W$ is shorthand notation for $(\rho_1)_W+(\rho_2)_W$.
These processes can be represented by the adiabatic-process matrix $q_{\rm adia}$,
\bg
q_{\rm adia} =
\bpm
1 & 0 & 1 & 0 & 0 & 0 & 0 & 0 & 0 & 0 \\
1 & 0 & 0 & 1 & 0 & 0 & 0 & 0 & 0 & 0 \\
0 & 2 & 1 & 1 & 0 & 0 & 0 & 0 & 0 & 0 \\
0 & 0 & 0 & 0 & 1 & 0 & 1 & 0 & 0 & 0 \\
0 & 0 & 0 & 0 & 1 & 0 & 0 & 1 & 0 & 0 \\
0 & 0 & 0 & 0 & 0 & 2 & 1 & 1 & 0 & 0 \\
-1 & 0 & 0 & 0 & -1 & 0 & 0 & 0 & 1 & 0 \\
0 & -1 & 0 & 0 & 0 & -1 & 0 & 0 & 1 & 0 \\
0 & 0 & -1 & 0 & 0 & 0 & -1 & 0 & 0 & 1 \\
0 & 0 & 0 & -1 & 0 & 0 & 0 & -1 & 0 & 1 \\
0 & 0 & 0 & 0 & 0 & 0 & 0 & 0 & -1 & -1
\epm,
\eg
which is an $N_{\rm orb} \times N_{\rm adia}$ matrix with $N_{\rm orb}=11$ and $N_{\rm adia}=10$.
Each column of $q_{\rm adia}$ denotes an adiabatic process as ordered in Eq.~\eqref{eq:sg212_adia}.
Also note that the row basis is ordered as
\bg
(A_1)_a, (A_2)_a, (E)_a, (A_1)_b, (A_2)_b, (E)_b, \nn
(A_1)_c, ({}^1 E {}^2 E)_c, (A)_d, (B)_d, (A)_e.
\label{eq:sg212_orb}
\eg
For a given column, each row represents how the multiplicity of a site-symmetry irrep changes by the corresponding adiabatic process.
For example, the first column indicates that the corresponding adiabatic process, $(A_1 + A_2)_a \eqv (A_1)_c$, changes the multiplicities of the site-symmetry irreps, $(m[(A_1)_a], m[(A_2)_a], m[(A_1)_c])$, by $(1,1,-1)$.

The SRSIs can be obtained by performing the Smith decomposition on the adiabatic-process matrix $q_{\rm adia}$.
Here, we briefly review the algorithm for generating the SRSIs, as detailed in Ref.~\cite{hwang2025stable}.
For this purpose, let us consider two atomic insulators, $AI_1$ and $AI_2$, with corresponding site-symmetry irrep multiplicity vectors $\bb m_1$ and $\bb m_2$.
If a series of adiabatic process encoded in $q_{\rm adia}$ deforms $AI_1$ to $AI_2$, or vice versa, then then we can write
\bg
\Delta \bb m = q_{\rm adia} \cdot \bb z,
\label{eq:deform1}
\eg
where $\Delta \bb m = \bb m_2 - \bb m_1$ for some integer-valued vector $\bb z \in \Z^{N_{\rm adia}}$.
Since $AI_{1,2}$ are topologically equivalent, we define the SRSIs to yield the same values for $AI_1$ and $AI_2$.
We denote the Smith decomposition of $q_{\rm adia}$ as $L \cdot \Lambda \cdot R$, where $L$ and $R$ are unimodular matrices, and the diagonal matrix $\Lambda$ has the form
\bg
{\rm diag} \, \Lambda = (\lambda_1, \lambda_2, \dots, \lambda_{r_{\rm adia}}, 0, \dots, 0),
\label{eq:lambda}
\eg
where $r_{\rm adia}$ is the rank of $q_{\rm adia}$ and $\lambda_{1,\dots,r_{\rm adia}}$ are nonnegative integers satisfying $\lambda_1 \le \lambda_2 \le \dots \le \lambda_{r_{\rm adia}}$.
Then, Eq.~\eqref{eq:deform1} can be written as $(L^{-1} \cdot \Delta \bb m)_i = \lambda_i \td{z}_i$ for $i=1,\dots, r_{\rm adia}$ and $(L^{-1} \cdot \Delta \bb m)_{i'} = 0$ for $i'=r_{\rm adia}+1, \dots, N_{\rm orb}$.
Here, $\td{z}_i = (R \cdot \bb z)_i$, which is also an integer vector.
For a given site-symmetry irrep multiplicity vector $\bb m$, we define $\Z_{\lambda_i}$-valued SRSI, $(L^{-1} \cdot \bb m)_i \pmod{\lambda_i}$, and $\Z$-valued SRSI $(L^{-1} \cdot \bb m)_{i'}$ for $i'$.
When $\lambda_i=1$ for some $i$, it defines a $\Z_1$-valued SRSI, which trivially takes 0.
Hence, we only define $\Z_{n \ge 2}$-valued and $\Z$-valued SRSIs, giving the same values for $AI_1$ and $AI_2$.

For convenience, we denote the list of all $\Z$-valued SRSIs, $(L^{-1} \cdot \bb m)_{i'}$ for $i'=r_{\rm adia}+1,\dots,N_{\rm orb}$, as $\bs \theta_\Z = \Delta_\Z \cdot \bb m$, with the integer-valued matrix $\Delta_\Z$.
Note that $\Delta_\Z$ can be always chosen such that its pseudoinverse $\Delta_\Z^\ddagger$ is integer-valued~\cite{hwang2025stable}.
This follows from the fact that $\Delta_\Z$ is composed of rows from the unimodular matrix $L^{-1}$~\cite{hwang2025stable}, though it can also be demonstrated as follows.
First, note that the defining property of $\Delta_\Z$ is
\bg
\Delta_\Z \cdot q_{\rm adia} = 0,
\label{eq:smith_zrsis}
\eg
ensured by the form of $\Lambda$.
We will demonstrate that there exists an integer-valued matrix $\Delta'_\Z$ satisfying the same condition and show that its pseudoinverse is also an integer-valued matrix.
Since $\Delta_\Z$ is integer-valued, its has the Smith decomposition $\Delta_\Z = L_\Delta \cdot \Lambda_\Delta \cdot R_\Delta$.
Then, Eq.~\eqref{eq:smith_zrsis} can be reduced to $\Lambda_\Delta \cdot R_\Delta \cdot q_{\rm adia}=0$.
Now, consider $\Lambda'_\Delta$, which has the same form as $\Lambda_\Delta$ but with all nonzero elements normalized to 1.
The diagonal form of $\Lambda'_\Delta$ also ensure that $\Lambda'_\Delta \cdot R_\Delta \cdot q_{\rm adia}=0$, meaning that $\Delta_\Z' = L_\Delta \cdot \Lambda'_\Delta \cdot R_\Delta$ also defines well-defined $\Z$-valued SRSIs, as $\Delta_\Z'$ satisfies the defining property of the matrix for $\Z$-valued SRSIs given in Eq.~\eqref{eq:smith_zrsis}, i.e., $\Delta'_\Z \cdot q_{\rm adia} = 0$.
As the pseudoinverse of $\Delta'_\Z$ is given by ${\Delta'_\Z}^\ddagger = R_\Delta^{-1} \cdot \Lambda'_\Delta \cdot L_\Delta^{-1}$, which is a product of integer-valued matrices, $\Delta'_\Z$ is also integer-valued.

In our example of SG $P4_332$, the Smith decomposition of $q_{\rm adia} = L \cdot \Lambda \cdot R$ are expressed with
\bg
L = \bpm 1 & 0 & 1 & 0 & 0 & 0 & 0 & 0 & 0 & 0 & 0 \\
1 & 0 & 0 & 1 & 0 & 0 & 0 & 0 & 0 & 0 & 0 \\
0 & 2 & 1 & 1 & 0 & -2 & 0 & 1 & 0 & 0 & 0 \\
0 & 0 & 0 & 0 & 1 & 1 & 1 & -1 & 0 & 0 & 0 \\
0 & 0 & 0 & 0 & 1 & 0 & 0 & 0 & 0 & 0 & 0 \\
0 & 0 & 0 & 0 & 0 & 3 & 1 & -2 & -1 & -1 & 1 \\
-1 & 0 & 0 & 0 & -1 & 1 & 0 & 0 & 0 & 0 & 0 \\
0 & -1 & 0 & 0 & 0 & 1 & 0 & 0 & 0 & 0 & 0 \\
0 & 0 & -1 & 0 & 0 & -1 & -1 & 1 & 0 & 0 & 0 \\
0 & 0 & 0 & -1 & 0 & 0 & 0 & 0 & -1 & 1 & 0 \\
0 & 0 & 0 & 0 & 0 & -1 & 0 & 0 & 1 & 0 & 0 \epm,
\nn
{\rm diag} \, \Lambda = (1,1,1,1,1,1,1,2,0,0),
\nn
R = \bpm 1 & 0 & 0 & 0 & 0 & 0 & 0 & -1 & 0 & 1 \\
0 & 1 & 0 & 0 & 0 & 1 & 0 & 0 & 0 & 1 \\
0 & 0 & 1 & 0 & 0 & 0 & 0 & 1 & 0 & -1 \\
0 & 0 & 0 & 1 & 0 & 0 & 0 & 1 & 0 & -1 \\
0 & 0 & 0 & 0 & 1 & 0 & 0 & 1 & 0 & 0 \\
0 & 0 & 0 & 0 & 0 & 0 & 0 & 0 & 1 & 1 \\
0 & 0 & 0 & 0 & 0 & -2 & 1 & -3 & 1 & 1 \\
0 & 0 & 0 & 0 & 0 & -1 & 0 & -1 & 1 & 1 \nn
0 & 0 & 0 & 0 & 0 & -1 & 0 & 0 & 0 & 0 \\
0 & 0 & 0 & 0 & 0 & -1 & 0 & -1 & 0 & 1 \epm.
\eg
From the diagonal elements of $\Lambda$, we can define a single $\Z_2$ valued SRSI, $(L^{-1})_{8,*} \cdot \bb m \pmod 2$, and three $\Z$-valued SRSIs, $(L^{-1})_{i,*} \cdot \bb m$ for $i=9,10,11$.
(Here, $A_{i,*}$ denotes the $i$-th row of matrix $A$.)
Thus, the SRSIs in Eq.~\eqref{eq:sg212_rsi} are defined as follows:
\bg
(L^{-1})_{8,*} \mod 2
= (1,1,1,0,0,\dots,0),
\nn
\bpm (L^{-1})_{9,*} \\ (L^{-1})_{10,*} \\ (L^{-1})_{11,*} \epm
= \bpm 1 & 0 & 0 & 1 & 0 & 0 & 1 & 0 & 1 & 0 & 1 \\
0 & 1 & 0 & 0 & 1 & 0 & 1 & 0 & 0 & 1 & 1 \\
0 & 0 & 1 & 0 & 0 & 1 & 0 & 2 & 1 & 1 & 2 \epm.
\eg
Note that, in general, the Smith decomposition of $q_{\rm adia}$ does not uniquely fix the unimodular matrices $L$ and $R$.
Consequently, different algorithms may yield different matrices $\Delta_\Z = (L^{-1})_{9-11,*}$ that define the $\Z$-valued SRSIs, equivalent up to taking linear combinations.
By reducing $\Delta_\Z$ to its Hermite normal form~\cite{cohen2013course}, following the convention used in Ref.~\cite{hwang2025stable}, one obtains a canonical representative, which is exactly the expression shown above.

One important property of SRSIs is that if two insulators, $AI_1$ and $AI_2$, have matching (both $\Z$- and $\Z_n$-valued) SRSIs, then $AI_1$ and $AI_2$ are deformable to each other with the inclusion of auxiliary fictitious site-symmetry irreps, thereby establishing the stable equivalence between $AI_1$ and $AI_2$~\cite{hwang2025stable}.
In other words, the following equation, which is a slight modification of Eq.~\eqref{eq:deform1}, holds:
\bg
(\bb m_{AI_2} + \bb m_{\rm aux}) - (\bb m_{AI_1} + \bb m_{\rm aux}) = q_{\rm adia} \cdot \bb z,
\eg
where $\bb m_{\rm aux}$ denotes the multiplicity vector for auxiliary site-symmetry irreps.
The necessity of $\bb m_{\rm aux}$ can be explained as follows: there may exist a solution $\bb z \in \Z^{N_{\rm adia}}$ for Eq.~\eqref{eq:deform1}.
However, without auxiliary site-symmetry irreps, the corresponding deformation processes may cause some site-symmetry irrep multiplicities to become negative integers during the processes, which are unphysical.
This does not happen with auxiliary site-symmetry irreps.
Let us provide a clear example of stable equivalence between two site-symmetry irrep configurations (or corresponding band representations).
The irreps $(A_1)_a$ and $(A_1)_b$ have the SRSIs, $\bb \theta = (\theta_1,\theta_2,\theta_3,\theta_4^{(2)})=(1,0,0,1)$ and $(1,0,0,0)$, respectively.
Since $\theta_4^{(2)}$ is defined modulo 2, $2(A_1)_a$ and $2(A_1)_b$ have the matching SRSIs with $\bb \theta=(2,0,0,0)$, which implies that $2(A_1)_a + \rho_{\rm aux}$ and $2(A_1)_b + \rho_{\rm aux}$ can be adiabatically deformable to each other for some auxiliary orbitals $\rho_{\rm aux}$.
Indeed, for $\rho_{\rm aux}=({}^1E{}^2E)_c$, we find the following sequence of adiabatic processes, $2(A_1)_a + ({}^1E{}^2E)_c \eqv 2(A_1 + E)_a
\eqv 2(A)_d \eqv 2(A_1 + E)_b \eqv 2(A_1)_b + 2(E)_b \eqv 2(A_1)_b + ({}^1E{}^2E)_c$, which are relevant to 2nd, 5th, 7th, and 4th processes in Eq.~\eqref{eq:sg212_adia}.
This demonstrates the stable equivalence between $2(A_1)_a$ and $2(A_1)_b$.

Finally, we present an explicit example illustrating that different choices of regularized bands $\mc B_{\rm reg}$ and auxiliary bands $\mc B_{\rm aux}$ can be used to realize the same physical band $\mc B_{\rm phys}$, even when the corresponding regularized bands are not topologically equivalent.
Consider the site-symmetry representations $\rho_{\rm reg} = (A_1)_a + (A_2)_b$ and $\rho_{\rm reg'} = (A_1)_b + (A_2)_a$.
These two configurations have identical SRSIs, $\theta = (1,1,0,1)$, and are therefore stably equivalent.
However, they are not adiabatically deformable into each other unless additional trivial irreps are included.
In particular, the deformation becomes possible only after adding the trivial irrep $(E)_a$, as illustrated by the sequence of adiabatic deformations shown below:
\bg
(A_1)_a + (A_2)_b + (E)_a \Leftrightarrow (A)_d + (A_2)_b
\nn
\Leftrightarrow (E)_b + (A_1)_b + (A_2)_b \Leftrightarrow (B)_d + (A_1)_b
\nn
\Leftrightarrow (A_2)_a + (A_1)_b + (E)_a
\eg
A physical band $\mc B_{\rm phys}$ with $\theta = (0,1,0,1)$ can be realized using a pair $(\mc B_{\rm reg}, \mc B_{\rm aux})$, where the site-symmetry representations correspond to $\rho_{\rm reg}$ and $\rho_{\rm aux} = (A_1)_b$, respectively.
The resulting symmetry-data vector of $\mc B_{\rm phys}$ coincides with $[(\rho_T)_\Gamma + \Gamma_1 + \Gamma_2, R_3, M_2 M_3 + M_5, X_1 + X_2]$ discussed in the main text.
Importantly, the stable equivalence between $\rho_{\rm reg}$ and $\rho_{\rm reg'}$ implies that the same physical band can equally be represented by the pair $(\mc B_{\rm reg'}, \mc B_{\rm aux})$, even though $\mc B_{\rm reg}$ and $\mc B_{\rm reg'}$ are not topologically equivalent.
This is possible because the regularization procedure allows the inclusion of additional auxiliary bands.
Explicitly, introducing additional bands $\mc B_{\rm add}$ corresponding to the trivial irrep $(E)_a$, the physical band can be formally expressed as
\bg
\mc B_{\rm reg} - \mc B_{\rm aux} =
(\mc B_{\rm reg} + \mc B_{\rm add}) - (\mc B_{\rm aux} + \mc B_{\rm add})
\nn
= (\mc B_{\rm reg'} + \mc B_{\rm add}) - (\mc B_{\rm aux} + \mc B_{\rm add})
= \mc B_{\rm reg'} - \mc B_{\rm aux},
\eg
and thus represented by both pairs, $(\mc B_{\rm reg}, \mc B_{\rm aux})$ and $(\mc B_{\rm reg'}, \mc B_{\rm aux})$.
In all cases, the associated SRSIs of $\mc B_{\rm phys}$ remain unchanged.
This example explicitly demonstrates that the SRSI characterization of physical photonic bands is independent of the specific choice of regularization, and depends only on the stable equivalence class of the underlying band configurations.
%%%%%%

%%%%%%
\subsection{Ab-initio calculation}
%%%%%%
We perform ab-initio simulations of photonic crystals using the MIT Photonic Bands (MPB) package~\cite{johnson2001block-iterative}. 
MPB solves the macroscopic Maxwell's equations for the frequency spectrum $\omega_{n \bk}$ (shown in Fig.~\ref{fig:sg212_band}\textbf{a}) and the magnetic eigenfields $\bb H_{n \bk} (\bb r) = e^{i \bk \cdot \bb r} \bb u_{n \bk} (\bb r)$ where $\bb u_{n \bk} (\bb r)$ is the cell-periodic field.
We start with the primitive unit cell of the SG $P4_332$ (No. 212) which is filled with a non-magnetic (i.e. $\mu = 1$), homogeneous, isotropic medium with a dielectric constant $\epsilon = 11$.
The lattice constant is set to $a=1$.
Then, we carve out spheres of radius $r = 0.2$, centered at the $24e$ Wyckoff positions (WPs).
For this, a representative position of $24e$ WP is set as $(0,-0.4,0.2)$.
Remaining 23 positions are determined by SG symmetries.
At all the 24 positions corresponding to the $24e$ WP, we added a sphere of a dielectric that has unit dielectric constant.
This structure is inspired by a class of photonic crystals called inverse opals which feature a complete photonic band gap~\cite{joannopoulos2008photonic}.
Due to the relatively large radius $r = 0.2$ of these spheres, there are substantial overlaps between them.
The default behavior of MPB is to set the dielectric constant in the overlap region to match that of the most recently placed object.
In our case, this means that dielectric constant in the regions of overlap is still set to 1.
We then direct MPB to solve for the five bands with smallest frequencies.
The resulting band structure is shown in Fig.~\ref{fig:sg212_band} and was discussed in the main text.
We confirmed that the lowest four bands are isolated from the fifth band (hence they are isolated from all higher bands).
Because of this, in Fig.~\ref{fig:sg212_band} we show only the lowest four bands.

To determine the symmetry-data vector associated with the bands, we used the MPBUtils library~\cite{christensen2023MPBUtilsjl} developed in Ref.~\cite{christensen2022location}.
These codes already take into account the irregular $\Gamma$-point symmetry content (little-group irreps at $\Gamma$) and provide the symmetry-data vector in Eq.~\eqref{eq:sg212_irrep_model}.
Let us elaborate on the assignment of little-group irreps in this work. 
When magneto-electric coupling is absent, the symmetry-data vector can be defined separately for electric and magnetic fields (See the SM~\cite{supple} for the details).
Note that different conventions for assigning little group irreps to electric and magnetic fields exist, depending on whether one chooses to keep track of the axial vector nature of the $H$ fields~\cite{morales2025transversality}.
That is, for orientation-reversing symmetries like mirror and inversion, the magnetic field transforms with an extra minus sign compared to the electric field.
Thus, vector, axial-vector, scalar, and pseudo-scalar representations of electric fields correspond to axial-vector, vector, pseudo-scalar, and scalar representations of magnetic fields, respectively~\cite{morales2025transversality}.
However, in the SGs $P4_332$ and $P432$, studied in the main text and the SM~\cite{supple} respectively, only (screw) rotations that preserve orientation exist.
Therefore, no distinction arises between electric and magnetic fields when little-group irreps are assigned.
As the permeability tensors of the photonic crystals that we study in this work are trivial, we have chosen to use the magnetic eigenfields to construct the Wilson loops as described below.
%%%%%%

%%%%%%
\subsection{Wilson loop method}
%%%%%%
The Wilson loop is defined for a path whose initial (base) and final points, $\bk_i$ and $\bk_f$, in momentum space are equal up to a reciprocal vector $\bb G$, i.e. $\bk_f = \bk_i + \bb G$.
For a given path $\mc{L}$ in momentum space, we discretize it by defining points $\bk_a$ along the path, where $a=0,1,\dots,N-1$ (Note that $\bk_0=\bk_i$ and $\bk_{N-1}=\bk_f$).
For photonic bands, we compute the Wilson loop as follows.
For magnetic (electric) Wilson loops, we construct the overlap matrices $(S_{\bk_a,\bk_b})_{nm} = \langle \bb u_{n \bk_a} | \bb u_{m \bk_b} \rangle$, where $\bb u_{n \bk}$ is the cell-periodic part of $n$-th magnetic (electric) eigenfields.
The inner product is defined as
\bg
\langle \bb u_{n \bk_a} | \bb u_{m \bk_b} \rangle
= \int_{\rm uc} d \bb r \, (\bb u_{n \bk_a}(\bb r))^* \cdot K(\bb r) \cdot \bb u_{m \bk_b}(\bb r),
\eg
where $K(\bb r)$ is the permittivity (permeability) tensor $\ep(\bb r)$ [$\mu(\bb r)$] for electric (magnetic) eigenfields~\cite{joannopoulos2008photonic,devescovi2024tutorial}, and $\int_{\rm uc}$ indicates that the integration is over a unit cell.
The Wilson loop associated with $\mc{L}$ is
\bg
(W_\mc{L})_{nm}
= (S_{\bk_{N-1},\bk_{N-2}} S_{\bk_{N-2},\bk_{N-3}} \cdots S_{\bk_1,\bk_0})_{nm},
\eg
where $n$ and $m$ are indices for the eigenfields of interest.

To obtain the eigenvalues of $W_\mc{L}$ in a gauge-independent way, we choose the periodic gauge $\bb H_{n \bk + \bb G} = \bb H_{n \bk}$ and thus $\bb u_{n \bk_{N-1}} = e^{-i \bb G \cdot \bb r} \, \bb u_{n \bk_0}$.
The Wilson loop spectrum is obtained by plotting the phases of the eigenvalues of $W_\mc{L}$, i.e. ${\rm Spec}(-i \log W_\mc{L})$, for different choices of $\mc{L}$ (chosen so that phases of Wilson loop eigenvalues change smoothly).
For further details, readers are referred to Refs.~\cite{blanco2020tutorial,devescovi2024tutorial}.

For tight-binding models, the Wilson loop is defined similarly.
The tight-binding Hamiltonian $H_{TB}(\bk)$ is defined with real-space basis orbitals whose positions are $\bx_\alpha$ ($\alpha=1,\dots,n_{tot}$) within the unit cell.
Then, $H_{TB}(\bk)$ satisfies $H_{TB}(\bk + \bb G) = V(-\bb G) H_{TB}(\bk) V(\bb G)$, where $V(\bb G)_{\alpha \beta} = e^{i \bb G \cdot \bb x_\alpha} \delta_{\alpha \beta}$.
In this case, for $n$-th energy eigenstate $\ket{u_{n \bk}}$, the periodic gauge is defined as $\ket{u_{n \bk + \bb G}} = V(-\bb G) \ket{u_{n \bk}}$.
The Wilson loop is then defined with the overlap matrix $\mc{S}_{\bk_a,\bk_b}$, which is given by $
(\mc{S}_{\bk_a,\bk_b})_{nm} = \brk{u_{n \bk_a}}{u_{m \bk_b}} = \sum_{\alpha=1}^{n_{tot}} \, (\ket{u_{n \bk_a}}_\alpha)^* \ket{u_{m \bk_b}}_\alpha$.
%%%%%%

%%%%%%
\section*{Acknowledgments}
The initial work of Y.H., V.G., and B.B. was supported by the Air Force Office of Scientific Research under award number FA9550-21-1-0131, and the National Science Foundation under grant no. DMR-1945058.
Y.H. received additional support from the US Office of Naval Research (ONR) Multidisciplinary University Research Initiative (MURI) grant N00014-20-1-2325 on Robust Photonic Materials with High-Order Topological Protection, and the UK Research and Innovation (UKRI) Future Leaders Fellowship MR/Y017331/1.
B.B. received additional support during the final stages of this work from the National Science Foundation under grant no. DMR-2510219. A.G.E., A.M.P, C.D. and M.G.V. acknowledge support from the Spanish Ministerio de Ciencia e Innovación (PID2022-142008NB-I00). A.G.E., and A.M.P, received funding from the  Basque Government Elkartek program (KK2025\_00058). A.G.E., A.M.P and M.G.V.  and from the IKUR Strategy under the collaboration agreement between Ikerbasque Foundation and DIPC on behalf of the Department of Science of the Basque Government, Programa de Ayuda de Apoyo a los agentes de la Red Vasca de Ciencia, Tecnologia e Innovacion acreditados en la categoria de Centros de Investigacion Basica y de Excelencia (Programa BERC) from the Departamento de Universidades e Investigacion del Gobierno Vasco and Centros Severo Ochoa AEI/CEX2024-0001491-S from the Spanish Ministerio de Ciencia e Innovacion. M.G.V acknowledge the support of the Canada Excellence Research Chairs Program for Topological Quantum Matter. The work of JLM has been partly supported by the Basque Government
Grant No. IT1628-22 and by grants  PID2021-123703NB-C21and PID2024-156016NB-I00 funded by MCIN/AEI/10.13039/501100011033/
and ERDF; “A way of making Europe”.
%%%%%%

%%%%%%Ref
\bibliography{Refs.bib}
%%%%%%

\end{document}

\title{Supplementary Information:\\ Building blocks of topological band theory for photonic crystals}
%%%%%%

%%%%%%Authors
\author{Yoonseok \surname{Hwang}}
\affiliation{Department of Physics, University of Illinois Urbana-Champaign, Illinois 61801, USA}
\affiliation{Anthony J. Leggett Institute for Condensed Matter Theory, University of Illinois Urbana-Champaign, Illinois 61801, USA}
\affiliation{Blackett Laboratory, Imperial College London, London SW7 2AZ, United Kingdom}

\author{Vaibhav \surname{Gupta}}
\affiliation{Department of Physics, University of Illinois Urbana-Champaign, Illinois 61801, USA}
\affiliation{Anthony J. Leggett Institute for Condensed Matter Theory, University of Illinois Urbana-Champaign, Illinois 61801, USA}

\author{Antonio Morales-P\'erez}
\affiliation{Donostia International Physics Center, Paseo Manuel de Lardizabal 4, 20018 Donostia-San Sebastian, Spain}
\affiliation{Material and Applied Physics Department, University of the Basque Country (UPV/EHU), Donostia-San Sebastián, Spain}

\author{Chiara Devescovi}
\affiliation{Donostia International Physics Center, Paseo Manuel de Lardizabal 4, 20018 Donostia-San Sebastian, Spain}
\affiliation{Institute for Theoretical Physics, ETH Zurich, Zurich, Switzerland}

\author{Mikel García-Díez}
\affiliation{Donostia International Physics Center, Paseo Manuel de Lardizabal 4, 20018 Donostia-San Sebastian, Spain}
\affiliation{Physics Department, University of the Basque Country (UPV/EHU), Bilbao, Spain}

\author{Juan L. Ma\~nes}
\affiliation{Physics Department, University of the Basque Country (UPV/EHU), Bilbao, Spain}
\affiliation{EHU Quantum Center, University of the Basque Country UPV/EHU, 48940 Leioa, Spain}

\author{Maia G. Vergniory}
\affiliation{Donostia International Physics Center, Paseo Manuel de Lardizabal 4, 20018 Donostia-San Sebastian, Spain}
\affiliation{Département de Physique et Institut Quantique, Université de Sherbrooke, Sherbrooke, QC J1K 2R1 Canada}

\author{Aitzol Garc\'ia-Etxarri}
\affiliation{Donostia International Physics Center, Paseo Manuel de Lardizabal 4, 20018 Donostia-San Sebastian, Spain}
\affiliation{IKERBASQUE, Basque Foundation for Science, Mar\'ia D\'iaz de Haro 3, 48013 Bilbao, Spain}

\author{Barry \surname{Bradlyn}}
\affiliation{Department of Physics, University of Illinois Urbana-Champaign, Illinois 61801, USA}
\affiliation{Anthony J. Leggett Institute for Condensed Matter Theory, University of Illinois Urbana-Champaign, Illinois 61801, USA}
%%%%%%

\maketitle
\tableofcontents
\hfill \\
%%%%%%

%%%%%%
\begin{center}
{\bf \large Supplementary Notes}
\end{center}
%%%%%%
In this Supplementary Information, we first present in Supplementary Note (SN)~\ref{sec:sn_sym} a comprehensive discussion of symmetry constraints and Wilson loops in photonic crystals, establishing their formal equivalence to tight-binding models.
Subsequently, in SN~\ref{sec:sn_sg212}, we provide a detailed analysis of photonic band structures in space group (SG) $P4_332$ (No. 212), including the allowed values of stable real-space invariants (SRSIs) derived from the Hilbert basis method, as well as a tight-binding model and Wilson loop calculations.
In SN~\ref{sec:sn_sg207}, we extend our approach to SG $P432$ (No. 207), demonstrating that the methods developed in the main text apply broadly to arbitrary space groups.
Throughout this Supplementary Information, we follow the conventions of the Bilbao Crystallographic Server (BCS)~\cite{aroyo2006bilbao1,aroyo2006bilbao2,vergniory2017graph,elcoro2017double} for group-theoretical notations, including Wyckoff positions (WPs), site-symmetry irreducible representations (irreps), momentum-space (little-group) irreps, and SGs.
%%%%%%

%%%%%%
\section{Symmetry and Wilson loops in photonic crystals and tight-binding models}
\label{sec:sn_sym}
%%%%%%
Although many of the theoretical foundations of photonic crystals are well-established~\cite{sakoda2005optical,joannopoulos2008photonic}, several important details remain scattered throughout the literature.
In particular, the general action of symmetries (including both unitary and antiunitary operations) away from maximal $k$-vectors or high-symmetry momenta are often implicitly assumed.
In order for our work to be self-contained, in this SN we systematically define sewing matrices and related symmetry relations for general symmetries acting on photonic crystals.
In particular, we present rigorous derivations showing how eigenfields transform under these symmetries, encoded through sewing matrices.
We also derive precise relations connecting the symmetry-data vectors of electric and magnetic fields.
Furthermore, while Wilson lines and loops are widely used in photonic band theory~\cite{blanco2020tutorial,devescovi2024tutorial}, the explicit ways in which they are constrained by sewing matrices have not been comprehensively summarized for photonic systems.
Here, we provide such a summary and compare the results to those in tight-binding (TB) models.
Our analysis demonstrates that the Wilson-loop formalism in photonic crystals and in TB models is fundamentally equivalent, provided that the inner product for eigenfields in photonic crystals and eigenstates in TB models is properly defined.
This SN compiles and clarifies these results, aiming to provide a unified formalism applicable to both photonic crystals and tight-binding models.
%%%%%%

%%%%%%
\subsection{Maxwell's equations in photonic crystals}
\label{sec:sn_sym_phc}
%%%%%%
We start from the Maxwell's equations in a (time-independent) medium:
\bg
\nabla \times \bb E (\bb r, t) = - \der_t \bb B (\bb r, t),
\quad
\nabla \times \bb H (\bb r, t) = \der_t \bb D (\bb r, t),
\quad
\nabla \cdot \bb D (\bb r, t) = 0,
\quad
\nabla \cdot \bb B (\bb r, t) = 0.
\eg
By assuming time-harmonic fields of the form $\bb E (\bb r, t) = \bb E (\bb r) e^{-i \om t}$ with frequency $\om$, and similarly for other fields, these equations become
\bg
\nabla \times \bb E (\bb r) = i \om \bb B (\bb r),
\quad
\nabla \times \bb H (\bb r) = - i \om \bb D (\bb r),
\quad
\nabla \cdot \bb D (\bb r) = 0,
\quad
\nabla \cdot \bb B (\bb r) = 0.
\label{seq:maxwell1}
\eg
The $\bb D$ and $\bb B$ fields are related to $\bb E$ and $\bb H$ fields through constitutive parameters $\ep (\bb r)$, $\mu (\bb r)$, $\xi (\bb r)$, and $\zeta (\bb r)$:
\bg
\bb D (\bb r) = \ep (\bb r) \cdot \bb E (\bb r) + \xi (\bb r) \cdot \bb H (\bb r),
\quad
\bb B (\bb r) = \mu (\bb r) \cdot \bb H (\bb r) + \zeta (\bb r) \cdot \bb E (\bb r).
\label{seq:constitutive1}
\eg
Here, $\ep (\bb r)$ and $\mu (\bb r)$ are permittivity and permeability tensors, respectively.
The tensors $\xi (\bb r)$ and $\zeta (\bb r)$ describe magneto-electric coupling.
In lossless media, $\ep (\bb r)$ and $\mu (\bb r)$ are Hermitian, i.e. $\ep (\bb r) = \ep (\bb r)^\dg$ and $\mu (\bb r) = \mu (\bb r)^\dg$, and $\xi (\bb r)$ and $\zeta (\bb r)$ are related via Hermitian conjugation, i.e. $\xi (\bb r) = \zeta (\bb r)^\dg$.
We assume this lossless condition throughout this work.
By introducing the 6-dimensional field $\Psi (\bb r)$ and the constitutive parameter tensor $\eta (\bb r)$,
\bg
\eta (\bb r) = \bpm \ep (\bb r) & \xi (\bb r) \\ \zeta (\bb r) & \mu (\bb r) \epm,
\quad
\Psi (\bb r) = \bpm \bb E (\bb r) \\ \bb H (\bb r) \epm,
\eg
Eqs.~\eqref{seq:maxwell1} and \eqref{seq:constitutive1} can be compactly written as
\bg\label{seq:maxwell_matrix}
\bpm & i \nabla \times \\ - i \nabla \times & \epm \, \Psi (\bb r) \equiv \mc{H} (\bb r) \, \Psi (\bb r) = \om \, \eta(\bb r) \Psi (\bb r),
\eg

For later convenience, we introduce the position basis states $\ket{\bb r}$.
Then we can define states $\ket{\Psi }$ such that $\Psi (\bb r) = \brk{\bb r}{\Psi}$.
We also define
\bg
\hat{\mc{H}} = \int d \bb r \, \ket{\bb r} \, \mc{H}(\bb r) \, \bra{\bb r},
\quad
\hat{\eta} = \int d \bb r \, \ket{\bb r} \, \eta (\bb r) \bra{\bb r}.
\eg
Then, the Maxwell's equations can be expressed as an eigenvalue problem:
\bg
\hat{\mc{H}} \, \ket{\Psi} = \om \, \hat{\eta} \, \ket{\Psi}.
\label{seq:maxwell2}
\eg
To discuss the properties of the operator $\hat{\mc{H}}$, we first define the inner product between two states $\Psi$ and $\Psi'$ as:
\bg
\brk{\Psi}{\Psi'}
= \int d \bb r \, \brk{\Psi}{\bb r} \cdot \brk{\bb r}{\Psi'}
= \int d \bb r \, \Psi(\bb r)^\dg \cdot \Psi'(\bb r)
= \int d \bb r \, \sum_{a=1}^6 \, (\Psi(\bb r)^*)_a (\Psi'(\bb r))_a
\eg
Then, we observe that $\mc{H}$ is self-adjoint, i.e.
\ba
\brk{\Psi}{\hat{\mc{H}} \Psi'}
=& \int d \bb r \, \brk{\Psi}{\bb r} \cdot \left[ \mc{H}(\bb r) \, \brk{\bb r}{\Psi'} \right]
= \int d \bb r \, \Psi(\bb r)^\dg \cdot \bpm & i \nabla \times \\ -i \nabla \times & \epm \Psi'(\bb r)
\nn
=& \int d \bb r \, \left[ \bpm & -i \nabla \times \\ i \nabla \times & \epm \Psi(\bb r)^\dg \right] \cdot \Psi'(\bb r)
= \int d \bb r \, \left[ \mc{H}(\bb r) \Psi(\bb r) \right]^\dg \cdot \Psi'(\bb r)
\nn
=& \brk{\hat{\mc{H}} \Psi}{\Psi'}
\ea
Consequently, $\om$ is real-valued, and eigenfields $\Psi$ and $\Psi'$ at different frequencies must be orthogonal under an inner product with a weight $\eta$:
\bg
\bra{\Psi} \hat{\eta} \ket{\Psi'}
= \int d \bb r \, \Psi(\bb r)^\dg \cdot \eta(\bb r) \cdot \Psi'(\bb r) = 0.
\eg
A precise orthonormality condition will be given after we discuss the Bloch theorem below.

In photonic crystals, the constitutive parameter tensor $\eta (\bb r)$ is periodic, satisfying $\eta (\bb r) = \eta (\bb r + \bb R)$ for any lattice vector $\bb R$.
In this case, the eigenfields can be labeled by the crystal momentum $\bk$.
We denote the $n$-th eigenfield with momentum $\bk$ and frequency $\om_{n \bk}$ as $\Psi_{n, \bk}$.
We adopt the following normalization condition:
\bg
\bra{\Psi_{n, \bk}} \hat{\eta} \ket{\Psi_{m, \bk'}} = N_{\rm cell} \, \delta_{nm} \delta_{\bk \bk'}
\label{seq:psi_norm}
\eg
Here, $N_{\rm cell}$ denotes the number of unit cells in the system.

The momentum $\bk$ parametrizes the eigenvalue for the translation operator $\hat{t}_{\bb R}$ as
\bg
\hat{t}_{\bb R} \, \ket{\Psi_{n \bk}} = e^{-i \bk \cdot \bb R} \ket{\Psi_{n \bk}}.
\label{seq:translation_def}
\eg
For the position basis $\ket{\bb r}$, the operator $\hat{t}_{\bb R}$ acts as $\hat{t}_{\bb R} \, \ket{\bb r} = \ket{\bb r + \bb R}$.
By the Bloch's theorem, the eigenfields can be expressed as
\bg
\ket{\Psi_{n \bk}} = e^{i \bk \cdot \hat{\bb r}} \, \ket{\bb u_{n \bk}},
\eg
where the cell-periodic part $\bb u_{n \bk} (\bb r) = \brk{\bb r}{\bb u_{n \bk}}$ satisfies the periodicity condition $\bb u_{n \bk} (\bb r) = \bb u_{n \bk} (\bb r + \bb R)$ for any lattice vector $\bb R$.

Since $\bb u_{n \bk}$ are cell-periodic, we define the $\eta$-weighted inner product within a single unit cell as
\bg
\bra{\bb u_{n \bk}} \hat{\eta} \ket{\bb u_{m \bk}}_{\rm uc}
:= \int_{\rm uc} d \bb r \, \bb u_{n \bk} (\bb r)^\dg \cdot \eta (\bb r) \cdot \bb u_{m \bk} (\bb r)
= \delta_{nm},
\label{seq:bloch_norm}
\eg
where $\int_{\rm uc}$ denotes integration over a single unit cell~\cite{joannopoulos2008photonic,devescovi2024tutorial}.
Note that the normalization in Eq.~\eqref{seq:bloch_norm} follows directly from the condition in Eq.~\eqref{seq:psi_norm}.
%%%%%%

%%%%%%
\subsection{Symmetries of photonic crystals}
\label{sec:sn_sym_sym}
%%%%%%
Now, let us consider a general symmetry operation $g = \{O_g | \bb v_g\}$, where $O_g$ is an $O(3)$ matrix representing the point group part, and $\bb v_g \in \Q^3$ denotes a (possibly fractional) translation.
Consequently, $g$ acts on real-space coordinates $\bb r$ as
\bg
g \circ \bb r = O_g \cdot \bb r + \bb v_g,
\quad
g^{-1} \circ \bb r = O_g^{-1} \cdot (\bb r - \bb v_g).
\eg
We then define the operator $\hat g$ that implements $g$ on states and operators (including possible time reversal).
To distinguish whether $\hat g$ is unitary or antiunitary, we introduce $\tau_g \in \{0,1\}$ defined as
\bg
\tau_g = \begin{cases}
0 & \text{if $\hat g$ is unitary},
\\
1 & \text{if $\hat g$ is antiunitary}.
\end{cases}
\eg
which equivalently determines whether $g$ reverses time.
In this notation, the action of $\hat{g}$ on the position-space basis $\ket{\bb r}$ is given by
\bg
\hat{g} \ket{\bb r} = \ket{g \circ \bb r} \, \mc{K}^{\tau_g} = \begin{cases}
\ket{g \circ \bb r} & \text{if $g$ is unitary}, \\
\ket{g \circ \bb r} \mc{K} & \text{if $g$ is antiunitary}.
\end{cases}
\label{seq:g_conj_def}
\eg
Here, $\mc{K}$ denotes complex conjugation, satisfying $\mc{K} c = c^* \mc{K}$ for any complex-valued quantity $c$.
We explicitly include $\mc{K}$ in Eq.~\eqref{seq:g_conj_def} for the antiunitary case to ensure consistent results when applying $\hat g$ to $\hat{g} \ket{\bb r} \, c$, where $c$ commutes with $\ket{\bb r}$.
Specifically, if $g$ is antiunitary, it must hold that $\hat{g} ( \ket{\bb r} \, c ) = \hat{g} ( c \, \ket{\bb r})$, leading to
\bg
\hat{g} ( \ket{\bb r} \, c ) = \ket{g \circ \bb r} \mc{K} \, c = \ket{g \circ \bb r} c^* \mc{K},
\quad
\hat{g} ( c \, \ket{\bb r} )= c^* \hat{g} \ket{\bb r} = c^* \ket{g \circ \bb r} \, \mc{K} = \ket{g \circ \bb r} \, c^* \, \mc{K}.
\eg
The notation adopted here for describing general symmetry operations is widely used (see, for example, Refs.~\cite{shiozaki2017topological,alexandradinata2018no}) because it allows both unitary and antiunitary operations to be treated within a unified framework.

Now, let us discuss the symmetry properties of Maxwell's equations.
First, to build some intuition, suppose that $g$ is a unitary symmetry operation and consider uniform, constant $\bb E$ and $\bb H$ fields.
These transform under $g$ as a vector and an axial-vector, respectively,
\bg
\bb E \to O_g \cdot \bb E,
\quad
\bb H \to {\rm Det} [O_g] \, O_g \cdot \bb H.
\eg
Moreover, under time-reversal symmetry, the fields transform as
\bg
\bb E \to \bb E,
\quad
\bb H \to - \bb H.
\eg
Since any antiunitary symmetry can be represented as a combination of a unitary symmetry and time-reversal, we define the real $6 \times 6$ matrix $\Delta_g$ and the operator $\hat{U}_g$ as
\bg
\Delta_g = \bpm O_g & \\ & (-1)^{\tau_g} {\rm Det} [O_g] \, O_g \epm,
\quad
\hat{U}_g = \Delta_g \, \hat{g} = \hat{g} \Delta_g.
\eg

We now show the condition for photonic crystals to possess a symmetry $g$.
Equation~\eqref{seq:maxwell2} can be written as
\bg
\hat{U}_g \, \hat{\mc{H}} \hat{U}_g^{-1} \, \left( \hat{U}_g \, \ket{\Psi_{n, \bk}} \right) = \om_{n, \bk} \, \hat{U}_g \, \hat{\eta} \, \hat{U}_g^{-1} \, \left( \hat{U}_g \, \ket{\Psi_{n, \bk}} \right).
\label{seq:maxwell_sym1}
\eg
One can directly verify that $\hat{U}_g \, \hat{\mc{H}} \hat{U}_g^{-1} = \hat{\mc{H}}$ for any rigid symmetry $g$.
Furthermore, by applying a translation operator $\hat{t}_{\bb R}$ to $\hat{U}_g \, \ket{\Psi_{n, \bk}}$, we find that $\hat{U}_g \, \ket{\Psi_{n, \bk}}$ carries momentum $(-1)^{\tau_g} \, O_g \cdot \bk$:
\ba
\hat{t}_{\bb R} \, \Delta_g \, \hat{g} \ket{\Psi_{n, \bk}}
=& \Delta_g \, (\hat{t}_{\bb R} \, \hat{g}) \, \ket{\Psi_{n, \bk}}
= \Delta_g \, (\hat{g} \, \hat{t}_{O_g^{-1} \cdot \bb R}) \, \ket{\Psi_{n, \bk}}
\nn
=& \Delta_g \, \hat{g} e^{-i \bk \cdot (O_g^{-1} \cdot \bb R)} \, \ket{\Psi_{n, \bk}}
= e^{-i (-1)^{\tau_g} (O_g \cdot \bk) \cdot \bb R} \, \Delta_g \, \hat{g} \ket{\Psi_{n, \bk}}
\nn
:=& e^{- i g\bk \cdot \bb R} \, [\Delta_g \, \hat{g} \ket{\Psi_{n, \bk}}]
= e^{- i g\bk \cdot \bb R} \, \hat{U}_g \, \ket{\Psi_{n, \bk}}.
\ea
In the fourth equality, we used the fact that $(O_g \cdot \bk) \cdot (O_g \cdot \bb r) = \bk \cdot \bb r$ since $O_g$ is an orthogonal matrix.
Therefore, the momentum of the transformed eigenfield is defined as
\bg
g\bk = (-1)^{\tau_g} \, O_g \cdot \bk.
\label{seq:gk}
\eg

Only if the relation $\hat{U}_g \, \hat{\eta} \, \hat{U}_g^{-1} = \hat{\eta}$ holds, Eq.~\eqref{seq:maxwell_sym1} reduces to the Maxwell's equation for eigenfields $\ket{\Psi_{n', g\bk}}$ with momentum $g \bk$ and frequency $\om_{n', g\bk} = \om_{n, \bk}$:
\bg
\hat{\mc H} \, \ket{\Psi_{n', g\bk}} = \om_{n, \bk} \hat{\eta} \, \ket{\Psi_{n', g\bk}},
\quad
\ket{\Psi_{n', g\bk}} = \hat{U}_g \, \ket{\Psi_{n, \bk}}
\label{seq:maxwell_sym2}
\eg
Thus, the symmetry condition on the constitutive parameter tensor $\eta (\bb r)$ can be expressed as~\cite{christensen2022location}
\bg
\eta(g \circ \bb r) = \Delta_g \cdot \mc{K}^{\tau_g} \, \eta(\bb r) \, \mc{K}^{\tau_g} \cdot \Delta_g^{-1}.
\eg
Equivalently, this leads to the following relations for the individual constitutive parameters:
\bg
\ep (g \circ \bb r) = O_g \cdot \mc{K}^{\tau_g} \, \ep(\bb r) \, \mc{K}^{\tau_g} \cdot O_g^{-1},
\quad
\mu (g \circ \bb r) = O_g \cdot \mc{K}^{\tau_g} \, \mu(\bb r) \, \mc{K}^{\tau_g} \cdot O_g^{-1},
\nn
\xi (g \circ \bb r) = (-1)^{\tau_g} {\rm Det} [O_g] \, O_g \cdot \mc{K}^{\tau_g} \, \xi (\bb r) \, \mc{K}^{\tau_g} \cdot O_g^{-1},
\quad
\zeta (g \circ \bb r) = (-1)^{\tau_g} {\rm Det} [O_g] \, O_g \cdot \mc{K}^{\tau_g} \, \zeta (\bb r) \, \mc{K}^{\tau_g} \cdot O_g^{-1}.
\label{seq:eta_sym}
\eg
As an example, consider time-reversal symmetry.
For time-reversal symmetry, $O_g = \mathds{1}_3$ (the identity matrix), $\bb v_g = \bb 0$, and $\tau_g=1$.
Thus, to preserve time-reversal symmetry, the constitutive parameters must satisfy $\ep(\bb r) = \ep(\bb r)^*$, $\mu(\bb r)= \mu(\bb r)^*$, $\xi(\bb r) = - \xi(\bb r)^*$, and $\zeta(\bb r) = - \zeta(\bb r)^*$.
In what follows, we will restrict our attention to the subgroup of symmetry operations satisfying Eq.~\eqref{seq:eta_sym}, i.e. to symmetries of the photonic crystal.
%%%%%%

%%%%%%
\subsection{Sewing matrices}
\label{sec:sn_sym_sewing}
%%%%%%
Now, from the relation in Eq.~\eqref{seq:maxwell_sym2} between the eigenfields $\ket{\Psi_{n, \bk}}$ and $\ket{\Psi_{n', g\bk}}$, we define the sewing matrix associated with the symmetry operation $g$.
Consider a set of bands labeled by $\mc{B}$.
For these bands, the sewing matrix $B_g(\bk)$ has components given by
\bg
[B_g(\bk)]_{nm}
= \frac{1}{N_{\rm cell}} \, \bra{\Psi_{n, g\bk}} \hat{\eta} \, \Delta_g \, \hat{g} \, \ket{\Psi_{m, \bk}}
= \frac{1}{N_{\rm cell}} \, \bra{\Psi_{n, g\bk}} \bpm \hat \ep & \hat \xi \\ \hat \zeta & \hat \mu \epm \cdot \bpm O_g & \\ & (-1)^{\tau_g} {\rm Det} [O_g] \, O_g \epm \cdot \hat{g} \, \ket{\Psi_{m, \bk}}.
\label{seq:sewing1}
\eg
Here, $n$ and $m$ denote band indices within the set $\mc{B}$.
The matrix $[B_g(\bk)]_{nm}$ is nonzero only if $\om_{n, g\bk} = \om_{m,\bk}$.
Note that the prefactor $N_{\rm cell}$ arises from our normalization condition in Eq.~\eqref{seq:psi_norm}.
It can be straightforwardly checked that for the identity operation $g= {\rm id}$, Eq.~\eqref{seq:sewing1} reduces to $\delta_{nm}$.

Importantly, the sewing matrix respects the group laws.
For symmetry operations $g_1$ and $g_2$, the sewing matrices satisfy
\bg
B_{g_1 g_2}(\bk) = B_{g_1}(g_2 \bk) \, B_{g_2} (\bk).
\label{seq:sewing_multi}
\eg
This relation can be shown by carefully defining the completeness relation with respect to the inner product weighted by $\eta$.
For the ket states $\ket{\Psi_{n, \bk}}$, the completeness relation reads $\frac{1}{N_{\rm cell}} \, \sum_{p,\bb q \in {\rm all}} \, \ket{\Psi_{p, \bb q}} \bra{\Psi_{p, \bb q}} \hat{\eta}$, and for the bra states $\bra{\Psi_{n,\bk}}$, it takes the form $\frac{1}{N_{\rm cell}} \, \sum_{p,\bb q \in {\rm all}} \, \hat{\eta} \ket{\Psi_{p, \bb q}} \bra{\Psi_{p, \bb q}}$.
Let us also comment on our notation.
The sewing matrix $B_g(\bk)$ is a unitary matrix when $g$ is unitary, and a unitary matrix combined with the complex conjugation operator $\mc{K}$ when $g$ is antiunitary.
If we denote $\hat{g} \ket{\Psi_{m,\bk}} = \ket{\hat{g} \Psi_{m,\bk}} \mc{K}^{\tau_g}$, then Eq.~\eqref{seq:sewing1} becomes
\bg
[B_g(\bk)]_{nm} = \frac{1}{N_{\rm cell}} \, \bra{\Psi_{n, g\bk}} \hat{\eta} \, \Delta_g \ket{\hat{g} \Psi_{m, \bk}} \, \mc{K}^{\tau_g}.
\label{seq:sewing_with_K}
\eg

In terms of the cell-periodic part of the eigenfields, Eq.~\eqref{seq:sewing1} can be expressed as
\bg
[B_g(\bk)]_{nm} = \bra{\bb u_{n, g\bk}} \hat{\eta} \, \Delta_g \, e^{-i g\bk \cdot \bb v_g} \, \hat{g} \ket{\bb u_{m, \bk}}_{\rm uc}.
\label{seq:sewing2}
\eg
By inserting the completeness relation for the position basis, the sewing matrix can also be written as
\ba
[B_g(\bk)]_{nm}
=& e^{-i g\bk \cdot \bb v_g} \, \int_{\rm uc} d \bb r \, \brk{\bb u_{n, g \bk}}{\bb r} \cdot \eta(\bb r) \cdot \Delta_g \cdot \bra{\bb r} \hat{g} \ket{\bb u_{m, \bk}}
\nn
=& e^{-i g\bk \cdot \bb v_g} \, \int_{\rm uc} d \bb r \, \bb u_{n, g \bk}(\bb r)^\dg \cdot \eta(\bb r) \cdot \Delta_g \cdot \mc{K}^{\tau_g} \, \bb u_{m, \bk}(g^{-1} \circ \bb r).
\ea
For each case depending on whether $g$ is unitary or antiunitary, the sewing matrix is given by
\bg
[B_g(\bk)]_{nm} = \begin{cases}
e^{-i (O_g \cdot \bk) \cdot \bb v_g} \, \int_{\rm uc} d \bb r \, \bb u_{n, O_g \cdot \bk}(\bb r)^\dg \cdot \eta(\bb r) \cdot \Delta_g \cdot \bb u_{m, \bk}(g^{-1} \circ \bb r) & \text{if $g$ is unitary},
\\
e^{i (O_g \cdot \bk) \cdot \bb v_g} \, \int_{\rm uc} d \bb r \, \bb u_{n, - O_g \cdot \bk}(\bb r)^\dg \cdot \eta(\bb r) \cdot \Delta_g \cdot [\bb u_{m, \bk}(g^{-1} \circ \bb r)]^* \, \mc{K} & \text{if $g$ is antiunitary}.
\end{cases}
\eg

In general, the sewing matrix is not gauge invariant.
A gauge transformation, which includes relabeling of the band indices and phases of the eigenfields, can be written as $\ket{\Psi_{n, \bk}} \to \sum_{m \in \mc{B}} \, \ket{\Psi_{m, \bk}} \, [\mc{G}(\bk)]_{mn}$, and similarly for $\ket{\bb u_{n, \bk}}$.
Under such a transformation, the sewing matrix transforms as $B_g(\bk) \to \mc{G}(g\bk)^{-1} \, B_g(\bk) \, \mc{G}(\bk)$, or explicitly,
\bg
[B_g(\bk)]_{nm} \to \sum_{n',m' \in \mc{B}} \, [\mc{G}(g\bk)^{-1}]_{n n'} \, [B_g(\bk)]_{n' m'} \, [\mc{G}(\bk)]_{m' m}.
\label{seq:sewing_gauge}
\eg

At high-symmetry momenta (HSM) or maximal $k$-vectors $\bk_*$, the eigenvalues and the trace of the sewing matrix $B_g(\bk_*)$ correspond to the eigenvalues and the group-theoretical character of the bands with respect to the unitary operation $g$.
To compute the sewing matrix in a gauge invariant way, we adopt the periodic gauge:
\bg
\ket{\Psi_{n, \bk + \bb G}} = \ket{\Psi_{n, \bk}},
\quad
\ket{\bb u_{n, \bk + \bb G}} = e^{- i \bb G \cdot \hat{\bb r}} \ket{\bb u_{n, \bk}},
\label{seq:periodic_gauge_phc}
\eg
for any reciprocal lattice vector $\bb G$.

For definiteness, let us suppose that $g\bk_* = (-1)^{\tau_g} \, O_g \cdot \bk_* = \bk_* + \bb G_g (\bk_*)$ for some reciprocal lattice vector $\bb G_g (\bk_*)$.
This means that $\bk_*$ is left invariant by $g$, up to a reciprocal lattice vector $\bb G_g (\bk_*)$.
Thus, at $\bk_*$, the sewing matrix takes the form
\ba
[B_g(\bk_*)]_{nm}
=& \bra{\bb u_{n, \bk_* + \bb G_g (\bk_*)}} \hat{\eta} \, \Delta_g \, e^{-i g\bk_* \cdot \bb v_g} \, \hat{g} \ket{\bb u_{m, \bk_*}}_{\rm uc}
\nn
=& \bra{\bb u_{n, \bk_*}} e^{i \bb G_g(\bk_*) \cdot \hat{\bb r}} \, \hat{\eta} \, \Delta_g \, e^{-i g\bk_* \cdot \bb v_g} \, \hat{g} \ket{\bb u_{m, \bk_*}}_{\rm uc}
\nn
=& e^{-i g\bk_* \cdot \bb v_g} \, \int_{\rm uc} d \bb r \, e^{i \bb G_g(\bk_*) \cdot \bb r} \, \bb u_{n, \bk_*}(\bb r)^\dg \cdot \eta(\bb r) \cdot \Delta_g \cdot \mc{K}^{\tau_g} \, \bb u_{m, \bk_*}(g^{-1} \circ \bb r).
\ea
Hence, the sewing matrix at $\bk_*$ is determined entirely from the eigenfields at $\bk_*$.

There are two advantages to adopt the periodic gauge.
First, the sewing matrix becomes periodic in the Brillouin zone (BZ):
\bg
B_g(\bk + \bb G) = B_g (\bk),
\label{seq:sewing_peri}
\eg
Second, at HSM $\bk_*$, the sewing matrix transforms under a gauge transform as
\bg
B_g (\bk_*) \to \mc{G} (\bk_*)^{-1} \, B_g (\bk_*) \, \mc{G} (\bk_*).
\eg
This implies that, if $g$ is unitary, the gauge transformation of $B_g (\bk_*)$ is simply a unitary similarity transformation.
Consequently, the eigenvalues and the trace of $B_g(\bk_*)$ are gauge invariant.
Specifically, for a unitary $g$, the sewing matrix is given by a unitary matrix whose components are
\bg
[B_g(\bk_*)]_{nm} = e^{-i (O_g \cdot \bk_*) \cdot \bb v_g} \, \int_{\rm uc} d \bb r \, e^{i (O_g \cdot \bk_* - \bk_*) \cdot \bb r} \, \bb u_{n, \bk_*}(\bb r)^\dg \cdot \eta(\bb r) \cdot \Delta_g \cdot \bb u_{m, \bk_*}(g^{-1} \circ \bb r).
\label{seq:sewing_uni_HSM}
\eg
Computing the sewing matrices for unitary operations at all HSM allows one to determine the multiplicities of momentum-space or little-group irreps, which defines the symmetry-data vector.
The formula for the sewing matrix for unitary operations in Eq.~\eqref{seq:sewing_uni_HSM} is well-known in the literature, for instance, as discussed in Ref.~\cite{sakoda2005optical} and the MPB Documentation~\cite{johnson2001block-iterative}, specifically in the absence of magneto-electric coupling, i.e. $\xi(\bb r) = 0$ and $\zeta(\bb r)=0$.
However, it should be noted that our derivation in Eq.~\eqref{seq:sewing2} generalizes to antiunitary operations and to arbitrary momenta $\bk$, as well as to cases where magneto-electric coupling is nonzero.
%%%%%%

%%%%%%
\subsection{Symmetry-data vectors of electric and magnetic fields}
\label{sec:sn_sym_vec}
%%%%%%
When magneto-electric coupling is absent, the symmetry-data vector can be defined separately for electric and magnetic fields.
For general SGs, little-group irreps may be assigned differently to the $\bb E$ and $\bb H$ fields, owing to their vectorial and axial-vectorial natures, respectively.
In particular, under orientation-reversing symmetries such as mirror or inversion, the $\bb H$ field transforms with an additional minus sign compared to the $\bb E$ field.
Practically, this means that the conversion rule between the symmetry-data vectors of $\bb E$ and $\bb H$ fields is as follows: vector, axial-vector, scalar, and pseudo-scalar representations of $\bb E$ fields correspond to axial-vector, vector, pseudo-scalar, and scalar representations of $\bb H$ fields, respectively, and vice versa~\cite{morales2025transversality}.
In fact, the precise conversion rule depends on how the symmetry character or the sewing matrix is defined.
Here, we present a rigorous derivation of how this conversion arises from Maxwell's equations and the definition of the sewing matrix.

In the absence of magneto-electric coupling, we can apply Eq.~\eqref{seq:maxwell_matrix} twice to see that the $\bb E$ and $\bb H$ eigenfields satisfy
\bg
\mc{H}_E (\bb r) \, \bb E_{n, \bk}(\bb r) \equiv \nabla \times [\mu(\bb r)^{-1} \cdot \nabla \times \bb E_{n, \bk} (\bb r)] = \om_{n,\bk}^2 \, \ep(\bb r) \, \bb E_{n, \bk} (\bb r),
\nn
\mc{H}_H (\bb r) \, \bb H_{n, \bk}(\bb r) \equiv \nabla \times [\ep(\bb r)^{-1} \cdot \nabla \times \bb H_{n, \bk} (\bb r)] = \om_{n,\bk}^2 \, \mu(\bb r) \, \bb H_{n, \bk} (\bb r).
\label{seq:maxwell_eh}
\eg
Since $\mc{H}_{E,H}$ are self-adjoint, we impose the following normalization conditions for the $\bb E$ and $\bb H$ fields with suitable inner products:
\bg
\bra{\bb E_{n, \bk}} \hat{\ep} \ket{\bb E_{m, \bk'}} = N_{\rm cell} \, \delta_{nm} \delta_{\bk \bk'},
\quad
\bra{\bb H_{n, \bk}} \hat{\mu} \ket{\bb H_{m, \bk'}} = N_{\rm cell} \, \delta_{nm} \delta_{\bk \bk'}.
\eg
Note that these conditions are not independent due to the Maxwell's equations: fixing the orthonormality condition (and any prefactors like $N_{\rm cell}$) for the $\bb H$ fields automatically determines it for the $\bb E$ fields, and vice versa.
In lossless media, for eigenfields with nonzero frequencies, the normalization of $\bb E$ fields follows from that of the $\bb H$ fields as follows:
\ba
\bra{\bb E_{n, \bk}} \hat{\ep} \ket{\bb E_{m, \bk'}}
=& \bra{\bb E_{n, \bk}} \hat{\ep}^\dg \ket{\bb E_{m, \bk'}}
= \int_{\rm all} d \bb r \, \bb D_{n, \bk} (\bb r)^\dg \cdot \bb E_{m, \bk'} (\bb r)
\nn
=& \int_{\rm all} d \bb r \, \left[ \frac{i}{\om_{n,\bk}} \nabla \times \bb H_{n, \bk} (\bb r) \right]^\dg \cdot \bb E_{m, \bk'}(\bb r)
= \frac{1}{i \om_{n,\bk}} \int_{\rm all} d \bb r \, \bb H_{n, \bk} (\bb r)^\dg \cdot [\nabla \times \bb E_{m, \bk'}(\bb r)]
\nn
=& \frac{\om_{m,\bk'}}{\om_{n,\bk}} \, \int_{\rm all} d \bb r \, \bb H_{n, \bk} (\bb r)^\dg \cdot \bb B_{m, \bk'}(\bb r)
= \frac{\om_{m,\bk'}}{\om_{n,\bk}} \, \bra{\bb H_{n,\bk}} \hat{\mu} \ket{\bb H_{m, \bk'}}
\nn
=& N_{\rm cell} \, \delta_{nm} \delta_{\bk \bk'}.
\label{seq:norm_convert}
\ea
Note that Maxwell’s equations are used in the second line, and the orthonormality condition for the $\bb H$ fields is applied in the last equality.

Now, let us define sewing matrices for $\bb E$ and $\bb H$ fields separately, for a symmetry operation $g$.
Based on the general definition of the sewing matrix in Eq.~\eqref{seq:sewing1}, we define
\ba
[B^{(E)}_g(\bk)]_{nm} =& \frac{1}{N_{\rm cell}} \, \bra{\bb E_{n, g\bk}} \hat{\ep} \cdot O_g \cdot \hat{g} \ket{\bb E_{m, \bk}},
\nn
[B^{(H)}_g(\bk)]_{nm} =& \frac{1}{N_{\rm cell}} \, \bra{\bb H_{n, g\bk}} \hat{\mu} \cdot \left[ (-1)^{\tau_g} {\rm Det}[O_g] \, O_g \right] \cdot \hat{g} \ket{\bb H_{m, \bk}}.
\label{seq:sewing_EH1}
\ea
Note that the sewing matrices for $\bb E$ and $\bb H$ fields are defined differently.
For the $\bb H$ fields, the definition of the sewing matrix already encodes the fact that $\bb E$ and $\bb H$ fields transforms differently under time-reversal and orientation-reversing symmetries.
Using a similar technique as in Eq.~\eqref{seq:norm_convert}, we can express $B^{(E)}_g(\bk)$ in terms of $B^{(H)}_g(\bk)$.
We have
\ba
[B^{(E)}_g(\bk)]_{nm}
=& \frac{1}{N_{\rm cell}} \, \int_{\rm all} d \bb r \, \bb E_{n, g\bk} (\bb r)^\dg \cdot \ep(\bb r) \cdot O_g \cdot \mc{K}^{\tau_g} \, \bb E_{m, \bk}(g^{-1} \circ \bb r)
\nn
=& \frac{1}{N_{\rm cell}} \, (-1)^{\tau_g} \, {\rm Det}[O_g] \, \frac{\om_{m,\bk}}{\om_{n,g\bk}} \, \int_{\rm all} d \bb r \, \bb H_{n, g\bk} (\bb r)^\dg \cdot O_g \cdot \mc{K}^{\tau_g} \, \bb B_{m, \bk}(g^{-1} \circ \bb r)
\nn
=& \frac{1}{N_{\rm cell}} \, (-1)^{\tau_g} \, {\rm Det}[O_g] \, \frac{\om_{m,\bk}}{\om_{n,g\bk}} \, \bra{\bb H_{n, g\bk}} O_g \hat g \ket{\bb B_{m, \bk}}.
\ea
The inner product in the last line can be further simplified by using the symmetry constraint on $\mu(\bb r)$, namely $\hat{g} \hat \mu \hat{g}^{-1} = O_g^{-1} \, \hat \mu \, O_g$, shown earlier in Eq.~\eqref{seq:eta_sym}.
Thus, we have
\bg
\bra{\bb H_{n, g\bk}} O_g \, \hat g \ket{\bb B_{m, \bk}}
= \bra{\bb H_{n, g\bk}} O_g \cdot \hat g \hat \mu \ket{\bb H_{m, \bk}}
= \bra{\bb H_{n, g\bk}} \hat \mu \cdot O_g \cdot \hat g \ket{\bb H_{m, \bk}}.
\eg
Hence, the sewing matrix for the $\bb E$ fields becomes
\ba
[B^{(E)}_g(\bk)]_{nm}
=& \frac{1}{N_{\rm cell}} \, (-1)^{\tau_g} \, {\rm Det}[O_g] \, \frac{\om_{m,\bk}}{\om_{n,g\bk}} \, \bra{\bb H_{n, g\bk}} \hat \mu \cdot O_g \hat g \ket{\bb H_{m, \bk}}
\nn
=& \frac{\om_{m,\bk}}{\om_{n,g\bk}} \, [B^{(H)}_g(\bk)]_{nm} = [B^{(H)}_g(\bk)]_{nm}.
\ea
In obtaining the last equality, we used the fact that $[B^{(H)}_g(\bk)]_{nm}=0$ unless $\om_{n, g\bk} = \om_{m, \bk}$.
In this case, if the little-group (irreducible) representations are assigned according to the trace or eigenvalues of the sewing matrices, the symmetry-data vectors for $\bb E$ and $\bb H$ fields coincide.
However, this can be confusing if, for example, the $\bb H$ fields transform as axial-vectors but are assigned to the vectorial representation, since ${\rm Det}[O_g]$ factor is already present in the sewing matrix definition.

To avoid such ambiguity, we can alternatively define the sewing matrices in the same way for both $\bb E$ and $\bb H$ fields as follows:
\bg
[\td{B}^{(E)}_g(\bk)]_{nm} = \frac{1}{N_{\rm cell}} \, \bra{\bb E_{n, g\bk}} \hat{\ep} \cdot O_g \cdot \hat{g} \ket{\bb E_{m, \bk}},
\quad
[\td{B}^{(H)}_g(\bk)]_{nm} = \frac{1}{N_{\rm cell}} \, \bra{\bb H_{n, g\bk}} \hat{\mu} \cdot O_g \cdot \hat{g} \ket{\bb H_{m, \bk}},
\label{seq:sewing_EH2}
\eg
Under this definition, they are related by $\td{B}^{(E)}_g(\bk) = (-1)^{\tau_g} \, {\rm Det}[O_g] \, \td{B}^{(H)}_g(\bk)$, thereby recovering the conventional conversion rules: vector, axial-vector, scalar, and pseudo-scalar representations of $\bb E$ fields correspond respectively to axial-vector, vector, pseudo-scalar, and scalar representations of $\bb H$ fields, and vice versa.

Both conventions in Eqs.~\eqref{seq:sewing_EH1} and \eqref{seq:sewing_EH2} for symmetry operations can be used in principle.
However, it is preferable to adopt a single convention consistently within a given work or analysis.
In the case of Eq.~\eqref{seq:sewing_EH1}, the representation $\girrep$ for transverse modes at the polarization singularity is assigned based on the difference between vectorial and trivial representations for both $\bb E$ and $\bb H$ fields, following Ref.~\cite{christensen2022location}.
On the other hand, using Eq.~\eqref{seq:sewing_EH2} again assigns $\girrep$ for the $\bb E$ fields based on the difference between vectorial and trivial representations, while for the $\bb H$ fields the representation is instead assigned based on the difference between {\it axial-vector} and {\it pseudoscalar} representations.
We note that for SG $P4_332$ (No. 212), discussed in the main text, and SG $P432$ (No. 207), which will be addressed in SN~\ref{sec:sn_sg207}, there are no orientation-reversing symmetries, and thus ${\rm Det}[O_g]$ is always trivial.
Furthermore, to identify the little-group irreps or symmetry-data vectors, the symmetry characters for unitary symmetries alone are computed, for which $(-1)^{\tau_g}=1$.
Thus for the examples considered in this work the two approaches Eqs.~\eqref{seq:sewing_EH1} and \eqref{seq:sewing_EH2} yield identical irrep assignments and symmetry-data vectors.
%%%%%%

%%%%%%
\subsection{Wilson lines and loops}
\label{sec:sn_sym_wilson}
%%%%%%
The Wilson line is defined for a path in momentum space with a basepoint (starting point) $\bk_b$ and an endpoint $\bk_f$.
When the endpoint differs from the basepoint by a reciprocal lattice vector $\bb G$, i.e. $\bk_f = \bk_b + \bb G$, the corresponding Wilson line is referred to as a Wilson loop~\cite{yu2011equivalent}.
For a given path $\mc{L}$ in momentum space, we discretize it by defining a series of points $\bk_\ell$ along the path, where $\ell=1, \dots, N$.
Note that $\bk_b = \bk_1$ and $\bk_f = \bk_N$.

To construct Wilson line or loop, we first define the overlap matrix,
\bg
[S_{\bk, \bk'}]_{nm} = \bra{\bb u_{n, \bk}} \hat \eta \ket{\bb u_{m, \bk'}}_{\rm uc}.
\label{seq:overlap_phc}
\eg
This definition applies to the generic case where the electric and magnetic fields are not decoupled, due to nonvanishing magneto-electric couplings.
The Wilson line or loop associated with a path $\mc{L}$ in momentum space is then defined as
\bg
(W_\mc{L})_{nm}
= (S_{\bk_N, \bk_{N-1}} \cdot S_{\bk_{N-1},\bk_{N-2}} \cdots S_{\bk_3,\bk_2} \cdot S_{\bk_2,\bk_1})_{nm},
\label{seq:wilson_def_discrete}
\eg
where $n$ and $m$ index the set $\mc{B}$ of eigenfields of interest.
In general, $W_\mc{L}$ is not gauge invariant.
Under a gauge transformation in which $\ket{\bb u_{n,\bk}}$ transforms as $\ket{\bb u_{n,\bk}} \to \sum_{m \in \mc{B}} \, \ket{\bb u_{m, \bk}} [\mc{G}(\bk)]_{mn}$, the Wilson line $W_\mc{L}$ transforms as
\bg
W_\mc{L} \to [\mc{G}(\bk_f)]^{-1} \cdot W_\mc{L} \cdot \mc{G}(\bk_b).
\eg
In the periodic gauge [Eq.~\eqref{seq:periodic_gauge_phc}], $\mc{G}(\bk)$ is periodic in the BZ.
This periodicity implies that, for Wilson loops where $\bk_f = \bk_b$ up to a reciprocal lattice vector, the eigenvalues of the Wilson loop are gauge invariant.
The Wilson loop spectrum is obtained by plotting the phases of the eigenvalues of $W_\mc{L}$, i.e. ${\rm Spec} [-i \log W_\mc{L}]$, for different choices of $\mc{L}$, ensuring that the phases of Wilson loop eigenvalues vary smoothly.

Moreover, the Wilson loop or line can also be defined without discretizing the path.
In the limit $N \to \infty$, Eq.~\eqref{seq:wilson_def_discrete} becomes
\bg
W_\mc{L} = \mc{P} \exp \left[ i \int_{\mc{L}} \, d \bk \cdot \bb A(\bk) \right].
\label{seq:wilson_def_cont}
\eg
Here, $\mc{P}$ is the path ordering symbol, and $\bb A(\bk)$ is the non-Abelian Berry connection.
The Berry connection is defined through a suitable inner product~\cite{devescovi2024tutorial}:
\bg
[\bb A(\bk)]_{nm} = i \bra{\bb u_{n, \bk}} \hat \eta \ket{ \nabla_\bk \bb u_{m, \bk}}_{\rm uc},
\eg
where $\ket{ \nabla_\bk \bb u_{m, \bk}} = \nabla_\bk \ket{\bb u_{m, \bk}}$ denotes the gradient of $\ket{\bb u_{m, \bk}}$ with respect to $\bk$.

The symmetry transformation of the Berry connection and Wilson line can be derived as in topological band theory for electronic systems (see, for example, Refs.~\cite{alexandradinata2014wilson,alexandradinata2016topological,benalcazar2017electric,wieder2018wallpaper,hwang2019fragile}).
For a generic symmetry operation $g = \{O_g | \bb v_g\}$, the Berry connection transforms according to\footnote{One can also derive the symmetry constraint for the Berry curvature.
The Berry curvature can be computed from the Berry connection as $[F_{ij}(\bk)]_{nm} = \der_{k_i} [A_j (\bk)]_{nm} - \der_{k_j} [A_i (\bk)]_{nm} - i \left[ A_i (\bk) \cdot A_j (\bk) - A_j (\bk) \cdot A_i (\bk) \right]_{nm}$, where $i,j=1,2,3$ and $[A_i (\bk)]_{nm}$ denotes the $i$-th component of $[\bb A(\bk)]_{nm}$.
Then, the symmetry constraint is given by $\sum_{i',j'} \, (O_g^{-1})_{i i'} (O_g^{-1})_{j j'} [F_{i' j'}(g\bk)]_{nm} = (-1)^{\tau_g} \, \left[ B_g(\bk) \cdot F_{ij}(\bk) \cdot B_g(\bk)^{-1} \right]_{nm}$.}
\bg
O_g^{-1} \cdot [\bb A(g\bk)]_{nm}
= \left[ B_g(\bk) \cdot \bb A(\bk) \cdot B_g(\bk)^{-1} \right]_{nm} + i (-1)^{\tau_g} \left[ B_g(\bk) \nabla_\bk B_g(\bk)^{-1} \right]_{nm} + ( O_g^{-1} \cdot \bb v_g) \, \delta_{nm}.
\eg
By combining this with Eq.~\eqref{seq:wilson_def_cont}, one can derive the symmetry transformation rule for Wilson lines.
Consider a Wilson line $W_\mc{L}$ defined along a path $\mc{L}$ starting at $\bk_b$ and ending at $\bk_f$.
A symmetry operation $g$ maps the path $\mc{L}$ to a new path $g \mc{L}$, whose basepoint and endpoint are $g \bk_b = (-1)^{\tau_g} \, O_g \cdot \bk_b$ and $g \bk_f = (-1)^{\tau_g} \, O_g \cdot \bk_f$, respectively.
Under this transformation, the Wilson line $W_\mc{L}$ and its transformed counterpart $W_{g \mc{L}}$ are related by the sewing matrices evaluated at the basepoint and endpoint of $\mc{L}$:
\ba
W_{g \mc{L}}
=& e^{i (-1)^{\tau_g}\int_\mc{L} \, d \bk \cdot [O_g]^{-1} \cdot \bb v_g} \, B_g(\bk_f) \cdot W_\mc{L} \cdot B_g(\bk_b)^{-1}
\nn
=& e^{i (-1)^{\tau_g} (\bk_f - \bk_b) \cdot [O_g]^{-1} \cdot \bb v_g} \, B_g(\bk_f) \cdot W_\mc{L} \cdot B_g(\bk_b)^{-1},
\label{seq:wilson_symmetry}
\ea
For Wilson loops, where the basepoint and endpoint differ only by a reciprocal lattice vector, the sewing matrices at $\bk_b$ and $\bk_f$ are equal under the periodic gauge, i.e. $B_g(\bk_b) = B_g(\bk_f)$ for any unitary or antiunitary symmetry $g$.
Note that Eq.~\eqref{seq:wilson_symmetry} can also be derived in discrete formalism by considering the symmetry transformation of the overlap matrix $S_{\bk, \bk'}$:
\bg
S_{g(\bk+\delta \bk), g\bk} = B_g (\bk + \delta \bk) \, [e^{i (-1)^{\tau_g} (g \delta \bk) \cdot \bb v_g} \, S_{\bk+\delta \bk, \bk}] \, B_g (\bk)^{-1}
\nn
= e^{i (-1)^{\tau_g} (O_g^{-1} \cdot \bb v_g) \cdot \delta \bk} \, B_g (\bk + \delta \bk) \cdot S_{\bk+\delta \bk, \bk} \cdot B_g (\bk)^{-1}.
\eg

When there is no magneto-electric coupling, the Berry connection and Wilson line and loop can be defined separately for $\bb E$ and $\bb H$ fields, using suitable inner products as discussed in SN~\ref{sec:sn_sym_vec}.
For example, the Berry connection for the $\bb E$ fields is defined as $[\bb A^{(E)}(\bk)]_{nm} = i \bra{\bb e_{n, \bk}} \hat \ep \ket{\nabla_\bk \bb e_{m, \bk}}_{\rm uc}$, where $\ket{\bb e_{n, \bk}} = e^{- i \bk \cdot \hat{\bb r}} \ket{\bb E_{n, \bk}}$ is the periodic part of the $\bb E$ fields.
In our ab-initio models of photonic crystals, we assume vacuum permeability $\mu(\bb r) = \mu_0$ and no magneto-electric coupling.
Hence, the inner product between $\bb H$ fields simplifies to the standard inner product up to a constant factor $\mu_0$.
Accordingly, we compute Wilson loops in our ab-initio models using the $\bb H$ fields.
One may wonder whether the Wilson loop spectrum for the $\bb E$ fields could differ quantitatively from that for the $\bb H$ fields.
We will discuss this issue in detail in SN~\ref{sec:sn_sg212_wl_cylinder} and \ref{sec:sn_sg212_wl_sphere}.
%%%%%%

%%%%%%
\subsection{Comparison between tight-binding models and photonic crystals}
\label{sec:sn_sym_compare}
%%%%%%
So far, we have presented a general symmetry analysis for photonic crystals
Here, we note that most of the results we derived, such as symmetry constraints on Wilson lines and loops and the definition of sewing matrices, can be translated to the TB theory, as long as one adjusts the definition of the inner product accordingly.

The TB Hamiltonian $H_{TB}(\bk)$ is defined in terms of real-space basis orbitals located at positions $\bx_\alpha$ ($\alpha=1,\dots,n_{\rm tot}$) within the unit cell.
Then, $H_{TB}(\bk)$ satisfies $H_{TB}(\bk + \bb G) = V(-\bb G) H_{TB}(\bk) V(\bb G)$, where $[V(\bb G)]_{\alpha \beta} = e^{i \bb G \cdot \bb x_\alpha} \delta_{\alpha \beta}$.
We denote the $n$-th energy eigenstate as $\ket{u_{n \bk}}$, which is a vector of length $n_{\rm tot}$.
Then, the periodic gauge condition for the TB models is defined as
\bg
\ket{u_{n \bk + \bb G}} = V(-\bb G) \ket{u_{n \bk}}.
\label{seq:periodic_gauge_TB}
\eg
The Berry connection, as well as Wilson lines and loops, is then defined using the standard inner product.
For example, the overlap matrix and the Berry connection are given by
\ba
[S_{\bk, \bk'}]_{nm}
=& \brk{u_{n, \bk}}{u_{m, \bk'}}
= \sum_{\alpha=1}^{n_{\rm tot}} \, (\ket{u_{n, \bk}}_\alpha)^* \ket{u_{m, \bk'}}_\alpha,
\nn
[\bb A (\bk)]_{nm}
=& i \brk{u_{n, \bk}}{\nabla_\bk u_{m, \bk}}
= i \sum_{\alpha=1}^{n_{\rm tot}} \, (\ket{u_{n, \bk}}_\alpha)^* (\nabla_\bk \ket{u_{m, \bk}}_\alpha).
\ea

As we have seen above, the overlap matrix and the Berry connection, and thus other quantities such as the Wilson line, are defined in the TB model using the standard inner product.
Because similar definitions apply to photonic crystals, with suitably modified inner products, the mathematical form of symmetry transformations, such as Eq.~\eqref{seq:wilson_symmetry}, remains the same in both frameworks.
With this straightforward correspondence between photonic crystals and TB models, we will perform symmetry analyses on Wilson loops in the later sections without explicitly differentiating between TB theory and photonic crystals, in most cases.
However, whenever our analysis is specific to photonic crystals, we will state it clearly.
%%%%%%

%%%%%%
\subsection{Long-wavelength limit and homogenization}
\label{sec:sn_sym_longwave}
%%%%%%
In this section, we review the long-wavelength limit description, also known as the homogenization, of photonic crystals.
Here, we consider photonic crystals without magneto-electric coupling and further assume that $\mu (\bb r) = \mu_0$.
We first review the homogenization of the $\bb H$ and $\bb D$ fields following Ref.~\cite{krokhin2002long}, and then discuss the $\bb E$ field, following Ref.~\cite{datta1993effective}.

First, let us consider the homogenization of the $\bb H$ field as described in Ref.~\cite{krokhin2002long}.
The $\bb H$ field satisfies the eigenvalue equation [a special case of Eq.~\eqref{seq:maxwell_eh} under the condition $\mu (\bb r) = \mu_0$] and the divergence-free condition:
\bg
\nabla \times [\ep(\bb r)^{-1} \cdot \nabla \times \bb H_\bk (\bb r)] = \mu_0 \, \om_\bk^2\, \bb H_\bk (\bb r),
\quad
\nabla \cdot \mu_0 \bb H_\bk (\bb r) = 0.
\label{seq:hfield_condition1}
\eg
To analyze the constraints placed on $\bb H$ in the long-wavelength limit, we define the Fourier transforms of the cell-periodic parts of $\bb H_\bk (\bb r)$ and $\ep (\bb r)^{-1}$ (both of which are cell-periodic functions) as follows:
\bg
\bb h_\bk (\bb r):= e^{-i \bk \cdot \bb r} \, \bb H_\bk (\bb r)= \sum_{\bb G} \, \bb H_{\bk, \bb G} \, e^{i \bb G \cdot \bb r},
\quad
\ep (\bb r)^{-1} = \sum_{\bb G} \, \td{\ep}_{\bb G} \, e^{i \bb G \cdot \bb r}.
\label{seq:hfield_fourier}
\eg
Inserting Eq.~\eqref{seq:hfield_fourier} into Eq.~\eqref{seq:hfield_condition1}, we obtain the following set of equations:
\bg
\mu_0 \, \om_\bk^2 \, \bb H_{\bk, \bb G} = - \sum_{\bb G'} \, (\bk + \bb G) \times \left[ \td{\ep}_{\bb G - \bb G'} \cdot (\bk + \bb G') \times \bb H_{\bk, \bb G'} \right],
\quad
(\bk + \bb G) \cdot \bb H_{\bk, \bb G} = 0.
\label{seq:hfield_condition2}
\eg

Let us first analyze the case where $\bb G = \bb 0$.
In this case, Eq.~\eqref{seq:hfield_condition2} reduces to
\bg
\mu_0 \, \om_\bk^2 \, \bb H_{\bk, \bb 0} = - \bk \times \left[ \td{\ep}_{\bb 0} \cdot \bk \times \bb H_{\bk, \bb 0} \right] - \sum_{\bb G' \ne \bb 0} \, \bk \times \left[ \td{\ep}_{-\bb G'} \cdot (\bk + \bb G') \times \bb H_{\bk, \bb G'} \right].
\eg
In the long-wavelength limit, where $|\bk| l \ll 1$ and $l$ is a characteristic system length scale such as the linear system size, we consider the lowest (acoustic) modes whose dispersion is linear in $\bk$, i.e. $\om_\bk \simeq v(\hat{\bk}) |\bk|$, with a possibly direction-dependent velocity $v(\hat{\bk})$.
In this limit, $\bk + \bb G'$ can be approximated as $\bb G'$ for $\bb G' \ne \bb 0$.
Therefore, the above equation becomes
\bg
\mu_0 v(\hat{\bk})^2 \, |\bk|^2 \, \bb H_{\bk, \bb 0} + \bk \times \left[ \td{\ep}_{\bb 0} \cdot \bk \times \bb H_{\bk, \bb 0} \right] \simeq - \sum_{\bb G' \ne \bb 0} \, \bk \times \left[ \td{\ep}_{-\bb G'} \cdot \bb G' \times \bb H_{\bk, \bb G'} \right].
\eg
Importantly, the left-hand side depends only on the $\bb H_{\bk, \bb 0}$ Fourier component, while the right-hand side involves only $\bb H_{\bk, \bb G' \ne \bb 0}$.
By comparing orders of $\bk$, we conclude that $\bb H_{\bk, \bb G' \ne \bb 0}$ is of higher order in $|\bk|$ compared to $\bb H_{\bk, \bb 0}$, i.e. $\bb H_{\bk, \bb 0} \sim \mc{O}(1)$ while $\bb H_{\bk, \bb G' \ne \bb 0} \sim \mc{O}(|\bk|)$.
Therefore, at leading order, the $\bb H$ field for the lowest frequency modes can be approximated as
\bg
\bb H_\bk (\bb r) \simeq \bb H_{\bk, \bb G = \bb 0} \, e^{i \bk \cdot \bb r}.
\label{seq:hfield_approx}
\eg
Furthermore, the vector $\bb H_{\bk, \bb 0}$ is transverse, satisfying $\bk \cdot \bb H_{\bk, \bb 0} = 0$, which follows from the divergence-free condition.
Hence, the field $\bb H_\bk (\bb r)$ is a approximately transverse plane wave:
\bg
\bk \cdot \bb H_\bk (\bb r) = 0.
\label{seq:hfield_transverse}
\eg
This condition is referred to in the main text as the emergent transversality.

It is also straightforward to verify that the $\bb D$ field is likewise a transverse plane wave in the long-wavelength limit.
From Maxwell's equation, we know that $\nabla \times \bb H_\bk (\bb r) = -i \om_\bk \bb D_\bk (\bb r)$.
Inserting Eq.~\eqref{seq:hfield_approx}, we obtain the displacement field $\bb D_\bk (\bb r)$ in the long-wavelength limit:
\bg
\bb D_\bk (\bb r) \simeq \frac{1}{v(\hat{\bk})} \, \left( \bb H_{\bk, \bb 0} \times \hat{\bk} \right) e^{i \bk \cdot \bb r},
\label{seq:dfield_approx}
\eg
which shows that $\bb D_\bk (\bb r)$ is a plane wave.
Moreover, it is clear that $\bk \cdot \bb D_\bk (\bb r) = 0$, confirming its transversality.

Now, let us turn to the long-wavelength limit of $\bb E$ field.
From Eq.~\eqref{seq:maxwell_eh}, the $\bb E$ fields satisfy
\bg
\nabla \times [\mu_0^{-1} \nabla \times \bb E_\bk (\bb r)] = \om_\bk^2 \, \ep(\bb r) \, \bb E_\bk (\bb r).
\label{seq:efield_condition1}
\eg
We express $\bb E_\bk (\bb r)$ and $\ep (\bb r)$ using the Fourier transformation as
\bg
\bb E_\bk (\bb r) = \sum_{\bb G} \, \bb E_{\bk, \bb G} \, e^{i (\bk + \bb G) \cdot \bb r},
\quad
\ep (\bb r) = \sum_{\bb G} \, \ep_{\bb G} \, e^{i \bb G \cdot \bb r}.
\label{seq:efield_fourier}
\eg
Substituting Eq.~\eqref{seq:efield_fourier} into Eq.~\eqref{seq:efield_condition1} leads to
\bg
(\bk + \bb G) \times (\bk + \bb G) \times \bb E_{\bk, \bb G} = - \mu_0 \, \om_\bk^2 \, \sum_{\bb G'} \, \ep_{\bb G - \bb G'} \cdot \bb E_{\bk, \bb G'}.
\label{seq:efield_condition2}
\eg

It is convenient to split $\bb E_{\bk, \bb G}$ into longitudinal and transverse components as $\bb E_{\bk, \bb G} = \bb E^{(T)}_{\bk, \bb G} + \bb E^{(L)}_{\bk, \bb G}$.
The longitudinal part $\bb E^{(L)}_{\bk, \bb G}$ is parallel to $\bk + \bb G$, while the transverse part $\bb E^{(T)}_{\bk, \bb G}$ is orthogonal to $\bk + \bb G$.
Following Ref.~\cite{datta1993effective}, for each $\bb E_{\bk, \bb G}$, let us introduce an orthonormal basis $\hat{\bb n}_{1,2,3} (\bb q)$, satisfying $\hat{\bb n}_1 (\bb q) = \hat{\bb n}_2 (\bb q) \times \hat{\bb n}_3 (\bb q)$, $\hat{\bb n}_3 (\bb q) = \bb q/|\bb q| = \hat{\bb q}$, where we denote $\bb q= \bk + \bb G$ for brevity.
Note that $\hat{\bb n}_{1,2} (\bb q)$ can be chosen arbitrarily as long as they are orthogonal to $\bb q$.
Thus, the longitudinal and transverse parts can be expanded as
\bg
\bb E^{(L)}_{\bk, \bb G} = E^{(L)}_{\bk, \bb G} \, \hat{\bb n}_3 (\bb q),
\quad
\bb E^{(T)}_{\bk, \bb G} = \sum_{\alpha=1,2} \, E^{(T,\alpha)}_{\bk, \bb G} \, \hat{\bb n}_\alpha (\bb q).
\eg
Rewriting Eq.~\eqref{seq:efield_condition2} gives
\bg
|\bk + \bb G|^2 \, \sum_{\alpha=1,2} \, E^{(T,\alpha)}_{\bk, \bb G} \, \hat{\bb n}_\alpha (\bb q)
= \mu_0 \, \om_\bk^2 \, \sum_{\bb G'} \, \ep_{\bb G - \bb G'} \cdot \left[ E^{(L)}_{\bk, \bb G'} \, \hat{\bb n}_3 (\bb q') + \sum_{\alpha=1,2} \, E^{(T,\alpha)}_{\bk, \bb G'} \, \hat{\bb n}_\alpha (\bb q') \right],
\label{seq:efield_condition3}
\eg
where $\bb q' = \bk + \bb G'$.
The left-hand side of Eq.~\eqref{seq:efield_condition3} involves only the transverse components.

By multiplying both sides of Eq.~\eqref{seq:efield_condition3} by $\hat{\bb n}_3 (\bb q)$, we find the following relation between the longitudinal and transverse components:
\bg
\sum_{\bb G'} \, \left[\hat{\bb n}_3 (\bb q) \cdot \ep_{\bb G - \bb G'} \cdot \hat{\bb n}_3 (\bb q') \right] \, E^{(L)}_{\bk, \bb G'} + \sum_{\bb G'} \, \sum_{\alpha=1,2} \, \left[ \hat{\bb n}_3 (\bb q) \cdot \ep_{\bb G - \bb G'} \cdot \hat{\bb n}_\alpha (\bb q') \right] \, E^{(T,\alpha)}_{\bk, \bb G'} = 0.
\label{seq:efield_condition4}
\eg
To express $E^{(L)}_{\bk, \bb G}$ in terms of $E^{(T,\alpha)}_{\bk, \bb G}$, we introduce a matrix notation using reciprocal lattice vectors as indices.
For instance, we write the Fourier components of $\ep$ and its inverse as
\bg
\ep_{\bG - \bG'} = \ep_{\bb q - \bb q'} := (\ep)_{\bb q, \bb q'},
\quad
\sum_{\bb q''} \, (\ep^{-1})_{\bb q, \bb q''} \, (\ep)_{\bb q'', \bb q} = (\ep^{-1} \cdot \ep)_{\bb q, \bb q'} = \delta_{\bb q, \bb q'}.
\eg
Then, Eq.~\eqref{seq:efield_condition4} becomes
\bg
E^{(L)}_{\bk, \bb G} = - \sum_{\alpha=1,2} \, \sum_{\bb q'} \, (\ep_{3,3}^{-1} \cdot \ep_{3,\alpha})_{\bb q, \bb q'} \, E^{(T,\alpha)}_{\bk, \bb G'}
\label{seq:efield_l}
\eg
where we define $(\ep_{\alpha, \alpha'})_{\bb q, \bb q'} = \hat{\bb n}_\alpha (\bb q) \cdot \ep_{\bb G - \bb G'} \cdot \hat{\bb n}_{\alpha'} (\bb q')$ and its inverse for $\alpha, \alpha' = 1,2,3$.
By projecting Eq.~\eqref{seq:efield_condition3} onto $\hat{\bb n}_{1,2} (\bb q)$, one obtains additional constraints.
Combining these with Eq.~\eqref{seq:efield_l}, one arrives at a self-consistent equation for the transverse components, as shown in Ref.~\cite{datta1993effective}:
\bg
|\bk + \bb G|^2 \, E^{(T,\alpha)}_{\bk, \bb G} = \mu_0 \, \om_\bk^2 \, \sum_{\bb G'} \, \sum_{\beta=1,2} \, M^{\alpha \beta}_{\bk + \bb G, \bk + \bb G'} \, E^{(T, \beta)}_{\bk, \bb G'},
\nn
M^{\alpha \beta}_{\bb q = \bk + \bb G, \bb q' = \bk+ \bb G'} = (\ep_{\alpha, \beta} - \ep_{\alpha,3} \cdot \ep_{3,3}^{-1} \cdot \ep_{3,\beta})_{\bb q, \bb q'}.
\label{seq:efield_condition5}
\eg

Now, we compare the relative scaling in $|\bk|$ between $E^{(T,\alpha)}_{\bk, \bb G = \bb 0}$ and $E^{(T,\alpha)}_{\bk, \bb G \ne \bb 0}$, based on Eq.~\eqref{seq:efield_condition5}.
For the lowest modes with $\om_\bk \to v(\hat{\bk}) |\bk|$ in the long-wavelength limit, Eq.~\eqref{seq:efield_condition5} reduces to
\bg
|\bk|^2 E^{(T,\alpha)}_{\bk, \bb 0} = \mu_0 v(\hat{\bk})^2 |\bk|^2 \, \sum_{\beta=1,2} \, M^{\alpha \beta}_{\bk, \bk} \, E^{(T, \beta)}_{\bk, \bb 0} + \mu_0 v(\hat{\bk})^2 |\bk|^2 \, \sum_{\bb G' \neq \bb 0} \sum_{\beta=1,2} \, M^{\alpha \beta}_{\bk, \bk + \bb G'} \, E^{(T, \beta)}_{\bk, \bb G'},
\label{seq:E_G_kdotp1}
\\
|\bb G|^2 E^{(T,\alpha)}_{\bk, \bb G \ne \bb 0} \simeq \mu_0 v(\hat{\bk})^2 |\bk|^2 \, \sum_{\beta=1,2} \, M^{\alpha \beta}_{\bk + \bb G, \bk} \, E^{(T, \beta)}_{\bk, \bb 0} + \mu_0 v(\hat{\bk})^2 |\bk|^2 \, \sum_{\bb G' \neq \bb 0} \sum_{\beta=1,2} \, M^{\alpha \beta}_{\bk + \bb G, \bk + \bb G'} \, E^{(T, \beta)}_{\bk, \bb G'},
\label{seq:E_G_kdotp2}
\eg
When $v (\hat \bk) \sim \mc{O}(1)$ and $M^{\alpha \beta}_{\bk + \bb G, \bk + \bb G'}$ remains bounded in the small $|\bk|$ limit (i.e. $|M^{\alpha \beta}_{\bk + \bb G, \bk + \bb G'}| \le C_M$), Eq.~\eqref{seq:E_G_kdotp2} implies that
\bg
|E^{(T,\alpha)}_{\bk, \bb G \ne \bb 0}| \le \mu_0 v(\hat \bk)^2 C_M \, \frac{|\bk|^2}{|\bb G|^2} \, \sum_{\beta=1,2} \, \left[ |E^{(T,\beta)}_{\bk, \bb 0}| + \sum_{\bb G' \ne \bb 0} \, |E^{(T,\beta)}_{\bk, \bb G'}| \right].
\label{seq:E_G_kdotp3}
\eg
This inequality determines the relative scaling between $E^{(T,\alpha)}_{\bk, \bb G = \bb 0}$ and $E^{(T,\alpha)}_{\bk, \bb G \ne \bb 0}$: when $E^{(T,\alpha)}_{\bk, \bb G = \bb 0} \sim \mc{O}(1)$, the nonzero-$\bb G$ components must vanish at least as fast as $|\bk|^2/|\bb G|^2$, i.e. $E^{(T,\alpha)}_{\bk, \bb G \ne \bb 0} \sim \mc{O}(|\bk|^2)$.
Hence, the transverse component of the field can be approximated as
\bg
\bb E^{(T)}_{\bk} (\bb r)
= \sum_{\bb G} \, \bb E^{(T)}_{\bk, \bb G} \, e^{i (\bk + \bb G) \cdot \bb r}
\simeq \sum_{\alpha=1,2} \, E^{(T,\alpha)}_{\bk, \bG = \bb 0} \, \hat{\bb n}_\alpha (\bk) \, e^{i \bk \cdot \bb r},
\label{seq:efield_approxT}
\eg
and satisfies $\bk \cdot \bb E^{(T)}_{\bk} (\bb r) = 0$ by definition.

For the longitudinal part, however, no such simplification is generally expected.
From Eq.~\eqref{seq:efield_l}, one sees that $E^{(T,\alpha)}_{\bk, \bb 0} \sim \mc{O}(1)$ may contribute significantly to $E^{(L)}_{\bk, \bb G}$ unless $(\ep^{-1}_{3,3} \cdot \ep_{3,\alpha})_{\bk + \bb G, \bk} = 0$, regardless of whether $\bb G = \bb 0$ or not.
Therefore, given that $\bb E^{(T)}_{\bk, \bb G}$ can be comparable to $E^{(T,\alpha)}_{\bk, \bb 0}$, we write the homogenized $\bb E$ field as
\bg
\bb E_\bk (\bb r) \simeq \sum_{\alpha=1,2} \, E^{(T,\alpha)}_{\bk, \bb 0} \, \hat{\bb n}_\alpha (\bk) \, e^{i \bk \cdot \bb r} + \sum_{\bb G} \, E^{(L)}_{\bk, \bb G} \, \hat{\bb n}_3 (\bk + \bb q) \, e^{i (\bk + \bb G) \cdot \bb r}.
\label{seq:efield_approx}
\eg
Only when $E^{(L)}_{\bk, \bb G \ne \bb 0}$ can be neglected at leading order, $\bb E_\bk (\bb r)$ reduce to a plane wave.
Even then, it is not necessarily transverse if $E^{(L)}_{\bk, \bb 0}$ survives at leading order.
Thus, $\nabla \cdot \bb E_\bk (\bb r)$ does not vanish in general, reflecting the existence of bound charges.

Finally, let us make a connection between the Fourier components of the $\bb E$ and $\bb H$ fields.
By inserting the long-wavelength description of $\bb E$ from Eq.~\eqref{seq:efield_approx} into Maxwell's equation, $\nabla \times \bb E_\bk (\bb r) = i \mu_0 \om_\bk \bb H_\bk (\bb r)$, we obtain 
\bg
\bb H_\bk (\bb r) = \frac{1}{\mu_0 v(\hat{\bk})} \left[ - E^{(T,2)}_{\bk, \bb 0} \, \hat{\bb n}_1 (\bk) +E^{(T,1)}_{\bk, \bb 0} \, \hat{\bb n}_2 (\bk) \right] \, e^{i \bk \cdot \bb r},
\label{seq:h_approx_e}
\eg
meaning that $\bb H_{\bk, \bb G = \bb 0} = \frac{1}{\mu_0 v(\hat{\bk})} [ - E^{(T,2)}_{\bk, \bb 0} \, \hat{\bb n}_1 (\bk) +E^{(T,1)}_{\bk, \bb 0} \, \hat{\bb n}_2 (\bk) ]$.
Comparing this with Eq.~\eqref{seq:dfield_approx}, we further find
\bg
\bb D_\bk (\bb r) = \frac{1}{\mu_0 v(\hat{\bk})^2} \left[ E^{(T,1)}_{\bk, \bb 0} \, \hat{\bb n}_1 (\bk) + E^{(T,2)}_{\bk, \bb 0} \, \hat{\bb n}_2 (\bk) \right] \, e^{i \bk \cdot \bb r}.
\label{seq:dfield_approx_e}
\eg
The long-wavelength expressions of $\bb E$, $\bb H$, and $\bb D$ fields in Eqs.~\eqref{seq:efield_approx}-\eqref{seq:dfield_approx_e} will be utilized in the analysis of the spherical Wilson loop presented in SN~\ref{sec:sn_sg212_wl_sphere}.
%%%%%%

%%%%%%
\section{Photonic band structures in SG $P4_332$ (No. 212)}
\label{sec:sn_sg212}
%%%%%%
In the main text, we identified the allowed SRSI values in photonic and electronic band structures of SG $P4_332$ by solving linear inequality constraints and visualizing the resulting solution space.
These solutions can be systematically obtained using the Hilbert basis method, which we outline in detail in this section.
In particular, we explain how the allowed SRSIs are constructed as lattice points in a polyhedron defined by constraints due to the symmetries and polarization singularity, and how the Hilbert basis method provides a finite basis generating this polyhedron.
We then provide a detailed description of the tight-binding model $H_{SG 212}(\bk)$ (referred to as $H_{TB}(\bk)$ in the main text) used to realize one such allowed set of photonic SRSIs, whose topological properties were also confirmed by ab-initio calculations in the main text.
We will discuss the construction of $H_{SG 212}(\bk)$, the symmetry analysis of the Wilson loop spectra, and the constraints imposed by the polarization singularity on the Wilson loop.
%%%%%%

\subsection{Hilbert basis method for allowed SRSIs}
\label{sec:sn_sg212_hilbert}
%%%%%%
To begin, we briefly recall the symmetries, WPs, site-symmetry irreps, and the definitions of SRSIs in the SG $P4_332$ (No. 212).
We follow the notation of the Bilbao Crystallographic server~\cite{aroyo2006bilbao1,aroyo2006bilbao2,aroyo2011crystallography}.
This SG is generated by the three-fold rotation $C_{3,111} = \{ 3_{111}| \bb 0 \}$, two-fold screw rotation $\td{C}_{2,110} = \{ 2_{110}| 1/4,3/4,1/4 \}$, four-fold screw rotation $\td{C}_{4,001} = \{ 4_{001}| 3/4,1/4,3/4 \}$, time-reversal symmetry $\mc{T}$, and lattice translations $\{ E|\bb a \in \Z^3 \}$.
The action of the point-group generators on positions $(x, y, z)$ is given by: $4_{001}$ maps to $(-y, x, z)$, $3_{111}$ to $(z, x, y)$, $2_{110}$ to $(y, x, -z)$, and the identity $E$ to $(x, y, z)$.
The Wyckoff positions (WPs) $4a$, $4b$, $8c$, $12d$, and $24e$ lie within a unit cell, with representative positions: $4a$: $(1/8, 1/8, 1/8)$, $4b$: $(5/8, 5/8, 5/8)$, $8c$: $(x, x, x)$, $12d$: $(y, 1/4 - y, 1/8)$, $24e$: $(x, y, z)$, where $x$, $y$, and $z$ are free real-valued parameters.
These WPs support the following site-symmetry irreps:
\bg
(A_1)_a, (A_2)_a, (E)_a, (A_1)_b, (A_2)_b, (E)_b, (A_1)_c, ({}^1 E {}^2 E)_c, (A)_d, (B)_d, (A)_e.
\eg
For notational convenience, we omit the multiplicity label of each WP.
The band representation (BR) multiplicity vector $\bb m$ is defined by the multiplicities of the site-symmetry irreps:
\bg
\bb m = \left( \nW{(A_1)_a}, \nW{(A_2)_a}, \dots, \nW{(B)_d}, \nW{(A)_e} \right).
\eg

In SG $P4_332$, there are three $\Z$-valued SRSIs, denoted by $\bs \theta_\Z = (\theta_1,\theta_2,\theta_3)^T$, and one $\Z_2$-valued SRSI $\theta_4^{(2)}$~\cite{hwang2025stable}.
These invariants are defined in terms of the BR multiplicity vector $\bb m$ as
\ba
\label{eq:SG212_SRSIs_SSG_multiplicities}
\theta_1 =& \nW{(A_1)_a} + \nW{(A_1)_b} + \nW{(A_1)_c} + \nW{(A)_d} + \nW{(A)_e}, \nn
\theta_2 =& \nW{(A_2)_a} + \nW{(A_2)_b} + \nW{(A_1)_c} + \nW{(B)_d} + \nW{(A)_e}, \nn
\theta_3 =& \nW{(E)_a} + \nW{(E)_b} + 2 \nW{({}^1E{}^2E)_c} + \nW{(A)_d} + \nW{(B)_d} + 2 \nW{(A)_e}, \nn
\theta^{(2)}_4 =& \nW{(A_1)_a} + \nW{(A_2)_a} + \nW{(E)_a} \pmod 2.
\ea
Note that $\theta_1 + \theta_2 + 2 \theta_3 = \nu/4$, where $\nu$ is the total number of bands $\mc{B}$ under consideration.

The symmetry-data vector $\bb v [\mc{B}]$ refers to the collection of little-group irrep multiplicities associated with a given band structure $\mc{B}$, evaluated at high-symmetry momenta (HSM), which in this space group are the points $\Gamma = (0,0,0)$, $R = (\pi,\pi,\pi)$, $M = (\pi,\pi,0)$, and $X = (0,\pi,0)$.
The full expression of $\bb v [\mc{B}]$ is
\ba
\bb v [\mc{B}]
=& [\nR{\Gamma_1}, \nR{\Gamma_2}, \nR{\Gamma_3}, \nR{\Gamma_4}, \nR{\Gamma_5}, \nR{R_1 R_2}, \nR{R_3}, \nR{M_1 M_4}, \nR{M_2 M_3}, \nR{M_5}, \nR{X_1}, \nR{X_2}]^T.
\ea
Compatibility relations between HSM reduce the number of independent components of $\bb v [\mc{B}]$.
In particular, the compatibility relations require that
\bg
\nR{\Gamma_3} = \nR{R_1 R_2}, \quad
\nR{\Gamma_4} = \nR{M_2 M_3}, \quad
\nR{\Gamma_5} = \nR{M_1 M_4}, \quad
\nR{M_5} = \nR{X_1} = \nR{X_2},
\label{seq:sg212_compatibility}
\eg
Using these relations, we define a reduced symmetry-data vector $\bb v' [\mc{B}]$ as
\ba
\bb v' [\mc{B}] &= [ \nR{\Gamma_1}, \nR{\Gamma_2}, \nR{\Gamma_3}, \nR{\Gamma_4}, \nR{\Gamma_5}, \nR{R_3}, \nR{M_5} ]^T.
\label{seq:sg212_symvec}
\ea
As detailed in the main text and originally proved in Ref.~\cite{hwang2025stable}, a one-to-one correspondence exists between the (reduced) symmetry-data vector $\bb v'$ and the $\Z$-valued SRSIs $\bs \theta_\Z$.
In particular, each component of $\bb v'$ can be expressed as a linear combination of $\bs \theta_\Z$:
\ba
\bb v' [\mc{B}]
&= (\theta_1, \theta_2, \theta_3, \theta_2 + \theta_3, \theta_1 + \theta_3, \theta_1 + \theta_2 + \theta_3, \theta_1 + \theta_2 + 2 \theta_3)^T \nn
&:= \mc{M} \cdot \bs \theta_\Z
\label{seq:sg212_map}
\ea
where the mapping matrix $\mc{M}$ is given by
\bg
\mc{M} = \bpm 1 & 0 & 0 & 0 & 1 & 1 & 1 \\ 0 & 1 & 0 & 1 & 0 & 1 & 1 \\
0 & 0 & 1 & 1 & 1 & 1 & 2 \epm^T.
\eg
An important advantage of this construction is that it automatically enforces the compatibility relations and integrality of the symmetry-data vector.
While a general symmetry-data vector $\bb v$ must satisfy both integrality and the compatibility relations among HSM irreps, the parameterization $\bb v' = \mc M \cdot \bs\theta_\Z$ guarantees these conditions by construction.
In particular, for any integer-valued $\bs \theta_\Z$, the resulting $\bb v'$ is automatically integer-valued, since $\mc M$ is an integer matrix, and it satisfies all compatibility relations.
This holds trivially for the reduced symmetry-data vector, and more generally for the unreduced case due to the way $\mc M$ is defined (see SN~\ref{sec:sn_sg207}).
This allows us to focus solely on the physical inequality constraints on $\bs \theta_\Z$ in the subsequent analysis.

By combining the mapping in Eq.~\eqref{seq:sg212_map} with the physical constraints imposed on the symmetry-data vector, we can determine the allowed values of SRSIs in photonic and electronic band structures.
In the main text, a graphical method was used to depict the allowed values of $\Z$-valued SRSIs in SG $P4_332$ (No. 212) on a two-dimensional plane spanned by two $\Z$-valued SRSIs $(\theta_1,\theta_2)$, under the condition that the total number of bands $\nu$ is fixed.
The quantity $\nu$ is related to the $\Z$-valued SRSIs by
\bg
\nu = 4 (\theta_1 + \theta_2 + 2 \theta_3).
\label{seq:sg212_filling}
\eg
Since SG $P4_332$ has only three $\Z$-valued SRSIs $\bs \theta_\Z$, it is straightforward to determine the allowed solutions directly using algebraic or graphical methods, as demonstrated in the main text.
However, for SGs with a larger number of $\Z$-valued SRSIs, such as SG $P432$ (No. 207) studied in detail in SN~\ref{sec:sn_sg207}, direct visualization or enumeration becomes impractical due to the high dimensionality.
To address this, we now introduce an algebraic approach for computing the allowed values of $\bs \theta_\Z$, based on polyhedral geometry and the Hilbert basis, which systematically identifies all integer solutions (lattice points) in the constrained polyhedron~\cite{bruns2009polytopes}.
%%%%%%

%%%%%%
\subsubsection{SRSIs allowed in photonic bands}
%%%%%%
We now examine the allowed values of $\bs \theta_\Z$ for photonic bands, by imposing the constraints on the symmetry-data vector arising from the presence of a polarization singularity.
Away from $\Gamma$, i.e. at all other HSM, each little-group irrep must appear with nonnegative multiplicity.
This leads to the conditions $( \bb v' [\mc{B}])_{6,7}\equiv v'_{6,7} \ge 0$, which correspond to
\ba
v'_6 =& \nR{R_3} = \theta_1 + \theta_2 + \theta_3 \ge 0,
\nn
v'_7 =& \nR{M_5} = \theta_1 + \theta_2 + 2 \theta_3 \ge 0 .
\label{seq:sg212_rsi_ph1}
\ea
At $\Gamma$, the two lowest bands, which correspond to the transverse $T$ modes, are assigned the representation $\girrep = \Gamma_4 - \Gamma_1$.
This implies the following constraints: $v'_1 = \nR{\Gamma_1} \ge -1$, $v'_4 = \nR{\Gamma_4} \ge 1$, and $v'_i \ge 0$ for $i=2,3,5$, explicitly:
\bg
v'_1 = \theta_1 \ge -1, \quad
v'_2 = \theta_2 \ge 0, \quad
v'_3 = \theta_3 \ge 0, \quad
v'_4 = \theta_2 + \theta_3 \ge 1, \quad
v'_5 = \theta_1 + \theta_3 \ge 0.
\label{seq:sg212_rsi_ph2}
\eg
Note that each component of the symmetry-data vector must be an integer, i.e. $v'_{1,\dots,7} \in \Z$.
Recall that this integrality condition is automatically satisfied since the variables $\theta_{1,2,3}$ are integers and the matrix $\mc{M}$ in Eq.~\eqref{seq:sg212_map} is integer-valued (We also discuss in SN~\ref{sec:sn_sg207} the case where a redefinition of $\bs \theta_\Z$ leads to a modified $\mc{M}$, in which the integrality condition is no longer guaranteed).
All inequalities in Eqs.~\eqref{seq:sg212_rsi_ph1} and \eqref{seq:sg212_rsi_ph2} can be compactly written as
\bg
\mc{M} \cdot \bs \theta_\Z \ge \bb b,
\label{seq:linear_prob_ph}
\eg
where $\bb b = (-1,0,0,1,0,0,0)^T$.
Note that we have ordered the inequalities such that the $i$\textsuperscript{th} component of $\bb b$ corresponds to the inequality satisfied by $v'_i$ in Eqs.~\eqref{seq:sg212_rsi_ph1} and \eqref{seq:sg212_rsi_ph2}.

We now demonstrate how the integral solutions for Eq.~\eqref{seq:linear_prob_ph} can be efficiently solved using the Hilbert basis~\cite{bruns2009polytopes}.
The inequality in Eq.~\eqref{seq:linear_prob_ph} defines a polyhedron $\{\bb x \in \R^3 | \mc{M} \cdot \bb x \ge \bb b\}$.
Then, the integral solutions $\bs \theta_\Z$ correspond to the lattice points within this polyhedron, i.e. $\{\bb x \in \Z^3 | \mc{M} \cdot \bb x \ge \bb b\}$.
These lattice points are generated by the finite Hilbert basis of the polyhedron, which is guaranteed to exist because the matrix $\mc{M}$ and the vector $\bb b$ are integer-valued (hence rational) and thus define a rational polyhedron~\cite{bruns2009polytopes}.
Specifically, any solution $\bs \theta_\Z$ can be written as
\bg
\bs \theta_\Z = \bb h_c + \sum_{A=1}^{D_r} \, n_A \bb h_{r,A},
\label{seq:hilbert}
\eg
where $\bb h_{r,A}$ ($A=1, \dots, D_r$) are the generators of the recession cone (representing unbounded directions of the polyhedron), and $\bb h_c$ belongs to the finite set $\{\bb h_c\}$ associated with the compact part of the polyhedron.
The coefficients $n_A$ are nonnegative integers, i.e. $n_A \in \N_0$.
The recession cone is determined independently of $\bb b$, and thus the same generators $\bb h_{r,A}$ can be used for polyhedra with the same $\mc{M}$~\cite{bruns2009polytopes}.
Then, the compact part is determined by the boundary condition $\bb b$ and the generators of the recession cone~\cite{chubarov2005basis}.
The components of the Hilbert basis can be computed efficiently using open-source tools such as Normaliz~\cite{bruns2010normaliz}, enabling systematic enumeration of all $\bs \theta_\Z$.
A brief summary of the Hilbert basis and its mathematical background is provided in SN~\ref{sec:sn_hilbert}.
For other applications of the Hilbert basis method, see Ref.~\cite{song2020fragile} for its use in classifying symmetry-indicated fragile topological phases, and Ref.~\cite{christensen2022location} for enumerating symmetry-data vectors in 3D photonic band structures.

\begin{figure*}[t]
\centering
\includegraphics[width=0.85\textwidth]{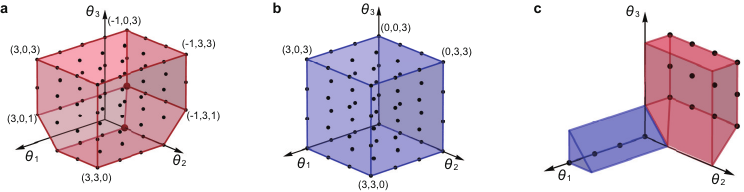}
\caption{{\bf Hilbert basis for $\Z$-valued SRSI $\bs \theta_\Z$ in SG $P4_332$ (No. 212).}
\textbf{a} Allowed values of $\bs \theta_\Z$ for photonic bands, determined by the inequality conditions in Eq.~\eqref{seq:linear_prob_ph}.
The red region indicates the allowed region by the inequalities, and the black dots denote the integer-valued solutions within the bounded range $\bs \theta_\Z \in [-1,3]^3$ (This finite range is chosen to illustrate the structure, since the full solution space is unbounded due to the recession cone).
The large red dots correspond to the compact part of the Hilbert basis, $\bb h_c = (-1, 0, 1)$ and $(0, 1, 0)$, while the remaining points are generated by adding the recession cone generators, $\bb h_{r,A} \in \{(1,0,0), (0,1,0), (0,0,1)\}$.
\textbf{b} Allowed values of $\bs \theta_\Z$ for electronic bands, defined by the same recession cone generators, but with a trivial compact part $\bb h_c = (0,0,0)$.
\textbf{c} Comparison of the photonic-only (red) and electronic-only (blue) regions of allowed $\bs \theta_\Z$.
Notably, $\bs \theta_\Z = (-1,b,c)$ with $b,c \in \N_0$ appear only in the photonic case, and $\bs \theta_\Z = (a,0,0)$ with $a \in \N_0$ appear only in the electronic case.}
\label{sfig:hilbert}
\end{figure*}

By using the Normaliz~\cite{bruns2010normaliz}, we obtain the Hilbert basis for Eqs.~\eqref{seq:sg212_rsi_ph1} and \eqref{seq:sg212_rsi_ph2} as
\ba
\{\bb h_c\} = \{& (-1,0,1)^T, (0,1,0)^T \},
\nn
\{\bb h_r \} = \{& (1,0,0)^T, (0,1,0)^T, (0,0,1)^T \},
\ea
where $\{\bb h_r\} = \{\bb h_{r,A}|A=1,\dots,D_r\}$ (thus, $D_r=3$ for SG $P4_332$).
Since there are two compact parts, the union of the following spans all solutions to Eqs.~\eqref{seq:sg212_rsi_ph1} and \eqref{seq:sg212_rsi_ph2}:
\ba
\bs \theta_{\Z,sol_1} =& (-1,0,1)^T + \sum_{A=1}^3 \, n_A \bb h_{r,A}, = (-1 + n_1, n_2, 1 + n_3)^T,
\nn
\bs \theta_{\Z,sol_2} =& (0,1,0)^T + \sum_{A=1}^3 \, n_A \bb h_{r,A} = (n_1, 1+ n_2, n_3)^T,
\label{seq:sg212_hilbert_ph}
\ea
where $n_A \in \N_0$.
Note that $\bb \theta_\Z$ generated with different compact parts, say $\bs \theta_{\Z,sol_i}$ and $\bs \theta_{\Z,sol_{j \ne i}}$, may have common points.
For instance, the point $\bs \theta_\Z = (0,1,1)$ can be generated either by $\bs \theta_{\Z,sol_1}$ with $(n_1,n_2,n_3)=(1,1,0)$, or by $\bs \theta_{\Z,sol_2}$ with $(0,0,1)$.
More generally, even for a fixed compact part $\bb h_c$, the recession cone part $\sum_A \, n_A \bb h_{r,A}$ need not be uniquely represented, since the generators of the recession cone may admit multiple nonnegative integer decompositions $\{n_A\}$.
Accordingly, when enumerating solutions, identical lattice points generated from different decompositions are removed to avoid double counting.
The structure of solutions in Eq.~\eqref{seq:sg212_hilbert_ph} is illustrated in Fig.~\ref{sfig:hilbert}\textbf{a}.
%

%%%%%%
%%%%%%

Since a set of bands is naturally characterized by its total number $\nu$, we now determine the values of $\bs \theta_\Z$ corresponding to a fixed $\nu$.
We can rewrite Eq.~\eqref{seq:sg212_hilbert_ph} more explicitly by using Eq.~\eqref{seq:sg212_filling}, which expresses $\nu$ in terms of $\bs \theta_\Z$.
For each solution branch $\bs \theta_{\Z,sol_i}$ ($i = 1,2$) in Eq.~\eqref{seq:sg207_hilbert_ph}, we can compute the corresponding value of $\nu$.
For example, $\bs \theta_{\Z,sol_1}$ yields
\bg
\nu = 4 + 4 n_1 + 4 n_2 + 8 n_3.
\label{seq:sg212_hilbert_filling}
\eg
Given a fixed $\nu$, this equation defines a Diophantine problem in the nonnegative integers $n_A \in \N_0$ ($A = 1,2,3$), which can be solved using standard linear programming techniques.
The positivity of the coefficients of $n_A$ reflects that the constraints on the symmetry-data vector enforce lower bounds, such as the nonnegativity of irrep multiplicities at all HSMs except $\Gamma$.

Repeating this process for the remaining solution set $\bs \theta_{\Z,sol_2}$, which also yields $\nu = 4 + 4 n_1 + 4 n_2 + 8 n_3$, we obtain the complete set of allowed solutions for $\bs \theta_\Z$.
For $\nu=4$ (no solutions exist for $\nu \le 3 $), there is a unique solution for $(n_1,n_2,n_3) = (0,0,0)$ for both $\bs \theta_{\Z, sol_{1,2}}$.
Substituting into the general forms $\bs \theta_{\Z, sol_{1,2}}$ gives two allowed SRSI vectors:
\bg
\{ \bs \theta_\Z \} = \{ (-1,0,1)^T, (0,1,0)^T \}.
\eg
From the mapping given in Eq.~\eqref{seq:sg212_map}, the corresponding (reduced) symmetry-data vectors $\bb v'$ are
\ba
\bs \theta_\Z = (-1,0,1)^T :& \quad
\bb v' = (-1,0,1,1,0,0,1)^T, \nn
\bs \theta_\Z = (0,1,0)^T :& \quad
\bb v' = (0,1,0,1,0,1,1)^T
\ea
Using the compatibility relations from Eq.~\eqref{seq:sg212_compatibility}, we translate these into the corresponding combinations of little-group irreps:
\ba
\bs \theta_\Z = (-1,0,1)^T :& \quad
( -\Gamma_1 + \Gamma_3 + \Gamma_4, R_1 R_2, M_2 M_3 + M_5, X_1 + X_2 ), \nn
\bs \theta_\Z = (0,1,0)^T :& \quad
( \Gamma_2 + \Gamma_4, R_3, M_2 M_3 + M_5, X_1 + X_2 ).
\ea
Recalling that the lowest two bands (the $T$ modes) at $\Gamma$ carry the representation $\girrep = \Gamma_4 - \Gamma_1$~\cite{christensen2022location}, we can rewrite the above expressions as
\ba
\label{eq:SG212_4_band_SRSIs}
\bs \theta_\Z = (-1,0,1)^T :& \quad
[\girrep + \Gamma_3, R_1 R_2, M_2 M_3 + M_5, X_1 + X_2], \nn
\bs \theta_\Z = (0,1,0)^T :& \quad
[\girrep+ \Gamma_1 + \Gamma_2, R_3, M_2 M_3 + M_5, X_1 + X_2].
\ea

The next allowed value of the number of bands is $\nu =8$, as can be directly inferred from Eq.~\eqref{seq:sg212_hilbert_filling}.
For this value, there are two solutions for $(n_1, n_2, n_3)$ for each of the solution sets $\bs \theta_{\Z, sol_1}$ and $\bs \theta_{\Z, sol_2}$: $\{ (n_1,n_2,n_3) \} = \{ (1,0,0), (0,1,0) \}$.
Substituting into the general forms yields
\bg
\{ \bs \theta_{\Z, sol_1} \} = \{ (0,0,1)^T, (-1,1,1)^T \}, \quad
\{ \bs \theta_{\Z, sol_2} \} = \{ (1,1,0)^T, (0,2,0)^T \}.
\eg
As there are no overlaps between the two sets, we obtain four distinct allowed values of $\bs \theta_\Z$ at $\nu = 8$:
\bg
\{ \bs \theta_\Z \} = \{ (0,0,1)^T, (-1,1,1)^T, (1,1,0)^T, (0,2,0)^T \}.
\eg
The corresponding representations at HSM are
\ba
\bs \theta_\Z = (0,0,1)^T :& \quad
[\girrep + \Gamma_1 + \Gamma_3 + \Gamma_5, R_1 R_2 + R_3, M_1 M_4 + M_2 M_3 + 2 M_5, 2 X_1 + 2 X_2], \nn
\bs \theta_\Z = (-1,1,1)^T :& \quad
[\girrep + \Gamma_2 + \Gamma_3 + \Gamma_4, R_1 R_2 + R_3, 2 M_2 M_3 + 2 M_5, 2 X_1 + 2 X_2], \nn
\bs \theta_\Z = (1,1,0)^T :& \quad
[\girrep + 2 \Gamma_1 + \Gamma_2 + \Gamma_5, 2 R_3, M_1 M_4 + M_2 M_3 + 2 M_5, 2 X_1 + 2 X_2], \nn
\bs \theta_\Z = (0,2,0)^T :& \quad
[\girrep + \Gamma_1 + 2 \Gamma_2 + \Gamma_4, 2 R_3, 2 M_2 M_3 + 2 M_5, 2 X_1 + 2 X_2].
\ea
%%%%%%

%%%%%%
\subsubsection{SRSIs allowed in electronic bands}
%%%%%%
For electrons, the physically allowed values of $\bs \theta_\Z$ are constrained by the requirement that each component of the symmetry-data vector must be a nonnegative integer.
This yields the same inequalities as in Eq.~\eqref{seq:sg212_rsi_ph1} for HSMs other than $\Gamma$, along with additional constraints at $\Gamma$:
\bg
v'_1 = \theta_1 \ge 0, \quad
v'_2 = \theta_2 \ge 0, \quad
v'_3 = \theta_3 \ge 0, \quad
v'_4 = \theta_2 + \theta_3 \ge 0, \quad
v'_5 = \theta_1 + \theta_3 \ge 0.
\label{seq:sg212_rsi_el}
\eg
All these inequalities can be collectively expressed as $\mc{M} \cdot \bs \theta_\Z \ge 0$, where the right-hand side is the zero vector.
This implies that the integral solutions $\bs \theta_\Z$ correspond to lattice points in a cone, whose Hilbert basis has a trivial compact part:
\bg
\{\bb h_c\} = \{ (0, 0, 0) \}.
\eg
Since $\mc{M}$ remains the same as in the photonic case, the generators of the recession cone ${\bb h_r}$ are also identical to those given in Eq.~\eqref{seq:sg212_hilbert_ph}.
Hence, the only difference from the photonic case lies in the absence of a nontrivial compact part $\{\bb h_c\}$ in the Hilbert basis.
The general solution of $\bs \theta_\Z$ is therefore
\bg
\bs \theta_{\Z, sol}
= \bb h_c + \sum_{A=1}^3 \, n_A \bb h_{r,A}
= \sum_A \, n_A \bb h_{r,A}
= (n_1, n_2, n_3)^T,
\label{seq:sg212_hilbert_el}
\eg
with $n_A \in \N_0$ and the corresponding number of bands
\bg
\nu = 4 n_1 + 4 n_2 + 8 n_3.
\eg
Because the compact part is trivial, the solution space is simpler than in the photonic case [Eq.~\eqref{seq:sg212_hilbert_ph}], consisting of a single family of solutions.
The structure of solutions in Eq.~\eqref{seq:sg212_hilbert_el} is illustrated in Fig.~\ref{sfig:hilbert}\textbf{b}.

For instance, when $\nu = 4$ (the minimal number of bands), the allowed solutions are
\bg
\{ \bs \theta_\Z \} = \{ (1,0,0)^T, (0,1,0)^T \},
\eg
which correspond to the following irrep content at HSM:
\ba
\bs \theta_\Z = (1,0,0)^T :& \quad
(\Gamma_1 + \Gamma_5, R_3, M_1 M_4 + M_5, X_1 + X_2), \nn
\bs \theta_\Z = (0,1,0)^T :& \quad
(\Gamma_2 + \Gamma_4, R_3, M_2 M_3 + M_5, X_1 + X_2).
\ea
In a similar way, the four distinct solutions for the next allowed number of bands, $\nu = 8$, can be found as follows:
\ba
\bs \theta_\Z = (0,2,0)^T :& \quad
(2 \Gamma_1 + 2 \Gamma_4, 2 R_3, 2 M_2 M_3 + 2 M_5, 2 X_1 + 2 X_2), \nn
\bs \theta_\Z = (1,1,0)^T :& \quad
(\Gamma_1 + \Gamma_2 + \Gamma_4 + \Gamma_5, 2 R_3, M_1 M_4 + M_2 M_3 + 2 M_5, 2 X_1 + 2 X_2), \nn
\bs \theta_\Z = (2,0,0)^T :& \quad
(2 \Gamma_1 + 2 \Gamma_5, 2 R_3, 2 M_1 M_4 + 2 M_5, 2 X_1 + 2 X_2), \nn
\bs \theta_\Z = (0,0,1)^T :& \quad
(\Gamma_3 + \Gamma_4 + \Gamma_5, R_1 R_2 + R_3, M_1 M_4 + M_2 M_3 + 2 M_5, 2 X_1 + 2 X_2).
\ea

Beyond specific values of $\nu$, the structure of allowed SRSIs for general $\nu$ can be systematically compared by examining Eqs.~\eqref{seq:sg212_hilbert_ph} and \eqref{seq:sg212_hilbert_el}, which characterize the Hilbert bases for photonic and electronic bands, respectively.
This comparison reveals the presence of SRSI values that are allowed exclusively in either photonic or electronic bands, as illustrated in Fig.~\ref{sfig:hilbert}\textbf{c}.
In particular, the photonic-only SRSIs take the form $\bs \theta_\Z = (-1, b, c)$, while the electronic-only SRSIs are of the form $(a, 0, 0)$, for all $a, b, c \in \N_0$.
%%%%%%

%%%%%%
\subsubsection{Remark on the Hilbert basis}
\label{sec:sn_hilbert}
%%%%%%
Here, we briefly summarize several standard results on the Hilbert basis~\cite{bruns2009polytopes} that underlie the construction used in this work.

(i) We consider integer solutions of a rational inequality system of the form $\mc M \cdot \bb x \ge \bb b$, where $\mc M$ and $\bs b$ are rational-valued (integer-valued with loss of generality).
Such a system defines a rational polyhedron $P \subset \R^d$.
Its associated recession cone is defined by ${\rm rec} (P) = \{\bb x \in \R^d | \mc M \cdot \bb x \ge \bb 0\}$.
That is, $\bb x \in {\rm rec} (P)$ if and only if $\bb x' + \lambda \bb x \in P$ for all $\bb x' \in P$ and all $\lambda \ge 0$.

A cone is said to be pointed if it contains no nontrivial lineality space, i.e. no nonzero vector $\bs \ell$ such that both $\bs \ell$ and $-\bs \ell$ belong to the cone.
Equivalently, the cone is pointed if ${\rm ker} \, \mc M = \{\bb 0\}$.
This condition is satisfied in our case due to the one-to-one mapping between $\bs \theta_\Z$ and symmetry-data vector $\bb v$.
For a pointed rational cone $C = \{\bb x \in \R^d| \mc M \cdot \bb x \ge \bb 0\}$, the set of lattice points in the cone, $\Z^d \cap C = \{\bb x \in \Z^d| \mc M \cdot \bb x \ge \bb 0 \}$ admits a finite Hilbert basis $\{\bb h_{r,A}|A=1,\dots,D_r\}$, which is minimal and unique as a set.

(ii) The existence of a Hilbert basis implies that any lattice point $\bb x \in \Z^d$ in the cone $C$ can be generated as a nonnegative integer combination of the basis elements.
That is, for any $\bb x \in \Z^d \cap C$, there exist nonnegative integers $n_A$ ($A=1,\dots,D_r$) such that $\bb x = \sum_{A=1}^{D_r} \, n_A \bb h_{r,A}$.
However, this decomposition need not be unique, since the generators themselves may satisfy linear relations.
In particular, the same lattice point $\bb x$ may admit distinct representations $\sum_A \, n_A \bb h_{r,A}$ and $\sum_A \, n'_A \bb h_{r,A}$ with different coefficient sets $\{n_A\}$ and $\{n'_A\}$.

(iii) For a general rational polyhedron $P = \{\bb x \in \R^d | \mc M \cdot \bb x \ge \bb b \}$, integer solutions can be analyzed by embedding the problem into a higher-dimensional cone by introducing an auxiliary variable $t$~\cite{chubarov2005basis}.
Specifically, we define the lifted cone, $\td C = \{(t,\bb x) \in \R^d| t \ge 0 \text{ and } \mc M \cdot \bb x \ge t \bb b\}$.
The defining conditions of $\td C$ can be rewritten as
\bg
\bpm 1 & \bb 0 \\ - \bb b & \mc M \epm \cdot \bpm t \\ \bb x \epm \ge \bb 0,
\eg
which makes it manifest that $\td C$ is a rational cone.
Each lattice point $(t,\bb x) \in \td C$ has a direct interpretation.
If $t=0$, then $\mc M \cdot \bb x \ge \bb 0$, so $\bb x \in {\rm rec}(P)$.
If $t=1$, then $\mc M \cdot \bb x \ge \bb b$, and $\bb x$ is an integer point of the original polyhedron $P$.

If $P$ is pointed, i.e. ${\rm ker} \, \mc M = \{\bb 0\}$, then the lifted cone $\td C$ is also pointed and therefore admits a finite Hilbert basis, which we denote by $\{ (\tau_\alpha, \td{\bb h}_\alpha) \}$.
Any lattice point in $\td C$ can then be expressed as a nonnegative integer combination of these generators.
Importantly, the grading by $t$ implies a natural separation of roles.
Generators with $\tau_\alpha = 0$ generate all lattice points with $t=0$ in the lifted cone $\td C$.
Projecting onto the $\bb x$-coordinates, the set $\{ \td{\bb h}_\alpha | \tau_\alpha=0 \}$ forms a Hilbert basis for the lattice points in the recession cone ${\rm rec}(P)$.
In contrast, any lattice point $(t, \bb x) \in \td C$ with $t=1$ must be generated by including exactly one generator with $\tau_\alpha = 1$, together with an arbitrary nonnegative integer combination of generators with $\tau_\alpha = 0$.
Generators with $\tau_\alpha > 1$ cannot appear in such a decomposition, since they would necessarily produce $t >1$.
As a result, generators with $\tau_\alpha = 1$ encode the compact part of the polyhedron $P$, while generators with $\tau_\alpha = 0$ encode its recession
directions.

Accordingly, the compact part $\{\bb h_c\}$ and the recession-cone generators $\{\bb h_r\}$ of the original polyhedron $P$ are obtained by projecting the Hilbert basis elements $(\tau_\alpha, \td{\bb h}_\alpha)$ of $\td C$ onto the $\bb x$-coordinates, with $\tau_\alpha=1$ and $\tau_\alpha=0$, respectively.
%%%%%%

%%%%%%
\subsection{Tight-binding model construction}
\label{sec:sn_sg212_tb}
%%%%%%
In the previous section, we derived the form of SRSIs allowed only in photonic systems.
Now we will explicitly construct a TB model for one such system.
Consider the 4-band photonic SRSI vector $\bs \theta_\Z = (-1,0,1)^T$ with the corresponding symmetry-data vector given in Eq.~\eqref{eq:SG212_4_band_SRSIs}.
From Eq.~\eqref{eq:SG212_SRSIs_SSG_multiplicities}, we find that one choice of site symmetry irrep multiplicities that realize $\theta_\mathbb{Z}$ is $m[(A_1)_a] = - m[(E)_a] = - 1$ and all other multiplicities being zero.
Accordingly, the physical photonic bands in this model admit an elementary band representation (EBR) decomposition $(E)_a \ua G \ominus (A_1)_a \ua G$, where $(\rho)_W \ua G$ denotes the EBR induced by the site-symmetry irrep $(\rho)_W$ at the WP $W$ in the SG $G$.
Note that we omit the multiplicity of $W$ for simplicity when writing $(\rho_W)$ (For example, $(E)_a$ refers the irrep $(E)$ at the WP $4a$).

We construct a tight-binding model that describes the physical photonic bands following Refs.~\cite{morales2025transversality,devescovi2024axion}.
The EBR decomposition indicates that the physical bands can be regularized into the EBR $(E)_a \ua G$ by introducing auxiliary trivial bands with symmetry-data vector coinciding with that of $(A_1)_a \ua G$.
Thus, we first construct the regularized bands of the EBR $(E)_a \ua G$.
This EBR is induced by the site-symmetry irrep $(E)_a$ at the WP $4a$, which will be the basis orbitals of tight-binding model.
The irrep $(E)_a$ defines the representation for the symmetry generators of the SG.
Recall that the SG $P4_332$ is generated by
\bg
\td C_{4,001} = \{ 4_{001}| 3/4, 1/4, 3/4 \}, \quad
C_{3,\rm 111} = \{ 3_{111}| \bb 0 \}, \quad
\td C_{2,110} = \{ 2_{110}| 1/4, 3/4, 1/4 \},
\eg
and time-reversal symmetry $\mc{T}$.
These symmetry operations transform the momentum $\bk$ as $\td C_{4,001} \bk = (-k_y,k_x,k_z)$, $C_{3,111} \bk = (k_z,k_x,k_y)$, $\td C_{2,110} \bk = (k_y,k_x,-k_z)$, and $\mc{T} \bk = -\bk$.

The basis orbitals corresponding to the irrep $(E)_a$ are located at the four positions $\bq_{4a,1}=(1/8,1/8,1/8)$, $\bq_{4a,2}=(3/8,7/8,5/8)$, $\bq_{4a,3}=(7/8,5/8,3/8)$, and $\bq_{4a,4}=(5/8,3/8,7/8)$ in the unit cell that comprise the WP $4a$.
Since the $(E)$ irrep at each position of WP $4a$ is two-dimensional, a total of 8 orbitals is required.
The matrix representation of symmetry operators can be constructed based on how the basis orbitals transform under these symmetries.
Using the induction procedure of Refs.~\cite{elcoro2017double,cano2018building}, we can write the explicit representation matrices as
\bg
U_{\td C_{4,001}}(\bk) = e^{-\frac{i}{4} (k_x - 3k_y + 3k_z)} \,
\bpm 0 & 0 & 0 & 0 & 0 & 1 & 0 & 0 \\ 0 & 0 & 0 & 0 & 1 & 0 & 0 & 0 \\
0 & 0 & 0 & 0 & 0 & 0 & \xi & 0 \\ 0 & 0 & 0 & 0 & 0 & 0 & 0 & \cm \xi \\
0 & 0 & 0 & 1 & 0 & 0 & 0 & 0 \\ 0 & 0 & 1 & 0 & 0 & 0 & 0 & 0 \\
\cm \xi & 0 & 0 & 0 & 0 & 0 & 0 & 0 \\ 0 & \xi & 0 & 0 & 0 & 0 & 0 & 0 \epm,
\quad
U_{C_{3,111}}(\bk) =
\bpm \xi & 0 & 0 & 0 & 0 & 0 & 0 & 0 \\ 0 & \cm \xi & 0 & 0 & 0 & 0 & 0 & 0 \\
0 & 0 & 0 & 0 & \xi & 0 & 0 & 0 \\ 0 & 0 & 0 & 0 & 0 & \cm \xi & 0 & 0 \\
0 & 0 & 0 & 0 & 0 & 0 & 0 & 1 \\ 0 & 0 & 0 & 0 & 0 & 0 & 1 & 0 \\
0 & 0 & 0 & \xi & 0 & 0 & 0 & 0 \\ 0 & 0 & \cm \xi & 0 & 0 & 0 & 0 & 0 \epm,
\nn
U_{\td C_{2,110}}(\bk) = e^{-\frac{i}{4} (3k_x + k_y - 3k_z)} \,
\bpm 0 & 0 & 0 & 1 & 0 & 0 & 0 & 0 \\ 0 & 0 & 1 & 0 & 0 & 0 & 0 & 0 \\
0 & 1 & 0 & 0 & 0 & 0 & 0 & 0 \\ 1 & 0 & 0 & 0 & 0 & 0 & 0 & 0 \\
0 & 0 & 0 & 0 & 0 & 1 & 0 & 0 \\ 0 & 0 & 0 & 0 & 1 & 0 & 0 & 0 \\
0 & 0 & 0 & 0 & 0 & 0 & 0 & \xi \\ 0 & 0 & 0 & 0 & 0 & 0 & \cm \xi & 0 \epm,
\quad
U_{\mc{T}} =
\bpm 0 & 1 & 0 & 0 & 0 & 0 & 0 & 0 \\ 1 & 0 & 0 & 0 & 0 & 0 & 0 & 0 \\
0 & 0 & 0 & 1 & 0 & 0 & 0 & 0 \\ 0 & 0 & 1 & 0 & 0 & 0 & 0 & 0 \\
0 & 0 & 0 & 0 & 0 & 1 & 0 & 0 \\ 0 & 0 & 0 & 0 & 1 & 0 & 0 & 0 \\
0 & 0 & 0 & 0 & 0 & 0 & 0 & 1 \\ 0 & 0 & 0 & 0 & 0 & 0 & 1 & 0 \epm,
\eg
where $\xi=e^{\frac{2\pi i}{3}}$ and $\cm \xi=e^{-\frac{2\pi i}{3}}$.

Now, we construct the tight-binding Hamiltonian $H_{SG212}(\bk)$ using these symmetry operators.
First, we introduce hopping parameters asymmetric to the SG and then perform a symmetrization process using the symmetry generators of the SG.
During this process, the intermediate Hamiltonians are simplified by introducing a sublattice embedding matrix.
Recall that the 8 basis orbitals are located at $\bx_{1,\dots,8}$, where $\bx_1 = \bx_2 = \bq_{a,1}$, $\bx_3 = \bx_4 = \bq_{a,2}$, $\bx_5 = \bx_6 = \bq_{a,3}$, and $\bx_7 = \bx_8 = \bq_{a,4}$.
This defines the sublattice embedding matrix $V(\bk)$, whose component is given by $V(\bk)_{\alpha \beta}=e^{i \bk \cdot \bx_\alpha} \delta_{\alpha \beta}$ ($\alpha,\beta=1,\dots,8$).
Next, we construct the Hamiltonian $H_0(\bk)$ symmetric under $\td C_{4,001}$.
This requires that $H_0(\bk)$ satisfies $H_0(\bk) = [U_{\td C_{4,001}}(\bk)]^{-1} \, H_0(\td C_{4,001} \bk) \, U_{\td C_{4,001}}(\bk)$.
For this, we define
\bg
H_0(\bk) = V(-\bk) \bpm 0_{4 \times 4} & h(\bk) \\ h(\bk)^\dg & 0_{4 \times 4} \epm V(\bk),
\\
h(\bk) = t_1 \bpm
0 & 0 & 0 & \cm \xi Q_1 Q_3 \\
0 & Q_1 Q_2 & 0 & 0
\\ 1 & 0 & 0 & 0
\\ 0 & 0 & \xi Q_2 & 0
\epm
+
t_2 \bpm
0 & 0 & 0 & Q_3 
\\ 0 & \cm \xi Q_1 & 0 & 0 \\ \xi Q_1 & 0 & 0 & 0
\\ 0 & 0 & t_2 & 0
\epm
+
t_3 \bpm
0 & 0 & 0 & 0 
\\ \xi Q_1 & 0 & 0 & \xi Q_3 \\ 0 & 0 & 0 & 0
\\ \cm \xi Q_1 & 0 & 0 & 1
\epm,
\eg
where $(Q_1,Q_2,Q_3)=(e^{-ik_x}, e^{-ik_y}, e^{-ik_z})$.
The parameters $(t_1,t_2,t_3)$ are real-valued and parametrize the nearest-neighbor hoppings.
After the symmetrization procedure described below, these three parameters span the most general nearest-neighbor hoppings consistent with the SG.

We then symmetrize $H_0(\bk)$ with respect to the remaining generators of the SG: $\td C_{2,110}$, $C_{3,111}$, and $\mc{T}$.
The symmetrization procedure is carried out as follows:
\bg
H_1(\bk) = H_0(\bk) + [U_{\td C_{2,110}}(\bk)]^{-1} \, H_0(\td C_{2,110} \bk) \, U_{\td C_{2,110}}(\bk),
\label{seq:h_sym_1}
\\
H_2(\bk) = H_1(\bk) + [U_{C_{3,111}}(\bk)]^{-1} \, H_1(C_{3,111}\bk) \, U_{C_{3,111}}(\bk) + U_{C_{3,111}}(\bk) \, H_1(C_{3,111}^{-1} \bk) \, [U_{C_{3,111}}(\bk)]^{-1},
\label{seq:h_sym_2}
\\
H_{SG 212}(\bk) = H_2(\bk) + [U_{\mc{T}}]^{-1} \, \cm{H_2(-\bk)} \, U_\mc{T} + \ep_0 \, \mathds{1}_8.
\label{seq:h_sym_3}
\eg
Note that $C_{3,111}^{-1} \bk = (k_y,k_z,k_x)$ and $\ep_0$ is the onsite energy.
Through Eqs.~\eqref{seq:h_sym_1}-\eqref{seq:h_sym_3}, $\td C_{2,110}$, $C_{3,111}$, and $\mc{T}$ are successively symmetrized.
After symmetrization, $H_{SG 212}(\bk)$ satisfies all symmetry constraints,
\bg
U_{\td C_{4,001}}(\bk) H_{SG 212}(\bk) [U_{\td C_{4,001}}(\bk)]^{-1} = H_{SG 212}(\td C_{4,001} \bk),
\quad
U_{C_{3,111}}(\bk) H_{SG 212}(\bk) [U_{C_{3,111}}(\bk)]^{-1} = H_{SG 212}(C_{3,111} \bk),
\nn
U_{\td C_{2,110}}(\bk) H_{SG 212}(\bk) [U_{\td C_{2,110}}(\bk)]^{-1} = H_{SG 212}(\td C_{2,110} \bk),
\quad
U_\mc{T} \cm{H_{SG 212}(\bk)} [U_{\mc{T}}]^{-1} = H_{SG 212}(-\bk),
\eg
meaning it is invariant under the full SG $P4_332$.
%

%%%%%%
\begin{figure*}[b]
\centering
\includegraphics[width=0.95\textwidth]{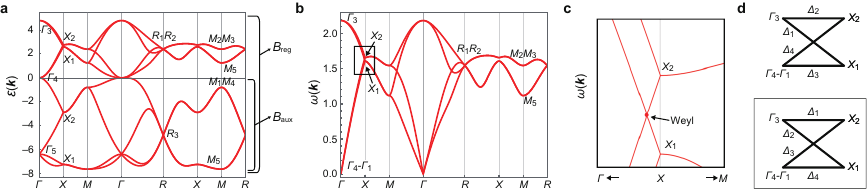}
\caption{{\bf Band structures of the model $H_{SG 212}(\bk)$.}
Band structure defined by (\textbf{a}) $\vep(\bk)$ and (\textbf{b}) $\om(\bk)$.
Note that $\Gamma=(0,0,0)$, $X=(0,\pi,0)$, $M=(\pi,\pi,0)$, and $R=(\pi,\pi,\pi)$.
\textbf{a} The regularized bands $\mc{B}_{\rm reg}$ corresponds to the EBR induced from $(E)_a$, while the auxiliary bands $\mc{B}_{\rm aux}$ are chosen to have a symmetry-data vector identical to that induced from $(A_1)_a$.
All irreps of auxiliary band except $\Gamma_1$ have negative $\vep(\bk)$.
\textbf{b} We plot the frequency spectrum $\om(\bk)$ for physical bands $\mc{B}_{\rm phys}$ corresponding to Eq.~\eqref{seq:sg212_ebr_phys}.
\textbf{c} The Weyl point along $\Gamma$-$X$ is enforced because of the compatibility relation shown in \textbf{d}.
The connectivity shown in the lower diagram of \textbf{d} is realized in our model.}
\label{sfig:sg212_bands}
\end{figure*}
%%%%%%

For convenience, we provide the full expression of $H_{SG 212}(\bk)$:
\bg
H_{SG 212}(\bk) =
V(-\bk) \bpm
\ep_0 \mathds{1}_2 & \eta_1(\bk) & \eta_2(\bk) & \eta_3(\bk) \\
h.c. & \ep_0 \mathds{1}_2 & \eta_4(\bk) & \eta_5(\bk) \\
h.c. & h.c. & \ep_0 \mathds{1}_2 & \eta_6(\bk) \\
h.c. & h.c. & h.c. & \ep_0 \mathds{1}_2
\epm V(\bk)
\label{seq:h_tb212}
\eg
where $h.c.$ denotes the Hermite conjugation.
The expressions of $\eta_{1,\dots,6}$ are given as
\bg
\eta_1(\bk) =
\bpm
Q_2(\lambda + \cm \lambda Q_3) & -t_3 Q_2(1+Q_3) \\
-t_3 Q_2(1+Q_3) & Q_2 (\cm \lambda + \lambda Q_3)
\epm,
\quad
\eta_2(\bk) =
\bpm
Q_1 (\lambda + \cm \lambda Q_2) & -\xi t_3 Q_1(1+Q_2) \\
-\cm \xi t_3 Q_1(1+Q_2) & Q_1 (\cm \lambda + \lambda Q_2)
\epm,
\\
\eta_3(\bk) =
\bpm
-t_3 Q_3(1+Q_1) & \cm \xi Q_3 (\lambda + \cm \lambda Q_1) \\
\xi Q_3 (\cm \lambda + \lambda Q_1) & -t_3 Q_3 (1+Q_1)
\epm,
\quad
\eta_4(\bk) =
\bpm
\cm \lambda + \lambda Q_1 & -\cm \xi t_3 (1+Q_1) \\
-\om t_3 (1+Q_1) & \lambda + \cm \lambda Q_1
\epm,
\nn
\eta_5(\bk) =
\bpm
-\cm \xi t_3 (1+\cm{Q_2}) & \cm \xi (\lambda + \cm \lambda \cm{Q_2}) \\
\xi (\cm \lambda + \lambda \cm{Q_2}) & -\xi t_3 (1+\cm{Q_2})
\epm,
\quad
\eta_6(\bk) =
\bpm
-\xi t_3 (1+Q_3) & \cm \xi (\cm \lambda + \lambda Q_3) \\
\xi (\lambda + \cm \lambda Q_3) & -\cm \xi t_3 (1+Q_3)
\epm,
\eg
where we define a complex-valued parameter, $\lambda = 2(t_1 + \xi t_2)$.

Since full Hilbert space is spanned by the irrep $(E)_a$, the full 8 bands must corresponds to the EBR $(E)_a \ua G$ for arbitrary values of the hopping parameters.
The EBR $(E)_a \ua G$ subduces the following little-group irreps at HSM:
\bg
\mc B_{\rm reg} = (E)_a \ua G:
(\Gamma_3 + \Gamma_4 + \Gamma_5, R_1 R_2 + R_3, M_1 M_4 + M_2 M_3 + 2 M_5, 2 X_1 + 2 X_2),
\label{seq:sg212_ebr_reg}
\eg
where $\Gamma=(0,0,0)$, $X=(0,\pi,0)$, $M=(\pi,\pi,0)$, and $R=(\pi,\pi,\pi)$.
To realize the physical photonic bands corresponding to the SRSI vector $\bs \theta_\Z = (-1,0,1)^T$, we compare the little-group irreps of the physical and regularized bands, $\mc B_{\rm phys}$ and $\mc B_{\rm reg}$, respectively.
The little-group irreps of $\mc B_{\rm phys}$ are given by
\bg
\mc B_{\rm phys}: \quad
[\girrep + \Gamma_3, R_1 R_2, M_2 M_3 + M_5, X_1 + X_2],
\label{seq:sg212_ebr_phys}
\eg
where $\girrep=\Gamma_4-\Gamma_1$, consistent with the irrep assignment for two transverse ($T$) modes, as discussed in the main text and Ref.~\cite{christensen2022location}.
This irregular linear combination of irreps in $\girrep$, marked by negative multiplicity of $\Gamma_1$, indicates the polarization singularity at $(\om,\bk)=(0,\bb 0)$.
We also note that the linear combinations of little-group irreps are regular at all high symmetry momenta except $\Gamma$.

By comparing Eqs.~\eqref{seq:sg212_ebr_reg} and \eqref{seq:sg212_ebr_phys}, the little-group irreps of $\mc B_{\rm aux}$ must be
\bg
\mc B_{\rm aux}: \quad
(\Gamma_1 + \Gamma_5, R_3, M_1 M_4 + M_5, X_1 + X_2).
\label{seq:sg212_ebr_aux}
\eg
As discussed earlier, the symmetry-data vector corresponding to Eq.~\eqref{seq:sg212_ebr_aux} is identical to that of the EBR $(A_1)_a \ua G$.
It is worth to noticing that both EBRs $(A_1)_a \ua G$ and $(A_1)_b \ua G$ have identical little-group irreps listed in Eq.~\eqref{seq:sg212_ebr_aux}.
For this reason, the physical bands $\mc B_{\rm phys}$ can be realized either as $(E)_a \ua G \ominus (A_1)_a \ua G$ or $(E)_a \ua G \ominus (A_1)_b \ua G$; however, both realizations yields the same symmetry-data vector for $\mc B_{\rm phys}$, and the same $\Z$-valued SRSIs.
Finally, as required by the definition of $(\rho_T)_\Gamma = \rho_V - \rho_{\rm triv}$, the vectorial (ir)rep $\rho_V = \Gamma_4$ is correctly included in $\mc{B}_{\rm reg}$, while the trivial irrep $\rho_{\rm triv} = \Gamma_1$ appears in $\mc{B}_{\rm aux}$, consistent with Eqs.~\eqref{seq:sg212_ebr_reg} and \eqref{seq:sg212_ebr_aux}.

To separate $\mc{B}_{\rm phys}$ and $\mc{B}_{\rm aux}$ energetically, we require that $\mc{B}_{\rm phys}$ ($\mc{B}_{\rm aux}$) has nonnegative (negative) energy at all maximal momenta.
The only exception is that $(\rho_T)_\Gamma$, corresponding to $T$ modes, must lie at zero energy.
For example, at $R$, the irreps $R_1R_2$ of $\mc{B}_{\rm phys}$ and $R_3$ of $\mc{B}_{\rm aux}$ must have positive and negative energy, respectively.
A similar condition is required at $\Gamma$ for all little-group irreps, except that $(\rho_T)_\Gamma$ has zero energy.
To find the parameter values $(t_1, t_2, t_3, \ep_0)$ that satisfy these energetic conditions, we analyzed the eigenvalues of $H_{SG 212} (\bk)$ at the HSM.
At $\Gamma$, for example, diagonalization yields the irreps $\Gamma_3$, $\Gamma_4$, and $\Gamma_5$ with energies $12 t_1 - 6 t_2 + \ep_0$, 
$-4 t_1 + 2 t_2 + 4 t_3 + \ep_0$, and $-4 t_1 + 2 t_2 - 4 t_3 + \ep_0$, respectively.
We must set $\Gamma_4$ to zero energy, which imposes $-4 t_1 + 2 t_2 + 4 t_3 + \ep_0 = 0$.
By requiring $\Gamma_3$ ($\Gamma_5$) to have positive (negative) energy, we obtain $12 t_1 - 6 t_2 + \ep_0 >0$ ($-4 t_1 + 2 t_2 - 4 t_3 + \ep_0 < 0$).
Similar constraints can be derived at the other HSM $M$ and $R$, while at $X$ and generic momenta the separation between the physical and auxiliary bands was verified numerically.
(This procedure can be performed purely numerically by scanning parameter sets at the HSM and checking energetic separation at general momenta.)
The required conditions for $H_{SG 212} (\bk)$ are satisfied when $\ep_0=-1.2$, $t_1=0.8$, $t_2=0.6$, $t_3=0.8$.
To define the photonic band structure, we denote the eigenvalues of $H_{SG 212}(\bk)$ as $\vep(\bk)=\om(\bk)^2$.
Since the physical bands have $\vep(\bk) \ge 0$, we define the frequency $\om(\bk)=\sqrt{\vep(\bk)} \ge 0$ for these bands.
In Figs.~\ref{sfig:sg212_bands}\textbf{a,b}, we plot the energy spectrum of the full, regularized bands with $\vep(\bk)$ and the frequency spectrum of the physical bands with $\om(\bk)$.

One interesting feature of the band structures is the appearance of Weyl points along the high-symmetry line (HSL) $\Gamma$-$X$, as shown in Figs.~\ref{sfig:sg212_bands}\textbf{b,c}.
This phenomenon can be explained by the compatibility relation.
The four physical bands carry little group representations $\Gamma_4-\Gamma_1$ and $\Gamma_3$ at $\Gamma$, and $X_1$ and $X_2$ at $X$.
On the HSL $\Gamma$-$X$, denoted as $\Delta=(0,k,0)$, these representations restrict to the following little-group representations at $\Delta$:
\bg
\Gamma_4 - \Gamma_1 \to \Delta_3 + \Delta_4,
\quad
\Gamma_3 \to \Delta_1 + \Delta_2,
\nn
X_1 \to \Delta_2 + \Delta_4,
\quad
X_2 \to \Delta_1 + \Delta_3.
\label{seq:sg212_compatibility_irrep}
\eg
These compatibility relations ensures the appearance of a Weyl point at some momentum, $(0,k_*,0)$, on the HSL $\Gamma$-$X$.
As shown in Fig.~\ref{sfig:sg212_bands}\textbf{d}, there must be a Weyl point regardless of the energetic ordering of irreps at $\Gamma$ and $X$.
In our model, with the chosen hopping parameters, the Weyl point is formed by the $\Delta_2$ and $\Delta_3$ irreps.
Note that, due to the symmetries of the system, there are 5 additional Weyl points located at $(0,-k_*,0)$, $(\pm k_*,0,0)$, $(0,0,\pm k_*)$.

To clarify the compatibility relations in Eq.~\eqref{seq:sg212_compatibility_irrep}, let us briefly explain the origin of the decomposition $X_1 \to \Delta_2 + \Delta_4$ along the HSL $\Delta = (0, k, 0)$.
Along $\Delta$, the little-group irreps $\Delta_{1,\dots,4}$ are distinguished by their eigenvalues of the four-fold screw rotation $\td C_{4,010} = \{4_{010}|1/4,3/4,3/4\}$.
In our convention, these eigenvalues take the form
\bg
e^{- i \frac{3}{4} k}, \quad
- e^{- i \frac{3}{4} k}, \quad
-i e^{- i \frac{3}{4} k}, \quad
+i e^{- i \frac{3}{4} k},
\label{seq:delta_irreps}
\eg
at momentum $(0,k,0)$.
At the $X$ point $(0,\pi,0)$, the irrep $X_1$ carries $\td C_{4,010}$ eigenvalues $e^{i \frac{\pi}{4}}$ and $e^{-i \frac{\pi}{4}}$.
When approaching $X$ from $\Gamma$ along $\Gamma$-$X$ line, these eigenvalues match continuously to those of $\Delta_2$ and $\Delta_4$, leading to the relation $X_1 \to \Delta_2 + \Delta_4$.

The remaining compatibility relations in Eq.~\eqref{seq:sg212_compatibility_irrep} follow from an analogous analysis.
Importantly, the specific decomposition depends on the path taken to approach the BZ boundary.
For instance, approaching $-X$ (which is equivalent to $X$ modulo a reciprocal lattice vector) from $\Gamma$ instead leads to $X_1 \to \Delta_2 + \Delta_3$.
This can be understood from the fact that $\Delta_2$ and $\Delta_3$ irreps carry $\td C_{4,010}$ eigenvalues $e^{-i \frac{\pi}{4}}$ and $e^{i \frac{\pi}{4}}$, respectively, at the momentum $-X$.
Finally, we note that in the BCS the compatibility
relation is written as $X_1 \to \Delta_2 + \Delta_3$ for the $\Gamma$–$X$ path.
This apparent difference originates from a convention choice.
Recall our definition of translation operator in Eq.~\eqref{seq:translation_def}, where a translation $t(\bb R)$ acts on the Bloch states to give $B_{t(\bb R)}(\bk) = e^{-i \bk \cdot \bb R}$ for a lattice vector $\bb R$.
In contrast, the BCS adopts the opposite sign convention, i.e. $e^{+i \bk \cdot \bR}$.
As a result, the BCS compatibility relation for the path $-X$-$\Gamma$ corresponds to our convention for the path $X$-$\Gamma$, and vice versa.
%%%%%%

%%%%%%
\subsection{Cylindrical Wilson loop}
\label{sec:sn_sg212_wl_cylinder}
\begin{figure*}[t]
\centering
\includegraphics[width=0.95\textwidth]{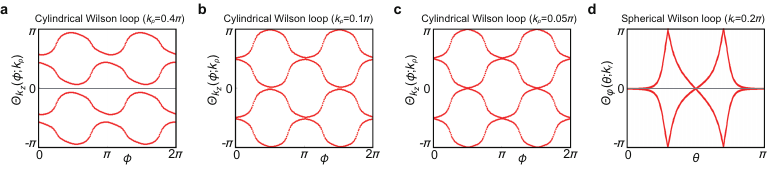}
\caption{{\bf Wilson loop spectra in the model $H_{SG 212}(\bk)$.}
\textbf{a–c} Spectra of the cylindrical Wilson loop for different values of $k_\rho$.
As $k_\rho \to 0$, the spectrum becomes gapless and exhibits a winding structure.
\textbf{d} Spectrum of the spherical Wilson loop, showing a winding structure corresponding to $|\mf{e}|=2$ for the Euler number $\mf{e}$.
As discussed in the main text, both cylindrical and spherical Wilson loops in the ab-initio model exhibit the same topological behavior.}
\label{sfig:sg212_wilson}
\end{figure*}

%%%%%%
In the main text, we discussed the connection between Wilson loops and the zero frequency polarization singularity.
Here, we focus on the key features of the cylindrical Wilson loop: first in the TB model $H_{SG 212}(\bk)$ [Eq.~\eqref{seq:h_tb212}], with results summarized in Fig.~\ref{sfig:sg212_wilson}, and second in the ab-initio photonic crystal model described in the main text (We will discuss the spherical Wilson loop separately in SN~\ref{sec:sn_sg212_wl_sphere}).
Specifically, we consider $k_z$-directed Wilson loops $W_{k_z}(\phi; k_\rho)$ computed along paths parallel to the $k_z$ axis in a cylindrical geometry centered at the $\Gamma$ point.
For each fixed azimuthal angle $\phi$ and fixed radius $k_\rho$, we evaluate the Wilson loop along the $k_z$-direction from a base (starting) point $\bk_b$ to $\bk_b + 2 \pi \hat{k}_z$.
An illustration of the integration path is shown in Fig.~\ref{sfig:wilson_path}\textbf{a}.
As $k_\rho \to 0$, we observe that the Wilson loop spectrum for the four physical bands becomes gapless and develops a winding structure, as shown in Figs.~\ref{sfig:sg212_wilson}\textbf{a}–\textbf{c}.
This section is devoted to explaining the origin of this winding behavior in the limiting case $k_\rho \to 0$.
%%%%%%

%%%%%%
%%%%%%

%%%%%%
\subsubsection{Summary of analysis}
\begin{figure*}[t]
\centering
\includegraphics[width=0.5\textwidth]{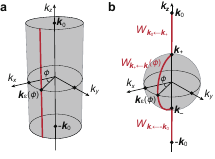}
\caption{{\bf Wilson loop paths.}
\textbf{a} Path for cylindrical Wilson loop $W_{k_z}(\phi; k_\rho)$.
Let $\bk_\ep (\phi) = r_\ep (\cos \phi, \sin \phi, 0)$ with $r_\ep = k_\rho$ denoting the radius parameter.
The Wilson loop is computed along a path parallel to $k_z$ axis that passes through $\bk_\ep (\phi)$.
\textbf{b} Path for almost-straight Wilson loop $W_{r_\ep}(\phi; \bk_b)$.
We define intermediate points $\bk_+ = (0,0,r_\ep)$ and $\bk_- = (0,0,-r_\ep)$ near the origin ($\Gamma$ point).
To avoid the polarization singularity at the origin, the path consists of three segments: a semicircular Wilson line $W_{\bk_+ \leftarrow \bk_-}(\phi)$ and two straight-line segments $W_{\bk_0 \leftarrow \bk_+}$ and $W_{\bk_- \leftarrow -\bk_0}$.
When the basepoint is $-\bk_0$, the almost-straight Wilson loop is defined as $W_{r_\ep}(\phi; -\bk_0) = W_{\bk_0 \leftarrow \bk_+} \cdot W_{\bk_+ \leftarrow \bk_-} (\phi) \cdot W_{\bk_- \leftarrow - \bk_0}$.}
\label{sfig:wilson_path}
\end{figure*}

%%%%%%
In the limit where the radius of the cylindrical Wilson loop approaches zero, we can instead consider the {\it almost-straight} Wilson loop $W_{r_\ep}(\phi; \bk_b)$, whose defining path mostly follows the $k_z$ axis, except in a small neighborhood around the $\Gamma$ point where the polarization singularity is located.
Here, $r_\ep$ denotes the radius parameter indicated in Fig.~\ref{sfig:wilson_path}, and $\phi$ is the azimuthal angle that determines how the Wilson loop path detours around the singularity.
The basepoint $\bk_b$ from which the Wilson loop is initiated can be chosen arbitrarily along the path.
Changing $\bk_b$ modifies the explicit analytic form or numerical values of the Wilson loop matrix itself, but does not affect its eigenvalue spectrum~\cite{alexandradinata2014wilson}.
In our analysis, however, we will mostly take $\bk_b$ to be either $-\bk_0$ or $\bk_\ep(\phi)$, both of which are shown in Fig.~\ref{sfig:wilson_path}.
See Figs.~\ref{sfig:wilson_path}\textbf{a} and \textbf{b} for the schematic comparison of cylindrical and almost-straight Wilson loops, respectively.
Throughout this section, we focus on the almost-straight Wilson loop, which captures the same topological features more conveniently in the $r_\ep \to 0$ limit.

To explain the winding structure of $W_{r_\ep}(\phi; \bk_b)$ in this limit, we must combine symmetry constraints with the effects of the polarization singularity.
Below, we summarize the main results of our symmetry analysis.
As a first step, we study symmetry constraints on $W_{r_\ep}(\phi; \bk_b)$ by considering different choices of basepoint $\bk_b$.
Importantly, we find the following:
\begin{itemize}
\item At $\phi = (2n+1) \frac{\pi}{4}$ for $n \in \Z$, the Wilson loop spectrum is fixed to
\bg
{\rm Spec} [-i \log W_{r_\ep} (\phi; \bk_b) ]
=\left\{ -\frac{3 \pi}{4}, -\frac{\pi}{4}, +\frac{\pi}{4}, +\frac{3 \pi}{4} \right\}.
\label{seq:sg212_wl_summary1}
\eg
Note that the eigenvalues of Wilson loop do not depend on the specific choice of basepoint $\bk_b$.
\item At other values of $\phi$, we find:
\ba
{\rm Spec} [ -i \log W_{r_\ep} (\phi; \bk_b) ]
&= \left\{ -\Theta_0, \, \Theta_0, \, -\Theta_0 + \pi, \, \Theta_0 + \pi \right\}
\quad \text{at } \phi = 0, \, \pi,
\nn
{\rm Spec} [ -i \log W_{r_\ep} (\phi; \bk_b) ]
&= \left\{ -\Theta_0 - \frac{\pi}{2}, \, \Theta_0 - \frac{\pi}{2}, \, -\Theta_0 + \frac{\pi}{2}, \, \Theta_0 + \frac{\pi}{2} \right\}
\quad \text{at } \phi = \frac{\pi}{2}, \, \frac{3\pi}{2}.
\label{seq:sg212_wl_summary2}
\ea
Here, $\Theta_0$ is a system-dependent angle parameter satisfying $\Theta_0 \in (-\pi, \pi]$.
\end{itemize}
These constraints must be satisfied by the Wilson loop $W_{r_\ep} (\phi; \bk_b)$ regardless of whether it is defined in an electronic or photonic system.

However, symmetry alone is not sufficient to explain the observed winding structure in the $r_\ep \to 0$ limit.
To resolve this, as a second step, we incorporate the role of the polarization singularity, which imposes additional constraints.
We show that $\Theta_0$ must approach either $0$ or $\pi/2$ as $r_\ep \to 0$, depending on the system details.
Note that the two choices of $\Theta_0$ differ by a $\pi/2$ rotation of the azimuthal angle $\phi$ in Eq.~\eqref{seq:sg212_wl_summary2}.
Consequently, in both cases, the Wilson loop spectrum exhibits a winding structure equivalent to that shown in Fig.~\ref{sfig:sg212_wilson}\textbf{c}, which corresponds to the case $\Theta_0 = \pi/2$.
%%%%%%

%%%%%%
%%%%%%

%%%%%%
\subsubsection{Basepoint $(0,0,-\pi)$ and azimuthal angle $\phi=\frac{\pi}{4}$}
%%%%%%
In this subsection, we focus on the basepoint $-\bk_0 = (0,0,-\pi)$ with azimuthal angle $\phi = \frac{\pi}{4}$.
Our goal is to derive the analytic expression of $W_{r_\ep}(\phi; -\bk_0)$ and to determine how symmetry constraints restrict its form.
To this end, we analyze the matrix representations of relevant symmetries at $-\bk_0$, acting on the four physical bands defined in Eq.~\eqref{seq:sg212_ebr_phys}.
These symmetry representation matrices, known as sewing matrices, characterize how the bands transform under symmetry operations.
Consider a symmetry operation $g$, which may be either unitary or antiunitary.
The operation $g$ maps $\bk$ to $g \bk$.
For a set of bands $\mc{B}$, we denote the eigenstates or eigenfields as $\ket{u_{n,\bk}}$ for $n \in \mc{B}$.
Then, the sewing matrix $[B_g (\bk)]_{nm}$ ($n,m \in \mc{B}$) of the symmetry $g$ for this set of bands can be defined via a suitable inner product between $\ket{u_{n,g\bk}}$ and $\ket{u_{m,\bk}}$, as defined in SN~\ref{sec:sn_sym_sewing}.
Recall that sewing matrices are periodic in the BZ under the periodic gauge [Eq~.~\eqref{seq:periodic_gauge_phc} and \eqref{seq:periodic_gauge_TB}], and that they respect the group multiplication law as defined in Eq.~\eqref{seq:sewing_multi}.

To determine the sewing matrices at $-\bk_0$, we begin by noting that the bands are labeled by two-dimensional irreps $X_1$ and $X_2$, ordered by increasing energy, at $X = (0,\pi,0)$, as shown in Fig.~\ref{sfig:sg212_bands}\textbf{b}.
Based on the definitions of these irreps, the corresponding sewing matrices can be constructed, for example by using the Representation tool~\cite{bilbao_Rep} provided by the BCS.
Next, observe that the point $X = (0,\pi,0)$ is mapped to $-\bk_0 = (0,0,-\pi)$ by the threefold rotation symmetry $C_{3,111} = \{3_{111} | \bb 0\}$, combined with a reciprocal lattice translation $\bb G_3 = 2\pi \hat{k}_z$, i.e.
\bg
C_{3,111}: \quad (k_x, k_y, k_z) \to (k_z, k_x, k_y) \quad \text{and} \quad X = (0, \pi, 0) \to (0,0,\pi) = -\bk_0 + \bb G_3.
\eg
Hence, the sewing matrices at $-\bk_0$ can be obtained from those at $X$ by conjugation with $C_{3,111}$\footnote{To see this, consider a symmetry operation $g$ and the associated sewing matrix $B_g (-\bk_0)$.
Since $-\bk_0 = C_{3,111} X - \bb G_3$, we have $B_g (-\bk_0) = B_g (C_{3,111} X - \bb G_3) = B_g (C_{3,111} X)$ under the periodic gauge.
Now, define $g' = C_{3,111}^{-1} \, g \, C_{3,111}$, i.e. the conjugation of $g$ by $C_{3,111}$.
Then, the group multiplication law imposes $B_{g'} (X) = B_{C_{3,111}^{-1}} (g \, C_{3,111} X) \cdot B_g (C_{3,111} X) \cdot B_{C_{3,111}} (X)$.
For any $g$ and $\bk$, we have $B_{g \, g^{-1}}(\bk) = \mathds{1} = B_g (g^{-1} \bk) \cdot B_{g^{-1}} (\bk)$, which yields $B_{g^{-1}} (\bk) = B_g (g^{-1} \bk)^{-1}$.
Applying this identity gives $B_{C_{3,111}^{-1}} (g \, C_{3,111} X) = B_{C_{3,111}} (g' X)^{-1}$.
Therefore, $B_g (-\bk_0) = B_{C_{3,111}} (g' X) \cdot B_{g'} (X) \cdot B_{C_{3,111}} (X)^{-1}$.
Since sewing matrices are not gauge invariant, we may choose a convenient gauge in which $B_{C_{3,111}} (g' X) = B_{C_{3,111}} (X) = \mathds{1}$.
In this gauge, we obtain $B_g (-\bk_0) = B_{g'} (X)$.
This means that the sewing matrix of $g$ at $-\bk_0$ can be derived directly from that of $g'$ at $X$, by properly identifying $g' = C_{3,111}^{-1} \, g \, C_{3,111}$.}.
Hence, we obtain
\bg
B_\mc{T} (-\bk_0) = -i \tau_0\sg_x \mc{K}
= -i \bpm 0 & 1 & 0 & 0 \\ 1 & 0 & 0 & 0 \\ 0 & 0 & 0 & 1 \\ 0 & 0 & 1 & 0 \epm \mc{K}, \quad
B_{\td{C}_{2,110}} (-\bk_0) = \tau_0 \sg_x, \quad
B_{\mc{T} \td{C}_{2,001}} (-\bk_0) = - i \tau_0 \sg_y \mc{K},
\nn
B_{\td{C}_{4,001}} (-\bk_0) = {\rm Diag} (e^{i \frac{\pi}{4}}, e^{- i \frac{\pi}{4}}, e^{- i \frac{3\pi}{4}}, e^{i \frac{3 \pi}{4}}),
\quad
B_{\mc{T} \td{C}_{2,100}} (-\bk_0) = {\rm Diag} (e^{- i \frac{3 \pi}{4}}, e^{-i \frac{\pi}{4}}, e^{i \frac{\pi}{4}}, e^{i \frac{3 \pi}{4}}) \mc{K},
\label{seq:sewing_sg212_k0}
\eg
where ${\rm Diag}$ is shorthand for a diagonal matrix.
Here, the relevant SG symmetries for deriving constraints on the Wilson loop are the little-group elements that leave $-\bk_0$ invariant up to a reciprocal lattice vector:
\ba
\td{C}_{2,110} = \{2_{110}|1/4,3/4,3/4\}:& \quad
(x,y,z) \to (y+1/4,x+3/4,-z+3/4),
\nn
\td{C}_{2,001} = \{2_{001}|1/2,0,1/2\}:& \quad
(x,y,z) \to (-x+1/2, -y, z+1/2),
\nn
\td{C}_{4,001} = \{4_{001}|3/4,1/4,3/4\}:& \quad
(x,y,z) \to (-y+3/4,x+1/4,z+3/4),
\nn
\td{C}_{2,100} = \{2_{100}|1/2,1/2,0\}:& \quad
(x,y,z) \to (x+1/2, -y+1/2,-z),
\ea
up to the possible action of time-reversal symmetry $\mc{T}$.
In all sewing matrices, the first two columns (or rows) correspond to the basis of the $X_1$ irrep, and the last two to the basis of the $X_2$ irrep.
Note that based on the group law, sewing matrices for other symmetries not listed in Eq.~\eqref{seq:sewing_sg212_k0} can be constructed by composing the listed ones.
For instance, the sewing matrix for $\td{C}_{2,001}$ is given by
\bg
B_{\td{C}_{2,001}} (-\bk_0)
= B_{\mc{T}} (\bk_0) \, B_{\mc{T} \td{C}_{2,001}} (-\bk_0)
= (- i \tau_0 \sg_x \mc{K}) (-i \tau_0 \sg_y \mc{K})
= (-i \tau_0 \sg_x)(i \tau_0 \sg_y) = i \tau_0 \sg_z.
\eg
Also note that $B_g(\bk_0) = B_g(-\bk_0)$ since $\bk_0 = -\bk_0 + \bb G_3$ and we are using the periodic gauge for sewing matrices (note that this is always possible, and is consistent with the conventions of Refs.~\cite{wieder2018wallpaper,vanderbilt2018berry,bradlyn2022lecture_notes,devescovi2024tutorial}.
In the following, we omit identity Pauli matrices such as $\tau_0$ or $\sg_0$ whenever it causes no ambiguity.

To examine how symmetry constraints affect Wilson loops in terms of sewing matrices, we apply the general symmetry transformation rule for Wilson loops~\cite{alexandradinata2014wilson,alexandradinata2016topological,benalcazar2017electric,wieder2018wallpaper,hwang2019fragile}.
For the reader's convenience, we restate Eq.~\eqref{seq:wilson_symmetry} here.
Consider a Wilson line $W_\mc{L}$ defined along a path $\mc{L}$ from a basepoint $\bk_b$ to an endpoint $\bk_f$.
A symmetry operation $g$ maps the path $\mc{L}$ to $g \mc{L}$, with corresponding basepoint $g \bk_b$ and endpoint $g \bk_f$.
The Wilson lines $W_\mc{L}$ and $W_{g \mc{L}}$ are related by the sewing matrices evaluated at the basepoint and endpoint of $\mc{L}$.
Let $g$ be defined by the data $O_g \in O(3)$, $\bb v_g \in \Q^3$, and $\tau_g \in \{0,1\}$ as in SN~\ref{sec:sn_sym_sym}.
Then the transformation rule is
\bg
W_{g \mc{L}} = e^{i (-1)^{\tau_g} \, (\bk_f - \bk_b) \cdot [O_g]^{-1} \cdot \bb v_g} \, B_g(\bk_f) \cdot W_C \cdot B_g(\bk_b)^{-1}.
\label{seq:wilson_symmetry1}
\eg

According to Eq.~\eqref{seq:wilson_symmetry1}, the almost-straight Wilson loop $W_{r_\ep}(\phi; -\bk_0)$ at $\phi = \frac{\pi}{4}$ satisfies the following symmetry constraints by $\td{C}_{2,110}$ and $\mc{T} \td{C}_{2,001}$:
\bg
[ W_{r_\ep} (\tfrac{\pi}{4}; -\bk_0) ]^{-1}
= e^{-i \frac{3\pi}{2}} B_{\td{C}_{2,110}} (\bk_0) \cdot W_{r_\ep} (\tfrac{\pi}{4}; -\bk_0) \cdot [B_{\td{C}_{2,110}} (\bk_0)]^{-1},
\label{seq:sg212_wilson_sewing1}
\\
[ W_{r_\ep} (\tfrac{\pi}{4}; -\bk_0) ]^{-1} = e^{-i \pi} B_{\mc{T} \td{C}_{2,001}} (\bk_0) \cdot W_{r_\ep} (\tfrac{\pi}{4}; -\bk_0) \cdot [B_{\mc{T} \td{C}_{2,001}} (\bk_0)]^{-1},
\label{seq:sg212_wilson_sewing2}
\eg
These constraints arise from how the symmetries $\td{C}_{2,110}$ and $\mc{T} \td{C}_{2,001}$ act on the Wilson loop path.
Let us first consider the symmetry operation $g = \td{C}_{2,110}$ and the Wilson loop path $\mc{L}$ associated with $W_\mc{L} = W_{r_\ep} (\tfrac{\pi}{4}; -\bk_0)$.
The operation $g$ maps the path $\mc{L}$, which runs from $-\bk_0$ to $\bk_0$, to the reversed path $g \mc{L}$, running from $\bk_0$ to $-\bk_0$.
Since the geometric shape of $\mc{L}$ is preserved, but the direction is reversed, the resulting Wilson line $W_{g \mc{L}}$ is the inverse of $W_\mc{L}$.
Next, we compute the phase factor in Eq.~\eqref{seq:wilson_symmetry1}.
The operation $g = \td{C}_{2,110} = \{2_{110} | 1/4, 3/4, 3/4 \}$ consists of a rotational part $O_g$, satisfying $O_g^{-1} \cdot (x,y,z) = (y,x,-z)$, and a fractional translation by $\bb v_g = (1/4, 3/4, 3/4)$.
Thus, the phase factor becomes
\bg
e^{i (2\bk_0) \cdot (3/4,1/4,-3/4)}
= e^{i (0,0, 2\pi) \cdot (3/4,1/4,-3/4)} = e^{-i \frac{3\pi}{2}}.
\eg
Therefore, by applying Eq.~\eqref{seq:wilson_symmetry1}, we find $[ W_{r_\ep} (\tfrac{\pi}{4}; -\bk_0) ]^{-1} = e^{-i \frac{3\pi}{2}} B_{\td{C}_{2,110}} (\bk_0) \cdot W_{r_\ep} (\tfrac{\pi}{4}; -\bk_0) \cdot [B_{\td{C}_{2,110}} (-\bk_0)]^{-1}$.
Finally, using the periodicity of sewing matrices in the BZ, we identify $B_{\td{C}_{2,110}} (\bk_0) = B_{\td{C}_{2,110}} (-\bk_0)$ and obtain Eq.~\eqref{seq:sg212_wilson_sewing1}.
A similar argument applies to the antiunitary symmetry $\mc{T} \td{C}_{2,001}$, leading to the second constraint in Eq.~\eqref{seq:sg212_wilson_sewing2}.

Now let us analyze what conditions are imposed by the sewing matrix constraints in Eqs.~\eqref{seq:sg212_wilson_sewing1}–\eqref{seq:sg212_wilson_sewing2}.
To this end, we focus on the constraints at $\phi = \frac{\pi}{4}$ and simplify the notation as follows: $W = W_{r_\ep} (\tfrac{\pi}{4}; -\bk_0)$, $U_1 = B_{\td{C}_{2,110}} (\bk_0) = \sg_x$, and $U_2 \mc{K} = B_{\mc{T} \td{C}_{2,001}} (\bk_0) = - i \sg_y \mc{K}$.
Note that both $U_1$ and $U_2$ are unitary matrices.
Under these definitions, the symmetry constraints become $W^{-1} = i U_1 \cdot W \cdot U_1^{-1}$ and $W^{-1} = - U_2 \mc{K} \, W \, \mc{K} U_2^{-1} = - U_2 \cdot \cm{W} \cdot U_2^{-1}$, where $\cm{x}$ denotes the complex conjugation of $x$.
Let $\ket{\Theta}$ be an eigenstate of $W$ with eigenvalue $e^{i\Theta}$: $W \ket{\Theta} = e^{i \Theta} \ket{\Theta}$.
Then, under the first constraint from $U_1$, we find
\bg
W [U_1^{-1} \ket{\Theta}] = -i U_1^{-1} \cdot W^{-1} \ket{\Theta} = e^{-i \Theta - i \frac{\pi}{2}} [U_1^{-1} \ket{\Theta}],
\eg
so $U_1^{-1} \ket{\Theta}$ is also an eigenstate of $W$, with eigenvalue $e^{-i \Theta - i \frac{\pi}{2}}$.
This implies that the spectrum of $W$,
\bg
{\rm Spec} [-i \log W] := \{\Theta\},
\eg
satisfies a spectral symmetry given by $\{\Theta\} = \{-\Theta - \frac{\pi}{2}\}$.
More precisely, for the eigenphase set $\{\Theta\} = \{\Theta_a | a=1,\dots,N_W\}$, where $N_W$ is the size of $W$, the symmetry constraint requires that
\bg
\exists \, b \text{ such that } \Theta_b = - \Theta_a - \frac{\pi}{2} \text{ for all } a.
\eg
Now consider the second constraint from $U_2$.
Starting again from $W \ket{\Theta} = e^{i \Theta} \ket{\Theta}$, we compute
\bg
W [\cm{U_2^{-1} \ket{\Theta}}] = (W \cm{U_2^{-1}}) \, \cm{\ket{\Theta}} = - (\cm{U_2^{-1} \cdot W^{-1}}) \, \cm{\ket{\Theta}} = - \cm{U_2^{-1} \cdot W^{-1} \ket{\Theta}} = e^{i \left( \Theta + \pi \right)} \cm{U_2^{-1} \ket{\Theta}},
\eg
where the overbar denotes complex conjugation.
This implies that the spectrum is invariant under a $\pi$-shift: $\{\Theta\} = \{\Theta + \pi\}$.
Combining the two constraints from $U_1$ and $U_2$, we conclude that the spectrum of Wilson loop $W_{r_\ep} (\pi/4; -\bk_0)$ satisfies
\bg
\{\Theta\} = \left\{-\Theta + \frac{\pi}{2} \right\} = \{\Theta+\pi\}.
\eg
Later, we will show that by choosing a different basepoint $\bk_b = \frac{1}{\sqrt{2}} (r_\ep, r_\ep, 0)$, the Wilson loop spectrum becomes quantized as
\bg
{\rm Spec} [-i \log W_{r_\ep} (\tfrac{\pi}{4}; \bk_b)] = \left\{ -\frac{3\pi}{4}, -\frac{\pi}{4}, +\frac{\pi}{4}, +\frac{3\pi}{4} \right\}
\eg
for arbitrary $r_\ep$ [See the discussion around Eq.~\eqref{seq:sg212_wilson_M_phases}].

Not only do the sewing constraints in Eqs.~\eqref{seq:sg212_wilson_sewing1} and \eqref{seq:sg212_wilson_sewing2} enforce symmetry on the Wilson loop spectrum, they also allow us to determine an analytic form of the Wilson loop matrix $W = W_{r_\ep} (\tfrac{\pi}{4}; -\bk_0)$.
Using the fact that $W^{-1} = W^\dg$ since $W$ is unitary, Eq.~\eqref{seq:sg212_wilson_sewing2} implies that $W^\dg = - \sg_y \cm{W} \sg_y$, i.e. $W^T = -\sg_y W \sg_y$.
Thus, we can express $W$ as
\ba
W =& (a_1 + i b_1) \sg_x + (a_2 + i b_2) \sg_y + (a_3 + i b_3) \sg_z + (a_4 + i b_4) \tau_x \sg_x + (a_5 + i b_5) \tau_x \sg_y 
\nn
&+ (a_6 + i b_6) \tau_x \sg_z + (a_7 + i b_7) \tau_y + (a_8 + i b_8) \tau_z \sg_x + (a_9 + i b_9) \tau_z \sg_y + (a_{10} + i b_{10}) \tau_z \sg_z
\ea
with real-valued parameters $a_{1,\dots,10}$ and $b_{1,\dots,10}$ which implicitly depend on $r_\ep$ and $\bk_0$.
Next, substituting this general form into Eq.~\eqref{seq:sg212_wilson_sewing1}, i.e. the constraint $W^\dg = i \sg_x W \sg_x$, we find that all coefficients not consistent with this relation must vanish.
The remaining nonzero parameters can be grouped into ten real parameters $t_{1,\dots,10}$, leading to the following simplified form:
\bg
W_{r_\ep} (\tfrac{\pi}{4}; -\bk_0)
= e^{i \frac{\pi}{4}} \, \bpm
t_1 & - i t_7 & t_4 & -i t_5 \\
-i t_8 & -t_1 & -i t_6 & -t_3 \\
t_3 & -i t_5 & t_2 & -i t_9 \\
-i t_6 & -t_4 & -i t_{10} & -t_2
\epm.
\label{seq:sg212_wilson_M}
\eg
Here, $t_{1,\dots,10}$ encode the dependence on the radius $r_\ep$ and the basepoint $\bk_0$, as before.
Moreover, the unitarity of $W$, namely $W^\dg \cdot W = \mathds{1}_4$, imposes additional constraints on these parameters.
From explicit computation, we obtain
\bg
t_1^2 + t_3^2 + t_6^2 + t_8^2 = t_1^2 + t_4^2 + t_5^2 + t_7^2
= t_2^2 + t_4^2 + t_6^2 + t_{10}^2 = t_2^2 + t_3^2 + t_5^2 + t_9^2 = 1,
\nn
t_1 t_7 + t_3 t_5 + t_1 t_8 + t_4 t_6 = t_1 t_4 + t_2 t_3 + t_6 t_8 + t_6 t_{10} = t_1 t_5 + t_2 t_6 + t_3 t_8 + t_3 t_9
\nn
= t_1 t_6 + t_2 t_5 + t_4 t_7 + t_4 t_{10} = t_1 t_3 + t_2 t_4 + t_5 t_7 + t_5 t_9 = t_3 t_6 + t_4 t_5 + t_2 t_9 + t_2 t_{10} =0.
\label{seq:sg212_wilson_M_cond}
\eg
%%%%%%

%%%%%%
\subsubsection{Basepoint $\frac{1}{\sqrt{2}} (r_\ep, r_\ep, 0)$ and azimuthal angle $\frac{\pi}{4}$}
%%%%%%
The goal of this subsection is to show that the eigenvalue spectrum of $-i \log W_{r_\ep} (\phi=\frac{\pi}{4}; \bk_b)$ is fixed to $\{-\frac{3\pi}{4}, -\frac{\pi}{4}, + \frac{\pi}{4}, +\frac{3\pi}{4}\}$, which is the result already summarized in Eq.~\eqref{seq:sg212_wl_summary1}.
For this purpose, we consider a basepoint in the $(k_x,k_y)$ plane, defined as $\bk_\ep (\phi) = r_\ep (\cos \phi, \sin \phi, 0)$ (see Fig.~\ref{sfig:wilson_path}\textbf{b}).
At $\phi = \frac{\pi}{4}$, this basepoint becomes $\bk_\ep (\frac{\pi}{4}) = \tfrac{1}{\sqrt{2}} (r_\ep, r_\ep, 0)$, and the sewing matrices can be deduced from the irreps at the $\Sigma = (u,u,0)$ line, where $u$ is a free real parameter.
The HSL connecting $\Gamma$ to $\Sigma$ involves the following irrep compatibility relations:
\bg
\Gamma_4 - \Gamma_1 \to \Sigma_2 + \Sigma_2,
\quad
\Gamma_3 \to \Sigma_1 + \Sigma_2.
\eg
In our model, the $\Sigma_1$ irrep that descends from $\Gamma_3$ lies lower in energy than the $\Sigma_2$ irrep coming from the same parent $\Gamma_3$ irrep.
Using this information, we obtain the sewing matrices as\footnote{Note that the sign of the exponent in the phase factor involving $r_\ep$ differs from that in the BCS, due to a difference in the convention for representing translation symmetry.
This point has been discussed in the context of the compatibility relations; see the discussion around Eq.~\eqref{seq:delta_irreps}.}
\bg
B_{\td{C}_{2,110}} \big( \bk_\ep(\tfrac{\pi}{4}) \big) = e^{-i \frac{r_\ep}{\sqrt{2}}} \, {\rm Diag} (-1, -1, +1, -1), \quad
B_{\mc{T} \td{C}_{2,001}} \big( \bk_\ep(\tfrac{\pi}{4}) \big) = - \mathds{1}_4 \mc{K}.
\label{seq:sewing_sg212_M}
\eg

The corresponding sewing matrix constraints are then given by
\bg
[ W_{r_\ep} \left( \tfrac{\pi}{4}; \bk_\ep (\tfrac{\pi}{4}) \right) ]^\dg
= e^{-i \frac{3\pi}{2}} B_{\td{C}_{2,110}} \left( \bk_\ep (\tfrac{\pi}{4}) \right) \cdot W_{r_\ep} \left( \tfrac{\pi}{4}; \bk_\ep (\tfrac{\pi}{4}) \right) \cdot [ B_{\td{C}_{2,110}} \left( \bk_\ep (\tfrac{\pi}{4}) \right) ]^{-1},
\label{seq:sg212_wilson_sewing3}
\\
[ W_{r_\ep} \left( \tfrac{\pi}{4}; \bk_\ep (\tfrac{\pi}{4}) \right) ]^\dg
= e^{-i \pi} B_{\mc{T} \td{C}_{2,001}} \left( \bk_\ep (\tfrac{\pi}{4}) \right) \cdot W_{r_\ep} \left( \tfrac{\pi}{4}; \bk_\ep(\tfrac{\pi}{4}) \right) \cdot [ B_{\mc{T} \td{C}_{2,001}} \left( \bk_\ep (\tfrac{\pi}{4}) \right) ]^{-1}.
\label{seq:sg212_wilson_sewing4}
\eg
These constraints imply that the Wilson loop matrix $W = W_{r_\ep} \left( \tfrac{\pi}{4}; \bk_\ep (\tfrac{\pi}{4}) \right)$ must take the form
\bg
\frac{1}{\sqrt{2}} \left[ (a_1 - i a_6) \sg_y + (a_2 - i a_4) \tau_x \sg_y + (a_3 + i a_5) \tau_y + (a_4 - i a_2) \tau_y \sg_x + (a_5 + i a_3) \tau_y \sg_z + (a_6 - i a_1) \tau_z \sg_y \right],
\eg
with real-valued parameters $a_{1,\dots,6}$.
We now impose the unitarity constraint $W^\dg \cdot W = \mathds{1}_4$.
Note that the determinant is given by ${\rm Det} W = (a_1^2 - a_2^2 - a_3^2 + a_4^2 +a_5^2 - a_6^2)^2\equiv v^2$.
Since $W$ is unitary, we must have ${\rm Det} W \in U(1)$.
Noting that the determinant is the square of $v=a_1^2 - a_2^2 - a_3^2 + a_4^2 +a_5^2 - a_6^2$ and that all $a_{1,\dots,6}$ are real, we must have $v\in\mathbb{R}$ and hence $\mathrm{Det} W$ is a positive real number.
Because $+1$ is the only positive real number in $U(1)$, we obtain ${\rm Det} W = v^2 = +1$, so $v$ must be either $+1$ or $-1$.
Then, the unitarity condition yields
\bg
(a_1 + a_6) v = a_1-a_6, \quad
(a_3 + a_5) v = a_5-a_3, \quad
(a_2 + a_4) v = a_4-a_2.
\eg

We solve the above constraints for the two cases $v = \pm 1$ separately.
(i) When $v=1$, $a_2=a_3=a_6=0$ and $v=a_1^2 + a_4^2 + a_5^2=1$.
Thus, $(a_1, a_4, a_5) \in S^2$, the unit 2-sphere, and can be parameterized as $a_1 = \cos \eta$, $a_4 = \cos \psi \sin \eta$, and $a_5 = \sin \psi \sin \eta$.
(ii) When $v=-1$, $a_1=a_4=a_5=0$ and $v=a_2^2 + a_3^2 + a_6^2=1$.
Now, $(a_2, a_3, a_6) \in S^2$ and can be parameterized as $a_2 = \cos \psi' \sin \eta'$, $a_3 = \sin \psi' \sin \eta'$, and $a_6 = \cos \eta'$.
For both cases $v=+1$ and $v=-1$, the spectrum is given by
\bg
{\rm Spec} [-i \log W_{r_\ep} \left( \tfrac{\pi}{4}; \bk_b \right)] = \left\{ -\frac{3\pi}{4}, -\frac{\pi}{4}, +\frac{\pi}{4}, +\frac{3\pi}{4} \right\},
\label{seq:sg212_wilson_M_phases}
\eg
independent of the values of the angle parameters $(\eta, \eta', \psi, \psi')$.
Here we have replaced the original basepoint $\bk_\ep (\frac{\pi}{4})$ by an arbitrary basepoint $\bk_b$ because the Wilson loop spectrum is invariant under the choice of basepoint.
%%%%%%

%%%%%%
\subsubsection{Basepoint $(0,0,-\pi)$ and azimuthal angle $\frac{\pi}{2}$}
%%%%%%
The goal of this subsection is to derive the analytic expression of $W_{r_\ep} (\phi;- \bk_0)$ at $\phi = \frac{\pi}{2}$ and to determine the resulting eigenvalue spectrum of $-i \log W_{r_\ep} (\phi;\bk_b)$, as summarized in Eq.~\eqref{seq:sg212_wl_summary2}.
For this purpose, we take the basepoint $-\bk_0 = (0,0,-\pi)$, for which we can use the sewing matrices provided in Eq.~\eqref{seq:sewing_sg212_k0}.
Specifically, we utilize
\bg
B_{\mc{T} \td{C}_{2,100}}(-\bk_0)
= {\rm Diag} (e^{-i \frac{3\pi}{4}}, e^{-i \frac{\pi}{4}}, e^{i \frac{\pi}{4}}, e^{i \frac{3 \pi}{4}}) \mc{K},
\quad
B_{\mc{T} \td{C}_{2,001}}(-\bk_0) = - i \sg_y \mc{K}.
\eg
These matrices impose the following constraints:
\bg
W_{r_\ep} (\tfrac{\pi}{2}; -\bk_0)
= B_{\mc{T} \td{C}_{2,100}} (\bk_0) \cdot W_{r_\ep} (\tfrac{\pi}{2}; -\bk_0) \cdot [ B_{\mc{T} \td{C}_{2,100}} (\bk_0) ]^{-1},
\label{seq:sg212_wilson_sewing5}
\\
[ W_{r_\ep} (\tfrac{\pi}{2}; -\bk_0) ]^\dg
= e^{-i \pi} B_{\mc{T} \td{C}_{2,001}} (\bk_0) \cdot W_{r_\ep}(\tfrac{\pi}{2}; -\bk_0) \cdot [ B_{\mc{T} \td{C}_{2,001}} (\bk_0) ]^{-1}.
\label{seq:sg212_wilson_sewing6}
\eg
(Recall that sewing matrices are periodic in the BZ in the periodic gauge.)
Using simplified notation where $W = W_{r_\ep}(\tfrac{\pi}{2}; -\bk_0)$, $U_3 = {\rm Diag} (e^{-i \frac{3\pi}{4}}, e^{-i \frac{\pi}{4}}, e^{i \frac{\pi}{4}}, e^{i \frac{3 \pi}{4}})$, and $U_4 = -i \sg_y$, these can be expressed as
\bg
W = U_3 \cdot \cm{W} \cdot U_3^{-1},
\quad
W^\dg = - U_4 \cdot \cm{W} \cdot U_4^{-1}.
\eg

Now consider an eigenstate $\ket{\Theta}$ satisfying $W \, \ket{\Theta} = e^{i \Theta}$.
Then, we can rewrite the above equations as $W \cdot U_3 \, \cm{\ket{\Theta}} = U_3 \cdot \cm{W \, \ket{\Theta}} = e^{-i \Theta} U_3 \cm{\ket{\Theta}}$ and $W \cdot \cm{U_4^{-1} \, \ket{\Theta}} = -\cm{U_4^{-1} \cdot W^\dg \, \ket{\Theta}} = e^{i \Theta + i \pi} \ket{\Theta}$.
Therefore, the Wilson loop spectrum must satisfy
\bg
{\rm Spec} [-i \log W_{r_\ep} (\tfrac{\pi}{2}, \bk_b)] = \{\Theta\} = \{-\Theta\} = \{ \Theta + \pi\}.
\eg
for any basepoint $\bk_b$.
At finite $r_\ep$, the numerically computed Wilson loop spectrum in Fig.~\ref{sfig:sg212_wilson}\textbf{a} indicates that
\bg
{\rm Spec} [- i \log W_{r_\ep} (\tfrac{\pi}{2}; \bk_b)]
= \{ -\Theta_0 - \frac{\pi}{2}, \Theta_0 - \frac{\pi}{2}, -\Theta_0 + \frac{\pi}{2}, \Theta_0 + \frac{\pi}{2} \}
\label{seq:sg212_wl_X_angle}
\eg
for some generic angle parameter $\Theta_0 \in (-\pi, \pi]$.
This coincides with Eq.~\eqref{seq:sg212_wl_summary2}.

To further constrain the structure of the Wilson loop matrix analytically, we now solve Eqs.~\eqref{seq:sg212_wilson_sewing5} and \eqref{seq:sg212_wilson_sewing6}.
These equations imply that the Wilson loop $W_{r_\ep} (\tfrac{\pi}{2}; -\bk_0)$ must take the following form:
\bg
W_{r_\ep} (\tfrac{\pi}{2}; -\bk_0) = \bpm
t'_1 & e^{-i \frac{\pi}{4}} \, t'_7 & -i t'_4 & e^{i \frac{\pi}{4}} \, t'_5 \\
-e^{i \frac{\pi}{4}} \, t'_8 & - t'_1 & -e^{-i \frac{\pi}{4}} \, t'_6 & -i t'_3 \\
i t'_3 & e^{i \frac{\pi}{4}} \, t'_5 & t'_2 & -e^{-i \frac{\pi}{4}} \, t'_9 \\
-e^{-i \frac{\pi}{4}} \, t'_6 & i t'_4 & e^{i \frac{\pi}{4}} \, t'_{10} & -t'_2
\epm,
\label{seq:sg212_wilson_X}
\eg
with real-valued parameters $t'_{1,\dots,10}$.
The unitarity condition $W^\dg \cdot W = \mathds{1}_4$ imposes further constraints:
\bg
{t'_1}^2 + {t'_3}^2 + {t'_6}^2 + {t'_8}^2 = {t'_1}^2 + {t'_4}^2 + {t'_5}^2 + {t'_7}^2
= {t'_2}^2 + {t'_4}^2 + {t'_6}^2 + {t'_{10}}^2 = {t'_2}^2 + {t'_3}^2 + {t'_5}^2 + {t'_9}^2 = 1,
\nn
t'_1 t'_7 + t'_3 t'_5 + t'_1 t'_8 + t'_4 t'_6 = t'_1 t'_4 + t'_2 t'_3 + t'_6 t'_8 + t'_6 t'_{10} = t'_1 t'_5 + t'_2 t'_6 + t'_3 t'_8 + t'_3 t'_9
\nn
= t'_1 t'_6 + t'_2 t'_5 + t'_4 t'_7 + t'_4 t'_{10} = t'_1 t'_3 + t'_2 t'_4 + t'_5 t'_7 + t'_5 t'_9 = t'_3 t'_6 + t'_4 t'_5 + t'_2 t'_9 + t'_2 t'_{10} =0.
\label{seq:sg212_wilson_X_cond}
\eg
These are identical in form to Eq.~\eqref{seq:sg212_wilson_M_cond}, except that the parameters $t_{1,\dots,10}$ are replaced by $t'_{1,\dots,10}$.
%%%%%%

%%%%%%
\subsubsection{Other important symmetry constraints}
%%%%%%
Here, we present additional symmetry information regarding the Wilson loop spectrum, which will be useful in later analyses.
%%%%%%

%%%%%%
\paragraph{Wilson loops at other $\phi$.}
%%%%%%
The Wilson loop $W_{r_\ep} (\phi; -\bk_0)$ at other high-symmetry azimuthal angles $\phi$, such as $0, \frac{3\pi}{4}, \pi$, etc., can be obtained from those at $\phi=\frac{\pi}{4}$ and $\frac{\pi}{2}$ by applying the symmetry $\td{C}_{4,001}$.
This symmetry imposes the constraint
\bg
W_{r_\ep} (\phi + \tfrac{\pi}{2}; -\bk_0) =
e^{i \frac{3 \pi}{2}} B_{\td{C}_{4,001}} (\bk_0) \cdot W_{r_\ep} (\phi; -\bk_0) \cdot [B_{\td{C}_{4,001}} (\bk_0)]^{-1},
\label{seq:sg212_sewing_c4}
\eg
where $B_{\td{C}_{4,001}} (\bk_0) = {\rm Diag} (e^{i \frac{\pi}{4}}, e^{- i \frac{\pi}{4}}, e^{- i \frac{3\pi}{4}}, e^{i \frac{3 \pi}{4}})$, as given in Eq.~\eqref{seq:sewing_sg212_k0}.
This implies that the Wilson loop spectrum, ${\rm Spec} [-i \log W_{r_\ep} (\phi; -\bk_0)] := \{\Theta(\phi)\}$, satisfies $\{\Theta(\phi + \tfrac{\pi}{2})\} = \left\{ \Theta(\phi) + \frac{3\pi}{2} \right\}$.
From this, we infer the following structure of the Wilson loop spectrum:
\ba
& {\rm Spec} [- i \log W_{r_\ep} (\phi; \bk_b)]
= \{ -\Theta_0, \Theta_0, -\Theta_0 + \pi, \Theta_0 + \pi \}
\quad \text{at $\phi = 0$ and $\pi$},
\nn
& {\rm Spec} [- i \log W_{r_\ep} (\phi; \bk_b)]
= \{ -\Theta_0 - \tfrac{\pi}{2}, \Theta_0 - \tfrac{\pi}{2}, -\Theta_0 + \tfrac{\pi}{2}, \Theta_0 + \tfrac{\pi}{2} \}
\quad \text{at $\phi = \tfrac{\pi}{2}$ and $\tfrac{3\pi}{2}$},
\nn
& {\rm Spec} [- i \log W_{r_\ep} (\phi; \bk_b)]
= \{ -\tfrac{3\pi}{4}, -\tfrac{\pi}{4}, +\tfrac{\pi}{4}, +\tfrac{3\pi}{4} \}
\quad \text{at $\phi = \tfrac{\pi}{4}, \tfrac{3\pi}{4}, \tfrac{5\pi}{4}$, and $\tfrac{7\pi}{4}$},
\ea
for any basepoint $\bk_b$.
Furthermore, applying Eqs.~\eqref{seq:sg212_sewing_c4} to \eqref{seq:sg212_wilson_X}, we obtain an explicit expression for the Wilson loop at $\phi=0$:
\ba
W_{r_\ep} (\phi=0; -\bk_0)
=& e^{- i \frac{3 \pi}{2}} \, [ B_{\td{C}_{4,001}} (\bk_0) ]^{-1} \cdot W_{r_\ep} (\phi = \tfrac{\pi}{2}; -\bk_0) \cdot B_{\td{C}_{4,001}} (\bk_0)
\nn
=& \bpm
i t'_1 & e^{-i \frac{\pi}{4}} \, t'_7 & -t'_4 & -e^{i \frac{\pi}{4}} \, t'_5 \\
e^{i \frac{\pi}{4}} \, t'_8 & -i t'_1 & -e^{-i \frac{\pi}{4}} \, t'_6 & -t'_3 \\
t'_3 & -e^{i \frac{\pi}{4}} \, t'_5 & i t'_2 & -e^{-i \frac{\pi}{4}} \, t'_9 \\
-e^{-i \frac{\pi}{4}} \, t'_6 & t'_4 & -e^{i \frac{\pi}{4}} \, t'_{10} & -i t'_2
\epm.
\label{seq:sg212_wilson_Xp}
\ea
%%%%%%

%%%%%%
\paragraph{Spectral symmetries of the Wilson loop.}
%%%%%%
Below, we provide a list of symmetries that enforce spectral symmetries in the Wilson loop spectrum.
\ba
\td{C}_{2,100} =& \{ 2_{100}| 1/2, 1/2, 0 \} : \,
\{ \Theta (-\phi) \} = \{ -\Theta(\phi) \},
\quad
\mc{T} : \,
\{ \Theta (\phi+ \pi) \} = \{ \Theta (\phi) \},
\nn
\td{C}_{2,010} =& \{ 2_{010}| 0, 1/2, 1/2 \} : \,
\{ \Theta (\pi - \phi) \} = \{ -\Theta (\phi) + \pi \},
\nn
\td{C}_{4,001} =& \{ 4_{001}| 3/4, 1/4, 3/4 \} : \,
\left\{ \Theta (\phi + \tfrac{\pi}{2}) \right\} = \left\{ \Theta(\phi) + \tfrac{3 \pi}{2} \right\},
\nn
\td{C}_{2,110} =& \{ 2_{110}| 1/4, 3/4, 3/4 \} : \,
\left\{ \Theta (\tfrac{\pi}{2} - \phi) \right\} = \left\{ -\Theta (\phi) - \tfrac{\pi}{2} \right\}.
\label{seq:sg212_wilson_sym}
\ea
%%%%%%

%%%%%%
\paragraph{Sewing matrices at $\bk_+ = (0,0,r_\ep)$.}
%%%%%%
To determine the sewing matrices at $\bk_+ = (0,0,r_\ep)$, we utilize the irrep information along the $\Delta$ line, parameterized by $\bk = (0,u,0)$ where $u$ is a continuous real parameter.
Since $\bk_+ = (0, 0, r_\ep)$ is related to the point $(0, r_\ep, 0)$ on the $\Delta$ line by a threefold rotation $C_{3,111}$, the sewing matrices at $\bk_+$ can be obtained from those at $(0, r_\ep, 0)$.
Because $r_\ep$ is infinitesimal (or smaller than the $k_y$ position of Weyl point illustrated in Fig.~\ref{sfig:sg212_bands}\textbf{d}), the four physical bands at $(0, r_\ep, 0)$ correspond to the irreps $\Delta_4$, $\Delta_3$, $\Delta_2$, and $\Delta_1$, ordered by increasing energy.
(See Fig.~\ref{sfig:sg212_bands}\textbf{d}.)

The sewing matrices at $\bk_+$ for symmetries that leave $\bk_+$ invariant modulo a reciprocal lattice vector are
\bg
B_{\td{C}_{4,001}} (\bk_+) = e^{-i \frac{3 r_\ep}{4}} \, {\rm Diag} (i, -i, -1, 1),
\quad
B_{\td{C}_{2,001}} (\bk_+) = e^{-i \frac{r_\ep}{2}} \, {\rm Diag} (-1, -1, 1, 1),
\nn
B_{\mc{T} \td{C}_{2,100}} (\bk_+) = {\rm Diag} (i, -i, 1, -1) \mc{K}.
\label{seq:sewing_sg212_kp1}
\eg
Here, the column and row bases follow the ordering of irreps $(\Delta_4, \Delta_3, \Delta_2, \Delta_1)$ (Note that the sewing matrix for $\mc{T} \td{C}_{2,100}$ does not follow the BCS convention.
This choice is made to simplify later calculations).
Next, we consider symmetries that exchange $\bk_+$ and $\bk_- = -\bk_+$.
For these symmetries, we choose the following gauge for the sewing matrices\footnote{Sewing matrices are gauge dependent [Eq.~\eqref{seq:sewing_gauge}], so multiple choices exist.
Here we fix all sewing matrices defined in this SN to ensure that all group multiplication relations [Eq.~\eqref{seq:sewing_multi}] are satisfied consistently.}:
\bg
B_\mc{T} (\bk_+) = {\rm Diag} (1, -1, 1, -1) \mc{K},
\quad
B_\mc{T} (\bk_-) = {\rm Diag} (1, -1, 1, -1) \mc{K},
\quad
B_{\td{C}_{2,100}} (\bk_+) = {\rm Diag} (-i, -i, 1, 1),
\nn
B_{\td{C}_{2,100}} (\bk_-) = {\rm Diag} (i, i, 1, 1),
\quad
B_{\td{C}_{2,110}} (\bk_+) = - e^{i \frac{3 r_\ep}{4}} \, \tau_0 \sg_z,
\quad
B_{\td{C}_{2,110}} (\bk_-) = - e^{-i \frac{3 r_\ep}{4}} \, \tau_0 \sg_z.
\label{seq:sewing_sg212_kp2}
\eg
The sewing matrices at $\bk_+$ will later be used to derive constraints from the polarization singularity.
%%%%%%

%%%%%%
\subsubsection{Angle dependence and polarization singularity}
%%%%%%
We now combine the symmetry analysis obtained so far with additional constraints arising from the polarization singularity.
To analyze the effect of the polarization singularity, it is useful to decompose the almost-straight Wilson loop $W_{r_\ep} (\bk; \bk_b)$ into a product of several Wilson lines.
As illustrated in Fig.~\ref{sfig:wilson_path}\textbf{b}, we define three Wilson lines: $W_{\bk_0 \leftarrow \bk_+}$, $W_{\bk_+ \leftarrow \bk_-} (\phi)$, and $W_{\bk_- \leftarrow -\bk_0}$.
The Wilson lines $W_{\bk_0 \leftarrow \bk_+}$ and $W_{\bk_- \leftarrow -\bk_0}$ follow straight paths along the $k_z$ direction.
The Wilson line $W_{\bk_+ \leftarrow \bk_-} (\phi)$ takes the semicircular arc that detours around the polarization singularity.
Thus, only this segment encodes the angle dependence on $\phi$.

For the basepoint $\bk_b = -\bk_0$, the Wilson loop takes the form:
\bg
W_{r_\ep} (\phi; -\bk_0) = W_{\bk_0 \leftarrow \bk_+} \cdot W_{\bk_+ \leftarrow \bk_-} (\phi) \cdot W_{\bk_- \leftarrow -\bk_0}.
\label{seq:wilson_decomp}
\eg
To relate Wilson loops for different azimuthal angles, observe that for a reference angle $\phi_0$, we can write $W_{\bk_- \leftarrow - \bk_0} = W_{\bk_+ \leftarrow \bk_-} (\phi_0)^{-1} \cdot {W_{\bk_0 \leftarrow \bk_+}}^{-1} \cdot W_{r_\ep} (\phi_0; -\bk_0)$.
Substituting into Eq.~\eqref{seq:wilson_decomp}, we obtain
\bg
W_{r_\ep} (\phi; -\bk_0)
= W_{\bk_0 \leftarrow \bk_+} \cdot [W_{\bk_+ \leftarrow \bk_-} (\phi) \cdot W_{\bk_+ \leftarrow \bk_-} (\phi_0)^{-1}] \cdot {W_{\bk_0 \leftarrow \bk_+}}^{-1} \cdot W_{r_\ep} (\phi_0; -\bk_0).
\label{seq:wilson_compare}
\eg
This relation shows that $W_{r_\ep} (\phi; -\bk_0)$ can be determined from $W_{r_\ep} (\phi_0; -\bk_0)$, given knowledge of $W_{\bk_0 \leftarrow \bk_+}$ and $W_{\bk_+ \leftarrow \bk_-} (\phi) \cdot W_{\bk_+ \leftarrow \bk_-} (\phi_0)^{-1}$.

Importantly, the straight segment $W_{\bk_0 \leftarrow \bk_+}$ is completely fixed up to four undetermined sign factors $s_{1,2,3,4}$ appearing in Eq.~\eqref{seq:wilson_sg212_straight}.
On the other hand, the relative contribution from the semicircular segments, expressed as $W_{\bk_+ \leftarrow \bk_-} (\phi) \cdot W_{\bk_+ \leftarrow \bk_-} (\phi_0)^{-1}$, cannot in general be determined solely from symmetry or irrep data.
Nevertheless, in the $r_\ep \to 0$ limit, the singular behavior induced by the polarization singularity allows us to constrain this term at specific angles $\phi = \frac{n\pi}{4}$ with $n \in \Z$.
Below, we first determine $W_{\bk_0 \leftarrow \bk_+}$ and then derive the approximate form of the angle-dependent segment $W_{\bk_+ \leftarrow \bk_-} (\phi) \cdot W_{\bk_+ \leftarrow \bk_-} (\phi_0)^{-1}$ in the $r_\ep \to 0$ limit.
%%%%%%

%%%%%%
\paragraph{Straight segment $W_{\bk_0 \leftarrow \bk_+}$.}
%%%%%%
The symmetries $\td{C}_{4,001} = \{4_{001}|3/4, 1/4, 3/4\}$ and $\mc{T} \td{C}_{2,100} = \mc{T} \{2_{100}|1/2,1/2,0\}$ constrain the form of the Wilson line $W_{\bk_0 \leftarrow \bk_+}$.
To apply the symmetry transformation rule in Eq.~\eqref{seq:wilson_symmetry1}, we need the sewing matrices of these symmetries at both the basepoint $\bk_+$ and endpoint $\bk_0$ of the Wilson line.
They are already given in Eqs.~\eqref{seq:sewing_sg212_k0} and \eqref{seq:sewing_sg212_kp1}, and are restated here for clarity:
\bg
B_{\td{C}_{4,001}} (\bk_0) = {\rm Diag} (e^{i \frac{\pi}{4}}, e^{- i \frac{\pi}{4}}, e^{- i \frac{3\pi}{4}}, e^{i \frac{3 \pi}{4}}),
\quad
B_{\td{C}_{4,001}} (\bk_+) = e^{-i \frac{3 r_\ep}{4}} \, {\rm Diag} (i, -i, -1, 1),
\nn
\quad
B_{\mc{T} \td{C}_{2,100}} (\bk_0) = {\rm Diag} (e^{- i \frac{3 \pi}{4}}, e^{-i \frac{\pi}{4}}, e^{i \frac{\pi}{4}}, e^{i \frac{3 \pi}{4}}) \mc{K},
\quad
B_{\mc{T} \td{C}_{2,100}} (\bk_+) = {\rm Diag} (i, -i, 1, -1) \mc{K}.
\label{seq:sg212_sewing_straight}
\eg

Applying Eq.~\eqref{seq:wilson_symmetry1}, we obtain the symmetry constraints:
\ba
W_{\bk_0 \leftarrow \bk_+}
=& e^{i \frac{3}{4} (\pi- r_\ep)} \, B_{\td{C}_{4,001}} (\bk_0) \cdot W_{\bk_0 \leftarrow \bk_+} \cdot B_{\td{C}_{4,001}} (\bk_+)^{-1}
\nn
=& B_{\mc{T} \td{C}_{2,100}} (\bk_0) \cdot W_{\bk_0 \leftarrow \bk_+} \cdot B_{\mc{T} \td{C}_{2,100}}^{-1}.
\ea
Solving these constraints yields
\bg
W_{\bk_0 \leftarrow \bk_+} = e^{-i \frac{3 \pi}{8}} \, \bpm 0 & 0 & s_3 & 0 \\ s_1 & 0 & 0 & 0 \\ 0 & 0 & 0 & s_4 \\ 0 & s_2 & 0 & 0 \epm,
\label{seq:wilson_sg212_straight}
\eg
where $s_{1,2,3,4} \in \{+1, -1\}$ are undetermined sign factors.

The form in Eq.~\eqref{seq:wilson_sg212_straight} encodes how the irreps change their energy ordering as we move along the HSL $(0,0,u)$ for $r_\ep \le u \le \pi$.
At momentum $(0,0,u)$, the $\Delta_{1,2,3,4}$ irreps have $\td{C}_{4,001}$ eigenvalues $e^{-i \frac{3}{4} u}$, $-e^{-i \frac{3}{4} u}$, $-i e^{-i \frac{3}{4} u}$, $i e^{-i \frac{3}{4} u}$, respectively.
The choice of $B_{\td{C}_{4,001}}(\bk_+)$ in Eq.~\eqref{seq:sg212_sewing_straight} matches this, as its basis is ordered as $(\Delta_4, \Delta_3, \Delta_2, \Delta_1)$ according to the increasing energy order in Fig.~\ref{sfig:sg212_bands}\textbf{d}.
In contrast, the sewing matrix $B_{\td{C}_{4,001}}(\bk_0)$ implies a basis ordering of $(\Delta_2, \Delta_4, \Delta_1, \Delta_3)$.
Thus, as the energy eigenstates evolve from $\bk_+$ to $\bk_0$ along $(0,0,u)$, the energy ordering of the $\Delta$ irreps changes.
This reflects that the compatibility relations [Eq.~\eqref{seq:sg212_compatibility_irrep} and Fig.~\ref{sfig:sg212_bands}\textbf{d}] enforce a permutation among the $\Delta_{1,2,3,4}$ irreps, which must be connected to the $X_1$ and $X_2$ irreps .
Since all four irreps have distinct $\td{C}_{4,001}$ eigenvalues, they do not mix, and the Wilson line $W_{\bk_0 \leftarrow \bk_+}$ records this permutation.
For example, the $\Delta_4$ irrep is represented by the vector $(1,0,0,0)$ at $\bk_+$ and by $(0,1,0,0)$ at $\bk_0$ in our chosen energy eigenstate basis.
Under Wilson line evolution, we find $W_{\bk_0 \leftarrow \bk_+} \cdot (1,0,0,0) = e^{-i \frac{3 \pi}{8}} \, (0, s_1 ,0 ,0)$, which correctly maps $\Delta_4$ at $\bk_+$ to the corresponding state at $\bk_0$, up to a phase.
%%%%%%

%%%%%%
\paragraph{Semicircular segments $W_{\bk_+ \leftarrow \bk_-} (\phi)$.}
%%%%%%
As shown in Eq.~\eqref{seq:wilson_compare}, the Wilson loop $W_{r_\ep} (\phi_0; -\bk_0)$ can be constructed from $W_{r_\ep} (\phi_0; -\bk_0)$ once we know the straight segment $W_{\bk_0 \leftarrow \bk_+}$ and the relative contribution from the semicircular segments,
\bg
W_{\bk_+ \leftarrow \bk_-} (\phi) \cdot W_{\bk_+ \leftarrow \bk_-} (\phi_0)^{-1}.
\eg
Here, we determine the form of this product for discrete angles $\phi= n \pi/4$ for $n \in \Z$, in the limit $r_\ep \to 0$.

Let us consider a small sphere of radius $r_\ep$ in the BZ centered at $\Gamma$, where the semicircular paths for $W_{\bk_+ \leftarrow \bk_-} (\phi)$ lie.
(See Fig.~\ref{sfig:wilson_path}\textbf{b}.)
In this region, the four physical bands split into two subsets, separated by a finite energy gap $E_g$: the two lower bands correspond to transverse modes.
In the absence of any polarization singularity, the Wilson loop $W_{\bk_+ \leftarrow \bk_-} (\phi) \cdot W_{\bk_+ \leftarrow \bk_-} (\phi_0)^{-1}$ would approach the identity matrix in the energy eigenbasis as $r_\ep \to 0$.
However, this may no longer hold when a polarization singularity is present.
In particular, we can write
\bg
W_{\bk_+ \leftarrow \bk_-}(\phi) \cdot W_{\bk_+ \leftarrow \bk_-}(\phi_0)^{-1}
= \bpm \mc{W}(\phi) \, \mc{W}(\phi_0)^{-1} & \\ & \sg_0
\epm + O(E_g^{-1}),
\label{seq:wilson_circle_correction}
\eg
where $\mc{W}(\phi)$ denotes the Wilson line defined from the two transverse modes, along the same semicircular path as $W_{\bk_+ \leftarrow \bk_-}(\phi)$.
Because the upper two bands are regular, the bottom-right (lower diagonal) block is approximated by $\sg_0$.
Furthermore, since they remain separated from the transverse modes by a finite energy gap, the off-diagonal blocks vanish to leading order\footnote{The small parameter here and in Eq.~\eqref{seq:wilson_circle_correction} is more precisely $\frac{v r_\ep}{E_g}$, where $v$ is a characteristic velocity such as group velocity.}.

Now we solve for $\mc{W}(\phi)$ at $\phi=0$ and $\pi/4$.
To obtain the analytic form of $\mc{W}(\phi)$, we use the sewing-matrix constraints.
The relevant high-symmetry momenta are $\bk_+=(0,0,r_\ep)$, $\bk_1 = (r_\ep,0,0)$, and $\bk_2 = (r_\ep/\sqrt{2},r_\ep/\sqrt{2},0)$.
Note that $\bk_1 = \bk_\ep (\phi = 0)$ and $\bk_2 = \bk_\ep (\phi = \tfrac{\pi}{4})$.

To set the sewing matrices at these momenta, we use the fact that the two transverse modes correspond to $\Delta_3$ and $\Delta_4$ irreps on the HSL $(0,u,0)$ for $0<u<\pi$ and two $\Sigma_2$ irreps on the HSL $(u',u',0)$ for $0<u'<\pi$.
The momentum $\bk_2$ lies on the HSL $(0,u',u')$, while $\bk_1$ lies on the HSL $(u,0,0)$, which is related to $(0,u,0)$ by the $C_{3,111}$ rotation.
Specifically, we need the sewing matrices of $\td{C}_{2,110}$ at $\bk_+$ and $\bk_2$, and of $\td{C}_{2,100}$ at $\bk_+$ and $\bk_1$.
The sewing matrices for the two transverse modes at $\bk_+$ are obtained from the four-band matrices in Eq.~\eqref{seq:sewing_sg212_kp2} by taking the $2 \times 2$ top-left block, while those at $\bk_{1,2}$ follow from the little-group irreps of the two transverse modes:
\bg
\mc{B}_{\td{C}_{2,110}} (\bk_\pm) = - e^{\pm i \frac{3 r_\ep}{4}} \sg_z,
\quad
\mc{B}_{\td{C}_{2,110}}(\bk_2) = - e^{-i \frac{r_\ep}{\sqrt{2}}} \sg_0.
\nn
\mc{B}_{\td{C}_{2,100}}(\bk_\pm) = \mp i \sg_0,
\quad
\mc{B}_{\td{C}_{2,100}}(\bk_1) = - e^{-i \frac{r_\ep}{2}} \, \sg_0,
\label{seq:sewing_sg212_circ}
\eg
Here, the choice of basis implies that $\Delta_4$ is energetically lower than $\Delta_3$ when $r_\ep$ is small, as shown in Fig.~\ref{sfig:sg212_bands}\textbf{d}.

The key idea is to decompose $\mc{W}(\phi=0, \pi/4)$ into two Wilson lines:
\bg
\mc{W}(0) = \mc{W}_{\bk_+ \leftarrow \bk_1} \cdot \mc{W}_{\bk_1 \leftarrow \bk_-},
\quad
\mc{W}(\pi/4) = \mc{W}_{\bk_+ \leftarrow \bk_2} \cdot \mc{W}_{\bk_2 \leftarrow \bk_-}.
\label{seq:sg212_wl_circ_decomp1}
\eg
Here, each Wilson line $\mc{W}_{\bk_+ \leftarrow \bk_{1,2}}$ and $\mc{W}_{\bk_{1,2} \leftarrow \bk_-}$ corresponds to a quarter-circle arc, with its basepoint and endpoint specified in the lower indices.
The Wilson line $\mc{W}_{\bk_+ \leftarrow \bk_1}$ is related to $\mc{W}_{\bk_1 \leftarrow \bk_-}$ by $\td C_{2,100}$, and $\mc{W}_{\bk_+ \leftarrow \bk_2}$ is related to $\mc{W}_{\bk_2 \leftarrow \bk_-}$ by $\td C_{2,110}$.
Using the symmetry transformation rule of Wilson lines together with the sewing matrices in Eq.~\eqref{seq:sewing_sg212_circ}, we obtain
\ba
\mc{W}_{\bk_1 \leftarrow \bk_-}^{-1}
=& e^{-i \frac{r_\ep}{2}} \, \mc{B}_{\td C_{2,100}}(\bk_+) \cdot W_{\bk_+ \leftarrow \bk_1} \cdot \mc{B}_{\td C_{2,100}}(\bk_1)^{-1}
= i \mc{W}_{\bk_+ \leftarrow \bk_1},
\nn
\mc{W}_{\bk_2 \leftarrow \bk_-}^{-1}
=& e^{-i \frac{3 r_\ep}{4} - i\frac{r_\ep}{\sqrt{2}}} \, \mc{B}_{\td C_{2,110}}(\bk_+) \cdot \mc{W}_{\bk_+ \leftarrow \bk_2} \cdot \mc{B}_{\td C _{2,110}}(\bk_2)^{-1}
= \sg_z \cdot \mc{W}_{\bk_+ \leftarrow \bk_2}
\label{seq:sg212_wl_circ_decomp2}
\ea
Combining Eqs.~\eqref{seq:sg212_wl_circ_decomp1} and \eqref{seq:sg212_wl_circ_decomp2}, we find
\ba
& \mc{W}(0)
= \mc{W}_{\bk_+ \leftarrow \bk_1} \cdot (-i \, \mc{W}_{\bk_+ \leftarrow \bk_1}^{-1})
= -i \sg_0,
\nn
& \mc{W}(\pi/4)
= \mc{W}_{\bk_+ \leftarrow \bk_2} \cdot (W_{\bk_+ \leftarrow \bk_2}^{-1} \cdot \sg_z)
= \sg_z.
\ea

Finally, the constraint from $\td C_{4,001}$ gives the form of $\mc{W}(\phi)$ at other angles $\phi = \tfrac{n \pi}{4}$ for $n \in \Z$.
For this, we use the sewing matrices for $\td C_{4,001}$ at $\bk=\bk_\pm$, which can be inferred from Eq.~\eqref{seq:sg212_sewing_straight}:
\bg
\mc{B}_{\td C_{4,001}}(\bk_+) = i e^{-i \frac{3r_\ep}{4}} \, \sg_z,
\quad
\mc{B}_{\td C_{4,001}}(\bk_+) = -i e^{i \frac{3r_\ep}{4}} \, \sg_z.
\eg
With these matrices, $\mc{W}(\phi)$ and $\mc{W}(\phi + \tfrac{\pi}{2})$ are related as
\bg
\mc{W}(\phi + \tfrac{\pi}{2})
= e^{i \frac{3 r_\ep}{2}} \, \mc{B}_{\td C_{4,001}}(\bk_+) \cdot \mc{W}(\phi) \cdot \mc{B}_{\td C_{4,001}}(\bk_-)^{-1}
= - \sg_z \cdot \mc{W}(\phi) \cdot \sg_z.
\label{seq:sg212_wl_c4}
\eg
For example, $\mc{W}(\tfrac{\pi}{2}) = -\sg_z \cdot \mc{W}(0) \cdot \sg_z = i \sg_0$.

By combining the expressions for $\mc{W}(\phi)$ at $\phi=\tfrac{\pi}{4}$ and $\tfrac{\pi}{2}$ with Eq.~\eqref{seq:wilson_circle_correction}, we obtain
\bg
W_{\bk_+ \leftarrow \bk_-}(\tfrac{\pi}{4}) \cdot W_{\bk_+ \leftarrow \bk_-}(0)^{-1}
= \bpm i \sg_z & \\ & \sg_0 \epm + O(E_g^{-1}),
\nn
W_{\bk_+ \leftarrow \bk_-}(\tfrac{\pi}{2}) \cdot W_{\bk_+ \leftarrow \bk_-}(0)^{-1}
= \bpm -\sg_0 & \\ & \sg_0 \epm + O(E_g^{-1}).
\label{seq:sg212_wl_circ_final}
\eg
%%%%%%

%%%%%%
\subsubsection{Winding structure in the limit $r_\ep \to 0$}
%%%%%%
By combining the results obtained in the previous subsections, we now show that the Wilson loop spectrum develops a winding structure as $r_\ep \to 0$ (see Fig.~\ref{sfig:sg212_wilson}\textbf{c}).
First recall Eq.~\eqref{seq:wilson_compare} with $\phi = \tfrac{\pi}{4}$ and $\phi_0 = 0$:
\bg
W_{r_\ep}(\tfrac{\pi}{4}; -\bk_0)
= W_{\bk_0 \leftarrow \bk_+} \cdot [W_{\bk_+ \leftarrow \bk_-} (\tfrac{\pi}{4}) \cdot W_{\bk_+ \leftarrow \bk_-} (0)^{-1}] \cdot W_{\bk_0 \leftarrow \bk_+}^{-1} \cdot W_{r_\ep} (0; -\bk_0)^{-1}.
\label{seq:wilson_compare_eval}
\eg
The analytic forms of $W_{r_\ep}(\frac{\pi}{4}; -\bk_0)$, $W_{r_\ep}(0; -\bk_0)$, and $W_{\bk_0 \leftarrow \bk_+}$ were given in Eqs.~\eqref{seq:sg212_wilson_M}, \eqref{seq:sg212_wilson_Xp}, and \eqref{seq:wilson_sg212_straight}, respectively, while $W_{\bk_+ \leftarrow \bk_-} (\tfrac{\pi}{4}) \cdot W_{\bk_+ \leftarrow \bk_-} (0)^{-1}$ was obtained in Eq.~\eqref{seq:sg212_wl_circ_final} in the limit $r_\ep \to 0$.
Substituting these results into Eq.~\eqref{seq:wilson_compare_eval}, one finds
\bg
e^{i \frac{\pi}{4}} \, \bpm
t_1 & - i t_7 & t_4 & -i t_5 \\
-i t_8 & -t_1 & -i t_6 & -t_3 \\
t_3 & -i t_5 & t_2 & -i t_9 \\
-i t_6 & -t_4 & -i t_{10} & -t_2
\epm
= \bpm i t'_1 & e^{- i \frac{\pi}{4}} t'_7 & -t'_4 & - e^{i \frac{\pi}{4}} t'_5 \\
i e^{i \frac{\pi}{4}} t'_8 & t'_1 & -i e^{- i \frac{\pi}{4}} t'_6 & -i t'_3 \\
t'_3 & - e^{i \frac{\pi}{4}} t'_5 & i t'_2 & - e^{- i \frac{\pi}{4}} t'_9 \\
i e^{- i \frac{\pi}{4}} t'_6 & -i t'_4 & i e^{i \frac{\pi}{4}} t'_{10} & -t'_2 \epm
\eg
valid to leading order in $E_g^{-1}$ as $r_\ep \to 0$.
Because $t_{1,\dots,10}$ and $t'_{1,\dots,10}$ are real, this implies
\bg
t_{1,\dots,6} = 0, \quad
t'_{1,\dots,6} = 0, \quad
(t'_7, t'_8, t'_9, t'_{10})
= (t_7, -t_8, -t_9, -t_{10}), \quad
\label{seq:sg212_wl_limit_cond}
\eg
as $r_\ep \to 0$.
Thus, the conditions in Eq.~\eqref{seq:sg212_wl_limit_cond} imply that
\bg
\lim_{r_\ep \to 0} \, W_{r_\ep} (0; -\bk_0)
= e^{-i \frac{\pi}{4}} \bpm 0 & t_7 & 0 & 0 \\
-i t_8 & 0 & 0 & 0 \\
0 & 0 & 0 & t_9 \\
0 & 0 & i t_{10} & 0 \epm,
\quad
\lim_{r_\ep \to 0} \, W_{r_\ep} (\tfrac{\pi}{4}; -\bk_0)
= e^{-i \frac{\pi}{4}} \bpm 0 & t_7 & 0 & 0 \\
t_8 & 0 & 0 & 0 \\
0 & 0 & 0 & t_9 \\
0 & 0 & t_{10} & 0 \epm.
\label{seq:sg212_wilson_limit}
\eg
Also, together with Eqs.~\eqref{seq:sg212_wilson_M_cond} and \eqref{seq:sg212_wilson_X_cond}, Eq.~\eqref{seq:sg212_wl_limit_cond} enforces $t_{7,8,9,10}^2 \to 1$, i.e. each of $t_{7,8,9,10}$ converges to either $+1$ or $-1$.
This yields 16 possible sign choices for $(t_7,t_8,t_9,t_{10})$.
However, not all 16 are allowed: we already derived that the eigenvalue spectrum of $W_{r_\ep}(\phi=\frac{\pi}{4}; \bk_b)$ is quantized as $\{-\frac{3\pi}{4}, -\frac{\pi}{4}, + \frac{\pi}{4}, +\frac{3\pi}{4}\}$, which leaves eight viable cases.
For four of them, ${\rm Spec} [-i \log W_{r_\ep}(\phi=0; \bk_b)] \to \{0,0,\pi,\pi\}$ as $r_\ep \to 0$; for the other four, ${\rm Spec} [-i \log W_{r_\ep}(\phi=0; \bk_b)] \to \{-\frac{\pi}{2},-\frac{\pi}{2},+\frac{\pi}{2},+\frac{\pi}{2}\}$:
\bg
t_{7,8,9,10} = (-1,+1,-1,-1), \, (-1,+1,+1,+1), \, (+1,-1,-1,-1), \, (+1,-1,+1,+1),
\label{seq:sg212_wl_limit_case1}
\eg
implies $\lim_{r_\ep \to 0} \, {\rm Spec} [-i \log W_{r_\ep}(\phi=0; \bk_b)] = \{0,0, \pi, \pi\}$, and
\bg
t_{7,8,9,10} = (-1,-1,-1,+1), \, (-1,-1,+1,-1), \, (+1,+1,-1,+1), \, (+1,+1,+1,-1),
\label{seq:sg212_wl_limit_case2}
\eg
implies $\lim_{r_\ep \to 0} \, {\rm Spec} [-i \log W_{r_\ep}(\phi=0; \bk_b)] = \{-\frac{\pi}{2},-\frac{\pi}{2},+\frac{\pi}{2},+\frac{\pi}{2}\}$.
We confirmed numerically that $(t_7, t_8, t_9, t_{10})=(+1,+1,-1,+1)$ as $r_\ep \to 0$ in our TB model $H_{SG\,212}(\bk)$.
Importantly, however, the emergence of the winding structure does not rely on this particular choice of $(t_7,t_8,t_9,t_{10})$, but follows generally from the symmetry and singularity constraints, as we discuss below.

By combining the sewing-matrix (symmetries) constraints and the effect of the polarization singularity, we determined the eigenvalue spectra of the Wilson loop $W_{r_\ep} (\phi; \bk_b)$ at $\phi=0$ and $\phi=\frac{\pi}{4}$ in the limit $r_\ep \to 0$.
More generally, $\td C_{4,001}$ symmetry quantizes the Wilson loop spectrum at $\phi=\frac{n\pi}{4}$ ($n \in \Z$):
\bg
{\rm Spec} [-i \log W_{r_\ep}(\phi; \bk_b)] =
\begin{cases}
 \{ -\Theta_0, \Theta_0, - \Theta_0 + \pi, \Theta_0 + \pi \} & \text{if } \phi=0,\pi, \\
\{-\frac{3\pi}{4}, -\frac{\pi}{4}, +\frac{\pi}{4}, +\frac{3\pi}{4}\} & \text{if } \phi = \frac{\pi}{4}, \frac{3\pi}{4}, \frac{5\pi}{4}, \frac{7\pi}{4}, \\
\{ -\Theta_0 - \frac{\pi}{2}, \Theta_0 - \frac{\pi}{2}, - \Theta_0 + \frac{\pi}{2}, \Theta_0 + \frac{\pi}{2} \} & \text{if } \phi = \frac{\pi}{2}, \frac{3\pi}{2}.
\end{cases}
\label{seq:sg212_wil_limit_quantized}
\eg
in the limit $r_\ep \to 0$.
Here, $\Theta_0 \in \{0,\tfrac{\pi}{2}\}$ is determined by $t_{7,8,9,10}$ according to Eqs.~\eqref{seq:sg212_wl_limit_case1} and \eqref{seq:sg212_wl_limit_case2}.

Having established the quantized Wilson loop eigenvalues at $\phi = \frac{n\pi}{4}$, we now combine these results with the general symmetry constraints to understand why the spectrum must exhibit the winding structure shown in Fig.~\ref{sfig:sg212_wilson}\textbf{c}.
Since the Wilson loop spectrum varies continuously with $\phi$ (because the Wilson loop path avoids all singular points, such as the polarization singularity), the eigenvalues at generic $\phi \neq \frac{n\pi}{4}$ interpolate between the quantized values at $\phi = \frac{n\pi}{4}$.
In addition, the entire spectrum must respect the spectral symmetries in Eq.~\eqref{seq:sg212_wilson_sym}.
To determine the allowed interpolation, we assume that the Wilson loop spectrum defined on the cylindrical path does not carry the chiral winding associated with a nonzero Chern number.
Otherwise, such winding would already manifest before taking the limit $r_\ep \to 0$.
Under this assumption, when $\Theta_0 = \frac{\pi}{2}$ the only interpolation consistent with Eqs.~\eqref{seq:sg212_wil_limit_quantized} and \eqref{seq:sg212_wilson_sym} is one that either coincides with or can be smoothly deformed into the winding structure shown in Fig.~\ref{sfig:sg212_wilson}\textbf{c}.
For $\Theta_0 = 0$, the structure is similar, but the entire spectrum is shifted by $\frac{\pi}{2}$.
Hence, our analysis explains the winding in the cylindrical Wilson loop spectrum for the TB model $H_{SG 212}(\bk)$ and the ab-initio photonic crystal model.
%%%%%%

%%%%%%
\subsubsection{Continuum description of transverse modes in isotropic limit}
\label{sec:sn_iso_wl}
%%%%%%
In the previous subsections, we established that at special azimuthal angles $\phi = \frac{n \pi}{4}$ ($n \in \Z$) the Wilson loop eigenvalues are quantized in the $r_\ep \to 0$ limit.
Taken together with the spectral symmetries of the Wilson loop, this necessarily leads to a winding structure in the $r_\ep \to 0$ limit.
To gain further insight, we now analyze the continuum description of the two transverse modes, from which the Wilson loop spectrum can be obtained analytically at arbitrary $\phi$.
In particular, we consider the isotropic limit, which is natural in our setting since the two lowest bands among the four bands corresponding to Eq.~\eqref{seq:sg212_ebr_phys} (serving as the transverse modes of interest) transform as $\Gamma_4 - \Gamma_1$ at the $\Gamma$ point, as assigned in Ref.~\cite{christensen2022location}.
This assignment is consistent with the $O(3)$ symmetry emergent in the isotropic limit~\cite{christensen2022location}, so it is justified to treat the transverse modes in this limit, up to smooth, symmetry-preserving deformation.

We introduce spherical coordinates on $S^2$ in momentum space, with unit vectors
\bg
\hat r = ( \cos \phi \sin \theta, \sin \phi \sin \theta, \cos \theta ),
\quad
\hat \theta = ( \cos \phi \cos \theta, \sin \phi \cos \theta, -\sin \theta),
\quad
\hat \phi = ( -\sin \phi, \cos \phi, 0).
\label{seq:sph_normal_vec}
\eg
By definition, $\hat r = \bk / |\bk|$ represents the polarization of the longitudinal mode.
The transverse polarizations are orthogonal to $\hat r$ and can therefore be written as linear combinations of $\hat \theta$ and $\hat \phi$.

In the isotropic limit, the two transverse modes are degenerate, so their polarization vectors may be chosen as any pair of independent linear combinations of $\hat \theta$ and $\hat \phi$.
One such choice is
\bg
\ket{T_1 (\theta, \phi)} = \frac{1}{\sqrt{2}} \left( \hat \theta (\theta, \phi) - i \hat \phi (\theta, \phi) \right),
\quad
\ket{T_2 (\theta, \phi)} = \frac{1}{\sqrt{2}} \left( \hat \theta (\theta, \phi) + i \hat \phi (\theta, \phi) \right).
\label{seq:iso_trans_vec}
\eg
[Here, we write their dependence on $(\theta, \phi)$ explicitly.]
However, this choice is not well defined at the north and south poles of $S^2$ ($\theta=0, \pi$).
Indeed, at these poles one finds
\bg
\ket{T_1 (\theta = 0, \phi)}
= \frac{1}{\sqrt{2}} e^{i \phi} (1, -i, 0),
\quad
\ket{T_2 (\theta = 0, \phi)}
= \frac{1}{\sqrt{2}} e^{- i \phi} (1, i, 0),
\nn
\ket{T_1 (\theta = \pi, \phi)}
= -\frac{1}{\sqrt{2}} e^{-i \phi} (1, i, 0),
\quad
\ket{T_2 (\theta = \pi, \phi)}
= -\frac{1}{\sqrt{2}} e^{i \phi} (1, -i, 0).
\eg
Since the azimuthal angle $\phi$ is not uniquely defined at the north and south poles of $S^2$, these eigenvectors are not single-valued at the poles.

Instead, we introduce two coordinate patches $N$ and $S$, covering $S^2$ except for the south and north poles, respectively.
On each patch the eigenmodes can be chosen to be single-valued:
\bg
\ket{T^{(N)}_{1} (\theta, \phi)} = \frac{1}{\sqrt{2}} e^{-i \phi} \left( \hat \theta (\theta, \phi) - i \hat \phi (\theta, \phi) \right),
\quad
\ket{T^{(N)}_2 (\theta, \phi)} = \frac{1}{\sqrt{2}} e^{i \phi} \left( \hat \theta (\theta, \phi) + i \hat \phi (\theta, \phi) \right),
\nn
\ket{T^{(S)}_1 (\theta, \phi)} = \frac{1}{\sqrt{2}} e^{i \phi} \left( \hat \theta (\theta, \phi) - i \hat \phi (\theta, \phi) \right),
\quad
\ket{T^{(S)}_2 (\theta, \phi)} = \frac{1}{\sqrt{2}} e^{-i \phi} \left( \hat \theta (\theta, \phi) + i \hat \phi (\theta, \phi) \right).
\label{seq:iso_trans_vec_poles}
\eg
By construction, $\ket{T^{(N)}_{1,2} (\theta, \phi)}$ ($\ket{T^{(S)}_{1,2} (\theta, \phi)}$) are now single-valued at the north (south) pole.
On the overlap between the $N$ and $S$ patches, the two sets of eigenmodes $\ket{T^{(N)}_{n} (\theta, \phi)}$ and $\ket{T^{(S)}_{m} (\theta, \phi)}$ ($n, m=1,2$) are related by a $2 \times 2$ transition function $g_{NS} (\theta,\phi)$:
\bg
\ket{T^{(N)}_n (\theta, \phi)} = \sum_{m=1}^2 \, \ket{T^{(S)}_m (\theta, \phi)} \, [g_{NS} (\theta, \phi)]_{mn}.
\eg
A direct comparison gives
\bg
g_{NS} (\theta, \phi) = {\rm Diag} (e^{2i \phi}, e^{-2i \phi}) = e^{2 i \phi \sg_z}.
\label{seq:iso_transitionf}
\eg

From these eigenmodes, we can compute the Berry connection.
For patch $P=N,S$, we define
\bg
[A^{(P)}_\phi (\theta, \phi)]_{nm} = \bra{T^{(P)}_n} i \der_\phi \ket{T^{(P)}_m},
\quad
[A^{(P)}_\theta (\theta, \phi)]_{nm} = \bra{T^{(P)}_n} i \der_\theta \ket{T^{(P)}_m},
\eg
which evaluates to
\bg
A^{(N)}_\phi (\theta, \phi) = (1 - \cos \theta) \sg_z,
\quad
A^{(S)}_\phi (\theta, \phi) = - (1 + \cos \theta) \sg_z,
\quad
A^{(N)}_\theta (\theta, \phi) = A^{(S)}_\theta (\theta, \phi) = 0.
\label{seq:iso_berry_con}
\eg

Now we are ready to compute the Wilson line $\mc{W} (\phi)$ along the semicircular path in Fig.~\ref{sfig:wilson_path}\textbf{b}.
For this purpose, we first define the Wilson lines within each patch:
\bg
\mc{W}^{(N)} (\phi) = \mc{P} \exp \left[ \int_\vep^\pi \, d \theta \, A^{(N)}_\theta (\theta, \phi) \right],
\quad
\mc{W}^{(S)} (\phi) = \mc{P} \exp \left[ \int_0^\vep \, d\theta \, A^{(S)}_\theta (\theta, \phi) \right]
\eg
for any $0 < \vep < \pi$.
On the overlap at $\theta = \vep$, the transition function $g_{NS}(\theta,\phi)$ connects the two patches, so that
\bg
\mc{W}(\phi)
= \mc{W}^{(N)} (\phi) \cdot g_{NS} (\vep, \phi) \cdot \mc{W}^{(S)} (\phi).
\eg
Since $A^{(P)}_\theta (\theta, \phi)=0$, we obtain
\bg
\mc{W}(\phi) = g_{NS} (\vep, \phi) = e^{2i \phi \sg_z}.
\label{seq:iso_wilson_circle}
\eg

\begin{figure*}[t]
\centering
\includegraphics[width=0.5\textwidth]{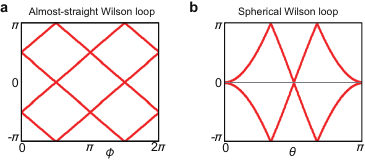}
\caption{{\bf Wilson loop spectra from the isotropic continuum approximation.}
\textbf{a} Wilson loop spectrum for $\lim_{r_\ep \to 0} \, W_{r_\ep} (\phi;-\bk_0)$ [Eq.~\eqref{seq:sg212_wilson_limit_iso}], i.e. the almost-straight Wilson loop obtained in the isotropic continuum description.
The eigenvalues evolve linearly with $\phi$ as in Eq.~\eqref{seq:iso_wilson_spec}, consistent with the winding structure observed in our TB and ab initio models for SG $P4_332$.
\textbf{b} Spherical Wilson loop spectrum computed from the two transverse modes in the isotropic continuum limit [Eq.~\eqref{seq:iso_wilson_euler}].
The spectrum shows winding corresponding to an Euler number with $|\mf{e}|=2$.}
\label{sfig:iso_wilson}
\end{figure*}

With this result, we can write the almost-straight Wilson loop $W_{r_\ep} (\phi; -\bk_0)$ for the four bands corresponding to Eq.~\eqref{seq:sg212_ebr_phys}.
By combining Eqs.~\eqref{seq:wilson_compare}, \eqref{seq:wilson_circle_correction}, and \eqref{seq:iso_wilson_circle}, we find
\bg
\lim_{r_\ep \to 0} \, W_{r_\ep} (\phi; -\bk_0) = W_{\bk_0 \leftarrow \bk_+} \cdot \bpm e^{2 i \phi \sg_z} & \\ & \sg_0 \epm \cdot W_{\bk_0 \leftarrow \bk_+}^{-1} \cdot \lim_{r_\ep \to 0} \, W_{r_\ep} (0, -\bk_0).
\eg
By substituting $W_{\bk_0 \leftarrow \bk_+}$ and $\lim_{r_\ep \to 0} \, W_{r_\ep}(0; -\bk_0)$ from Eqs.~\eqref{seq:wilson_sg212_straight} and \eqref{seq:sg212_wilson_limit}, we obtain
\bg
\lim_{r_\ep \to 0} \, W_{r_\ep} (\phi; -\bk_0)
= e^{- \frac{i \pi}{4}} \, \bpm
0 & t_7 & 0 & 0 \\
-i e^{2 i \phi} t_8 & 0 & 0 & 0 \\
0 & 0 & 0 & t_9 \\
0 & 0 & i e^{-2 i \phi} \, t_{10} & 0 \\
\epm
= \bpm 0 & e^{- \frac{i \pi}{4}} & 0 & 0 \\
-e^{\frac{i \pi}{4} + 2 i \phi} & 0 & 0 & 0 \\
0 & 0 & 0 & - e^{- \frac{i \pi}{4}} \\
0 & 0 & e^{\frac{i \pi}{4} - 2 i \phi} & 0 \epm.
\label{seq:sg212_wilson_limit_iso}
\eg
where in the last equality we set $(t_7, t_8, t_9, t_{10}) = (+1,+1,-1,+1)$, as confirmed numerically in our TB model $H_{SG 212} (\bk)$ in the limit $r_\ep \to 0$.
Thus, we derive the expression previously given in the main text.
The eigenvalue spectrum then follows as
\bg
\lim_{r_\ep \to 0} \, {\rm Spec}[-i \log W_{r_\ep} (\phi; -\bk_0)]
= \left( -\phi-\frac{\pi}{2}, -\phi+\frac{\pi}{2}, \phi-\frac{\pi}{2}, \phi+\frac{\pi}{2} \right).
\label{seq:iso_wilson_spec}
\eg
This shows that, in the isotropic approximation near the polarization singularity, the Wilson loop eigenvalues evolve linearly with $\phi$, with slopes $\pm 1$, as illustrated in Fig.~\ref{sfig:iso_wilson}\textbf{a}.
Moreover, the winding structure in Eq.~\eqref{seq:iso_wilson_spec} is topologically equivalent to those obtained in both our TB and ab-initio models (compare Fig.~\ref{sfig:iso_wilson}\textbf{a} with Fig.~\ref{sfig:sg212_wilson}\textbf{c} and Fig.~5\textbf{b} in the main text).
%%%%%%

%%%%%%
%%%%%%

%%%%%%
\subsection{Spherical Wilson loop}
\label{sec:sn_sg212_wl_sphere}
%%%%%%
We now examine the spherical Wilson loop spectra obtained in our TB and ab-initio models for SG $P4_332$ (No. 212).
The spherical Wilson loop is defined on a sphere $S^2$ of radius $k_r$ centered at $\Gamma$: for each polar angle $\theta$, we evaluate the Wilson loop $W_\phi(\theta;k_r)$ along the azimuthal direction $\phi$, and plot the eigenvalues of $-i \log[W_\phi(\theta;k_r)]$ as $\theta$ varies from 0 to $\pi$.
As shown in Fig.~\ref{sfig:sg212_wilson}\textbf{d} for the TB model and Fig.~5\textbf{c} of the main text for ab-initio model, the two transverse modes exhibit a characteristic winding corresponding to an Euler number\footnote{The number $|\mf{e}|$ is obtained by counting the number of eigenvalue crossings at $\pi$.
In our case in Fig.~\ref{sfig:sg212_wilson}\textbf{d}, two crossings correspond to $|\mf{e}|=2$.
Note that the sign of $\mf{e}$ depends on the global orientation convention of the real two-band bundle (or wavefunctions)~\cite{milnor1974characteristic,ahn2019failure} and hence only $|\mf{e}|$ has a well-defined meaning that does not depend on the orientation choice.} $|\mf{e}|=2$.
Ordinarily, such a winding is protected by the combined inversion–time-reversal symmetry $I\mc{T}$ satisfying $(I\mc{T})^2=\mathds{1}$.
However, SG $P4_332$ possesses time-reversal symmetry $\mc{T}$ but lacks inversion $I$, and hence does not have $I\mc{T}$.
Below, we analyze why the same winding nevertheless emerges in this non-centrosymmetric system and how it can be consistently interpreted within our TB and ab-initio models.
In the TB model, it arises from an emergent $I\mc{T}$ symmetry, while in the ab-initio model it originates from emergent transversality.
%%%%%%

%%%%%%
\subsubsection{Emergent $I \mc{T}$ symmetry in the tight-binding model}
%%%%%%
In our TB model [$H_{SG 212}(\bk)$ in Eq.~\eqref{seq:h_tb212}], the effective $k \cdot p$ Hamiltonian can be made almost real-valued by an appropriate unitary transformation.
In this case, the combined inversion–time-reversal operation $I \mc{T}$ is represented simply by complex conjugation $\mc{K}$ and becomes an emergent symmetry.

To construct the effective Hamiltonian $H_{\rm eff}(\bk)$, we apply the L\"owdin partitioning~\cite{lowdin1951note,winkler2003spin} to the three {\it active} bands connected to the polarization singularity (see Fig.~\ref{sfig:sg212_bands}\textbf{a}).
These comprise two transverse modes from the four physical bands and one longitudinal mode originating from the auxiliary bands.
Expanding $H_{\rm eff}(\bk)$ up to quadratic order in $\bk$, we find that a suitable unitary basis renders all matrix elements nearly real: the maximal imaginary components of the coefficients are on the order of $10^{-9}$, whereas the typical real components are $\sim 10^{-1}$.
Hence, to a very high numerical accuracy,
\bg
\mc{K} \, H_{\rm eff}(\bk) \, \mc{K} = (H_{\rm eff}(\bk))^* = H_{\rm eff}(\bk),
\eg
indicating an emergent $I \mc{T}$ symmetry represented by $\mc{K}$ at small momenta.
This emergent symmetry explains why the Euler number, which is normally well defined only for $I \mc{T}$-symmetric systems, is also quantized here.
It also accounts for the $\pi$-crossings in the spherical Wilson loop spectrum of the TB model, which underlie the observed winding structure with $|\mf{e}|=2$.
Below, we briefly review the L\"owdin partitioning procedure and explicitly demonstrate the near-reality of $H_{\rm eff}(\bk)$ in the transformed basis.

To derive, $H_{\rm eff}(\bk)$, let the full Hamiltonian $H (\bk)$ satisfy $H(\bk) \ket{n, \bk} = E_n (\bk) \ket{n, \bk}$.
We divide the Hilbert space into the active subspace $S$ and its complement $S_c$, with the corresponding projectors
\bg
P(\bk) = \sum_{\mu \in S} \, \ket{\mu, \bk} \bra{\mu, \bk},
\quad
Q(\bk) = \sum_{M \in S_c} \, \ket{M, \bk} \bra{M, \bk},
\eg
which satisfy $P(\bk) + Q(\bk) = \mathds{1}$, $P(\bk)^2 = P(\bk)$, $Q(\bk)^2 = Q(\bk)$, and $P(\bk) Q(\bk) = Q(\bk) P(\bk) = 0$.
To expand $H(\bk)$ around a reference momentum $\bk_0$, we write
\bg
\ket{n, \bk} = P(\bk_0) \ket{n, \bk} + Q(\bk_0) \ket{n, \bk}.
\eg
Substituting this decomposition into the eigenvalue equation $H(\bk) \ket{n,\bk} = E_n(\bk) \ket{n, \bk}$ yields two coupled equations,
\bg
Q(\bk_0) H(\bk) P(\bk_0) \ket{n, \bk} + Q(\bk_0) H(\bk) Q(\bk_0) \ket{n, \bk} = E_n(\bk) Q(\bk_0) \ket{n, \bk},
\label{seq:partition_evp1}
\\
P(\bk_0) H(\bk) P(\bk_0) \ket{n, \bk} + P(\bk_0) H(\bk) Q(\bk_0) \ket{n, \bk} = E_n(\bk) P(\bk_0) \ket{n, \bk}.
\label{seq:partition_evp2}
\eg
From Eq.~\eqref{seq:partition_evp1}, we can express
\bg
Q(\bk_0) \ket{n, \bk} = [ E_n(\bk) - Q(\bk_0) H(\bk) Q(\bk_0) ]^{-1} \, Q(\bk_0) H(\bk) P(\bk_0) \ket{n, \bk}.
\eg
Inserting this into Eq.~\eqref{seq:partition_evp2} leads to an effective eigenvalue problem for the projected state $P(\bk_0)\ket{n,\bk}$,
\bg
H_{\rm eff} (\bk) [P(\bk_0) \ket{n, \bk}] = E_n(\bk) [P(\bk_0) \ket{n, \bk}],
\eg
where the effective Hamiltonian acting on the active subspace is defined as
\bg
H_{\rm eff} (\bk) = P(\bk_0) H(\bk) P(\bk_0) + P(\bk_0) H(\bk) Q(\bk_0) \, [ E_n(\bk) - Q(\bk_0) H(\bk) Q(\bk_0) ]^{-1} \, Q(\bk_0) H(\bk) P(\bk_0).
\label{seq:partition_eff1}
\eg

We next expand $H_{\rm eff}(\bk)$ around $\bk_0$.
The full Hamiltonian can be expressed as
\bg
H(\bk) = H(\bk_0) + (\der_i H)(\bk_0) \delta k_i + \frac{1}{2} (\der_i \der_j H)(\bk_0) \delta k_i \delta k_j + \mc{O}(\delta \bk^3),
\label{seq:hfull_expand}
\eg
where $\delta \bk = \bk - \bk_0$ and $(\der_i H)(\bk_0)$ denotes the derivative of $H(\bk)$ with respect to $k_i$ evaluated at $\bk_0$.
We consider the case where all active bands are degenerate at $\bk_0$ with a common energy $E_0$, so that $E_n(\bk) = E_0 + \mc{O}(\delta \bk)$ for all $\mu \in S$.
In this situation, $H_{\rm eff}(\bk)$ provides a valid low-energy approximation around $E_0$ and $\bk_0$.
Substituting Eq.~\eqref{seq:hfull_expand} into Eq.~\eqref{seq:partition_eff1}, we obtain
\ba
H_{\rm eff}(\bk) =&
P(\bk_0) \left[ H(\bk_0) + (\der_i H)(\bk_0) \delta k_i + \frac{1}{2} (\der_i \der_j H)(\bk_0) \delta k_i \delta k_j \right] P(\bk_0).
\nn
&+ [P(\bk_0) (\der_i H)(\bk_0) Q(\bk_0)] \, [E_0 - Q(\bk_0) H(\bk_0) Q(\bk_0)]^{-1} \, [Q(\bk_0) (\der_j H)(\bk_0) P(\bk_0)] \, \delta k_i \delta k_j + \mc{O}(\delta \bk^3).
\ea
In the active-band basis, $H_{\rm eff}(\bk)$ is then given by
\ba
[H_{\rm eff}(\bk)]_{\mu \nu}
:=& \bra{\mu, \bk_0} H_{\rm eff}(\bk) \ket{\nu, \bk_0}
\nn
=& \bra{\mu, \bk_0} H(\bk_0) \ket{\nu, \bk_0} + \bra{\mu, \bk_0} (\der_i H)(\bk_0) \ket{\nu, \bk_0} \delta k_i + \frac{1}{2} (\der_i \der_j H)(\bk_0) \delta k_i \delta k_j
\nn
& + \sum_{M \in S_c} \, \frac{\bra{\mu, \bk_0} (\der_i H)(\bk_0) \ket{M, \bk_0} \bra{M, \bk_0} (\der_j H)(\bk_0) \ket{\nu, \bk_0} }{E_0 - E_M (\bk_0)} \delta k_i \delta k_j + \mc{O}(\delta \bk^3).
\label{seq:heff_final}
\ea

We apply the above formalism to the TB model $H_{SG 212}(\bk)$.
In this model, we take $E_0=0$ and $\bk_0 = \bb 0$.
The energy spectrum of $H(\bb 0)$ [$=H_{SG 212}(\bb 0)$] is ${\rm Spec}[H(\bb 0)] =\{-6.4, -6.4, -6.4, 0, 0, 0, 4.8, 4.8\}$.
We identify the active subspace $S=\{4,5,6\}$ and the remaining subspace $S_c=\{1,2,3,7,8\}$.
(In our terminology, the fourth band corresponds to the longitudinal mode, whereas the fifth and sixth bands correspond to the transverse modes.)
Expanding up to second order in $\bk$, we obtain the effective Hamiltonian,
\bg
H_{\rm eff} (\bk) = M_1 k_x^2 + M_2 k_y^2 + M_3 k_z^3 + M_4 k_x k_y + M_5 k_y k_z + M_6 k_z k_x + \mc{O}(\bk^2),
\eg
where the complex-valued coefficient matrices $M_{1,\dots,6}$ are listed below:
\bg
M_1 = \bpm -0.10955 & -0.03338 - 0.12019i & 0.25786 + 0.24225i \\
-0.03338 + 0.12019i & 0.34288 & 0.07785 - 0.04728i \\
0.25786 - 0.24225i & 0.07785 + 0.04728i & 0.11667 \epm,
\nn
M_2 = \bpm 0.20739 & 0.23844 + 0.04065i & -0.08801 - 0.18856i \\
0.23844 - 0.04065i & 0.02594 & 0.17093 + 0.24689i \\
-0.08801 + 0.18856i & 0.17093 - 0.24689i & 0.11667 \epm,
\nn
M_3 = \bpm 0.25216 & -0.20506 + 0.07954i & -0.16985 - 0.05369i \\
-0.20506 - 0.07954i & -0.01883 & -0.24878 - 0.19962i \\
-0.16985 + 0.05369i & -0.24878 + 0.19962i & 0.11667 \epm,
\nn
M_4 = \bpm 0.10110 & -0.04889 - 0.05912i & -0.06507 + 0.07370i \\
-0.04889 + 0.05912i & -0.04276 & 0.08561 + 0.00434i \\
-0.06507 - 0.07370 & 0.08561 - 0.00434i & -0.05833 \epm,
\nn
M_5 = \bpm -0.01942 & 0.01012 + 0.10422i & 0.00274 + 0.011345i \\
0.01012 - 0.10422i & -0.03891 & 0.09611 - 0.08741i \\
0.00274 - 0.01135i & 0.09611 + 0.08741i & 0.05833 \epm,
\nn
M_6 = \bpm 0.09099 & 0.05125 - 0.06790i & 0.06781 - 0.06235i \\
0.05125 + 0.06790i & -0.03265 & 0.01049 - 0.09175i \\
0.06781 + 0.06235i & 0.01049 + 0.09175i & -0.05833 \epm.
\eg
Note that the magnitudes of the imaginary components are comparable to those of the real components in these matrices.

To minimize the imaginary parts, we perform a unitary transformation with $U(3)$ matrix, $U(\bs \lambda)$.
This matrix is parameterized by 9 real variables, $\bs \lambda=(\lambda_1, \dots, \lambda_9) \in \R^9$:
\bg
U(\bs \lambda) = \exp \left[ i \bpm
\lambda_1 & \lambda_2 + i \lambda_3 & \lambda_4 + i \lambda_5 \\
\lambda_2 - i \lambda_3 & \lambda_6 & \lambda_7 - i \lambda_8 \\
\lambda_4 - i \lambda_5 & \lambda_7 - i \lambda_8 & \lambda_9
 \epm \right],
\eg
Under this unitary transformation, $H_{\rm eff} (\bk)$ is transformed to $U(\bs \lambda) H_{\rm eff} (\bk) U(\bs \lambda)^{-1}$, which corresponds to transforming each coefficient matrix $M_a$ as $M_a \to U(\bs \lambda) M_a U(\bs \lambda)^{-1}$.
To quantify the overall imaginary weight, we define a cost function,
\bg\label{eq:cost}
{\rm cost}(\bs \lambda) = \sum_{a=1}^6 \, || {\rm Im}(U(\bs \lambda) M_a U(\bs \lambda)^{-1}) ||_F,
\eg
where $||A||_F = \sqrt{\sum_{ij} |A_{ij}|^2}$ is the Frobenius norm of a matrix $A$.
We then determine the optimal parameters $\bs \lambda_*$ such that the cost function becomes minimal:
\bg
\bs \lambda_* = \arg \min_{\bs \lambda \in \R^9} \, {\rm cost}(\bs \lambda).
\eg
Using the Nelder–Mead algorithm~\cite{nelder1965simplex} to numerically minimize Eq.~\eqref{eq:cost}, we find
\bg
U_* = U(\bs \lambda_*) = \bpm
-0.03575 - 0.29553i & 0.13200 - 0.67723i & 0.43107 - 0.49949i \\
-0.75382 + 0.23954i & 0.46103 + 0.26513i & 0.02800 - 0.30125i \\
-0.35495 + 0.39971i & -0.45566 - 0.18297i & 0.55981 + 0.39969i
\epm.
\eg
The transformed matrices $M_{a,*}:= U_* M_a U_*^{-1}$ are given by
\bg
M_{1,*} = \bpm 0.35479 & -0.06036 & -0.10776 \\
-0.06036 & 0.19475 & -0.32181 \\
-0.10776 & -0.32181 & -0.19954 \epm + \mc{O}(10^{-10}),
\quad
M_{2,*} = \bpm 0.34849 & -0.12072 & 0.07259 \\
-0.12072 & -0.17473 & 0.33055 \\
0.07259 & 0.33055 & 0.17624 \epm + \mc{O}(10^{-10}),
\nn
M_{3,*} = \bpm -0.35328 & 0.18107 & 0.03517 \\
0.18107 & 0.32998 & -0.00874 \\
0.03517 & -0.00874 & 0.37330 \epm + \mc{O}(10^{-10}),
\quad
M_{4,*} = \bpm 0.01045 & 0.03941 & 0.01356 \\
0.03941 & 0.14216 & 0.08416 \\
0.01356 & 0.08416 & -0.15261 \epm + \mc{O}(10^{-10}),
\nn
M_{5,*} = \bpm 0.06275 & 0.13508 & -0.08743 \\
0.13508 & -0.07105 & 0.01446 \\
-0.08743 & 0.01446 & 0.00839 \epm + \mc{O}(10^{-9}),
\quad
M_{6,*} = \bpm -0.05479 & -0.07500 & -0.14474 \\
-0.07500 & 0.04068 & 0.04027 \\
-0.14474 & 0.04027 & 0.01411 \epm + \mc{O}(10^{-9}).
\eg
Their imaginary parts are reduced below $10^{-9}$, meaning that $H_{\rm eff} (\bk)$ becomes effectively real-valued.
Consequently, the system exhibits an emergent $I \mc{T}$-symmetry represented simply by complex conjugation $\mc{K}$, consistent with the presence of an Euler number $|\mf{e}|=2$ in the spherical Wilson loop spectrum.
%%%%%%

%%%%%%
\subsubsection{Emergent transversality in the ab-initio model}
%%%%%%
In the ab-initio calculations, the spherical Wilson loop was evaluated using the magnetic ($\bb H$) field eigenmodes, as shown in Fig.~5\textbf{c} of the main text (See also the discussion in SN~\ref{sec:sn_sym_wilson} regarding our choice of using $\bb H$ fields to compute the Wilson loops).
In this case, the origin of the observed winding structure (corresponding to the Euler number $|\mf{e}|=2$) can be traced to the emergent transversality of the $\bb H$ field in the long-wavelength limit.
As discussed in SN~\ref{sec:sn_sym_longwave} and Eq.~\eqref{seq:hfield_approx}, the lowest (acoustic) modes of the $\bb H$ field can be approximated as $\bb H_\bk (\bb r) \simeq \bb h_\bk \, e^{i \bk \cdot \bb r}$, subject to the transversality condition $\bb h_\bk \cdot \bk = 0$.

Recall that the spherical Wilson loop is defined on a small sphere $S^2$ of radius $k_r$ centered at the $\Gamma$ point.
On this sphere, the transversality condition implies that the two transverse $\bb H$-field modes span the tangent bundle of $S^2$ in momentum space.
The Euler number of the tangent bundle to $S^2$ is given by the Euler characteristic of the sphere, which is 2~\cite{milnor1974characteristic}.
Consequently, the Wilson-loop spectrum of the $\bb H$ field naturally exhibits the winding corresponding to $|\mf{e}|=2$.

Alternatively, the same result follows directly from the two-patch construction of the isotropic transverse modes introduced in SN~\ref{sec:sn_iso_wl}.
There, the transverse modes $\ket{T^{(P)}_{1,2} (\theta, \phi)}$ defined on the two patches $P = N,S$ can be identified with the cell-periodic parts $\bb h_\bk$ of the two transverse $\bb H$ eigenfields\footnote{The cell-periodic fields $\bb h_\bk$ and the modes $\ket{T^{(P)}_{n=1,2} (\theta, \phi)}$ are equivalent up to normalization.
The $\bb h_\bk$ fields are normalized using an inner product weighted by $\mu_0$, whereas $\ket{T^{(P)}_n (\theta, \phi)}$ are normalized with respect to the standard (unweighted) inner product.}.
Using these eigenmodes and associated Berry connections [see Eqs.~\eqref{seq:iso_trans_vec_poles} and \eqref{seq:iso_berry_con}], the Wilson line along $\phi$ at fixed $\theta$ is
\bg
W_\phi (\theta; k_r) = e^{-2 i \pi \cos \theta \, \sigma_z},
\label{seq:iso_wilson_phi}
\eg
whose eigenvalue spectrum is
\bg
{\rm Spec} [-i \log W_\phi (\theta; k_r) ] = \{ +2\pi \cos \theta,\, -2\pi \cos \theta \}.
\label{seq:iso_wilson_euler}
\eg
The eigenvalue flow of $-i \log[W_\phi (\theta; k_r)]$ as $\theta$ varies from $0$ to $\pi$ produces the winding pattern shown in Fig.~\ref{sfig:iso_wilson}\textbf{b}, corresponding to $|\mf{e}|=2$.

As we increase the radius of the sphere, we expect gaps to open up in the Wilson loop spectrum which destroys the winding and trivializes it.
Surprisingly, for our specific structure we find that the Wilson loop winding is robust to within the precision of our calculations even for relatively large spheres.
Even for space groups with no inversion symmetry like SG $P4_332$, the Wilson loops show the $|\mf{e}|=2$ winding pattern when the sphere has radii comparable to the length of the reciprocal lattice vectors.
We have explicitly checked that at these radii, the fields themselves explicitly break inversion symmetry at moderate radii.
We expect that more intensive simulations with much higher resolutions are needed to conclusively observe the gapped Wilson loop winding in the MPB calculations.
As this is outside the scope of this paper, we leave this analysis to a future work.
%%%%%%

%%%%%%
\subsubsection{Comparison between $\bb H$-field and $\bb E$-field Wilson loops}
%%%%%%
We previously showed that the spherical Wilson loop based on the magnetic field $\bb H$ exhibits a winding of $|\mf{e}|=2$, which arises from the emergent transversality of the long-wavelength modes.
Since the Wilson loop can equally be defined using the electric field $\bb E$, it is natural to ask whether the same topological winding persists in that formulation and, if not, to what extent the $\bb E$- and $\bb H$-field Wilson loops differ quantitatively.
Unlike the $\bb H$ field, the $\bb E$ field may contain small longitudinal components and therefore fail to satisfy the transversality condition, as discussed in SN~\ref{sec:sn_sym_longwave}.
Below, we examine how these differences manifest in the spherical Wilson loop, starting from the long-wavelength relations between $\bb E$ and $\bb H$ and identifying the conditions under which their Wilson-loop spectra coincide.

We begin by considering the overlap matrices that enter the discrete formalism of the Wilson loop.
In the absence of magneto-electric coupling, separate overlap matrices can be defined for the $\bb E$ and $\bb H$ fields using the corresponding inner products weighted by $\ep(\bb r)$ and $\mu(\bb r)$, respectively\footnote{We defined the general overlap matrix in Eq.~\eqref{seq:overlap_phc} for the case with magneto-electric coupling.
Here, as the magneto-electric coupling is absent, we separately define the overlap matrices for $\bb E$ and $\bb H$ fields using the inner products weighted by $\ep(\bb r)$ and $\mu(\bb r)$.}:
\bg
[S^{(E)}_{\bk',\bk}]_{nm} = \langle \bb e_{n, \bk'} | \hat \ep(\bb r) |\bb e_{m, \bk} \rangle_{\rm uc},
\quad
[S^{(H)}_{\bk',\bk}]_{nm} = \langle \bb h_{n, \bk'} | \hat \mu(\bb r) |\bb h_{m, \bk} \rangle_{\rm uc},
\eg

Using the Maxwell equations, $S^{(E)}_{\bk',\bk}$ can be rewritten as
\ba
[S^{(E)}_{\bk',\bk}]_{nm}
=& \frac{1}{N_{\rm cell}} \int_{\rm all} d \bb r \, \bb e_{n, \bk'}(\bb r)^\dg \cdot \ep(\bb r) \cdot \bb e_{m,\bk} (\bb r)
\nn
=& \frac{1}{N_{\rm cell}} \frac{i}{\om_{m, \bk}} \int_{\rm all} d \bb r \, \bb e_{n, \bk'}(\bb r)^\dg \cdot \ep(\bb r) \cdot [ \left( \nabla_{\bb r} + i \bk' - i (\bk' - \bk) \right) \times \bb h_{m, \bk} (\bb r)]
\nn
=& \frac{1}{N_{\rm cell}} \frac{\om_{n, \bk'}}{\om_{m, \bk}} \int_{\rm all} d \bb r \, \bb h_{n, \bk'} (\bb r)^\dg \cdot \mu(\bb r) \cdot \bb h_{m, \bk} (\bb r)
+ \frac{1}{N_{\rm cell} \, \om_{m, \bk}} \int_{\rm all} d \bb r \, \bb e_{n, \bk'}(\bb r)^\dg \cdot [(\bk'-\bk) \times \bb h_{m, \bk} (\bb r)]
\nn
=& \frac{\om_{n, \bk'}}{\om_{m, \bk}} \langle \bb h_{n, \bk'} | \hat{\mu} | \bb h_{m, \bk} \rangle_{\rm uc} + \frac{1}{\om_{m, \bk}} \langle \bb e_{n, \bk'} | (\bk'-\bk) \times \bb h_{m, \bk} \rangle_{\rm uc}.
\label{seq:e_overlap_1}
\ea
Introducing the frequency matrix $[\Omega_\bk]_{nm} = \om_{n, \bk} \delta_{nm}$, Eq.~\eqref{seq:e_overlap_1} can be compactly written as
\bg
[S^{(E)}_{\bk',\bk}]_{nm}
= \sum_{n', m'} \, (\Omega_{\bk'})_{n n'} \, \left[[S^{(H)}_{\bk',\bk}]_{nm} + \frac{1}{\om_{n', \bk'}} \langle \bb e_{n', \bk'} | (\bk'-\bk) \times \bb h_{m', \bk} \rangle_{\rm uc} \right] \, (\Omega_\bk^{-1})_{m' m},
\label{seq:overal_e_h_compare}
\eg
where $n'$ and $m'$ are restricted to the band indices within the subset of interest.

We now consider, as before, a momentum-space sphere $S^2$ of small radius $k_r$, on which the spherical Wilson loop $W^{(a)}_\phi(\theta; k_r)$ ($a = H, E$) is defined for the $\bb H$ and $\bb E$ fields.
In the discretized formalism, the Wilson loop for the $\bb H$ field is expressed as
\bg
[W^{(H)}_\phi (\theta; k_r)]_{nm}
= \left[ \mc{S}^{(H)}_{2\pi, 2\pi - \Delta \phi} (\theta) \cdot \mc{S}^{(H)}_{2\pi - \Delta \phi, 2\pi - 2\Delta \phi} (\theta) \cdot \mc{S}^{(H)}_{2\pi - 2\Delta \phi, 2\pi - 3\Delta \phi} (\theta) \cdots \mc{S}^{(H)}_{\Delta \phi, 0} (\theta) \right]_{nm}.
\label{seq:wl_H_sph}
\eg
Here, $\Delta \phi$ denotes a small discretization step along the azimuthal direction on the sphere.
The matrix $\mc{S}^{(H)}_{\phi', \phi} (\theta) = S^{(H)}_{\bk_{\theta, \phi'}, \bk_{\theta, \phi}}$ is defined with $\bk_{\theta,\phi} = k_r \, \hat r (\theta, \phi)$ on the sphere [Eq.~\eqref{seq:sph_normal_vec}].
The definition of $W^{(E)}_\phi (\theta; k_r)$ is analogous, with $\mc{S}^{(E)}_{\phi', \phi} (\theta)$ replacing $\mc{S}^{(H)}_{\phi', \phi} (\theta)$.

In the long-wavelength limit with $\mu(\bb r) = \mu_0$ and periodic $\ep(\bb r)$, the $\bb E$ and $\bb H$ fields are approximated by Eqs.~\eqref{seq:efield_approx} and \eqref{seq:h_approx_e}.
Since the two lowest $\bb H$ eigenfields are transverse plane waves, we introduce two orthonormal tangent vectors $\bb t_{1,2} (\theta, \phi)$ on the momentum-space sphere, defined as
\bg
\bb t_1 (\theta, \phi) = \frac{1}{\sqrt{\mu_0 \, V_{\rm uc}}} \, \hat \theta (\theta, \phi),
\quad
\bb t_2 (\theta, \phi) = \frac{1}{\sqrt{\mu_0 \, V_{\rm uc}}} \, \hat \phi (\theta, \phi),
\label{seq:t_field_sph}
\eg
where $V_{\rm uc}$ is the unit-cell volume, chosen such that $\langle \bb t_n (\theta, \phi) | \mu_0 | \bb t_m (\theta, \phi) \rangle_{\rm uc} = \delta_{nm}$ for $n,m=1,2$.
On the sphere, the cell-periodic parts of the $\bb H$ and $\bb D$ fields can then be expanded as
\bg
\bb h_{n, \bk_{\theta,\phi}} = \sum_{m=1,2} \, \bb t_m (\theta,\phi) \, [M_h (\theta,\phi)]_{mn},
\quad
\bb d_{n, \bk_{\theta,\phi}} = \frac{k_r}{\om_{n, \bk_{\theta,\phi}}} \, \sum_{m=1,2} \, \bb t_m (\theta,\phi) \, [M_d (\theta,\phi)]_{mn},
\label{seq:h_small_sph}
\eg
where $M_h (\theta, \phi)$ and $M_d (\theta, \phi)$ are unitary matrices encoding the system-specific mixing of $\bb t_{1,2}(\theta, \phi)$.
The Maxwell equation $\nabla_{\bb r} \times \bb H_{n, \bk} (\bb r) = - i \om_{n, \bk} \bb D_{n, \bk}(\bb r)$ imposes the constraint
\bg
M_d (\theta, \phi) = i \sg_y M_h (\theta, \phi).
\label{seq:mixing_t_relation}
\eg
Combining this with Eq.~\eqref{seq:efield_approx}, the cell-periodic parts of the two lowest $\bb E$ eigenfields can be expressed as
\bg
\bb e_{n, \bk_{\theta, \phi}} (\bb r) = \bb e^{(T)}_{n, \bk_{\theta, \phi}} + \delta \bb e_{n, \bk_{\theta, \phi}} (\bb r),
\nn
\bb e^{(T)}_{n, \bk_{\theta, \phi}} = \mu_0 \, \frac{\om_{n, \bk_{\theta, \phi}}}{k_r} \, \sum_{m=1,2} \, \bb t_m (\theta, \phi) \, [M_d (\theta,\phi)]_{mn},
\quad
\delta \bb e_{n, \bk_{\theta, \phi}} (\bb r) = \sum_{\bb G} \, \bb E^{(L)}_{n, \bk_{\theta, \phi}, \bb G} \, e^{i \bb G \cdot \bb r},
\label{seq:e_small_sph}
\eg
where $\bb E^{(L)}_{n, \bk, \bb G}$ are longitudinal Fourier components of $\bb E_{n, \bk} (\bb r)$ satisfying $\bb E_{n, \bk, \bb G} \parallel (\bk + \bb G)$.

Now, using the explicit forms of $\bb h_{n, \bk_{\theta, \phi}}$ and $\bb e_{n, \bk_{\theta, \phi}} (\bb r)$, we revisit Eq.~\eqref{seq:overal_e_h_compare}.
On the momentum-space sphere, it can be written as
\bg
\mc{S}^{(E)}_{\phi + \Delta \phi, \phi} (\theta)
= \Omega_{\bk_{\theta,\phi'}} \cdot \left[ \mc{S}^{(H)}_{\phi + \Delta \phi, \phi} (\theta) + \mc{M}_{\phi+ \Delta \phi, \phi} (\theta) \right] \cdot \Omega_{\bk_{\theta, \phi}}^{-1},
\label{seq:overal_e_h_comp1}
\eg
where we have introduced, for notational simplicity, a mixing term $\mc{M}_{\phi + \Delta \phi, \phi} (\theta)$ that couples the $\bb E$ and $\bb H$ fields:
\bg
\left[ \mc{M}_{\phi + \Delta \phi, \phi} (\theta) \right]_{nm}
= \frac{1}{\om_{n, \bk_{\theta, \phi + \Delta \phi}}} \langle \bb e_{n, \bk_{\theta, \phi + \Delta \phi}} | (\bk_{\theta, \phi + \Delta \phi}-\bk_{\theta, \phi}) \times \bb h_{m, \bk_{\theta, \phi}} \rangle_{\rm uc}.
\label{seq:overal_mixing}
\eg
To compare the $\bb H$- and $\bb E$-field Wilson loops constructed from $\mc{S}^{(H,E)}_{\phi + \Delta \phi, \phi} (\theta)$, we next expand both $\mc{S}^{(H)}_{\phi + \Delta \phi, \phi} (\theta)$ and $\mc{M}_{\phi + \Delta \phi, \phi} (\theta)$ in powers of $\Delta \phi$.
%%%%%%

%%%%%%
\paragraph{Expansion of $\mc{S}^{(H)}_{\phi + \Delta \phi, \phi} (\theta)$.}
%%%%%%
From Eq.~\eqref{seq:h_small_sph}, $\mc{S}^{(H)}_{\phi + \Delta \phi, \phi} (\theta)$ can be expressed as
\ba
\mc{S}^{(H)}_{\phi + \Delta \phi, \phi}
=& M_h (\theta, \phi + \Delta \phi)^\dg \cdot \mc{S}^{(t)}_{\phi + \Delta \phi, \phi} (\theta) \cdot M_h (\theta, \phi),
\label{seq:overal_H_t}
\ea
where $\mc{S}^{(t)}_{\phi + \Delta \phi, \phi} (\theta)$ is the overlap matrix between $\bb t_n (\theta, \phi)$ vectors.
Using Eq.~\eqref{seq:t_field_sph}, we obtain
\bg
[\mc{S}^{(t)}_{\phi+\Delta \phi, \phi} (\theta)]_{nm}
= \langle \bb t_n (\theta, \phi+\Delta \phi) | \mu_0 | \bb t_m (\theta, \phi) \rangle_{\rm uc}
= \left[ \sg_0 + i \Delta \phi \, A^{(t)}_\phi (\theta, \phi) \right]_{nm} + \mc{O}(\Delta \phi^2),
\eg
where $A^{(t)}_\phi (\theta, \phi) = \cos \theta \, \sg_y$ is the Berry connection along the $\phi$ direction computed from $\bb t_n$.
Combining Eqs.~\eqref{seq:wl_H_sph} and \eqref{seq:overal_H_t}, the spherical Wilson loop for the $\bb H$ field becomes
\ba
W^{(H)}_\phi (\theta; k_r)
=& M_h (\theta, 2\pi)^\dg \cdot \left[ \mc{S}^{(t)}_{2\pi, 2\pi-\Delta \phi} (\theta) \cdot \mc{S}^{(t)}_{2\pi-\Delta \phi, 2\pi-2\Delta \phi} (\theta) \cdots \mc{S}^{(t)}_{\Delta \phi, 0} \right] \cdot M_h(\theta, 0)
\nn
=& M_h (\theta, 0)^\dg \cdot \mc{P} \exp \left[ i \oint d \phi \, A^{(t)}_\phi (\theta, \phi) \right] \cdot M_h (\theta, 0)
\nn
=& M_h (\theta, 0)^\dg \cdot e^{2 i \pi \cos \theta \, \sg_y} \cdot M_h (\theta, 0)
\ea
in the limit $\Delta \phi \to 0$, where we used the periodicity $M_h (\theta, 2\pi) = M_h (\theta, 0)$ in the second line.
Thus, $W^{(H)}_\phi (\theta; k_r)$ is unitarily equivalent to $e^{2 i \pi \cos \theta \, \sg_y}$.
This form is also equivalent to $e^{-2 i \pi \cos \theta \, \sg_z}$ derived in Eq.~\eqref{seq:iso_wilson_phi}, which explains the winding corresponding to the Euler number $|\mf{e}|=2$ in the $\bb H$-field Wilson loop spectrum.
%%%%%%

%%%%%%
\paragraph{Expansion of $\mc{M}_{\phi + \Delta \phi, \phi} (\theta)$.}
%%%%%%
The above discussion implies that if the contribution of $\mc{M}_{\phi + \Delta \phi, \phi} (\theta)$ is of order $\mc{O} (\Delta \phi^2)$, then the $\bb H$- and $\bb E$-field Wilson loops, $W^{(H,E)}_\phi (\theta; k_r)$, have identical spectra.
This follows directly from Eq.~\eqref{seq:overal_e_h_comp1} and the definition of the Wilson loops:
\bg
\Omega_{\bk_{\theta, 2\pi}}^{-1} \cdot W^{(E)}_\phi (\theta; k_r) \cdot \Omega_{\bk_{\theta, 0}}
= \Omega_{\bk_{\theta, 0}}^{-1} \cdot W^{(E)}_\phi (\theta; k_r) \cdot \Omega_{\bk_{\theta, 0}}
\nn
= \left( \mc{S}^{(H)}_{2\pi, 2\pi - \Delta \phi} (\theta) + \mc{M}_{2\pi, 2\pi - \Delta \phi} (\theta) \right) \cdots \left( \mc{S}^{(H)}_{2\Delta \phi, \Delta \phi} (\theta) + \mc{M}_{2\Delta \phi, \Delta \phi} (\theta) \right) \cdot \left( \mc{S}^{(H)}_{\Delta \phi, 0} (\theta) + \mc{M}_{\Delta \phi, 0} (\theta) \right).
\label{seq:overal_e_h_comp2}
\eg
If $\mc{M}_{\phi + \Delta \phi, \phi} (\theta) \sim \mc{O} (\Delta \phi^2)$, then $W^{(E)}_\phi (\theta; k_r)$ is related to $W^{(H)}_\phi (\theta; k_r)$ by a similarity transformation with $\Omega_{\bk_{\theta, 0}}$.
Since similar matrices share the same eigenvalues, the two Wilson loops have identical spectra.
To determine whether this condition holds, we expand $\mc{M}_{\phi + \Delta \phi, \phi} (\theta)$ in powers of $\Delta \phi$ and show that, in general, a first-order correction may arise from the longitudinal components of the $\bb E$ field.

We decompose $\mc{M}_{\phi + \Delta \phi, \phi} (\theta)$ into two contributions:
\ba
\left[ \mc{M}^{(T)}_{\phi + \Delta \phi, \phi} (\theta) \right]_{nm}
=& \frac{1}{\om_{n, \bk_{\theta, \phi + \Delta \phi}}} \langle \bb e^{(T)}_{n, \bk_{\theta, \phi + \Delta \phi}} | (\bk_{\theta, \phi + \Delta \phi}-\bk_{\theta,\phi}) \times \bb h_{m, \bk_{\theta, \phi}} \rangle_{\rm uc},
\nn
\left[ \mc{M}^{(L)}_{\phi + \Delta \phi, \phi} (\theta) \right]_{nm}
=& \frac{1}{\om_{n, \bk_{\theta, \phi + \Delta \phi}}} \langle \bb e^{(L)}_{n, \bk_{\theta, \phi + \Delta \phi}} | (\bk_{\theta, \phi + \Delta \phi}-\bk_{\theta, \phi}) \times \bb h_{m, \bk_{\theta, \phi}} \rangle_{\rm uc},
\ea
based on Eq.~\eqref{seq:e_small_sph}.
We first simplify $\mc{M}^{(T)}_{\phi + \Delta \phi, \phi} (\theta)$ and show that it contributes only at $\mc{O} (\Delta \phi^2)$.
Using Eqs.\eqref{seq:t_field_sph} and \eqref{seq:e_small_sph}, we obtain
\bg
\mu_0 \sum_{n', m'=1,2} \, [M_d (\theta, \Delta + \phi)^\dg]_{n n'} \, \int_{\rm uc} d \bb r \, \bb t_{n'} (\theta, \Delta + \phi)^\dg \cdot \, [\hat r (\theta, \phi + \Delta \phi) - \hat r (\theta, \phi)] \times \bb t_{m'} (\theta, \phi) \, [M_h (\theta, \phi)]_{m'm}
\nn
= \frac{1}{2} (\Delta \phi)^2 \sin^2 \theta \, \left[ M_d (\theta, \phi + \Delta \phi)^\dg \cdot \sg_z \cdot M_d (\theta, \phi) \right]_{nm} + \mc{O}(\Delta \phi^4),
\eg
where we used the Taylor expansion of the unit vectors $\{\hat r, \hat \theta, \hat \phi\}$ for small $\Delta \phi$, together with the relation between $M_h (\theta, \phi)$ and $M_d (\theta, \phi)$ given in Eq.~\eqref{seq:mixing_t_relation}.
Being of order $\mc{O} (\Delta \phi^2)$, $\mc{M}^{(T)}_{\phi + \Delta \phi, \phi} (\theta)$ introduces no spectral deviation between $W^{(E)}_\phi (\theta; k_r)$ and $W^{(H)}_\phi (\theta; k_r)$ to leading order.

We next examine $\mc{M}^{(L)}_{\phi + \Delta \phi, \phi} (\theta)$, which originates from the longitudinal components of the $\bb E$ field.
From Eq.~\eqref{seq:e_small_sph},
\ba
[\mc{M}^{(L)}_{\phi + \Delta \phi, \phi} (\theta)]_{nm}
=& \frac{1}{N_{\rm cell} \, \om_{n, \bk_{\theta, \phi + \Delta \phi}}} \, \int_{\rm all} d \bb r \, \sum_{\bb G} \, e^{-i \bb G \cdot \bb r} \, {\bb E^{(L)}_{n, \bk_{\theta, \phi + \Delta \phi}, \bb G}}^\dg \cdot (\bk_{\theta, \phi + \Delta \phi} - \bk_{\theta, \phi}) \times \bb h_{m, \bk_{\theta, \phi}}
\nn
=& \frac{V_{\rm uc}}{\om_{n, \bk_{\theta, \phi + \Delta \phi}}} \bb E^{(L)}_{n, \bk_{\theta, \phi + \Delta \phi}, \bb 0} \cdot (\bk_{\theta, \phi + \Delta \phi} - \bk_{\theta, \phi}) \times \bb h_{m, \bk_{\theta, \phi}},
\ea
where $\int_{\rm all} d \bb r \, e^{i \bb G \cdot \bb r} = N_{\rm cell} V_{\rm uc} \delta_{\bb G, \bb 0}$ is used.
Writing the longitudinal Fourier component as
\bg
\bb E^{(L)}_{n, \bk_{\theta, \phi + \Delta \phi}, \bb 0}
= s_{\theta, \phi + \Delta \phi} \, |\bb E^{(L)}_{n, \bk_{\theta, \phi}, \bb 0}| \, \hat r (\theta, \phi + \Delta \phi),
\eg
where $s_{\theta, \phi + \Delta \phi} = \pm 1$ represents the sign of the longitudinal component, we see that $[\mc{M}^{(L)}_{\phi + \Delta \phi, \phi} (\theta)]_{nm}$ is equal to
\bg
s_{\theta, \phi + \Delta \phi} \, \frac{V_{\rm uc} \, k_r}{\om_{n, \bk_{\theta, \phi + \Delta \phi}}} |\bb E^{(L)}_{n, \bk_{\theta, \phi}, \bb 0}| \hat r (\theta, \phi + \Delta \phi) \cdot (\hat r (\theta, \phi + \Delta \phi) - \hat r (\theta, \phi)) \times \sum_{m=1,2} \, \bb t_m (\theta, \phi) [M_h (\theta, \phi)]_{mn}.
\eg
The term $\hat r (\theta, \phi + \Delta \phi) \cdot (\hat r (\theta, \phi + \Delta \phi) - \hat r (\theta, \phi))$ is generally of order $\mc{O} (\Delta \phi)$.

To summarize, when the longitudinal Fourier component of the $\bb E$ field is subleading in the long-wavelength limit, the $\bb E$- and $\bb H$-field Wilson loops become equivalent up to a similarity transformation.
Otherwise, the longitudinal correction can appear already at first order in $\Delta \phi$, leading to quantitative differences between the two Wilson loops.
A systematic analytical and numerical investigation of these deviations, as well as the possible topological (in)equivalence of more general Wilson loops beyond the spherical case, is left for future work.
Indeed, even for the bands above the lowest gap, which are free from the polarization singularity, 
the proof of the equivalence between the $\bb E$- and $\bb H$-field Chern numbers requires a rigorous mathematical treatment based on functional analysis and operator theory~\cite{de2020equivalence}, emphasizing that such topological equivalence is not entirely trivial.
%%%%%%

%%%%%%
\section{Photonic band structures in SG $P432$ (No. 207)}
\label{sec:sn_sg207}
%%%%%%
In this section, we extend our analysis of photonic band structures to a different space group, SG $P432$ (No. 207), to demonstrate the general applicability of our SRSI-based framework.
The methods developed for SG $P4_332$ (No. 212), which include the construction of SRSIs, their mapping to symmetry-data vectors, and the classification of physically allowed SRSI values using the Hilbert basis method, can be systematically applied to any SG.
Here, we explicitly work out the case of SG $P432$ to illustrate this process in a more complex setting, where the number of $\Z$-valued SRSIs increases and graphical or brute-force enumeration becomes impractical.

We begin by defining the SRSIs and constructing their mapping to the symmetry-data vector.
We then classify the allowed values of SRSIs in photonic and electronic band structures using the Hilbert basis method, supplemented by graphical visualization in selected fixed-SRSI slices.
Finally, we analyze specific photonic bands corresponding to one such allowed SRSI value using both ab-initio and tight-binding models, providing detailed matching of symmetry content and topological features.
%%%%%%

%%%%%%
\subsection{Stable real-space invariants (SRSIs)}
\label{sec:sn_sg207_rsi}
%%%%%%
The symmorphic SG $P432$ is generated by fourfold $C_{4,001}$, threefold $C_{3,111}$, twofold $C_{2,110}$ rotations, and three-dimensional translations $\{E | \bb v \in \Z\}$.
The rotation symmetries $C_{4,001}$, $C_{3,111}$, and $C_{2,110}$ transform $(x,y,z)$ to $(-y,x,z)$, $(z,x,y)$, and $(y,x,-z)$, respectively.
In this SG, there are four maximal WPs, with representative positions as follows: $1a(0,0,0)$, $1b(1/2,1/2,1/2)$, $3c(0,1/2,1/2)$, and $3d(1/2,0,0)$.
The nonmaximal WPs includes $6e(x,0,0)$, $6f(x,1/2,1/2)$, $8g(x,x,x)$, $12h(x,1/2,0)$, $12i(0,y,y)$, $12j(1/2,y,y)$, and $24k(x,y,z)$.
The full list of site-symmetry irreps at all WPs is:
\bg
(A_1)_a, (A_2)_a, (E)_a, (T_1)_a, (T_2)_a, (A_1)_b, (A_2)_b, (E)_b, (T_1)_b, (T_2)_b, (A_1)_c, (B_1)_c,
\nn
(A_2)_c, (B_2)_c, (E)_c, (A_1)_d, (B_1)_d, (A_2)_d, (B_2)_d, (E)_d, (A)_e, (B)_e, ({}^1 E {}^2 E)_e,
\nn
(A)_f, (B)_f, ({}^1 E {}^2 E)_f, (A_1)_g, ({}^1 E {}^2 E)_g, (A)_h, (B)_h, (A)_i, (B)_i, (A)_j, (B)_j, (A)_k.
\label{seq:sg207_irrep_full}
\eg

By considering adiabatic deformation processes between site-symmetry irreps, one can derive 9 $\Z$-valued SRSIs, following the algorithm explained in the Methods of the main text and originally proposed in Ref.~\cite{hwang2025stable}.
Explicit expressions for the SRSIs for all 230 SGs are also provided in Ref.~\cite{hwang2025stable}.
Then, the $\Z$-valued SRSIs, denoted by $\bs \theta^{(\rm full)}_\Z = (\theta^{(\rm full)}_1, \dots, \theta^{(\rm full)}_9)$, are given by
\bg
\bs \theta^{(\rm full)}_\Z =
\bpm
\begin{smallmatrix}
1 & 0 & 0 & 0 & 0 & 0 & 0 & -1 & 0 & 1 & 0 & 0 & -1 & 0 & 1 & 1 & 0 & 0 & 0 & 0 & 1 & 0 & 0 & -1 & 0 & 2 & 1 & 0 & 1 & 0 & 1 & 0 & 1 & 0 & 1
\\
0 & 1 & 0 & 0 & 0 & 0 & 0 & -1 & 0 & 1 & 0 & 1 & -1 & -1 & 1 & 1 & 1 & -1 & 0 & 0 & 0 & 1 & 0 & -1 & 0 & 2 & 1 & 0 & 2 & -1 & 0 & 1 & 1 & 0 & 1
\\
0 & 0 & 1 & 0 & 0 & 0 & 0 & 1 & 0 & 0 & 0 & 0 & 1 & 1 & 0 & 0 & 0 & 1 & 1 & 0 & 1 & 1 & 0 & 1 & 1 & 0 & 0 & 2 & 0 & 2 & 1 & 1 & 1 & 1 & 2
\\
0 & 0 & 0 & 1 & 0 & 0 & 0 & 0 & 0 & 1 & 0 & 1 & 0 & 0 & 1 & 1 & 0 & 0 & 0 & 1 & 1 & 0 & 2 & 0 & 1 & 2 & 1 & 2 & 2 & 1 & 1 & 2 & 2 & 1 & 3
\\
0 & 0 & 0 & 0 & 1 & 0 & 0 & 0 & 0 & 1 & 0 & 0 & 0 & 1 & 1 & 0 & 0 & 0 & 1 & 1 & 0 & 1 & 2 & 0 & 1 & 2 & 1 & 2 & 1 & 2 & 2 & 1 & 2 & 1 & 3
\\
0 & 0 & 0 & 0 & 0 & 1 & 0 & 1 & 0 & -1 & 0 & -1 & 2 & 1 & -1 & -1 & -1 & 1 & 1 & 0 & 0 & 0 & 0 & 2 & 0 & -2 & 0 & 0 & -2 & 2 & 0 & 0 & 0 & 0 & 0
\\
0 & 0 & 0 & 0 & 0 & 0 & 1 & 1 & 0 & -1 & 0 & 0 & 1 & 1 & -1 & -1 & 0 & 1 & 0 & 0 & 0 & 0 & 0 & 1 & 1 & -2 & 0 & 0 & -1 & 1 & 0 & 0 & -1 & 1 & 0
\\
0 & 0 & 0 & 0 & 0 & 0 & 0 & 0 & 1 & -1 & 0 & -1 & 1 & 0 & 0 & -1 & 0 & 1 & 0 & 0 & 0 & 0 & 0 & 1 & -1 & 0 & 0 & 0 & -1 & 1 & 0 & 0 & -1 & 1 & 0
\\
0 & 0 & 0 & 0 & 0 & 0 & 0 & 0 & 0 & 0 & 1 & 1 & -1 & -1 & 0 & 1 & 1 & -1 & -1 & 0 & 0 & 0 & 0 & 0 & 0 & 0 & 0 & 0 & 2 & -2 & 0 & 0 & 0 & 0 & 0
\end{smallmatrix}
\epm \cdot \bb m^{(\rm full)},
\eg
where the full site-symmetry irrep multiplicity vector $\bb m^{(\rm full)}$ is defined by
\bg
\bb m^{(\rm full)} = \left( \nW{(A_1)_a}, \nW{(A_2)_a}, \dots, \nW{(B)_j}, \nW{(A)_k} \right)^T,
\eg
where $\nW{(\rho)_W}$ denotes the multiplicity of the irrep $(\rho)_W$, and the basis is ordered according to Eq.~\eqref{seq:sg207_irrep_full}.
It collects the multiplicities of all site-symmetry irreps at all WPs, including nonmaximal WPs.
Accordingly, the superscript $(\rm full)$ on $\bs \theta^{(\rm full)}_\Z$ indicates that these $\Z$-valued SRSIs are computed from the full multiplicity vector $\bb m^{(\rm full)}$.

For simplicity, we assume that all site-symmetry irreps at nonmaximal WPs have zero multiplicity, as they can always be adiabatically deformed into those at connected maximal WPs.
(Note that this assumption is not essential, and the following analysis remains valid without it.) 
With this simplification, the $\Z$-valued SRSIs reduce to
\ba
\theta_1 =& \nW{(A_1)_a} - \nW{(E)_b} + \nW{(T_2)_b} - \nW{(A_2)_c} + \nW{(E)_c} + \nW{(A_1)_d},
\nn
\theta_2 =& \nW{(A_2)_a} - \nW{(E)_b} + \nW{(T_2)_b} - \nW{(A_2)_c} + \nW{(B_1)_c} - \nW{(B_2)_c} + \nW{(E)_c} \nn
& + \nW{(A_1)_d} - \nW{(A_2)_d} + \nW{(B_1)_d}
\nn
\theta_3 =& \nW{(E)_a} + \nW{(E)_b} + \nW{(A_2)_c} + \nW{(B_2)_c} + \nW{(A_2)_d} + \nW{(B_2)_d}
\nn
\theta_4 =& \nW{(T_1)_a} + \nW{(T_2)_b} + \nW{(B_1)_c} + \nW{(E)_c} + \nW{(A_1)_d} + \nW{(E)_d}
\nn
\theta_5 =& \nW{(T_2)_a} + \nW{(T_2)_b} + \nW{(B_2)_c} + \nW{(E)_c} + \nW{(B_2)_d} + \nW{(E)_d}
\nn
\theta_6 =& \nW{(A_1)_b} + \nW{(E)_b} - \nW{(T_2)_b} + 2\nW{(A_2)_c} - \nW{(B_1)_c} + \nW{(B_2)_c} - \nW{(E)_c} \nn
& - \nW{(A_1)_d} + \nW{(A_2)_d} - \nW{(B_1)_d} + \nW{(B_2)_d}
\nn
\theta_7 =& \nW{(A_2)_b} + \nW{(E)_b} - \nW{(T_2)_b} + \nW{(A_2)_c} + \nW{(B_2)_c} - \nW{(E)_c} - \nW{(A_1)_d} + \nW{(A_2)_d}
\nn
\theta_8 =& \nW{(T_1)_b} - \nW{(T_2)_b} + \nW{(A_2)_c} - \nW{(B_1)_c} - \nW{(A_1)_d} + \nW{(A_2)_d}
\nn
\theta_9 =& \nW{(A_1)_c} -\nW{(A_2)_c} + \nW{(B_1)_c} - \nW{(B_2)_c} + \nW{(A_1)_d} - \nW{(A_2)_d} + \nW{(B_1)_d} - \nW{(B_2)_d}.
\label{seq:sg207_rsi0}
\ea
Here, each $\theta^{(\rm full)}_i$ ($i=1,\dots,9$) reduces to $\theta_i$.
Accordingly, we distinguish $\bs \theta^{(\rm full)}_\Z$, computed from the full multiplicity vector $\bb m^{(\rm full)}$, from the simplified SRSIs $\bs \theta_\Z$, which are defined under the assumption that all site-symmetry irreps are shifted to maximal WPs.
Note that the total dimension $\nu$ of all site-symmetry irreps per unit cell or total number of bands that are being considered is expressed in terms of the SRSIs as
\bg
\nu = \theta_1 + \theta_2 + 2 \theta_3 + 3 \theta_4 + 3 \theta_5 + \theta_6 + \theta_7 + 3 \theta_8 + 3 \theta_9.
\label{seq:sg207_filling}
\eg

We can rewrite Eq.~\eqref{seq:sg207_rsi0} in a compact form using the reduced irrep multiplicity vector
\bg
\bb m = \left( \nW{(A_1)_a}, \nW{(A_2)_a}, \nW{(E)_a}, \dots, \nW{(A_2)_d}, \nW{(B_2)_d}, \nW{(E)_d} \right)^T,
\label{seq:sg207_irrep}
\eg
which includes only the irrep multiplicities associated with the maximal WPs $1a$, $1b$, $3c$, and $3d$.
The vector $\bb m$ has a length $N_{EBR}=20$.
Then, we express $\bs \theta_\Z$ as
\bg
\bs \theta_\Z
= \bpm \begin{smallmatrix}
1 & 0 & 0 & 0 & 0 & 0 & 0 & -1 & 0 & 1 & 0 & 0 & -1 & 0 & 1 & 1 & 0 & 0 & 0 & 0
\\
0 & 1 & 0 & 0 & 0 & 0 & 0 & -1 & 0 & 1 & 0 & 1 & -1 & -1 & 1 & 1 & 1 & -1 & 0 & 0
\\
0 & 0 & 1 & 0 & 0 & 0 & 0 & 1 & 0 & 0 & 0 & 0 & 1 & 1 & 0 & 0 & 0 & 1 & 1 & 0
\\
0 & 0 & 0 & 1 & 0 & 0 & 0 & 0 & 0 & 1 & 0 & 1 & 0 & 0 & 1 & 1 & 0 & 0 & 0 & 1
\\
0 & 0 & 0 & 0 & 1 & 0 & 0 & 0 & 0 & 1 & 0 & 0 & 0 & 1 & 1 & 0 & 0 & 0 & 1 & 1
\\
0 & 0 & 0 & 0 & 0 & 1 & 0 & 1 & 0 & -1 & 0 & -1 & 2 & 1 & -1 & -1 & -1 & 1 & 1 & 0
\\
0 & 0 & 0 & 0 & 0 & 0 & 1 & 1 & 0 & -1 & 0 & 0 & 1 & 1 & -1 & -1 & 0 & 1 & 0 & 0
\\
0 & 0 & 0 & 0 & 0 & 0 & 0 & 0 & 1 & -1 & 0 & -1 & 1 & 0 & 0 & -1 & 0 & 1 & 0 & 0
\\
0 & 0 & 0 & 0 & 0 & 0 & 0 & 0 & 0 & 0 & 1 & 1 & -1 & -1 & 0 & 1 & 1 & -1 & -1 & 0
\end{smallmatrix} \epm \cdot \bb m
:= \Delta_\Z \cdot \bb m,
\label{seq:sg207_rsi}
\eg
where the matrix $\Delta_\Z$ defines the linear map from irrep multiplicities at maximal WPs to the corresponding $\Z$-valued SRSIs.
%%%%%%

%%%%%%
\subsection{SRSI-based classification of photonic and electronic bands}
\label{sec:sn_sg207_map}
%%%%%%
We now investigate how photonic and electronic bands can be classified in terms of $\Z$-valued SRSIs.
We focus on topologically trivial bands, which are characterized by integer-valued multiplicities of site-symmetry irreps.
Consequently, the associated $\Z$-valued SRSIs also take integer values.
To enable this classification, we first establish the mapping between the symmetry-data vector $\bb v$ and the SRSIs $\bs \theta_\Z$.

In SG $P432$, the little-group irreps are defined at HSM $\Gamma = (0,0,0)$, $M = (\pi,\pi,0)$, $R = (\pi,\pi,\pi)$, and $X = (0,\pi,0)$, and are labeled as
\bg
\Gamma_1, \Gamma_2, \Gamma_3, \Gamma_4, \Gamma_5,
M_1, M_2, M_3, M_4, M_5,
R_1, R_2, R_3, R_4, R_5,
X_1, X_2, X_3, X_4, X_5.
\label{seq:sg207_kirrep}
\eg
Following this ordering, we define the symmetry-data vector as
\bg
\bb v = [ \nR{\Gamma_1}, \nR{\Gamma_2}, \nR{\Gamma_3}, \dots, \nR{X_3}, \nR{X_4}, \nR{X_5} ]^T,
\label{seq:sg207_svec}
\eg
which has total length $N_{BZ}=20$.
This vector $\bb v$ is determined by the site-symmetry irrep multiplicity vector $\bb m$ introduced in Eq.~\eqref{seq:sg207_irrep}, through a linear transformation defined by the elementary band representation (EBR) matrix:
\bg
\bb v = EBR \cdot \bb m,
\eg
where $EBR$ is an $N_{BZ}$ by $N_{EBR}$ matrix.
Each column of $EBR$ represents the symmetry-data vector associated with an EBR induced from a specific site-symmetry irrep.
This information can be obtained from the BCS~\cite{bilbao_EBR}, and the matrix $EBR$ is explicitly given by
\bg
EBR = \bpm
\begin{smallmatrix}
1 & 0 & 0 & 0 & 0 & 1 & 0 & 0 & 0 & 0 & 1 & 0 & 0 & 0 & 0 & 1 & 0 & 0 & 0 & 0
\\
0 & 1 & 0 & 0 & 0 & 0 & 1 & 0 & 0 & 0 & 0 & 1 & 0 & 0 & 0 & 0 & 1 & 0 & 0 & 0
\\
0 & 0 & 1 & 0 & 0 & 0 & 0 & 1 & 0 & 0 & 1 & 1 & 0 & 0 & 0 & 1 & 1 & 0 & 0 & 0
\\
0 & 0 & 0 & 1 & 0 & 0 & 0 & 0 & 1 & 0 & 0 & 0 & 1 & 0 & 1 & 0 & 0 & 1 & 0 & 1
\\
0 & 0 & 0 & 0 & 1 & 0 & 0 & 0 & 0 & 1 & 0 & 0 & 0 & 1 & 1 & 0 & 0 & 0 & 1 & 1
\\
1 & 0 & 1 & 0 & 0 & 0 & 0 & 0 & 0 & 1 & 0 & 0 & 0 & 1 & 1 & 1 & 0 & 1 & 1 & 0
\\
0 & 1 & 1 & 0 & 0 & 0 & 0 & 0 & 1 & 0 & 0 & 0 & 1 & 0 & 1 & 0 & 1 & 1 & 1 & 0
\\
0 & 0 & 0 & 1 & 0 & 0 & 1 & 1 & 0 & 0 & 0 & 1 & 1 & 1 & 0 & 0 & 0 & 1 & 0 & 1
\\
0 & 0 & 0 & 0 & 1 & 1 & 0 & 1 & 0 & 0 & 1 & 0 & 1 & 1 & 0 & 0 & 0 & 0 & 1 & 1
\\
0 & 0 & 0 & 1 & 1 & 0 & 0 & 0 & 1 & 1 & 1 & 1 & 0 & 0 & 2 & 1 & 1 & 0 & 0 & 2
\\
1 & 0 & 0 & 0 & 0 & 0 & 1 & 0 & 0 & 0 & 0 & 0 & 0 & 1 & 0 & 0 & 0 & 1 & 0 & 0
\\
0 & 1 & 0 & 0 & 0 & 1 & 0 & 0 & 0 & 0 & 0 & 0 & 1 & 0 & 0 & 0 & 0 & 0 & 1 & 0
\\
0 & 0 & 1 & 0 & 0 & 0 & 0 & 1 & 0 & 0 & 0 & 0 & 1 & 1 & 0 & 0 & 0 & 1 & 1 & 0
\\
0 & 0 & 0 & 1 & 0 & 0 & 0 & 0 & 0 & 1 & 0 & 1 & 0 & 0 & 1 & 1 & 0 & 0 & 0 & 1
\\
0 & 0 & 0 & 0 & 1 & 0 & 0 & 0 & 1 & 0 & 1 & 0 & 0 & 0 & 1 & 0 & 1 & 0 & 0 & 1
\\
1 & 0 & 1 & 0 & 0 & 0 & 0 & 0 & 1 & 0 & 1 & 0 & 0 & 0 & 1 & 1 & 1 & 1 & 0 & 0
\\
0 & 1 & 1 & 0 & 0 & 0 & 0 & 0 & 0 & 1 & 0 & 1 & 0 & 0 & 1 & 1 & 1 & 0 & 1 & 0
\\
0 & 0 & 0 & 1 & 0 & 1 & 0 & 1 & 0 & 0 & 1 & 1 & 1 & 0 & 0 & 1 & 0 & 0 & 0 & 1
\\
0 & 0 & 0 & 0 & 1 & 0 & 1 & 1 & 0 & 0 & 1 & 1 & 0 & 1 & 0 & 0 & 1 & 0 & 0 & 1
\\
0 & 0 & 0 & 1 & 1 & 0 & 0 & 0 & 1 & 1 & 0 & 0 & 1 & 1 & 2 & 0 & 0 & 1 & 1 & 2
\end{smallmatrix}
\epm.
\eg

The mapping between the symmetry-data vector $\bb v$ and the $\Z$-valued SRSI vector $\bs \theta_\Z$ can be constructed using the pseudoinverse method:
\bg
\bb v = EBR \cdot \Delta_\Z^\ddagger \cdot \bs \theta_\Z := \mc{M} \cdot \bs \theta_\Z,
\label{seq:sg207_map}
\eg
where $A^\ddagger$ denotes the Moore-Penrose pseudoinverse of a matrix $A$.
The pseudoinverse of $\Delta_\Z$ is given by
\bg
\Delta_\Z^\ddagger = \bpm
1 & 0 & 0 & 0 & 0 & 0 & 0 & 0 & 0 & 0 & 0 & 0 & 0 & 0 & 0 & 0 & 0 & 0 & 0 & 0
\\
0 & 1 & 0 & 0 & 0 & 0 & 0 & 0 & 0 & 0 & 0 & 0 & 0 & 0 & 0 & 0 & 0 & 0 & 0 & 0
\\
0 & 0 & 1 & 0 & 0 & 0 & 0 & 0 & 0 & 0 & 0 & 0 & 0 & 0 & 0 & 0 & 0 & 0 & 0 & 0
\\
0 & 0 & 0 & 1 & 0 & 0 & 0 & 0 & 0 & 0 & 0 & 0 & 0 & 0 & 0 & 0 & 0 & 0 & 0 & 0
\\
0 & 0 & 0 & 0 & 1 & 0 & 0 & 0 & 0 & 0 & 0 & 0 & 0 & 0 & 0 & 0 & 0 & 0 & 0 & 0
\\
0 & 0 & 0 & 0 & 0 & 1 & 0 & 0 & 0 & 0 & 0 & 0 & 0 & 0 & 0 & 0 & 0 & 0 & 0 & 0
\\
0 & 0 & 0 & 0 & 0 & 0 & 1 & 0 & 0 & 0 & 0 & 0 & 0 & 0 & 0 & 0 & 0 & 0 & 0 & 0
\\
0 & 0 & 0 & 0 & 0 & 0 & 0 & 0 & 1 & 0 & 0 & 0 & 0 & 0 & 0 & 0 & 0 & 0 & 0 & 0
\\
0 & 0 & 0 & 0 & 0 & 0 & 0 & 0 & 0 & 0 & 1 & 0 & 0 & 0 & 0 & 0 & 0 & 0 & 0 & 0
\epm^T,
\eg
which leads to the mapping matrix $\mc{M}$:
\bg
\mc{M} =EBR \cdot \Delta_\Z^\ddagger
= \bpm
1 & 0 & 0 & 0 & 0 & 1 & 0 & 0 & 0 & 0 & 1 & 0 & 0 & 0 & 0 & 1 & 0 & 0 & 0 & 0
\\
0 & 1 & 0 & 0 & 0 & 0 & 1 & 0 & 0 & 0 & 0 & 1 & 0 & 0 & 0 & 0 & 1 & 0 & 0 & 0
\\
0 & 0 & 1 & 0 & 0 & 1 & 1 & 0 & 0 & 0 & 0 & 0 & 1 & 0 & 0 & 1 & 1 & 0 & 0 & 0
\\
0 & 0 & 0 & 1 & 0 & 0 & 0 & 1 & 0 & 1 & 0 & 0 & 0 & 1 & 0 & 0 & 0 & 1 & 0 & 1
\\
0 & 0 & 0 & 0 & 1 & 0 & 0 & 0 & 1 & 1 & 0 & 0 & 0 & 0 & 1 & 0 & 0 & 0 & 1 & 1
\\
1 & 0 & 0 & 0 & 0 & 0 & 0 & 0 & 1 & 0 & 0 & 1 & 0 & 0 & 0 & 0 & 0 & 1 & 0 & 0
\\
0 & 1 & 0 & 0 & 0 & 0 & 0 & 1 & 0 & 0 & 1 & 0 & 0 & 0 & 0 & 0 & 0 & 0 & 1 & 0
\\
0 & 0 & 0 & 1 & 0 & 0 & 1 & 0 & 0 & 1 & 0 & 0 & 0 & 0 & 1 & 1 & 0 & 0 & 0 & 1
\\
1 & 0 & 1 & 0 & 0 & 0 & 0 & 0 & 1 & 1 & 0 & 0 & 0 & 0 & 1 & 1 & 0 & 1 & 1 & 0
\epm^T.
\label{seq:sg207_map_matrix}
\eg
Thus, the symmetry-data vector $\bb v$ is expressed in terms of $\Z$-valued SRSIs $\bs \theta_\Z$ by combining Eqs.~\eqref{seq:sg207_map} and \eqref{seq:sg207_map_matrix}.
Explicitly, each component of $\bb v$ is given by
\bg
v_1 = \nR{\Gamma_1} = \theta_1 + \theta_6 + \theta_9,
\quad
v_2 = \nR{\Gamma_2} = \theta_2 + \theta_7,
\quad
v_3 = \nR{\Gamma_3} = \theta_3 + \theta_9,
\quad
v_4 = \nR{\Gamma_4} = \theta_4 + \theta_8,
\nn
v_5 = \nR{\Gamma_5} = \theta_5,
\quad
v_6 = \nR{M_1} = \theta_1 + \theta_3,
\quad
v_7 = \nR{M_2} = \theta_2 + \theta_3 + \theta_8,
\quad
v_8 = \nR{M_3} = \theta_4 + \theta_7,
\nn
v_9 = \nR{M_4} = \theta_5 + \theta_6 + \theta_9,
\quad
v_{10} = \nR{M_5} = \theta_4 + \theta_5 + \theta_8 + \theta_9,
\quad
v_{11} = \nR{R_1} = \theta_1 + \theta_7,
\nn
v_{12} = \nR{R_2} = \theta_2 + \theta_6,
\quad
v_{13} = \nR{R_3} = \theta_3,
\quad
v_{14} = \nR{R_4} = \theta_4,
\quad
v_{15} =\nR{R_5} = \theta_5 + \theta_8 + \theta_9,
\nn
v_{16} = \nR{X_1} = \theta_1 + \theta_3 + \theta_8 + \theta_9,
\quad
v_{17} = \nR{X_2} = \theta_2 + \theta_3,
\quad
v_{18} = \nR{X_3} = \theta_4 + \theta_6 + \theta_9,
\nn
v_{19} = \nR{X_4} = \theta_5 + \theta_7 + \theta_9,
\quad
v_{20} = \nR{X_5} = \theta_4 + \theta_5 + \theta_8,
\label{seq:sg207_map_svec}
\eg
where $v_i=(\bb v)_i$ for $i=1,\dots,N_{BZ}$.
Importantly, the same relations hold if one uses $\bs \theta^{(\rm full)}_\Z$ instead of $\bs \theta_\Z$, since the mapping between $\bb v$ and the $\Z$-valued SRSIs does not depend on the simplifying assumption regarding the absence of site-symmetry irreps at nonmaximal WPs.
In addition to this property, the present construction of $\mc M$ has the important technical advantage that both integrality and compatibility relations of the symmetry-data vector are automatically enforced.
Specifically, $\mc M$ is defined as the product $\EBR \cdot \Delta_\Z^\ddagger$, where $\EBR$ is integer-valued by definition, and $\Delta_\Z^\ddagger$ is integer-valued in any space group~\cite{hwang2025stable}.
As a result, any integer-valued $\bs\theta_\Z$ produces an integer-valued symmetry-data vector $\bb v$.
Moreover, compatibility relations can be written as $\mc C \cdot \bb v = \bb 0$ with a matrix $\mc C$.
Since $\mc C \cdot \EBR = \bb 0$ holds identically in any space
group~\cite{po2017symmetry,elcoro2020application}, any $\bb v$ obtained from Eq.~\eqref{seq:sg207_map} automatically satisfies all compatibility relations.
This explains why, unlike in SN~\ref{sec:sn_sg212}, we can work directly with the full symmetry-data vector here without the need to impose compatibility relations separately.

By combining the mapping established in Eq.~\eqref{seq:sg207_map_svec} with the physical constraints imposed on the symmetry-data vector, we can determine the allowed values of SRSIs in photonic and electronic band structures.
In the main text, a graphical method was used to depict the allowed values of $\Z$-valued SRSIs in SG $P4_332$ (No. 212) on a two-dimensional plane defined by two $\Z$-valued SRSIs $(\theta_1,\theta_2)$.
For SG $P432$, however, the number of $\Z$-valued SRSIs increases to nine [see $\bs \theta_\Z$ in Eq.~\eqref{seq:sg207_rsi0}], requiring 9-dimensional space for complete visualization.
This makes a direct graphical representation infeasible.
To address this, we first apply the Hilbert basis method, introduced in SN~\ref{sec:sn_sg212_hilbert}, to enumerate all allowed values of $\bs \theta_\Z$.
We then demonstrate that a graphical representation is still possible by fixing seven independent linear combinations of $\bs \theta_\Z$, thereby reducing the number of free integer variables to two.
This reduction enables a two-dimensional visualization of the allowed values of the remaining SRSI variables.
%%%%%%

%%%%%%
\subsubsection{SRSIs allowed in photonic bands}
%%%%%%
We first examine the allowed values of $\bs \theta_\Z$ for photonic bands by considering constraints on the symmetry-data vector imposed by the polarization singularity.
All little-group irreps at HSM except $\Gamma$ must appear with nonnegative multiplicity in the absence of singularities.
This leads to the conditions $v_i \ge 0$ for $i = 6, \dots, N_{BZ} = 20$, which can be written explicitly as
\bg
\theta_1 + \theta_3 \ge 0,
\quad
\theta_2 + \theta_3 + \theta_8 \ge 0,
\quad
\theta_4 + \theta_7 \ge 0,
\quad
\theta_5 + \theta_6 + \theta_9 \ge 0,
\quad
\theta_4 + \theta_5 + \theta_8 + \theta_9 \ge 0,
\nn
\theta_1 + \theta_7 \ge 0,
\quad
\theta_2 + \theta_6 \ge 0,
\quad
\theta_{3,4} \ge 0,
\quad
\theta_5 + \theta_8 + \theta_9 \ge 0,
\quad
\theta_1 + \theta_3 + \theta_8 + \theta_9 \ge 0,
\quad
\theta_2 + \theta_3 \ge 0,
\nn
\theta_4 + \theta_6 + \theta_9 \ge 0,
\quad
\theta_5 + \theta_7 + \theta_9 \ge 0,
\quad
\theta_4 + \theta_5 + \theta_8 \ge 0.
\label{seq:sg207_rsi_ph1}
\eg
At $\Gamma$, the little-group representation for two lowest bands, which correspond to two $T$ modes, is taken to be $\girrep = \Gamma_4 - \Gamma_1$~\cite{christensen2022location}.
This condition implies $v_1 = \nR{\Gamma_1} \ge -1$ and $v_4 = \nR{\Gamma_4} \ge 1$, with the remaining irreps at $\Gamma$ satisfying $v_i \ge 0$ for $i=2,3,5$:
\bg
\theta_1 + \theta_6 + \theta_9 \ge -1,
\quad
\theta_2 + \theta_7 \ge 0,
\quad
\theta_3 + \theta_9 \ge 0,
\quad
\theta_4 + \theta_8 \ge 1,
\quad
\theta_5 \ge 0.
\label{seq:sg207_rsi_ph2}
\eg
All the inequalities in Eqs.~\eqref{seq:sg207_rsi_ph1} and \eqref{seq:sg207_rsi_ph2} can be collectively expressed in matrix form as
\bg
\mc{M} \cdot \bs \theta_\Z \ge \bb b
\eg
where the vector $\bb b$ is given by $\bb b = (-1,0,0,1,0, 0,0,0,0,0, 0,0,0,0,0, 0,0,0,0,0)^T$.
{Note that we have ordered the inequalities such that the $i$\textsuperscript{th} component 
of $\bb b$ corresponds to the inequality satisfied by $v_i$ in 
Eqs.~\eqref{seq:sg207_rsi_ph1} and \eqref{seq:sg207_rsi_ph2}.}

For the given $\mc{M}$ and $\bb b$, the Hilbert basis of the inequality system $\mc{M} \cdot \bs \theta_\Z \ge \bb b$, computed by using Normaliz~\cite{bruns2010normaliz}, consists of the compact part $\bb h_c$ and the recession cone part $\bb h_r$:
\ba
\{\bb h_c\} = \{& (-1,-1,1,0,0,1,1,1,-1)^T, (0,0,0,0,0,0,0,1,0)^T, (0,0,0,1,0,0,0,0,0)^T, (0,1,0,1,1,-1,0,0,0)^T \},
\nn
\{\bb h_r \} = \{& (-1,-1,1,0,0,1,1,0,0)^T, (-1,-1,1,0,0,2,1,1,-1)^T, (0,-1,1,0,0,1,1,1,-1)^T, (0,-1,1,0,1,1,1,
\nn
& 0,-1)^T, (0,0,0,0,0,0,0,0,1)^T, (0,0,0,0,0,0,0,1,0)^T, (0,0,0,0,0,0,1,0,0)^T, (0,0,0,0,0,1,0,0,0)^T,
\nn
& (0,0,0,0,1,0,0,0,0)^T, (0,0,0,1,0,0,0,0,0)^T, (0,0,1,0,0,0,0,0,0)^T, (0,0,1,0,1,1,0,0,-1)^T, (0,1,
\nn
& 0,0,0,-1,0,0,1)^T, (0,1,0,0,0,0,0,0,0)^T, (0,1,0,1,0,-1,0,-1,1)^T, (1,0,0,0,0,0,0,0,0)^T, (1,1,0,
\nn
& 1,0,-1,-1,-1,1)^T, (1,1,0,1,1,-1,-1,-1,0)^T \},
\ea
where $\{\bb h_r\} = \{\bb h_{r,A}|A=1,\dots,D_r\}$ with $D_r = 18$.
Since there are four vectors in the compact part $\{\bb h_c\}$, the complete solution space to Eqs.~\eqref{seq:sg207_rsi_ph1} and \eqref{seq:sg207_rsi_ph2} is given by the union of the following solution families:
\bg
\bs \theta_{\Z,sol_1} = (-1,-1,1,0,0,1,1,1,-1)^T + \sum_{A=1}^{18} \, n_A \bb h_{r,A},
\quad
\bs \theta_{\Z,sol_2} = (0,0,0,0,0,0,0,1,0)^T + \sum_{A=1}^{18} \, n_A \bb h_{r,A},
\nn
\bs \theta_{\Z,sol_3} = (0,0,0,1,0,0,0,0,0)^T + \sum_{A=1}^{18} \, n_A \bb h_{r,A},
\quad
\bs \theta_{\Z,sol_4} = (0,1,0,1,1,-1,0,0,0)^T + \sum_{A=1}^{18} \, n_A \bb h_{r,A},
\label{seq:sg207_hilbert_ph}
\eg
where $n_A \in \N_0$.

Since a set of bands is naturally characterized by the total number of bands $\nu$, we now determine the allowed values of $\bs \theta_\Z$ at fixed $\nu$.
This can be done by evaluating the filling formula in Eq.~\eqref{seq:sg207_filling} for each solution family $\bs \theta_{\Z, sol_i}$ ($i=1,\dots,4$) from Eq.~\eqref{seq:sg207_hilbert_ph}.
For example, for $\bs \theta_{\Z, sol_1}$, the filling becomes:
\ba
\nu =& 2 + 2 n_{1} + 3 n_{2} + 3 n_{3} + 3 n_{4} + 3 n_{5} + 3 n_{6} + n_{7} + n_{8} + 3 n_{9} \nn
&+ 3 n_{10} + 2 n_{11} + 3 n_{12} + 3 n_{13} + n_{14} + 3 n_{15} + n_{16} + 3 n_{17} + 3 n_{18}.
\ea
For a fixed $\nu$, this Diophantine equation can be solved for $n_A \in \N_0$ ($A=1,\dots,18$) using standard integer linear programming techniques.
By performing the same analysis for the remaining solution families $\bs \theta_{\Z, sol_{2,3,4}}$, one obtains the complete set of allowed $\bs \theta_\Z$ values consistent with the given filling $\nu$.

For instance, for $\nu=2$ (there is no solution for $\nu=1$), there is a unique solution,
\bg
\theta_\Z = (-1,-1,1,0,0,1,1,1,-1)^T.
\eg
Using the mapping in Eq.~\eqref{seq:sg207_map}, the corresponding symmetry-data vector is
\bg
\bb v = (-1, 0, 0, 1, 0, 0, 1, 1, 0, 0, 0, 0, 1, 0, 0, 0, 0, 0, 0, 1)^T.
\eg
This vector corresponds to the following combination of little-group irreps:
\bg
[\girrep, M_2 + M_3, R_3, X_5],
\eg
where $\girrep=\Gamma_4 - \Gamma_1$.
For $\nu = 3$, six distinct solutions are found:
\ba
\{\bs \theta_\Z\} =& \{ (0, 0, 0, 0, 0, 0, 0, 1, 0)^T,
(0, 0, 0, 1, 0, 0, 0, 0, 0)^T,
(-1, -1, 1, 0, 0, 2, 1, 1, -1)^T,
\nn
& (-1, 0, 1, 0, 0, 1, 1, 1, -1)^T,
(-1, -1, 1, 0, 0, 1, 2, 1, -1)^T,
(0, -1, 1, 0, 0, 1, 1, 1, -1)^T \}.
\ea
Each solution corresponds to the following sets of little-group irreps:
\ba
& [\girrep + \Gamma_1, M_2 + M_5, R_5, X_1 + X_5],
\quad
[\girrep + \Gamma_1, M_3 + M_5, R_4, X_3 + X_5],
\nn
& [\girrep + \Gamma_1, M_2 + M_3 + M_4, R_2 + R_3, X_3 + X_5],
\quad
[\girrep + \Gamma_2, 2 M_2 + M_3, R_2 + R_3, X_2 + X_5],
\nn
& [\girrep + \Gamma_2, M_2 + 2 M_3, R_1 + R_3, X_4 + X_5],
\quad
[\girrep + \Gamma_1, M_1 + M_2 + M_3, R_1 + R_3, X_1 + X_5].
\ea
%%%%%%

%%%%%%
\subsubsection{SRSIs allowed in electronic bands}
%%%%%%
For electrons, the physically allowed values of $\bs \theta_\Z$ are constrained by the requirement that every component of the symmetry-data vector must be a nonnegative integer.
Consequently, the inequalities for $\bs \theta_\Z$ include those in Eq.~\eqref{seq:sg207_rsi_ph1}, together with the additional constraints associated with irreps at $\Gamma$:
\bg
\theta_1 + \theta_6 + \theta_9 \ge 0,
\quad
\theta_2 + \theta_7 \ge 0,
\quad
\theta_3 + \theta_9 \ge 0,
\quad
\theta_4 + \theta_8 \ge 0,
\quad
\theta_5 \ge 0.
\label{seq:sg207_rsi_el}
\eg
Together, the inequalities in Eqs.~\eqref{seq:sg207_rsi_ph1} and \eqref{seq:sg207_rsi_el} can be compactly expressed as $\mc{M} \cdot \bs \theta_\Z \ge 0$, where the right-hand side is the zero vector.
This form indicates that the space of solutions corresponds to the set of integer lattice points in a polyhedral cone.
Accordingly, the Hilbert basis of this cone consists solely of generators of the recession cone, with a trivial compact part:
$\{ \bb h_c \} = \{ (0, 0, 0, 0, 0, 0, 0, 0, 0)^T \}$.
Note that, since the matrix $\mc{M}$ is identical to that in the photonic case, the Hilbert basis of the recession cone remains unchanged.
That is, the same set of generators $\{\bb h_r\}$ given in Eq.~\eqref{seq:sg207_hilbert_ph} applies here as well.
Accordingly, the general solution for $\bs \theta_\Z$ takes the form
\bg
\bs \theta_\Z = \bb h_c + \sum_{A=1}^{18} \, n_A \bb h_{r,A}
= \sum_A \, n_A \bb h_{r,A},
\eg
where $n_A \in \N_0$.
Thus, all physically allowed values of $\bs \theta_\Z$ for electronic band structures are given by integer nonnegative combinations of the cone generators ${\bb h_r}$.

For instance, when the total number of bands is $\nu = 1$, there are four solutions:
\bg
\{ \bs \theta_\Z \} = \{ (0, 0, 0, 0, 0, 1, 0, 0, 0)^T,
(0, 1, 0, 0, 0, 0, 0, 0, 0)^T,
(0, 0, 0, 0, 0, 0, 1, 0, 0)^T,
(1, 0, 0, 0, 0, 0, 0, 0, 0)^T \}.
\eg
Each of these corresponds to the following little-group irrep contents:
\bg
(\Gamma_1, M_4, R_2, X_3), \quad
(\Gamma_2, M_2, R_2, X_2), \quad
(\Gamma_2, M_3, R_1, X_4), \quad
(\Gamma_1, M_1, R_1, X_1).
\eg
%%%%%%

%%%%%%
\subsubsection{Graphical method for solving the inequality conditions}
%%%%%%
We now solve for the allowed $\Z$-valued SRSIs in photonic and electronic bands using a graphical approach.
To make the 9-dimensional lattice of SRSIs more accessible, we reduce its dimensionality by imposing 7 constraints that fix either individual SRSIs or linear combinations of them.
This leaves only 2 SRSIs as free integer variables.
With this reduction, the inequality constraints can be visualized on a two-dimensional plane, where the allowed regions and lattice points for the remaining SRSIs are represented geometrically.
To enhance clarity and to better capture a broader set of allowed SRSI configurations under these constraints, we redefine the basis of the $\Z$-valued SRSI vector $\bs \theta_\Z$.
This redefinition modifies the visual representation without altering the intrinsic lattice structure of the SRSIs.
We also discuss how this redefinition influences the procedure used to obtain the allowed SRSI values.
%

%%%%%%
\begin{figure}[b]
\centering
\includegraphics[width=0.49\textwidth]{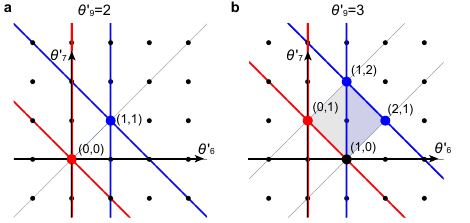}
\caption{{\bf SRSIs of photonic and electronic bands allowed in SG $P432$ (No. 207).}
Three $\Z$-valued SRSIs, $(\theta'_6, \theta'_7, \theta'_9)$, are shown.
These are subject to the inequality conditions given in Eq.~\eqref{seq:sg207_ineqs}.
Panels \textbf{a} and \textbf{b} depict the allowed values of $(\theta'_6, \theta'_7)$ for total number of bands $\nu=2$ and $\nu=3$, respectively, where $\nu = \theta'_9$.
All other $\Z$-valued SRSIs are fixed as $(\theta'_{1,\dots,5},\theta'_8)=(-1,0,0,0,0,0)$.
The gray, red, and blue regions represent $M_{co}$, $M_{ph}$, and $M_{el}$, which are defined by inequality sets $I_{co}$, $I_{ph}$, and $I_{el}$.
The boundaries where these inequalities are saturated are indicated by gray, red, and blue lines.
Red dots in $M_{ph}$ and blue dots in $M_{el}$ represent the allowed SRSIs that correspond to physically valid, integer-valued symmetry-data vectors.
Larger black dots indicate SRSIs that are allowed in both photonic and electronic band structures.}
\label{sfig:sg207_rsi}
\end{figure}
%%%%%%

To begin, we redefine the $\Z$-valued SRSIs $\bs \theta_\Z$ as a new set $\bs \theta'_\Z$:
\bg
\theta'_1 = \theta_1,
\quad
\theta'_2 = -\theta_1 + \theta_2,
\quad
\theta'_3 = \theta_2 + \theta_3,
\quad
\theta'_4 = \theta_4,
\quad
\theta'_5 = \theta_5,
\quad
\theta'_6 = \theta_6 + \theta_9,
\quad
\theta'_7 = \theta_7 + \theta_9,
\nn
\theta'_8 = \theta_8 + \theta_9,
\quad
\theta'_9 = \theta_1 + \theta_2 + 2 \theta_3 + 3 \theta_4 + 3 \theta_5 + \theta_6 + \theta_7 + 3 \theta_8 + 3 \theta_9 = \nu.
\eg
This redefinition is expressed compactly as
\bg
\bs \theta'_\Z =
\bpm
1 & 0 & 0 & 0 & 0 & 0 & 0 & 0 & 0 \\
-1 & 1 & 0 & 0 & 0 & 0 & 0 & 0 & 0 \\
0 & 1 & 1 & 0 & 0 & 0 & 0 & 0 & 0 \\
0 & 0 & 0 & 1 & 0 & 0 & 0 & 0 & 0 \\
0 & 0 & 0 & 0 & 1 & 0 & 0 & 0 & 0 \\
0 & 0 & 0 & 0 & 0 & 1 & 0 & 0 & 1 \\
0 & 0 & 0 & 0 & 0 & 0 & 1 & 0 & 1 \\
0 & 0 & 0 & 0 & 0 & 0 & 0 & 1 & 1 \\
1 & 1 & 2 & 3 & 3 & 1 & 1 & 3 & 3
\epm
\cdot \bs \theta_\Z
:= A \cdot \bs \theta_\Z.
\label{seq:sg207_rsi_redef}
\eg
The last component, $\theta'_9$, corresponds to the total number of bands $\nu$, as in Eq.~\eqref{seq:sg207_filling}.
This basis change rescales the lattice volume by ${\rm Det}(A) = -2$, so some integer points in the transformed polyhedron may correspond to fractional values of the original symmetry-data vector $\bb v$.
Therefore, we must discard any $\bs \theta'_\Z$ that does not yield an integer-valued $\bb v$.
Under the redefinition of $\Z$-valued SRSIs, the mapping from $\bs \theta'_Z$ to $\bb v$ becomes
\bg
\bb v = \mc{M}' \cdot \bs \theta'_\Z = \mc{M} \cdot A^{-1} \cdot \bs \theta_\Z,
\label{seq:sg207_map_reduced}
\eg
where $\mc{M}'$ is derived from the original map in Eq.~\eqref{seq:sg207_map} and the redefinition matrix $A$ in Eq.~\eqref{seq:sg207_rsi_redef}.
Since $|{\rm Det}(A)| = 2$ is an integer greater than one, the inverse matrix $A^{-1}$, and hence the redefined mapping matrix $\mc{M}' = \mc{M} \cdot A^{-1}$, generally contain fractional elements.
As a result, the mapping in Eq.~\eqref{seq:sg207_map_reduced} may send integer-valued vectors $\bs \theta'_\Z$ to fractional symmetry-data vectors $\bb v$.
However, since $\bb v$ must be integer-valued to correspond to physical symmetry-data vector, not all integer points $\bs \theta'_\Z$ in the redefined lattice are valid.
To account for this, we impose an additional constraint: only those $\bs \theta'_\Z$ that yield integer-valued $\bb v$ are retained.
Below, we detail the graphical method procedures, incorporating this additional requirement.

First, we impose dimensional-reduction constraints by fixing six $\Z$-valued SRSIs as
\bg
(\theta'_1, \theta'_2, \theta'_3, \theta'_4, \theta'_5, \theta'_8) = (-1,0,0,0,0,0),
\eg
leaving 3 free SRSI variables, $(\theta'_6, \theta'_7, \theta'_9=\nu)$.
Under this setting, the mapping in Eq.~\eqref{seq:sg207_map_reduced} simplifies to
\ba
(v_1,\dots,v_5) = 
& \left( -1 + \theta'_6, -1 + \nu/2 - \theta'_6/2 + \theta'_7/2, 1 - \nu/2 + \theta'_6/2 + \theta'_7/2, \nu/2 - \theta'_6/2 - \theta'_7/2, 0 \right),
\nn
(v_6,\dots,v_{10}) = 
& \left(0, \nu/2 - \theta'_6/2 - \theta'_7/2, \nu/2 - \theta'_6/2 + \theta'_7/2, \theta'_6, 0 \right),
\nn
(v_{11},\dots,v_{15}) =
& \left(-1 + \nu/2 - \theta'_6/2 + \theta'_7/2, -1 + \nu/2 + \theta'_6/2 - \theta'_7/2, 1, 0, 0 \right),
\nn
(v_{16},\dots,v_{20}) =
& \left( 0, 0, \theta'_6, \theta'_7, \nu/2 - \theta'_6/2 - \theta'_7/2 \right),
\label{seq:sg207_sym_vec_reduced}
\ea
where $v_i = (\bb v)_i$ for $i = 1, \dots, 20$.
The inequality conditions for photonic and electronic bands, originally given in Eqs.~\eqref{seq:sg207_rsi_ph1}, \eqref{seq:sg207_rsi_ph2}, and \eqref{seq:sg207_rsi_el}, are then translated into the following forms:
\ba
& \text{Photonic:} \quad \theta'_6 \ge 0, \quad \theta'_7 \ge 0, \quad \nu-2 \ge \theta'_6 - \theta'_7 \ge 2-\nu, \quad \nu -2 \ge \theta'_6 + \theta'_7 \ge \nu-2,
\nn
& \text{Electronic:} \quad \theta'_6 \ge 1, \quad \theta'_7 \ge 0, \quad \nu-2 \ge \theta'_6 - \theta'_7 \ge 2-\nu, \quad \nu \ge \theta'_6 + \theta'_7 \ge \nu-2.
\label{seq:sg207_ineqs}
\ea
The inequality conditions above can be grouped into three categories:
\ba
I_{co}:& \quad \theta'_7 \ge 0, \quad \nu-2 \ge \theta'_6 - \theta'_7 \ge 2-\nu, \quad \theta'_6 + \theta'_7 \ge \nu-2,
\nn
I_{ph,only}:& \quad \theta'_6 \ge 0, \quad \nu-2 \ge \theta'_6 + \theta'_7,
\nn
I_{el,only}:& \quad \theta'_6 \ge 1, \quad \nu \ge \theta'_6 + \theta'_7.
\ea
Here, $I_{co}$ denotes the set of inequalities satisfied by both photonic and electronic bands.
The conditions $I_{ph,only}$ and $I_{el,only}$ are additionally satisfied only by photonic and electronic bands, respectively.
In other words, photonic (electronic) bands are determined by $I_{co}$ and $I_{ph,only}$ ($I_{el,only}$).
The inequalities $I_{co}$ defines a region $M_{co}$ in the space of $(\theta'_6,\theta'_7,\theta'_9)$.
The subregion $M_{ph}$ ($M_{el}$) is further constrained by $I_{ph,only}$ ($I_{el,only}$).
For fixed total number of bands with $\nu= \theta'_9=2$ and $3$, the regions $M_{ph,el,co}$ are shown in Fig.~\ref{sfig:sg207_rsi}.

We can now classify trivial photonic bands by defining the lattice of allowed SRSI values as the intersection between the photonic region $M_{ph}$ and an integer sublattice $Z$:
\bg
Lat_{\bs \theta_\Z', ph} = M_{ph} \cap Z.
\label{seq:lat_ph_triv}
\eg
Here, $Z$ is defined as
\bg
Z = \{(\theta'_6,\theta'_7,\theta'_9) \in \Z^3 | \mc{M}' \cdot \bs \theta'_\Z \in \Z^{N_{BZ}} \},
\label{seq:Z_lattice}
\eg
where $\bs \theta'_\Z = (-1,0,0,0,0,\theta'_6,\theta'_7,0,\theta'_9)$ reflects the dimensional-reduction constraints imposed earlier.
Unlike the sublattice $Z$ introduced in the main text, the set in Eq.~\eqref{seq:Z_lattice} includes an additional requirement: the mapped symmetry-data vector $\bb v = \mc{M}' \cdot \bs \theta'_\Z$ must be integer-valued.
This condition is necessary because the redefinition of the SRSI variables via the matrix $A$, with ${\rm Det} (A) \ne \pm 1$, implies that not every integer-valued $\bs \theta'_\Z$ corresponds to an integer vector $\bb v$.
As a result, only those $\bs \theta'_\Z$ that satisfy both the inequality conditions and this integrality constraint are physically allowed.
Similarly, the lattice of SRSI values allowed for electronic bands is defined by
\bg
Lat_{\bs \theta_\Z', el} = M_{el} \cap Z.
\label{seq:lat_el_triv}
\eg

We now compare the SRSIs and symmetry-data vectors allowed in photonic and electronic band structures.
For the case of $\nu=2$, the allowed SRSI lattices are $Lat_{\bs \theta_\Z', ph} = \{(0,0,2)\}$ and $Lat_{\bs \theta_\Z', el} = \{(1,1,2)\}$, which are clearly distinguishable with no overlap.
These are illustrated in Fig.~\ref{sfig:sg207_rsi}\textbf{a}.
The corresponding symmetry-data vectors are:
\ba
(0,0,2) \in Lat_{\bs \theta_\Z', ph}:& \,
[\girrep, M_2 + M_3, R_3, X_5],
\nn
(1,1,2) \in Lat_{\bs \theta_\Z', el}:& \,
(\Gamma_3, M_3 + M_4, R_3, X_3 + X_4).
\label{seq:sg207_symvec_nu2}
\ea
In SN~\ref{sec:sn_sg207_TB}, we construct a tight-binding model corresponding to the photonic point $(0,0,2) \in Lat_{\bs \theta_\Z', ph}$.

For $\nu=3$, our procedure yields $Lat_{\bs \theta_\Z', ph} = \{(1,0,3),(0,1,3)\}$ and $Lat_{\bs \theta_\Z', el} = \{(1,0,3),(1,2,3),(2,1,3)\}$, as illustrated in Fig.~\ref{sfig:sg207_rsi}\textbf{b}.
Among them, $(\theta'_6, \theta'_7, \theta'_9)=(0,1,3)$ is allowed only for photonic bands and does not appear in electronic ones.
The corresponding symmetry-data vectors are:
\ba
(1,0,3) \in Lat_{\bs \theta_\Z', ph}:& \,
[\girrep + \Gamma_1, M_2 + M_3 + M_4, R_2 + R_3, X_3 + X_5],
\nn
(0,1,3) \in Lat_{\bs \theta_\Z', ph}:& \,
[\girrep + \Gamma_2, M_2 + 2 M_3, R_1 + R_3, X_4 + X_5],
\nn
(1,0,3) \in Lat_{\bs \theta_\Z', el}:& \,
(\Gamma_4, M_2 + M_3 + M_4, R_2 + R_3, X_3 + X_5),
\nn
(1,2,3) \in Lat_{\bs \theta_\Z', el}:& \,
(\Gamma_2 + \Gamma_3, 2 M_3 + M_4, R_1 + R_3, X_3 + 2 X_4),
\nn
(2,1,3) \in Lat_{\bs \theta_\Z', el}:& \,
(\Gamma_1 + \Gamma_3, M_3 + 2 M_4, R_2 + R_3, 2 X_3 + X_4).
\label{seq:sg207_symvec_nu3}
\ea
%%%%%%

%%%%%%
\subsection{Ab-initio and tight-binding models}
\label{sec:sn_sg207_TB}
%%%%%%
Earlier, we demonstrated that bands with $(\theta'_6, \theta'_7, \theta'_9) = (0,0,2)$ in SG $P432$ (No. 207) can only be realized in photonic band structure when the number of bands is $\nu=2$.
The corresponding little-group irrep content is:
\bg
\mc{B}_{\rm phys}: [(\rho_T)_\Gamma = \Gamma_4 - \Gamma_1, M_2 + M_3, R_3, X_5].
\label{seq:sg207_irrep_model}
\eg
Like the SG $P4_332$ (No. 212) example in SN~\ref{sec:sn_sg212}, we construct ab-initio and TB models that realize this band structure.
%%%%%%

%%%%%%
\paragraph{Ab-initio model.}
%%%%%%
Let us first describe our ab-initio model, whose construction is analogous to that for SG $P4_332$.
We built a photonic crystal in SG $P432$ and computed the eigenspectra using MPB~\cite{johnson2001block-iterative}.
The structure is similar to the SG $P4_332$ model (see Methods in the main text).
We set the lattice constant to 1 and fill the entire cubic unit cell with a non-magnetic, homogeneous, and isotropic medium of dielectric constant $\epsilon=11$.
An arbitrary representative position $(0.4, 0.2, -0.3)$ is chosen.
This position and its 23 symmetry-related counterparts form the general WP $24k$ of SG $P432$.
We place a sphere of radius $r=0.2$ and $\epsilon=1$ at each of these 24 points.
This ensures that the resulting structure respects all symmetries of SG $P432$.
The unit cell is shown in Fig.~\ref{sfig:unitcell}\textbf{a}.

Note the comparison between the unit cell structures of the ab initio models in SGs $P432$ and $P4_332$.
In SG $P432$ all rotation generators are symmorphic (no fractional translations), whereas in SG $P4_332$ the fourfold and twofold generators are screw rotations.
For example, the operation $\{4_{001}| 3/4, 1/4, 3/4 \}$ in $P4_332$ and $\{4_{001}|\bb 0\}$ in $P432$ map $(x,y,z)$ to $(-y+3/4, x+ 1/4, z + 3/4)$ and $(-y,x,z)$, respectively.
This difference accounts for the different parameterizations of the general WPs ($24k$ in $P432$ and $24e$ in $P4_332$) and for the different visual appearances of the two unit cells in Fig.~\ref{sfig:unitcell}.

\begin{figure*}[t]
\centering
\includegraphics[width=0.75\textwidth]{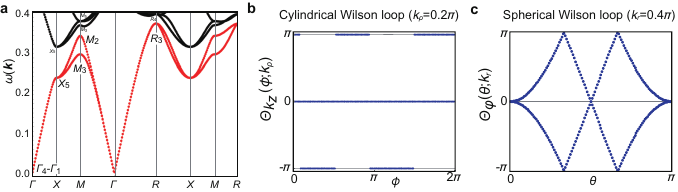}
\caption{{\bf Frequency and Wilson loop spectra of the ab-initio model in SG $P432$.}
\textbf{a} Frequency spectrum.
We used MPB to solve for the 6 lowest frequency bands to show that the lowest two bands are isolated from the higher bands.
Note that $\Gamma=(0,0,0)$, $X=(0,\pi,0)$, $M=(\pi,\pi,0)$, and $R=(\pi,\pi,\pi)$.
The irrep assignment at HSM confirms that the lowest two bands correspond to $\mc{B}_{\rm phys}$ in Eq.~\eqref{seq:sg207_irrep_model}.
\textbf{b} Spectrum of the cylindrical Wilson loop $W_{k_z} (\phi; k_\rho)$ for $k_\rho=0.2 \pi$.
\textbf{d} Spectrum of the spherical Wilson loop $W_\phi (\theta; k_r)$, showing a winding corresponding to the Euler number $|\mf{e}|=2$.}
\label{sfig:sg207_PhC}
\end{figure*}

\begin{figure*}[t]
\centering
\includegraphics[width=0.6\textwidth]{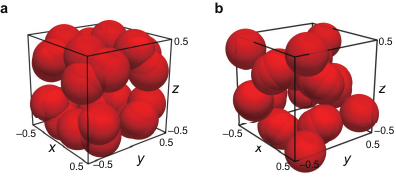}
\caption{{\bf Comparison between the unit cells in the ab-initio models for SGs $P432$ and $P4_332$.}
\textbf{a} Unit cell in SG $P432$.
The spheres with radius 0.2 are located at $24k$ WP with a representative position $(0.4,0.2,-0.3)$.
\textbf{b} Unit cell in SG $P4_332$.
The spheres with radius 0.2 are located at $24e$ WP with a representative position $(0,0.4,-0.2)$.}
\label{sfig:unitcell}
\end{figure*}

The frequency spectrum is shown in Fig.~\ref{sfig:sg207_PhC}\textbf{a}, where the lowest two bands are separated from higher bands by the lowest frequency gap.
The irreps of these bands at HSM match Eq.~\eqref{seq:sg207_irrep_model}.
For the two bands below the lowest frequency gap, we then computed the cylindrical and spherical Wilson loop spectra [$W_{k_z}(\phi; k_\rho)$ and $W_\phi (\theta; k_r)$] as defined in SN~\ref{sec:sn_sg212_wl_cylinder} and \ref{sec:sn_sg212_wl_sphere}, shown in Figs.~\ref{sfig:sg207_PhC}\textbf{b} and \textbf{c}.
The cylindrical Wilson loop spectrum does not exhibit a winding structure even when the radius $k_\rho \to 0$.
The spherical Wilson loop spectrum has winding with Euler number $|\mf{e}|=2$ due to emergent transversality in the long-wavelength limit, as explained in SN~\ref{sec:sn_sg212_wl_sphere}.
%%%%%%

%%%%%%
%%%%%%

%%%%%%
%%%%%%

%%%%%%
\paragraph{Tight-binding model.}
%%%%%%
We also confirmed the same Wilson loop features in a TB model.
We built the TB model $H_{SG 207}(\bk)$ as follows.
First, note that $\mc{B}_{\rm phys}$ can be expressed formally as the difference between little-group irreps of two EBRs, $(A_2)_d \ua G$ and $(A_1)_a \ua G$:
\bg
(A_2)_d \ua G \ominus (A_1)_a \ua G: \quad
[\girrep, M_2+M_3, R_3, X_5].
\eg
This is seen by comparing the HSM irreps of the two EBRs:
\ba
(A_2)_d \ua G:& \quad
(\Gamma_4, M_1 + M_2 + M_3, R_1 + R_3, X_1 + X_5),
\nn
(A_1)_a \ua G:& \quad
(\Gamma_1, M_1, R_1, X_1).
\ea
Note that for the symmetry-data vector in this example, once the regularized bands are fixed to $(A_2)_d \ua G$, the auxiliary bands are uniquely fixed to $(A_1)_a \ua G$ in order to reproduce the target symmetry-data vector for physical bands.
Thus the physical bands $\mc{B}_{\rm phys}$ can be realized by regularizing to $B_{\rm reg}=(A_2)_d \ua G$ and treating $B_{\rm aux}=(A_1)_a \ua G$ as an auxiliary band\footnote{While the same physical symmetry-data vector may admit alternative realizations with different $\Z_n$-valued SRSIs, such as $(A_2)_c-(A_1)_b$, here we restrict to the representative choice $(A_2)_d-(A_1)_a$, since the present analysis focuses on properties determined solely by the symmetry-data vector.}.

Based on this, we construct $H_{SG 207} (\bk)$ with basis orbitals $(A_2)_d$, so the full set of bands realizes $B_{\rm reg}$.
The basis orbits correspond to the site-symmetry irrep $(A_2)$ at WP $3d$ with $\bq_{d,1} = (0,0,1/2)$, $\bq_{d,2} = (1/2, 0, 0)$, and $\bq_{d,3} = (0, 1/2, 0)$.
This defines the sublattice embedding matrix $V(\bk)_{\alpha \beta} = e^{i \bk \cdot \bq_{d, \alpha}} \delta_{\alpha \beta}$ ($\alpha, \beta=1,2,3$).
Using the transformation of these orbitals under $C_{4,001}$, $C_{3,111}$, $C_{2,110}$ and time-reversal $\mc T$, and the induction procedure of Refs.~\cite{elcoro2017double,cano2018building}, we obtain the matrix representation of symmetry operators:
\bg
U_{C_{4,001}}= \bpm 1 & 0 & 0 \\ 0 & 0 & -1 \\ 0 & 1 & 0 \epm,
\quad
U_{C_{3,111}} = \bpm 0 & 0 & -1 \\ -1 & 0 & 0 \\ 0 & 1 & 0 \epm,
\quad
U_{C_{2,110}} = \bpm -1 & 0 & 0 \\ 0 & 0 & 1 \\ 0 & 1 & 0 \epm,
\quad
U_{\mc{T}} = \bpm 1 & 0 & 0 \\ 0 & 1 & 0 \\ 0 & 0 & 1 \epm.
\eg
We first construct $H_0(\bk) = V(- \bk) h(\bk) V(\bk)$, which we choose to be symmetric under $C_{4,001}$:
\bg
h(\bk) = \frac{1}{4} \bpm
h_{11}(\bk) & t_1 (1-Q_1) & t_1 (1-Q_2) \\
t_1 (1-\cm{Q_1}) & h_{22}(\bk) & - t_1 (1-\cm{Q_1})(1-Q_2)
\\ t_1 (1-\cm{Q_2}) & - t_1 (1-Q_1)(1-\cm{Q_2}) & h_{33}(\bk)
\epm,
\eg
where the diagonal elements are given by
\bg
h_{11}(\bk) = t_3 (1 + Q_1 Q_2)(\cm{Q_1} + \cm{Q_2})(Q_3 + \cm{Q_3}) + \frac{\ep_0}{3}, \nn
h_{22}(\bk) = t_2 (Q_2 + \cm{Q_2} + Q_3 + \cm{Q_3}) + \frac{\ep_0}{3}, \quad
h_{33}(\bk) = t_2 (Q_1 + \cm{Q_1} + Q_3 + \cm{Q_3}) + \frac{\ep_0}{3},
\eg
with $(Q_1,Q_2,Q_3)=(e^{-ik_x}, e^{-ik_y}, e^{-ik_z})$ and real-valued parameters ($t_1, t_2, t_3, \ep_0$).
Here, $t_1$ describes nearest-neighbor hoppings between different sublattices, while $t_2$ and $t_3$ correspond to hoppings within the same sublattice types.
Among them, $t_2$ is a second nearest-neighbor hopping, whereas $t_3$ corresponds to a fourth nearest-neighbor one.
Although including a fourth nearest-neighbor term may look unusual, we keep $t_3$ (with a small magnitude) because without it the auxiliary bands become flat along the $\Gamma$–$X$ HSL.
$t_3$ lifts this accidental flatness by pushing the auxiliary band's energy downward away from $\Gamma$.

The diagonal elements contain intra-sublattice hoppings (between sites of the same sublattice type) up to next-nearest neighbors, whereas the off-diagonal elements, representing inter-sublattice hoppings, include the nearest-neighbor contributions.
The onsite energy is given by $\ep_0 \in \R$.
The Hamiltonian $H_0(\bk)$ satisfies the symmetry relation
\bg
H_0(\bk) = [U_{C_{4,001}}(\bk)]^{-1} \, H_0(C_{4,001} \bk) \, U_{C_{4,001}}(\bk).
\eg
Here, $C_{4,001} \bk = (-k_y, k_x, k_z)$, and we also define $C_{3,111} \bk = (k_z, k_x, k_y)$ and $C_{2,110} \bk = (k_y, k_x, -k_z)$.
We then symmetrize $H_0(\bk)$ with respect to $C_{2,110}$, $C_{3,111}$, and $\mc{T}$:
\bg
H_1(\bk) = H_0(\bk) + [U_{C_{2,110}}(\bk)]^{-1} \, H_0(C_{2,110} \bk) \, U_{C_{2,110}}(\bk),
\nn
H_2(\bk) = H_1(\bk) + [U_{C_{3,111}}(\bk)]^{-1} \, H_1(C_{3,111}\bk) \, U_{C_{3,111}}(\bk) + U_{C_{3,111}}(\bk) \, H_1(C_{3,111}^{-1} \bk) \, [U_{C_{3,111}}(\bk)]^{-1},
\nn
H_{SG 207}(\bk) = H_2(\bk) + [U_{\mc T}]^{-1} \, \cm{H_2(-\bk)} \, U_{\mc T}.
\label{seq:sg207_h_sym}
\eg
%

%%%%%%
\begin{figure*}[t]
\centering
\includegraphics[width=0.95\textwidth]{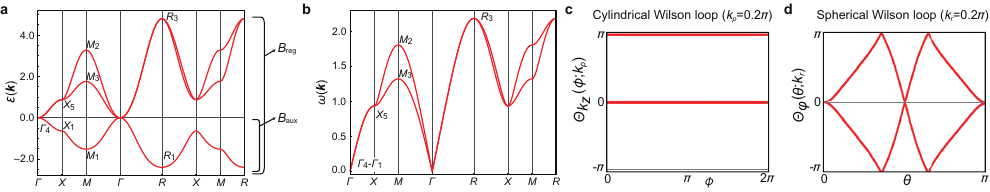}
\caption{{\bf Band structure and Wilson loop spectra of the TB model $H_{SG 207}(\bk)$.}
Band structure defined by \textbf{a} $\vep(\bk)$ and \textbf{b} $\om(\bk)$.
HSM are $\Gamma=(0,0,0)$, $X=(0,\pi,0)$, $M=(\pi,\pi,0)$, and $R=(\pi,\pi,\pi)$.
\textbf{a} Physical and auxiliary bands correspond to $(A_2)_d \ua G$ and $(A_1)_a \ua G$ respectively.
All irreps of auxiliary band except $\Gamma_1$ have negative $\ep(\bk)$.
\textbf{b} We plot the frequency spectrum $\om(\bk)$ for physical bands corresponding to $(A_2)_d \ua G \ominus (A_1)_a \ua G$.
\textbf{c} Cylindrical Wilson loop spectrum does not exhibit a winding structure.
 \textbf{d} Spherical Wilson loop spectrum shows the winding corresponding to the Euler number $|\mf{e}|=2$.}
\label{sfig:sg207_bands}
\end{figure*}
%%%%%%

To energetically separate $\mc{B}_{\rm phys}$ and $\mc{B}_{\rm aux}$, we impose that $\mc{B}_{\rm phys}$ ($\mc{B}_{\rm aux}$) has positive (negative) energy at all momenta, exception that $\girrep$ must be pinned at zero energy.
This condition can be efficiently enforced by diagonalizing $H_{SG 207} (\bk)$ at HSM $\Gamma=(0,0,0)$, $X=(0,\pi,0)$, $M=(\pi,\pi,0)$, and $R=(\pi,\pi,\pi)$ and matching the resulting eigenvalues to the corresponding irreps.
For instance, at $\Gamma$, the three eigenvalues are degenerate and equal to $8 t_2 + 8 t_3 + \ep_0$, corresponding to $\Gamma_4 = \girrep + \Gamma_1$.
Pinning the this irrep at zero gives $\ep_0=-8(t_2+t_3)$.
Repeating the same procedure at $X$, $M$, and $R$ yields conditions that isolate the physical bands from the auxiliary band.
A representative parameter set satisfying these constraints is $\ep_0=0.88$, $t_1=0.3$, $t_2=-0.15$, and $t_3=0.04$.
We also numerically confirmed that these parameter values satisfy the required energy separation between the physical and auxiliary bands at all momenta.

We denote the eigenvalues of $H_{SG 207}(\bk)$ by $\vep(\bk)=\om(\bk)^2$.
The frequency spectrum $\om(\bk) \ge 0$ is defined on the physical bands.
Figures~\ref{sfig:sg207_bands}\textbf{a,b} show the total energy spectrum $\vep(\bk)$ and the frequency spectrum $\om(\bk)$.
We also compute the cylindrical and spherical Wilson loop spectra, $W_{k_z}(\phi; k_\rho)$ and $W_\phi(\theta; k_r)$, shown in Fig.~\ref{sfig:sg207_bands}.
Consistent with the ab-initio model, the cylindrical Wilson loop spectrum is gapped (i.e. exhibits no winding) as $k_\rho \to 0$, whereas the spherical Wilson loop shows a winding with Euler number $|\mf{e}|=2$.
The origin of this behavior is discussed below.
%%%%%%

%%%%%%
\subsection{Wilson loop analysis}
\label{sec:sn_sg207_wilson}
%%%%%%
Here, we analyze the Wilson loop spectra of both the TB model $H_{SG 207}(\bk)$ and the ab-initio model.
Our gold is to show that (i) the spherical Wilson loop spectrum in the TB model\footnote{As discussed in SN~\ref{sec:sn_sg207_TB}, in the ab-initio model the protected winding with $|\mf{e}|=2$ can be be attributed to the emergent transversality in the long-wavelength limit of photonic crystals.} exhibits a protected winding characterized by the Euler number $|\mf e|=2$ and (ii) the cylindrical Wilson loop spectrum remains non-winding (gapped) in both models, in the limit $k_\rho \to0$.
%%%%%%

%%%%%%
\subsubsection{Spherical Wilson loop}
%%%%%%
To demonstrate that the spherical Wilson loop $W_\phi (\theta;k_r)$ indicates a quantized Euler number $|\mf e|=2$, we show that, in the
TB model $H_{SG 207}(\bk)$, the real structure of its effective Hamiltonian emerges in the small $k_r$ limit.
Since the TB model is a three-band model, which corresponds to two transverse modes and one longitudinal mode near the $\Gamma$ point, we can obtain the effective Hamiltonian near $\Gamma$ for these modes through a Taylor expansion, without using the more general L\"owdin partitioning introduced in SN~\ref{sec:sn_sg212_wl_sphere}.
The resulting effective Hamiltonian is given by
\bg
H_{\rm eff}(\bk) = \bpm
0.22 k_x^2 + 0.22 k_y^2 - 0.16 k_z^2 & 0.6 k_z k_x & 0.6 k_y k_z \\
0.6 k_z k_x & 0.22 k_y^2 + 0.22 k_z^2 - 0.16 k_x^2 & -0.6 k_x k_y \\
0.6 k_y k_z & -0.6 k_x k_y & 0.22 k_z^2 + 0.22 k_x^2 - 0.16 k_y^2
\epm.
\label{seq:sg207_heff}
\eg
Because $H_{\rm eff}(\bk)$ is real-valued, i.e. $H_{\rm eff} (\bk) = (H_{\rm eff} (\bk))^* = \mc K \, H_{\rm eff} (\bk) \, \mc{K}$, it realizes an emergent $I \mc{T}$ symmetry with $(I \mc{T})^2=+\mathds{1}$.
This ensures that the Euler number $\mf e$ is well-defined.
(Recall that $I$ denotes inversion, which is absent in SG $P432$.)
Therefore, the winding in the spherical Wilson loop spectrum of the TB model $H_{SG 207}(\bk)$ is protected in the limit $k_\rho \to 0$.
%%%%%%

%%%%%%
\subsubsection{Cylindrical Wilson loop}
%%%%%%
We next analyze the cylindrical Wilson loop $W_{k_z} (\phi; k_\rho)$ in the limit $k_\rho \to 0$.
In this limit, we can instead study the almost-straight Wilson loop $W_{r_\ep} (\phi;\bk_b)$, following the analytic technique introduced in SN~\ref{sec:sn_sg212_wl_cylinder}.

We first consider $\phi=0$ with basepoint $\bk_b = -\bk_0 = (0,0,-\pi)$.
The almost-straight Wilson loop is defined along the line $-\bk_0 \to \bk_1 =(r_\ep,0,0) \to \bk_0$.
(See Fig.~\ref{sfig:wilson_path}.)
Decomposing the Wilson loop as $W_{r_\ep} (0; -\bk_0) = W_{\bk_0 \leftarrow \bk_1} \cdot W_{\bk_1 \leftarrow -\bk_0}$, the two segments can be related by the $C_{2,100}$ rotation:
\bg
W_{\bk_0 \leftarrow \bk_1}^{-1}
= B_{C_{2,100}} (\bk_1) \cdot W_{\bk_1 \leftarrow -\bk_0} \cdot B_{C_{2,100}} (-\bk_0)^{-1}.
\eg
The sewing matrices $B_{C_{2,100}} (\bk_1)$ and $B_{C_{2,100}} (-\bk_0)$ can be determined from the information about representation of bands at relevant momenta.
At $\pm \bk_0 = (0,0, \pm \pi)$, the lowest two bands transform as the irrep $X_5$, giving $B_{C_{2,100}} (\pm \bk_0) = \sg_x$.
At $\bk_1$, the representation is $\Delta_3 + \Delta_4$ and $B_{C_{2,100}} (\bk_1) = -\sg_0$.
(Note that $X=(0,\pi,0)$ and $\Delta = (0,k,0)$, and owing to the $C_{3,111}$ symmetry, the sewing matrices at $(0,0,\pi)$ and $(r_\ep,0,0)$ can be fixed from the representations at $X$ and $\Delta$, as explained in SN~\ref{sec:sn_sg212_wl_cylinder}.)
Hence,
\bg
W_{\bk_0 \leftarrow \bk_1}^{-1}
= - W_{\bk_1 \leftarrow -\bk_0} \cdot \sg_x,
\quad
W_{r_\ep} (0; -\bk_0) = W_{\bk_0 \leftarrow \bk_1} \cdot \left( - W_{\bk_0 \leftarrow \bk_1}^{-1} \cdot \sg_x \right) = -\sg_x,
\eg
so that ${\rm Spec} [-i \log W_{r_\ep} (0; -\bk_0)]$ is $\{0, \pi\}$.

A similar analysis applies for $\phi = \frac{\pi}{4}$.
For $\bk_2 = \frac{1}{\sqrt 2} (r_\ep, r_\ep, 0)$, the expression $W_{r_\ep}(\frac{\pi}{4}; -\bk_0) = W_{\bk_0 \leftarrow \bk_2} \cdot W_{\bk_2 \leftarrow -\bk_0}$ is constrained by the $C_{2,110}$ rotation:
\bg
W_{\bk_0 \leftarrow \bk_2}^{-1}
= B_{C_{2,110}} (\bk_2) \cdot W_{\bk_2 \leftarrow -\bk_0} \cdot B_{C_{2,110}} (-\bk_0)^{-1}.
\eg
From the $X_5$ irrep, $B_{C_{2,110}} (\pm \bk_0) = \sg_z$, and on $\Sigma=(k,k,0)$, where the bands correspond to $\Sigma_2 + \Sigma_2$, we have $B_{C_{2,110}} (\bk_2) = - \sg_0$.
It follows that
\bg
W_{\bk_0 \leftarrow \bk_2}^{-1}
= -W_{\bk_2 \leftarrow -\bk_0} \cdot \sg_z,
\quad
W_{r_\ep}(\tfrac{\pi}{4}; -\bk_0)
= -\sg_z,
\eg
and hence ${\rm Spec} [-i \log W_{r_\ep} (\tfrac{\pi}{4}; -\bk_0)] = \{0, \pi\}$.
Therefore, as $\phi$ varies in steps of $\tfrac{\pi}{4}$, the Wilson loop eigenvalues at these angles can be connected smoothly without any winding.
This behavior is consistent with the numerically flat spectra shown in Figs.~\ref{sfig:sg207_PhC}\textbf{b} and \ref{sfig:sg207_bands}\textbf{c}.

Next we confirm that the polarization singularity at $\Gamma$ does not induce a winding structure in the limit $r_\ep \to 0$.
Let $\bk_\pm = (0,0, \pm r_\ep)$.
Recall that the almost-straight Wilson loops at different angles are related by
\bg
W_{r_\ep} (\phi; -\bk_0)
= W_{\bk_0 \leftarrow \bk_+} \cdot \left[ W_{\bk_+ \leftarrow \bk_-} (\phi) \cdot W_{\bk_+ \leftarrow \bk_-}(0)^{-1} \right] \cdot W_{\bk_0 \leftarrow \bk_+}^{-1} \cdot W_{r_\ep} (0; -\bk_0).
\label{seq:sg207_wl_compose}
\eg
[We rewrite Eq.~\eqref{seq:wilson_compare} for convenience; see Fig.~\ref{sfig:wilson_path} for the definition of the semicircular segment $W_{\bk_+ \leftarrow \bk_-} (\phi)$.]
The first segment $W_{\bk_0 \leftarrow \bk_+}$ is constrained by $C_{4,001}$ and $\mc T C_{2,100}$.
From the $X_5$ and $\Delta_3 + \Delta_4$ representations, we determine the sewing matrices
\bg
B_{C_{4,001}} (\bk_0) = i \sg_y,
\quad
B_{C_{4,001}} (\bk_+) = i \sg_z,
\quad
B_{\mc T C_{2,100}} (\bk_0) = \sg_x \mc K,
\quad
B_{\mc T C_{2,100}} (\bk_+)= -i \sg_0 \mc K.
\eg
Solving the two symmetry constraints from $C_{4,001}$ and $\mc T C_{2,100}$,
\bg
W_{\bk_0 \leftarrow \bk_+}
= B_{C_{4,001}} (\bk_0) \cdot W_{\bk_0 \leftarrow \bk_+} \cdot B_{C_{4,001}} (\bk_+)^{-1}
= \sg_y \cdot W_{\bk_0 \leftarrow \bk_+} \cdot \sg_z,
\nn
W_{\bk_0 \leftarrow \bk_+}
= B_{\mc T C_{2,100}} (\bk_0) \cdot W_{\bk_0 \leftarrow \bk_+} \cdot B_{\mc T C_{2,100}} (\bk_+)^{-1}
= i \sg_x \cdot \cm{W_{\bk_0 \leftarrow \bk_+}},
\eg
for a unitary matrix $W_{\bk_0 \leftarrow \bk_+}$, we find that
\bg
W_{\bk_0 \leftarrow \bk_+}
= \pm \frac{1}{\sqrt 2} \bpm 1 & i s \\ i & s \epm,
\label{seq:wilson_sg207_straight}
\eg
where $s \in \{+1, -1\}$ is a sign factor.

The remaining step is to evaluate $W_{\bk_+ \leftarrow \bk_-} (\phi)$.
At $\phi=0$ and $\frac{\pi}{4}$, this can be determined from symmetry constraints.
For $\phi=0$ ($\frac{\pi}{4}$), the relevant symmetry is $C_{2,100}$ ($C_{2,110}$), which exchanges $\bk_+ \leftrightarrow \bk_-$ and leaves $\bk_1$ ($\bk_2$) invariant.
From the $\Delta_3 + \Delta_4$ and $\Sigma_2 + \Sigma_2$ representations, we determine the sewing matrices as $B_{C_{2,100}} (\bk_+) = \sg_0$ and $B_{C_{2,110}} (\bk_+) = -i \sg_z$.
(Recall that we have already fixed $B_{C_{2,100}} (\bk_1) = - \sg_0$ and $B_{C_{2,110}} (\bk_2) = - \sg_0$, respectively.)
Writing $W_{\bk_+ \leftarrow \bk_-} (0) = W_{\bk_+ \leftarrow \bk_1} \cdot W_{\bk_1 \leftarrow \bk_-}$, we apply a symmetry constraint
\bg
W_{\bk_1 \leftarrow \bk_-}^{-1}
= B_{C_{2,100}} (\bk_+) \cdot W_{\bk_+ \leftarrow \bk_1} \cdot B_{C_{2,100}} (\bk_1)^{-1}
= - W_{\bk_+ \leftarrow \bk_1},
\eg
which gives $W_{\bk_+ \leftarrow \bk_-} (0) = -\sg_0$.

We repeat a similar procedure for $\phi = \frac{\pi}{4}$.
By writing $W_{\bk_+ \leftarrow \bk_-} (\frac{\pi}{4}) = W_{\bk_+ \leftarrow \bk_2} \cdot W_{\bk_2 \leftarrow \bk_-}$, we apply the constraint
\bg
W_{\bk_2 \leftarrow \bk_-}^{-1}
= B_{C_{2,110}} (\bk_+) \cdot W_{\bk_+ \leftarrow \bk_2} \cdot B_{C_{2,110}} (\bk_2)^{-1}
= i \sg_z \cdot W_{\bk_+ \leftarrow \bk_2}.
\eg
Consequently, we find that $W_{\bk_+ \leftarrow \bk_-} (\frac{\pi}{4}) = -i \sg_z$.
Substituting these results into Eq.~\eqref{seq:sg207_wl_compose} verifies that both sides are consistent, confirming the internal consistency of all symmetry constraints.

Finally, in the isotropic limit around $\Gamma$, we use the result derived in Eq.~\eqref{seq:iso_wilson_circle} of SN~\ref{sec:sn_sg212_wl_cylinder} for isotropic transverse modes.
This gives $W_{\bk_+ \leftarrow \bk_-}(\phi) \cdot W_{\bk_+ \leftarrow \bk_-}(0)^{-1} = e^{2 i \phi \sg_z}$.
Combining this with $W_{\bk_0 \leftarrow \bk_+}$ in Eq.~\eqref{seq:wilson_sg207_straight} and
$W_{r_\ep} (0;-\bk_0) = -\sg_x$, we obtain
\bg
W_{r_\ep} (\phi; -\bk_0)
= \bpm -\sin 2\phi & -\cos 2\phi \\
-\cos 2\phi & \sin 2\phi \epm.
\eg
The spectrum ${\rm Spec} [-i \log W_{r_\ep} (\phi; -\bk_0)]$ is therefore $\{0, \pi\}$ for all $\phi$.
Hence the Wilson-loop spectrum is completely flat and gapped.
This demonstrates that the cylindrical Wilson loop spectra in the SG $P432$ models realizing the irrep content of Eq.~\eqref{seq:sg207_irrep_model} do not host any winding structure.

In summary, unlike the four-band cylindrical Wilson-loop spectrum in the SG $P4_332$ (No. 212) models, which exhibits a winding structure in the limit $k_\rho \to 0$ due to the polarization singularity, the two-band spectrum in the SG $P432$ (No. 207) models shows no winding.
Extending our analytic framework to general SGs and photonic band structures to determine whether and how polarization singularities affect Wilson-loop observables will be an important direction for future work.
In addition, when such effects are present, developing a systematic way to identify the Wilson-loop paths that can reveal them remains an open problem.
%%%%%%

%%%%%%Ref
\bibliography{Refs.bib}
%%%%%%